\documentclass{cernrep}
\usepackage{a4} 
\usepackage[bookmarks, colorlinks=true, linktoc=page, linkcolor=black, citecolor=black, urlcolor=blue]{hyperref}  
\usepackage{fancyhdr}
\fancyhfoffset{4 mm}
\fancypagestyle{ARTTITLE}{%
\fancyhf{} 
\lhead{\footnotesize{Proceedings of the General Introductory CAS course on
\it{Accelerator Physics \& Technologies}, 2019 and beyond}}
\lfoot{Available online at \url{https://cas.web.cern.ch/previous-schools}}
\rfoot{\thepage\hspace*{3mm}}

}

\begin{document}

\title{Beam Instrumentation and Diagnostics} 
\author{Peter Forck}
\institute{ GSI Helmholtz-Zentrum f\"ur Schwerionenforschung, Darmstadt, Germany}

\begin{abstract}
{The determination of beam parameters is essential for the operation and development of any accelerator facility. The working principle of frequently used beam instruments for electron and proton beams is discussed. The article comprises of the beam instrumentation for beam current determination, the usage of beam position monitors for bunched beams, methods for transverse profile and emittance diagnostics, bunch shape measurement and the usage of beam loss monitors.}
\end{abstract}


\maketitle
\thispagestyle{ARTTITLE}

\section{Demands for beam diagnostics}
Beam diagnostics is an essential constituent of any accelerator as it is the
sensory organ showing the~properties and the behaviour of the beam. It deals
with the real beam including all possible imperfections of a real technical
installation. There is a vast range of applications for the diagnostics 
resulting in quite different demands for the beam instruments and the related display of the measurement results: 
\begin{itemize}
\item Reliable, quick measurements to determine the basic parameters of a
machine setting are used as a fast check of the general functionality of the accelerator.
The readings from the instrumentation give a single number or simple plots.
These devices should be non-destructive for the beam, yielding online information. A prominent example is a current measurement by a current transformer, as discussed in Section~\ref{kap-current}.
\item Instruments are installed for a daily check of performance and stability or the control of necessary parameter changes in the accelerator setting. They are also used for solving simpler machine problems in case of any device malfunction. An example is the transverse profile determination, in most cases performed by a destructive device, as discussed in Section~\ref{kap-profile}.
\item Complex instrumentation for the commissioning of a new accelerator component, for the development of higher performance and for solving more serious problems in case of a malfunction must be available. The devices can be more complex to use and might be destructive for the beam. The~determination of the beam emittance is an example as discussed in Section~\ref{kap-emittance}.
\item An additional application is the active control of device settings after
the detection of the beam's properties where the beam instrumentation serves as the sensor. This is generally called feedback of a given
parameter to yield an improved beam delivery. An example is the determination of the~beam position inside a synchrotron and the correction of the orbit to its nominal value.
\end{itemize}
General articles and books on beam instrumentation and its usage for beam diagnostics are Refs.~\cite{schmick_cas2018,smaluk_book,brandt_cas2007,strehl_book,minty_book,kuro_book,forck_juas}. 

The outline of this contribution is oriented on the beam quantities. Section 2 describes the measurement of beam current by current transformers in a non-interfering manner and Faraday Cup as an~invasive device. Section 3 covers the frequent usage of beam position monitors which delivers the transverse centre-of-mass of a bunched beam. In nearly all facilities these instruments are used for the daily beam alignment as well as the monitor for advanced diagnostics. Section 4 describes several methods for transverse profile measurements. There are methods which can be operated for all beams, but also quite different methods are used for electrons (which are in most cases relativistic) and protons (which require a large facility to reach relativistic velocities). Some of these methods are destructive for the~beam and cannot, therefore, be applied at a synchrotron to monitor the beam properties of the circulating beam. Section 5 deals with the determination of transverse emittance in transfer lines, where the statistical description of the ensemble of beam particles is relevant. Section 6 deals with the determination of longitudinal parameters, with the focus on bunch shape determination at circular and linear light sources. In the last section, monitors detecting the lost beam are discussed. Generally, the basic functionality of the~beam instrumentation is discussed, including some examples of related measurements; more details and applications can be found in the references.

\section{Measurement of beam current}
\label{kap-current}
The total electrical current is one of the most important parameters for the
operation of a particle accelerator. In the daily operation, a first check
concerns the current in almost all accelerator laboratories; in most cases, it
is measured with a beam current transformer. Several types of current transformers are 
available to determine the current at 
electron and proton
LINACs and synchrotrons, even for short pulses, such as for the transfer between
synchrotrons, as well as for coasting beams, such as for storage rings. These devices
are commercially available \cite{berg-com}, even though quite different types
are used, a general overview of the current measurement device is presented in
\cite{belo-dipac11}. Current transformers are generally non-intercepting devices. Their principle is the detection of
the magnetic field carried by the beam. For currents below about 1 $\mu$A,
transformers cannot be used owing to noise limitations. 

From the first days of accelerators, Faraday cups were used. The cup gives a
direct measurement of the particle beam's charge because the particles are stopped in the cup. For high current, this destructive method can not be
applied, because the total energy carried by the beam can destroy the
intercepting material. For higher energetic particles with energies above some 100 MeV/u for ions, the penetration depth
reaches more than several cm, and Faraday cups are not useful any more.

\subsection{Current transformer for pulsed beams}
\label{chapter-passive-trafo}
\begin{figure}
\centering \includegraphics*[width=50mm,angle=0]
{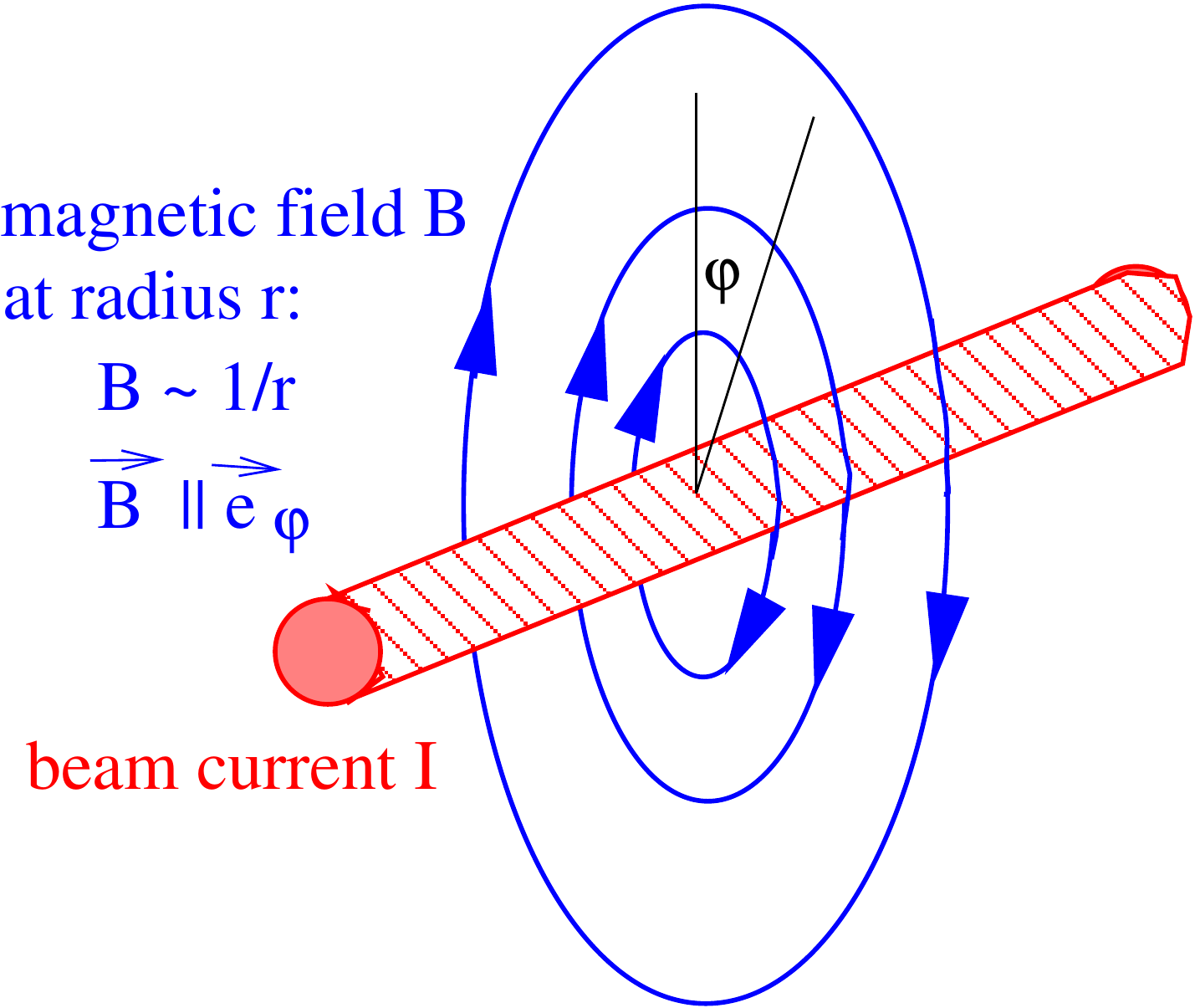}
\caption{The magnetic field of a current.}
\label{magionenstrahl}
\end{figure}

In an accelerator the current is formed by $N_{part}$ particles of charge state $q$
per unit of time $t$ or unit of length $l$ and velocity $\beta=v/c$. The
electrical current passing a given location is
\begin{equation}
I_{beam} = \frac{qeN_{part}}{t} = \frac{qeN_{part}}{l} \cdot \beta c
\end{equation}
with $e$ being the elementary charge. The magnetic field $B$ of a current can
be calculated according to the~Biot-Savart law
\begin{equation}
\mbox{d}\vec{B}(\vec{r})= \mu_{0} I_{beam} \cdot \frac{\mbox{d}\vec{l} \times \vec{r}}{4\pi r^{3}}
\end{equation}
with $\mu_{0}=4\pi \cdot 10^{-7}$ Vs/Am is the permeability of the vacuum,
$d\vec{l}$ the length in direction of the beam and $\vec{r}$ the distance
between the centre of the beam and the location of the field determination. Due to the~cylindrical symmetry outside of the beam only the azimuthal component 
has to be considered along
the~unitary vector $\vec{e}_{\varphi}$ as shown in Fig.~\ref{magionenstrahl}
\begin{equation} 
\vec{B} = \mu_{0} \frac{I_{beam}}{2\pi r} \cdot \vec{e}_{\varphi} .
\end{equation} 
For a beam current of 1 $\mu$A and a distance of 10 cm the magnetic field has
a value of only 2 pT. To put this into perspective, the Earth magnetic field has a value of about 50 $\mu$T.

\begin{figure}
\parbox{0.55\linewidth}{ \includegraphics*[width=65mm,angle=0]
{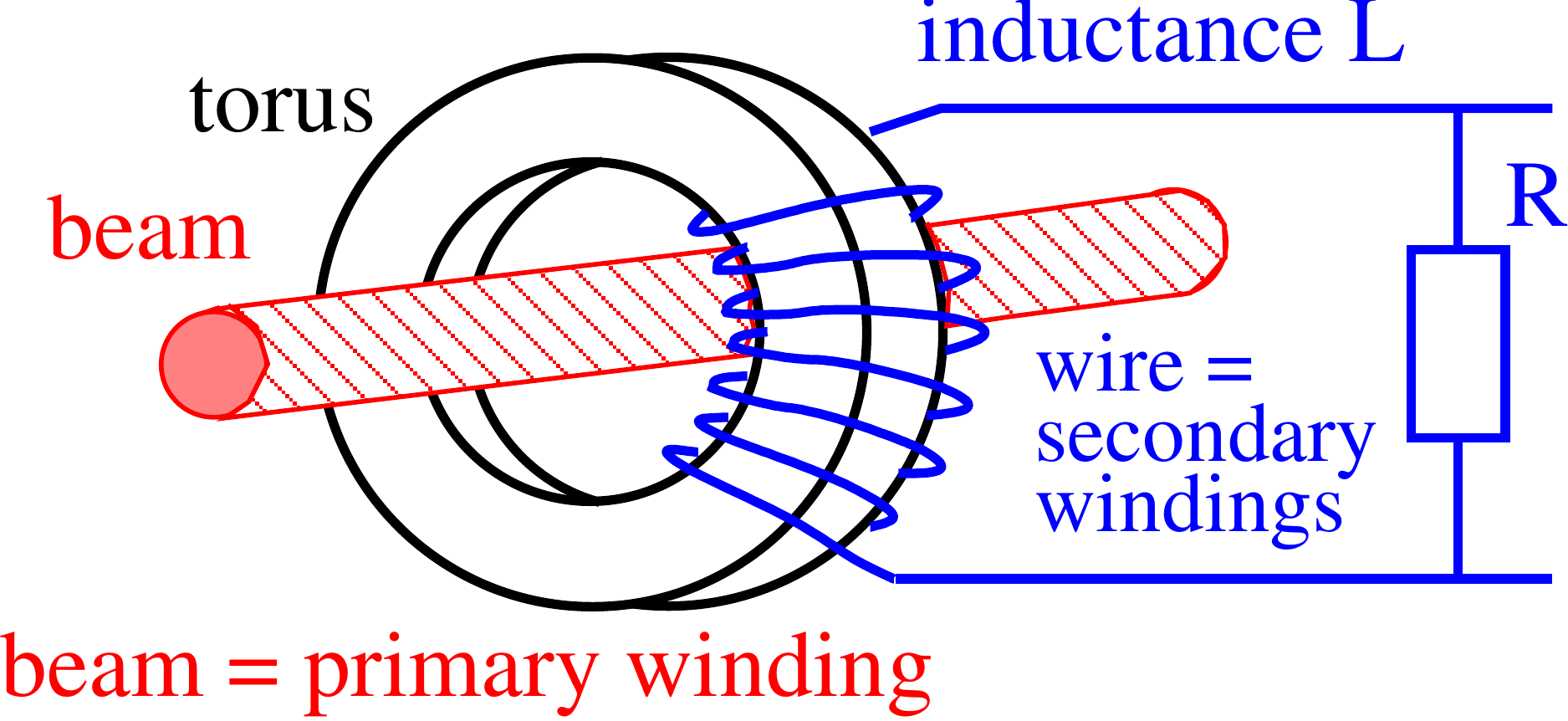} }
\parbox{0.40\linewidth}{ \includegraphics*[width=63mm,angle=0]
{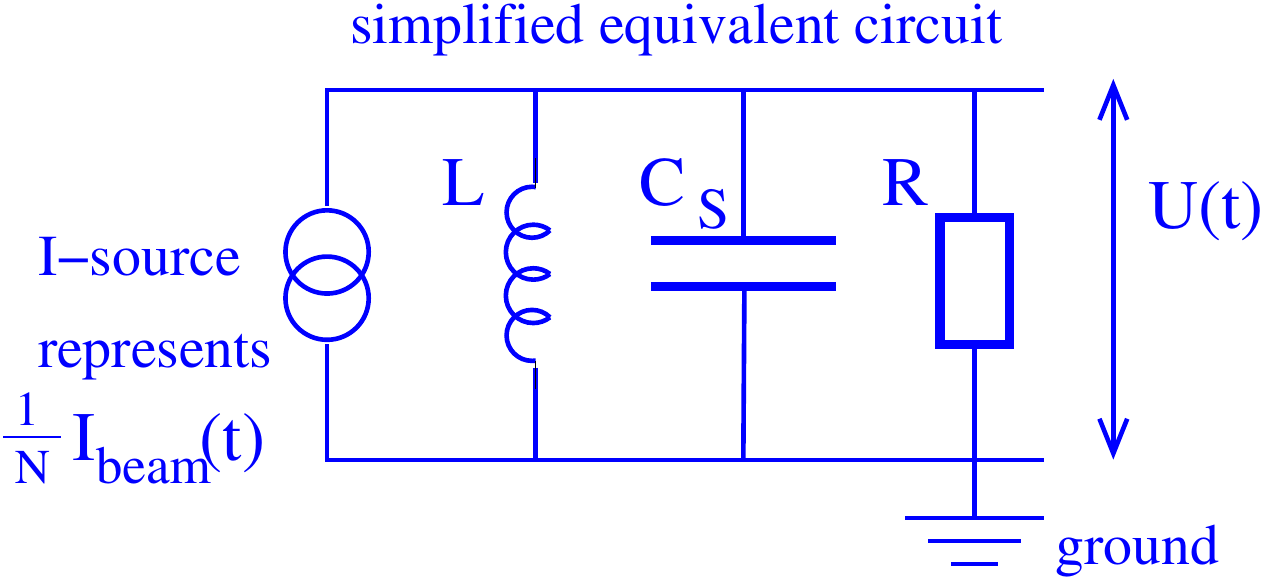} }
\caption{Left: Scheme of a current transformer built as a ring-core (torus) around 
the beam. Right: Simplified equivalent circuit.}
\label{trafo-prinz-start}
\end{figure}

The beam current can be determined by monitoring the accompanying magnetic
field with a current transformer schematically shown in
Fig.~\ref{trafo-prinz-start}. The beam passes through a highly permeable torus
as the 'primary winding'. An insulated wire wound around the torus with $N$
turns serve as the 'secondary winding' of the transformer with the
inductance $L$.
The inductance $L$ for a torus material of length $l$ in beam direction, inner
radius $r_{i}$ and outer radius $r_{o}$ having a relative permeability
$\mu_{r}$ and $N$ windings is given by
\begin{equation}
L = \frac{\mu_{0} \mu_{r}}{2\pi} \cdot l N^{2} \cdot \ln\frac{r_{o}}{r_{i}} .
\label{eq-trans-induct}
\end{equation}
The value of the induction can be chosen by the core size (as given by $r_{i}$ and $r_{o}$) and the number of windings $N$. A further reason for the torus is to guide the field-lines, so only the azimuthal
component is measured and the signal strength is nearly independent of the
beam position inside the vacuum pipe.

Generally, for an ideal current transformer loaded with a low value of ohmic
resistance $R$ the~ratio between the primary current $I_{prim}$ conducted in $N_{prim}$ windings and the secondary current $I_{sec}$ conducted in $N_{sec}$ is given by
\begin{equation}
I_{sec} = \frac{N_{prim}}{N_{sec}} \cdot I_{prim}
~~~\Longrightarrow ~~~
I_{sec} = \frac{1}{N} \cdot I_{prim} 
~~~~ \mbox{due to} ~~ N_{prim}=1
\label{eq-current-trans-I}
\end{equation}
$N_{prim}=1$ is one due
to the single pass of the beam through the torus. $N_{sec}$ is the winding
number on the secondary side and for simplicity it is called $N$ further-on.

For most practical cases a measurement of a voltage $U$ is preferred,
therefore the resistance $R$ is introduced leading to
\begin{equation}
U = R \cdot I_{sec} = \frac{R}{N} \cdot I_{beam}~~.
\label{eq-current-trans-U}
\end{equation}
The ratio between the usable signal voltage $U$ and the beam current
$I_{beam}$ is called sensitivity $S$ (or transfer impedance, for a more
stringent discussion see Section~\ref{kap-pickup-treatment})
\begin{equation}
U = S \cdot I_{beam} ~~~.
\end{equation}

The properties of a transformer can be influenced by different external
electrical elements to match the response to a given time structure
of the beam. We first consider the
characteristics of a so-called passive transformer or \textbf{F}ast {\bf
C}urrent \textbf{T}ransformer \textbf{FCT}, 
where the voltage at a 50
$\Omega$ resistor is recorded. The equivalent circuit of the secondary
transformer side is depicted in Fig.~\ref{trafo-prinz-start} on the right
side. The~beam current is modelled by a current source with a reduction given
by the number of windings $N$ according to Eq.~\ref{eq-current-trans-I}. One
has to take also some stray capacitance $C_{S}$ into account, 
which are caused
by the~capacitance between the windings, the windings and the torus and along
the shielded cable to the resistor $R$. Generally, the voltage $U(t)$ of the parallel shunt of the three elements is measured. Depending on the~choice of elements the response is influenced as discussed in the general textbooks on beam diagnostics the review article on transformers 
\cite{smaluk_book,brandt_cas2007,strehl_book,forck_juas,belo-dipac11}; here we state only the results.

\begin{figure}
\parbox{0.59\linewidth}{ \centering \includegraphics*[width=90mm,angle=0]
{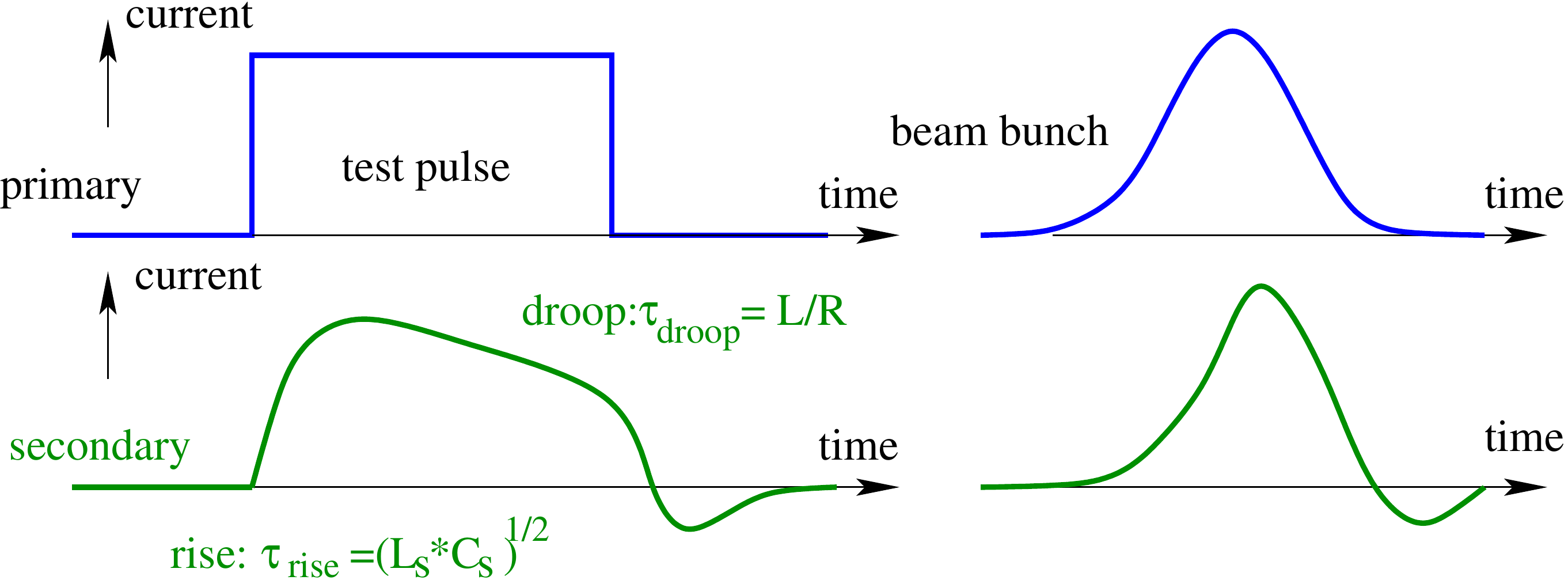}}
\parbox{0.40\linewidth}{ \centering \includegraphics*[width=40mm,angle=90]
	{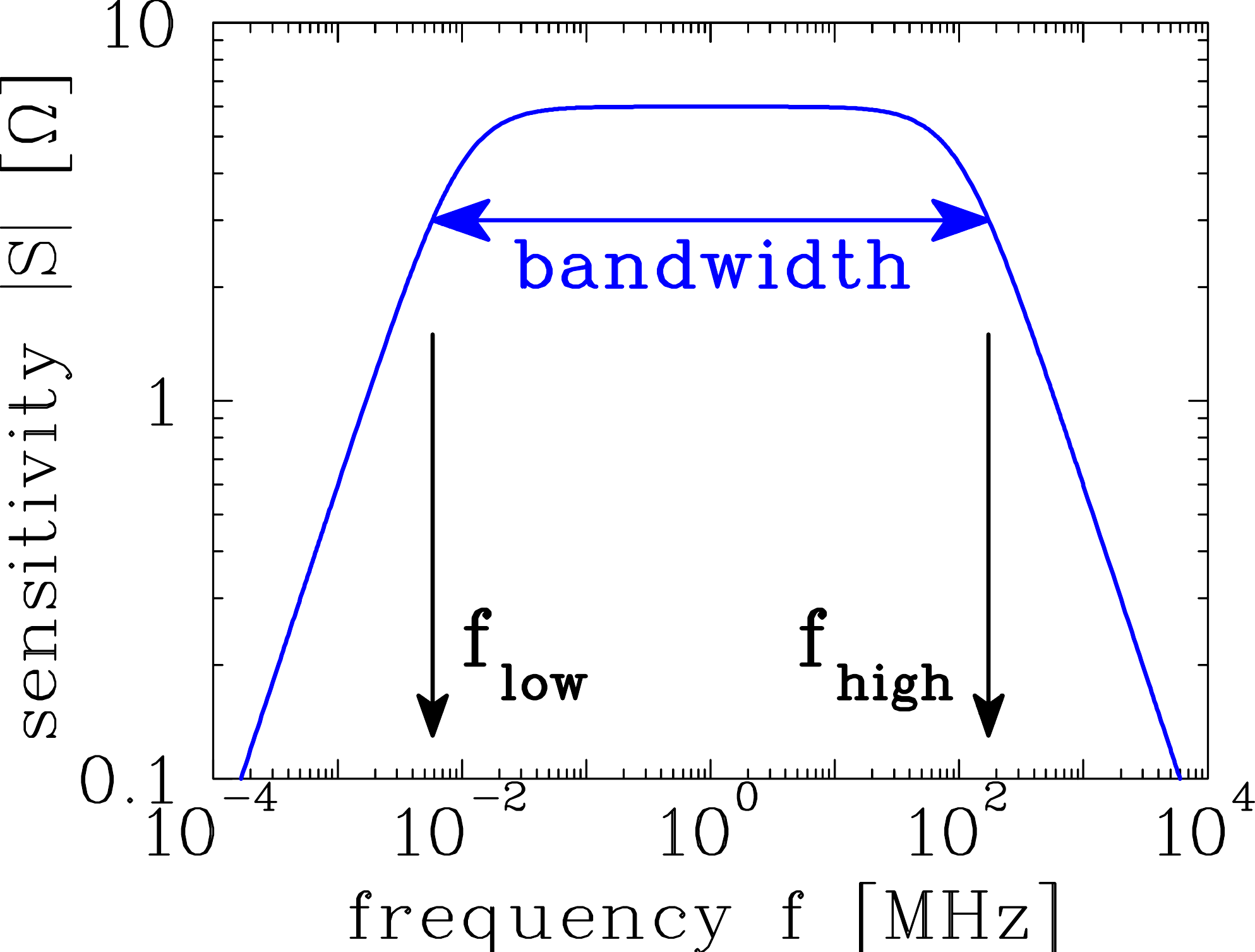}} 
\caption{Left: The response of a transformer to a rectangular pulse and a more
realistic beam pulse. Right: The~corresponding frequency domain plot for the sensitivity as a function of frequency.}
\label{rise-droop-koziol}
\end{figure}

For the following discussion, we are interested in the time response of the
measurement device to a given beam pulse. Therefore, one uses frequently the
rise time constant $\tau_{rise}$ and the droop time constant $\tau_{droop}$ as
depicted in Fig.~\ref{rise-droop-koziol}: If the excitation is given by a step
function, the signal amplitude $A$ increases as $A\propto
(1-e^{-t/\tau_{rise}})$ and $\tau_{rise}$ corresponds to the time for 
an increase by $e^{-1}=37$ \%.
It is linked to the upper cut-off frequency by
\begin{equation}
\tau_{rise} = \frac{1}{\omega_{high}} 
= \frac{1}{2 \pi f_{high}} ~~~.
\end{equation} 
Correspondingly, the droop time constant is linked to the lower cut-off
frequency as
\begin{equation}
\tau _{droop} = \frac{1}{2 \pi f_{low}} ~~~.
\end{equation} 
Both equations have a general meaning and will be used to transfer
quantities like bandwidth, as given in the frequency domain, into the corresponding
description in the time domain 

For the
passive current transformer as described by the equivalent circuit of
Fig.~\ref{trafo-prinz-start} the rise and droop time constant are given by
\begin{equation}
\tau_{rise} = RC_{S} ~~~\mbox{and} ~~~~ \tau_{droop}= \frac{L}{R} ~~~~.
\end{equation}

\begin{figure}
\parbox{0.49\linewidth}{ \centering \includegraphics*[width=70mm,angle=0]
{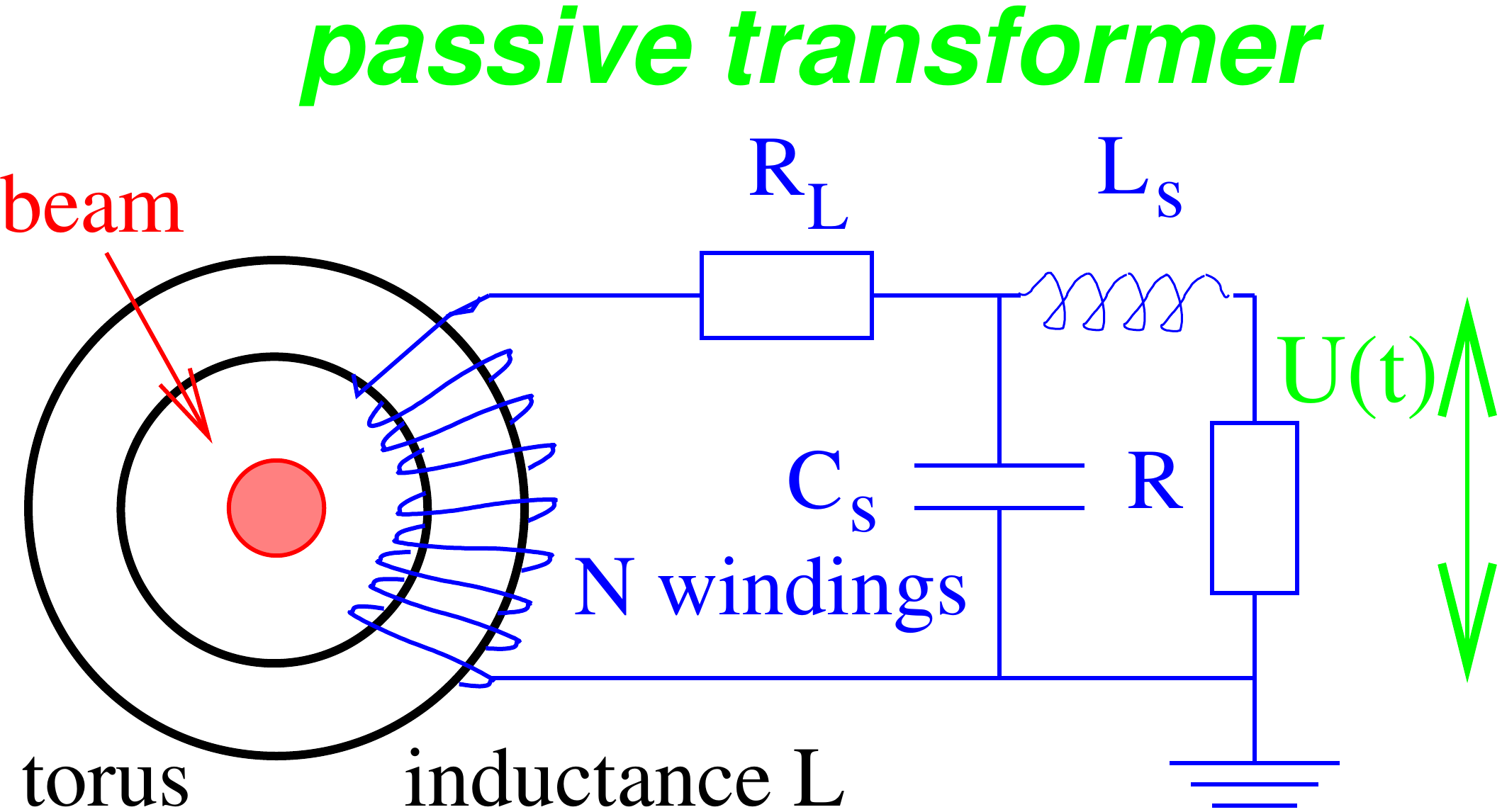}
} \parbox{0.49\linewidth}{ \centering \includegraphics*[width=32mm,angle=0]
{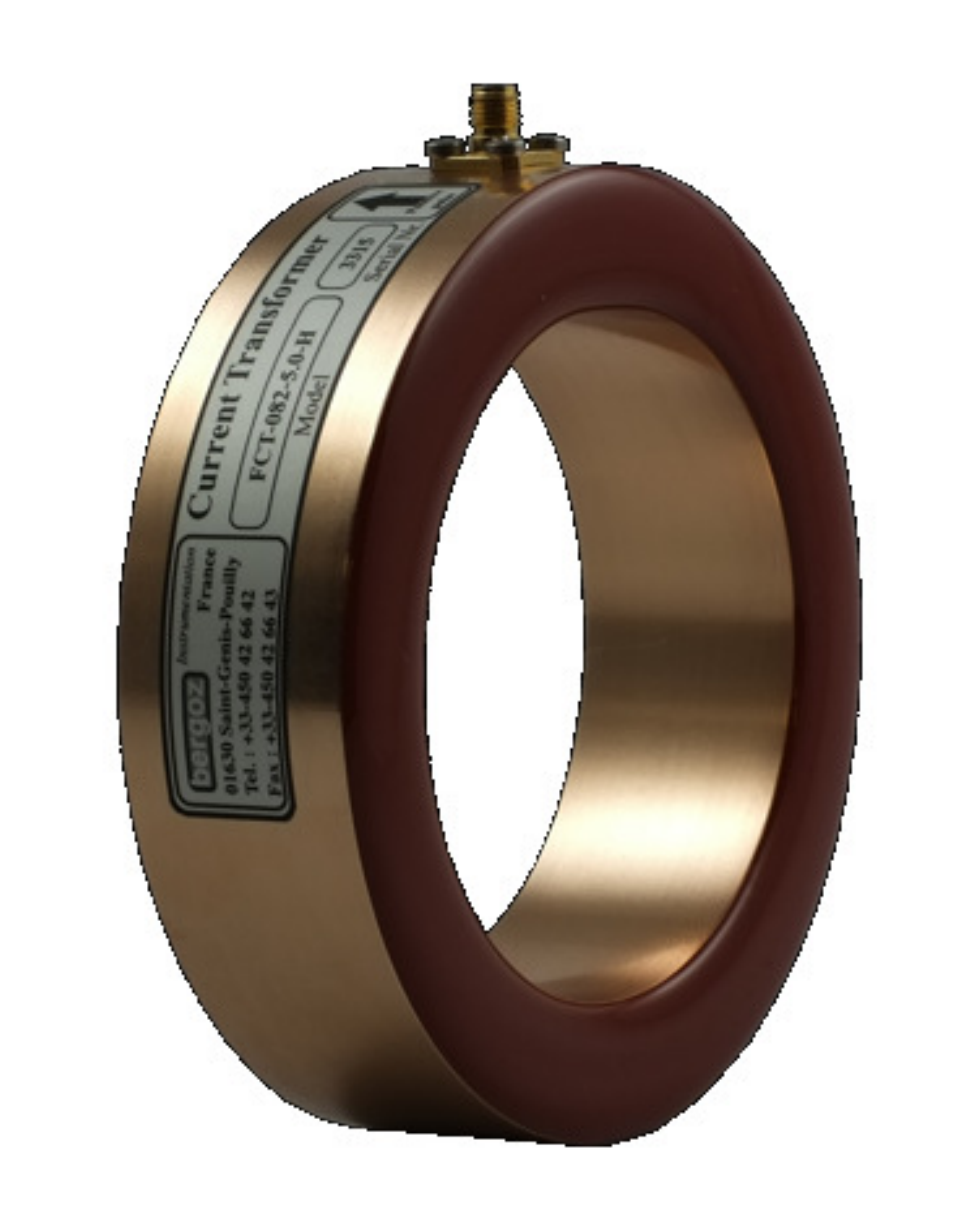} }
\caption{Left: Equivalent circuits of a passive beam transformer. Right: Photo of a commercially available passive transformer \cite{berg-com}.}
\label{scheme-passive-transformer}
\end{figure}

A more realistic schematic diagram of the passive transformer is shown in
Fig.~\ref{scheme-passive-transformer}. The main difference is the additional
loss resistivity in the cables, which is represented by a serial resistor
$R_{L}$. Additionally, there is a stray inductance between the windings, which
is best modelled by a serial insertion of an inductance $L_{S}$. With these two
modifications the rise and droop times are modified to yield
\begin{equation}
\tau_{rise} = \sqrt{L_{S} C_{S}} ~~~~~~\mbox{and} ~~~~ 
\tau_{droop}= \frac{L}{R+R_{L}} ~~~~.
\end{equation} 
The specifications of the GSI device are listed in
Table~\ref{tab-HEST-passive-trafo}. Careful matching is necessary between the~torus and the 50 $\Omega$ resistor $R$, where the voltage drop is measured.

\begin{table}
\centering
\caption{Basic specification of the GSI passive transformer \cite{reeg-dipac01}.}
\begin{tabular}{ll} \hline \hline
\textbf{Parameter} & \textbf{Value} \\ \hline
Torus radii & $r_{i}=70 $ mm, $r_{o}= 90$ mm\\
Torus thickness $l$ & $16$ mm\\
Torus material & Vitrovac 6025: (CoFe)$_{70\%}$(MoSiB)$_{30\%}$\\
Torus permeability $\mu_{r}$ & $\mu_{r}\simeq 10^{5}$ for $f< 100$ kHz, $\mu_{r}\propto
1/f$ above\\
Number of windings $N$ & $10$\\
Sensitivity $S$ & 4 V/A at $R= 50 ~\Omega$ ($10^{4}$ V/A with amplifier)\\
Current resolution $I_{min}$  & 40 $\mu$A$_{rms}$ for full bandwidth\\
Droop time constant $\tau _{droop} =L/R$ & 0.2 ms, corresponding to 5\% per 10 $\mu$s pulse length\\
Rise time constant $\tau _{rise} =\sqrt{L_{S} C_{S}}$ & 1 ns\\
Bandwidth & $f_{low}= 0.75$ kHz to $f_{high}= 660$ MHz\\
\hline \hline
\end{tabular}
\label{tab-HEST-passive-trafo}
\end{table}

\begin{figure}
\parbox{0.49\linewidth}{
\centering \includegraphics*[width=50mm,angle=0]
{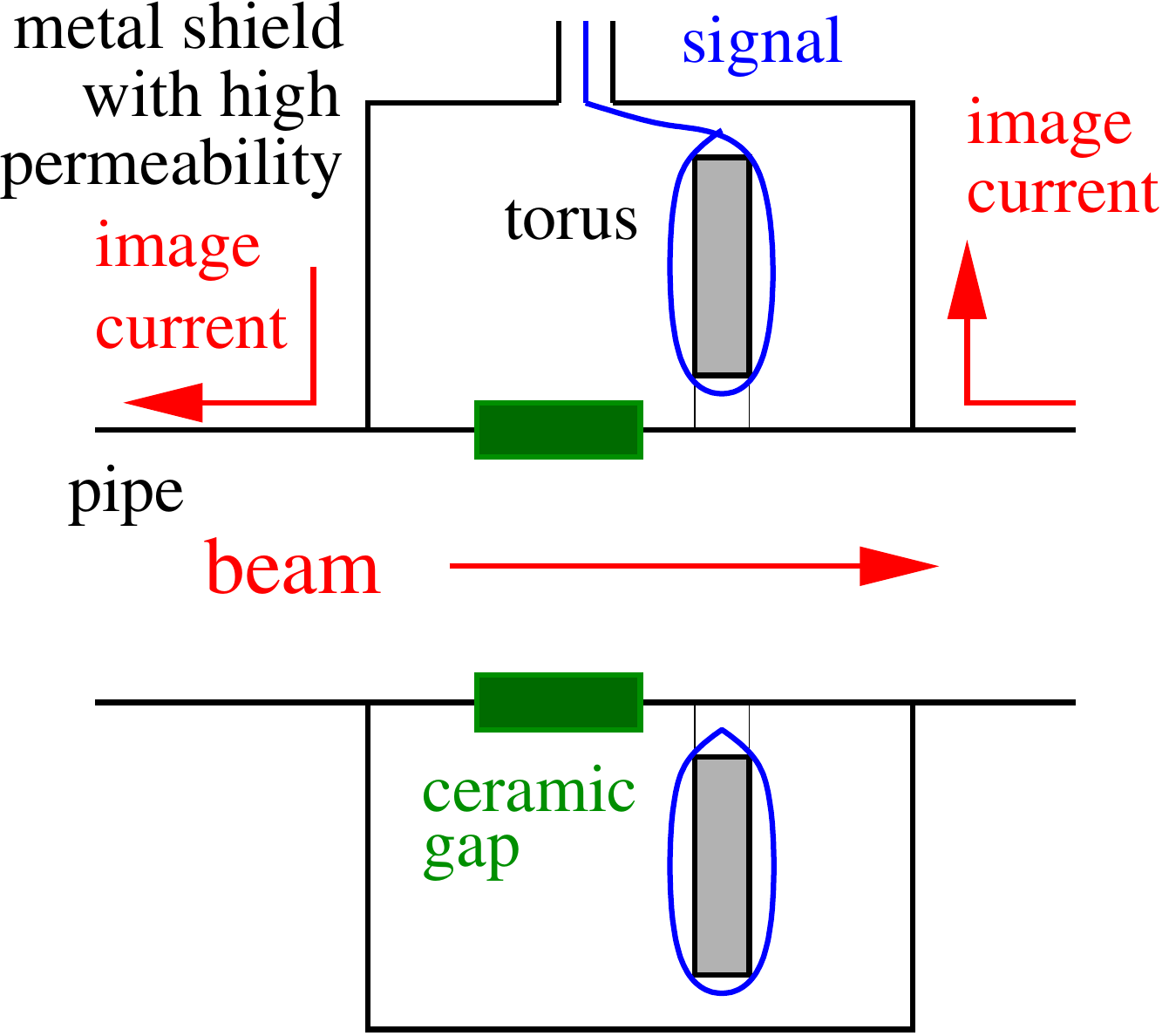}
}
\parbox{0.49\linewidth}{
\centering \includegraphics*[width=55mm,angle=0]
{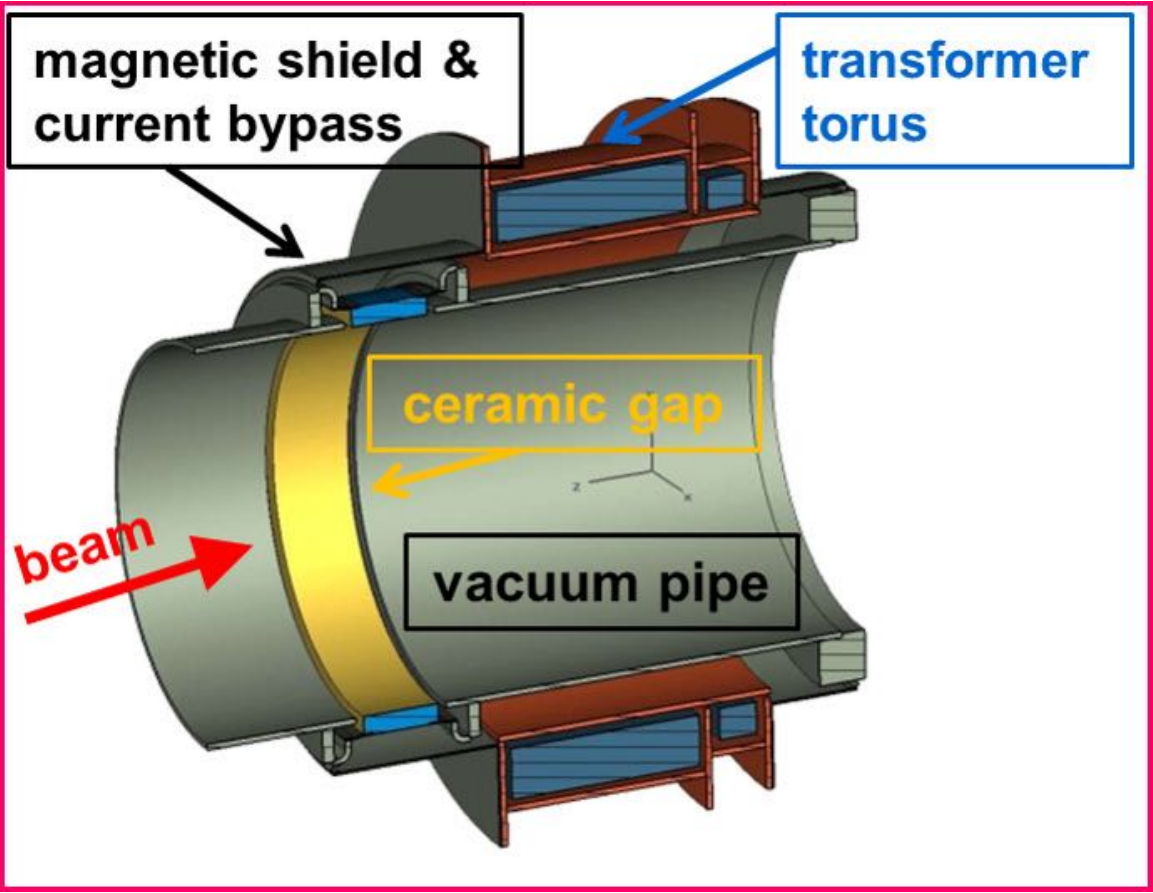}
}
\caption{Left: The general 
scheme beam pipe intersected by a ceramic gap and a transformer
housing as used as an image current pass 
and shielding against external magnetic fields is shown. Right:
The technical drawing for a~transformer arrangement surrounding a beam
pipe including the gap and magnetic shield is shown.}
\label{scheme-trafo-abschirmung}
\end{figure}

An additional general point is that close to the transformer the electrical
conductivity of the beam pipe has to be interrupted as schematically shown in
Fig.~\ref{scheme-trafo-abschirmung}. This is done with an
insulator, either a~ceramic gap or a plastic vacuum seal. The reason is to
prevent a flow of image current inside of the~transformer
torus. This image current has the opposite sign and without the gap, the
fields of the image current and beam current add up to zero. The image current
has to be bypassed outside of the transformer torus by some metallic housing.
It is surrounded by high permeability $\mu$-metal, also used for the shielding
of the transformer against external magnetic fields. A general review on
transformers is presented in \cite{belo-dipac11, webber-biw94}.

\begin{figure}
\parbox{0.49\linewidth}{ \centering \includegraphics*[width=75mm,angle=0]
{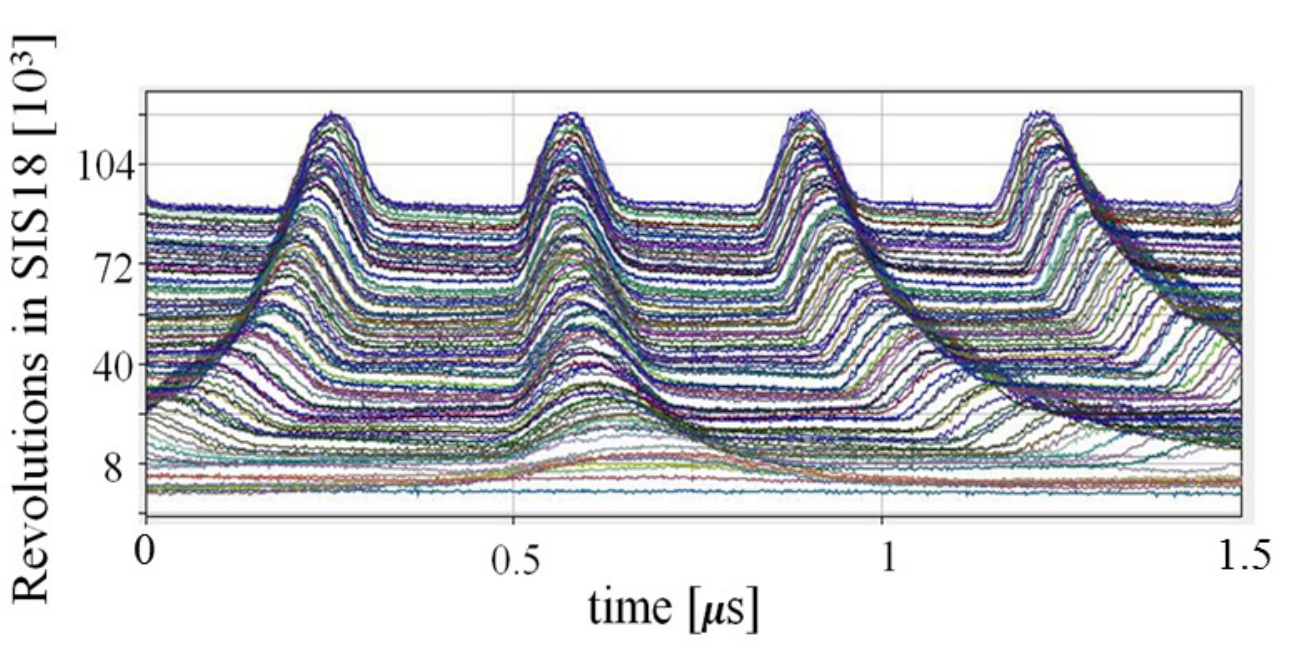}
}
\parbox{0.49\linewidth}{ \centering \includegraphics*[width=75mm,angle=0]
{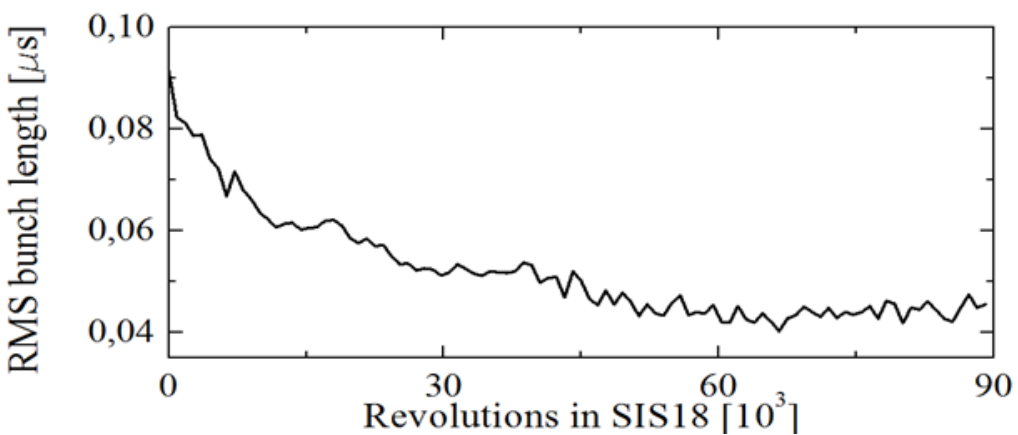}
}
\vspace*{-0.2cm}
\caption{Left: Signals from a passive transformer installed in GSI
synchrotron during the acceleration of a U$^{73+}$ beam. The beam is
injected at 11
MeV/u, corresponding to a velocity of $\beta = 15$ \% then four
bunches are build and 
accelerated to 350 MeV/u, corresponding to $\beta = 69$ \% with an
accelerating frequency swing from 0.85 MHz to 3.80 MHz within 0.3
s. Each 0.15 ms a trace of the circulating bunches are
shown. Right: The width (one standard deviation) of the bunches
during acceleration is shown.}
\label{FCT-sis-acceleration-width}
\end{figure}

Passive transformers or so-called \textbf{F}ast \textbf{C}urrent \textbf{T}ransformers \textbf{FCT} are mainly used when beam pulses in a time range from ns 
to $\mu$s have to be observed. The observables are the actual bunch shape, the~arrival time with respect to an external reference (e.g. the acceleration frequency) and the total amount of particles gained by the integration of the signal. The bunch structure can be observed with a~bandwidth typically 1 GHz, corresponding to a rise time constant 
of $\tau_{rise} =160 
$ ps \cite{berg-com, reeg-dipac01}. As an~example 
the observation of the bunch structure during acceleration inside a 
synchrotron is shown in Fig.~\ref{FCT-sis-acceleration-width} for 
the acceleration of an ion beam injected at a non-relativistic velocity of $
\beta =15$ \% of velocity of light and accelerated to $\beta = 69$ \%. Due to the increase of the velocity, the bunches approach each other and get smaller as expected from the conservation of the normalized longitudinal emittance, see Section~\ref{kap-longi}. Generally, the observation of the bunches during acceleration and possible bunch merging or splitting is 
observed by the FCT and eventually a feedback system for the control of the acceleration frequency, amplitude and phase is generated from the FCT signal. An innovative diagnostics method has been developed \cite{hancock-prab,hancock-www-note} to determine the longitudinal phase space
distribution and hence the longitudinal emittance from a measurement of the bunch shape on a turn-by-turn basis. The idea is
based on the~tomographic reconstruction used as an imaging
technique in medicine for X-ray tomography.

Further types of current transformers exist to adapt the signal output to the time dependent beam structure by dedicate electrical processing: For very short beam pulses in much below 1 ns the signal is integrated by special transformer arrangements by the so-called \textbf{I}ntegrating \textbf{C}urrent \textbf{T}ransformer \textbf{ICT}, while for long pulses above 10 $\mu$s the so-called \textbf{A}lternating \textbf{C}urrent \textbf{T}ransformer \textbf{ACT} are used; their proerties are discussed in Refs. \cite{brandt_cas2007,strehl_book,forck_juas,belo-dipac11,webber-biw94} 

\subsection{The dc-transformer}
The transformers discussed in the previous sections only work
for a pulsed beam. An essential task for beam diagnostics is the measurement of a coasting beam current (so-called direct current dc beam). The~application is either a
LINAC producing beam permanently (so-called continuous-wave cw-mode) or a~synchrotron with storage times from seconds to many hours. 

\begin{figure}
\centering
\includegraphics*[width=85mm,angle=0]
{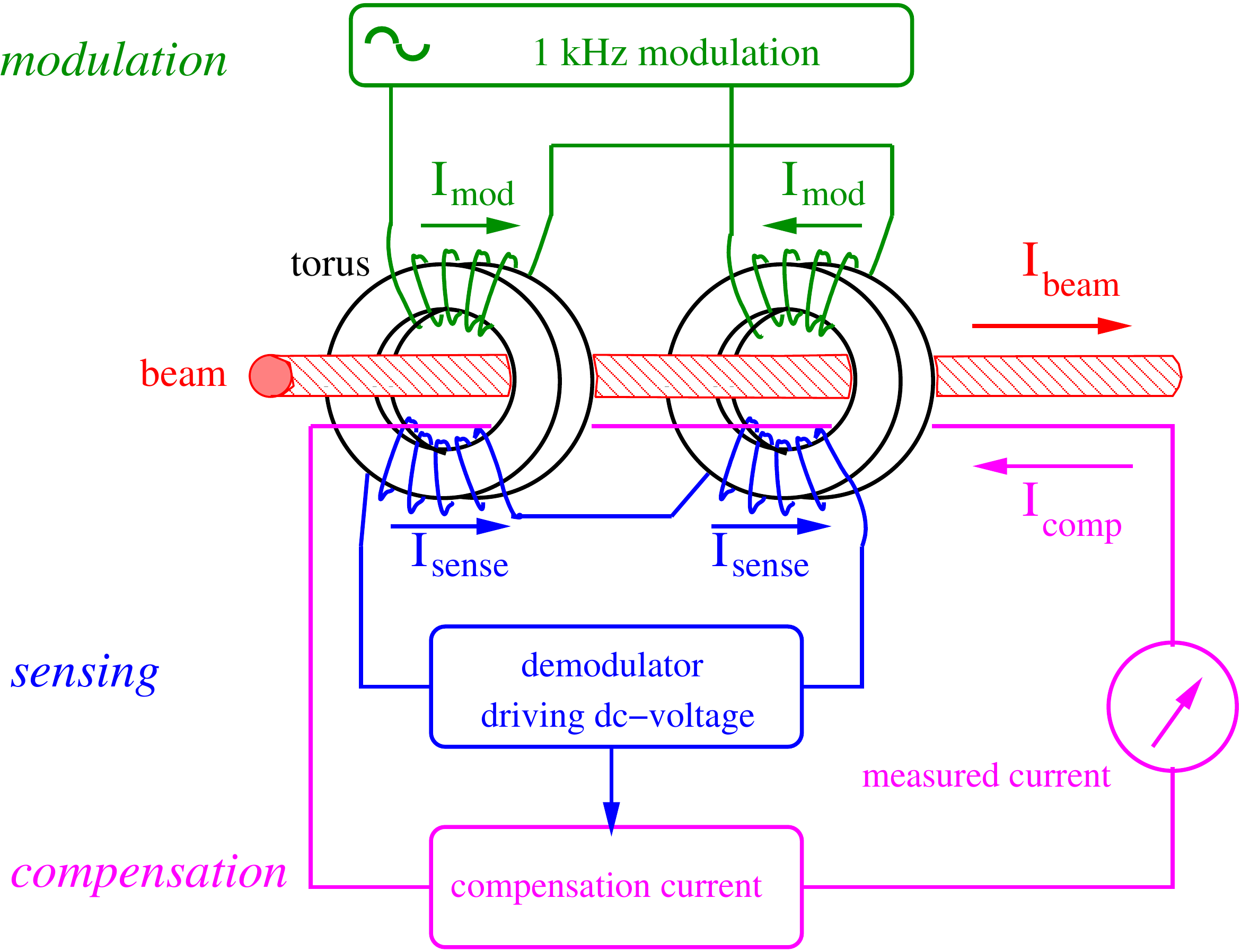}
\caption{Schematics of a dc-transformer, see text.}
\label{dc-prinz-koziol}
\end{figure}

\begin{figure}
\centering \includegraphics*[width=110mm,angle=0]
{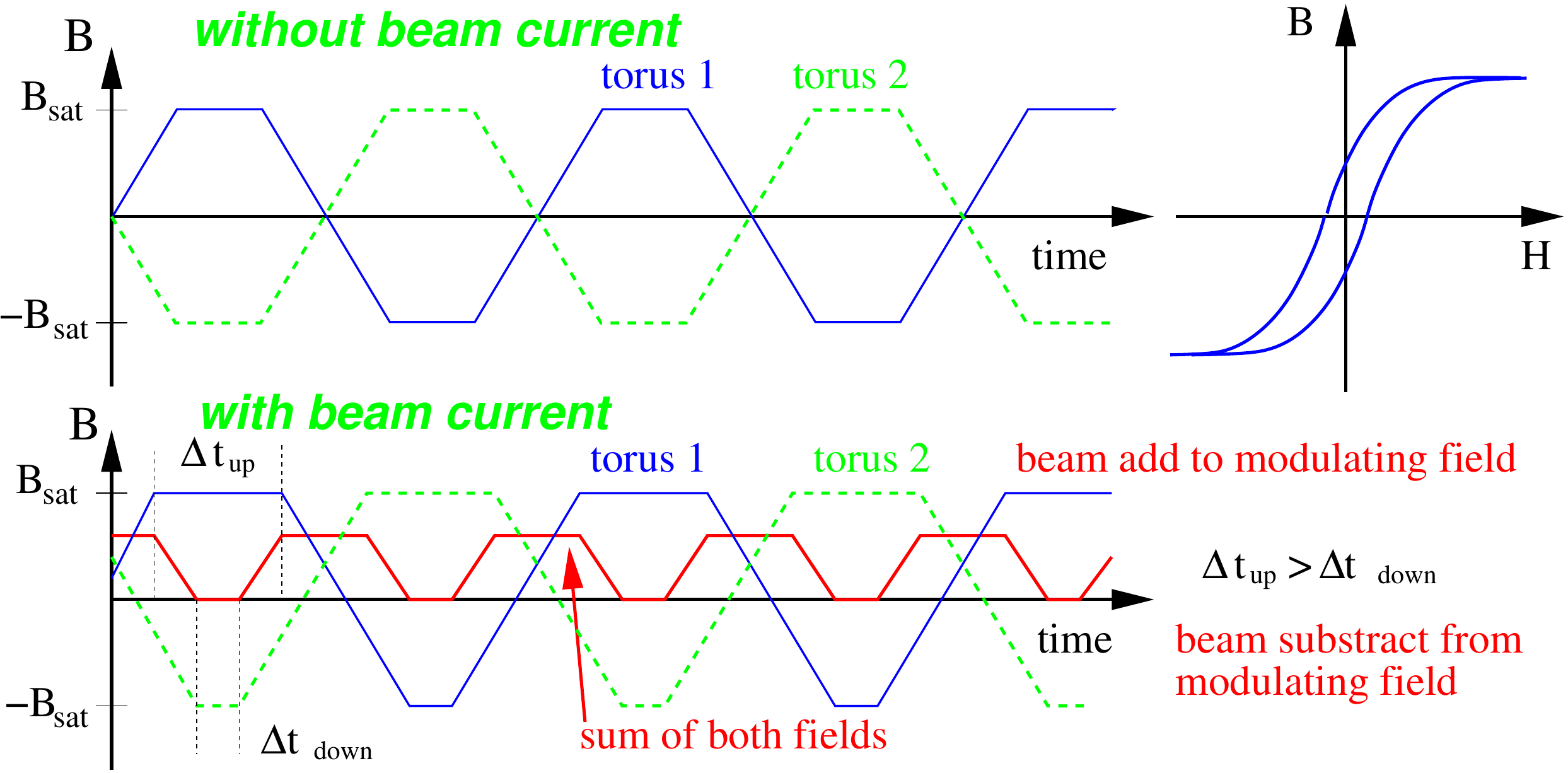}
\vspace*{-.4cm}
\caption{The fluxes in the two tori of a dc-transformer without and with
a beam. The magnetic field of the beam adds to the modulated field for one
modulation phase and 180$^{o}$ degree later it is subtracted. The sum of the
maximum magnetization gives the measured signal. The scheme is drawn for a
triangular-modulation, for a realistic sine-modulation the edges are
smoother.}
\label{flux-dc-trafo}
\end{figure}

\begin{figure}
\parbox{0.49\linewidth}{ 
\centering \includegraphics*[width=55mm,angle=0]
{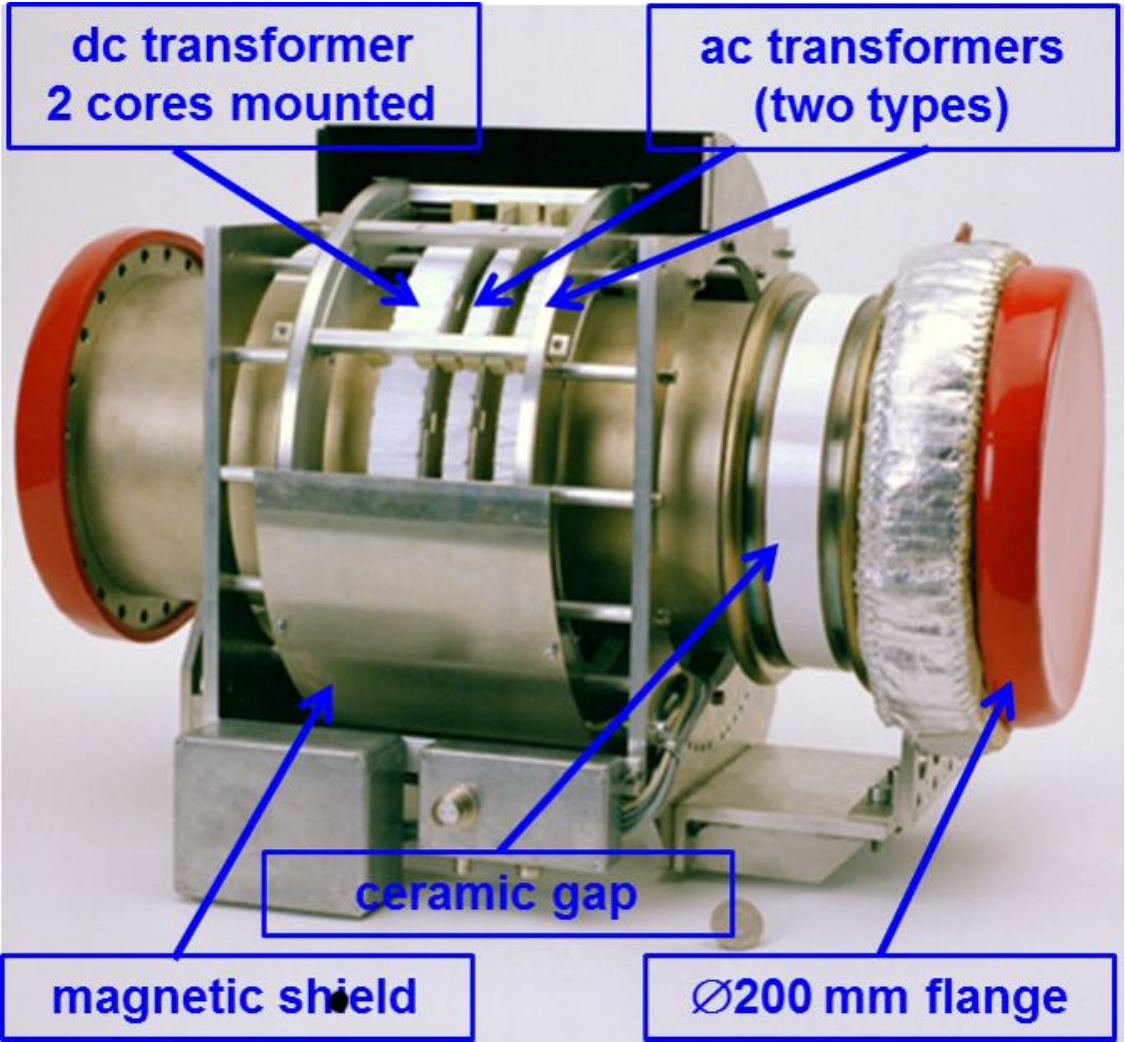} }
\parbox{0.49\linewidth}{ 
\centering \includegraphics*[width=52mm,angle=90]
{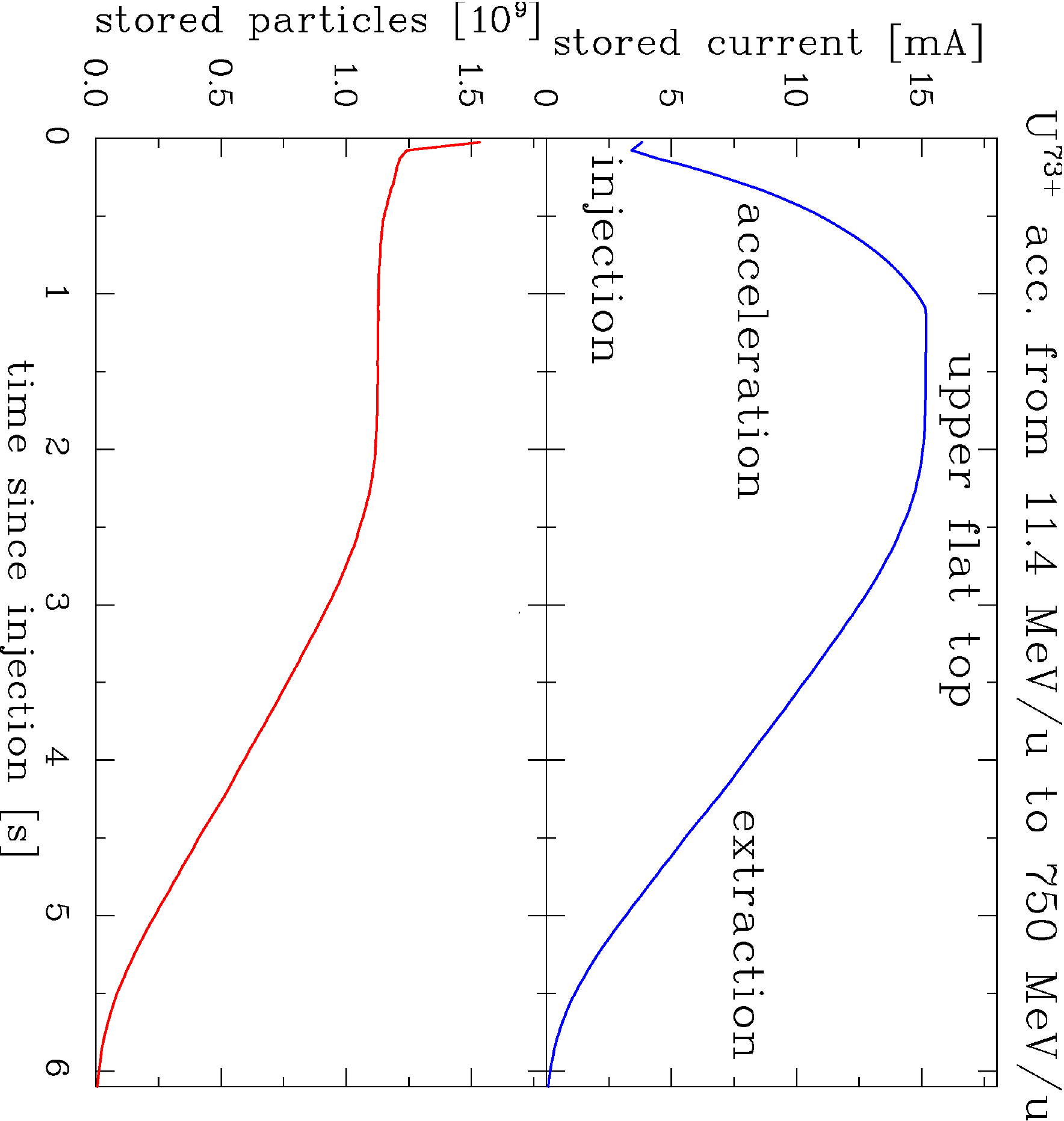} }
\caption{Left: The dc-transformer installed at the GSI synchrotron. The left torus 
is the dc-transformer (two tori mounted closely together), further types transformers are installed in this chamber. 
Right: The signal of a dc-transformer for a $^{238}$U$^{73+}$ beam at the
GSI synchrotron. The electrical current is shown at the top and the number
of stored particles at the bottom. The acceleration needs about 1 s. After
a delay of about 0.5 s, the slow extraction of 4 s length follows. }
\label{sis-trafo-foto}
\end{figure}

\begin{table}
\centering
\caption{The specification of the dc-transformer installed at the GSI
synchrotron \cite{reeg-dipac01}.} 
\begin{tabular}{ll} \hline \hline
\textbf{Parameter} & \textbf{Value} \\ \hline
Torus radii & $r_{i}=135$ mm, $r_{o}=145$mm \\
Torus thickness $l$ & 10 mm\\
Torus material & Vitrovac 6025: (CoFe)$_{70\%}$(MoSiB)$_{30\%}$\\
Torus permeability & $\mu_{r}\simeq 10^{5}$\\
Isolating gap & Al$_{2}$O$_{3}$\\
Number of windings $N$& 16 for modulation and sensing\\
& 12 for feedback\\
Ranges for beam current& 300 $\mu$A to 1 A\\
Current resolution $I_{min}$  & 1 $\mu$A\\
Bandwidth & dc to 20 kHz\\
Rise time constant $\tau_{rise}$ & 20 $\mu$s\\
Offset compensation & $\pm 2.5$ $\mu$A in auto mode\\
& $<15 $ $\mu$A/day in free run\\
Temperature coeff. & 1.5 $\mu$A/$^{o}$C\\
\hline \hline
\end{tabular}
\label{tab-dc-transformer}
\end{table}

The principle of a dc transformer, the so-called \textbf{D}irect {\bf
C}urrent \textbf{C}urrent \textbf{T}ransformer \textbf{DCCT} 
\cite{berg-com,belo-dipac11,webber-biw94,reeg-dipac01} is shown
schematically in Fig.~\ref{dc-prinz-koziol}. It consists of two tori with three
types of 
windings: The first windings of each torus with opposite orientation are
used as a modulator. The modulation frequency is typically 1-10~kHz. The
amplitude of the modulation current is high enough to force the torus into
magnetic saturation of $B_{sat}\simeq 0.6 $ T, 
for positive and negative azimuthal orientation each period. The secondary windings with equal
orientation act as a detector for the modulated signal, see
Fig.~\ref{flux-dc-trafo}. Assuming perfectly identical magnetic
characteristics of both tori, the detector signal, as shown in the scheme,
should be exactly zero if there is no beam current flowing through the tori.
However, an asymmetric shifting of the hysteresis curve results if a dc-beam
is fed through the toroids, because of the additional magnetic field from the~beam. The sum signal $U_{S}$ is different from zero with a modulation twice
the modulation frequency. In the demodulator stage, this signal is rectified.
The dc-current is measured utilizing the~current generated in the feedback
compensation circuit, which forces the output signal back to zero flowing through the third winding of both tori. The detector
can even be used at even harmonics of the~modulation frequency, which results
in higher sensitivity and improvement of the signal-to-noise ratio.

The specification of a typical dc-transformer developed for the heavy-ion
synchrotron at GSI is given in Table \ref{tab-dc-transformer} and a photo in
Fig.~\ref{sis-trafo-foto}. The resolution is about 1 $\mu$A. The offset
drift, mainly caused by the magnetic properties of the tori, is of the order
of 20 $\mu$A per day. The offset can be improved by an~automatic zero
compensation at times the synchrotron does not contain any beam, e.g., after
the ramp down phase of the magnets. For the parameters of the commercially available type see \cite{berg-com}.

An example of a dc-transformer measurement is shown in
Fig.~\ref{sis-trafo-foto}. The electrical current and the~number of
stored particles are shown. A $^{238}$U$^{73+}$ beam is injected in the
synchrotron, and the acceleration starts shortly after the injection. A loss of
particles is seen during the rf bunching process due to some misalignment.
During the acceleration the slope of the magnetic field $dB/dt= 1.3$
T/s is constant, resulting in a constant rise of of the particle momentum.
The current grows linearly only for non-relativistic velocities due to the
decrease of the revolution time, inversely proportional to the~velocity. For
the maximal energy of 750 MeV/u, corresponding to $\beta = 84$ \%, the
relativistic increase of the mass starts to be significant. After reaching
the maximum energy, a $\sim 0.5$ s flat top is seen for the~de-bunching phase followed by the slow extraction.

\subsection{Energy loss and ranges of particles in matter}
\label{kap-current-energyloss}
\subsubsection{Material interaction of protons and ions}
\begin{figure}
\centering \includegraphics*[width=65mm,angle=90]
{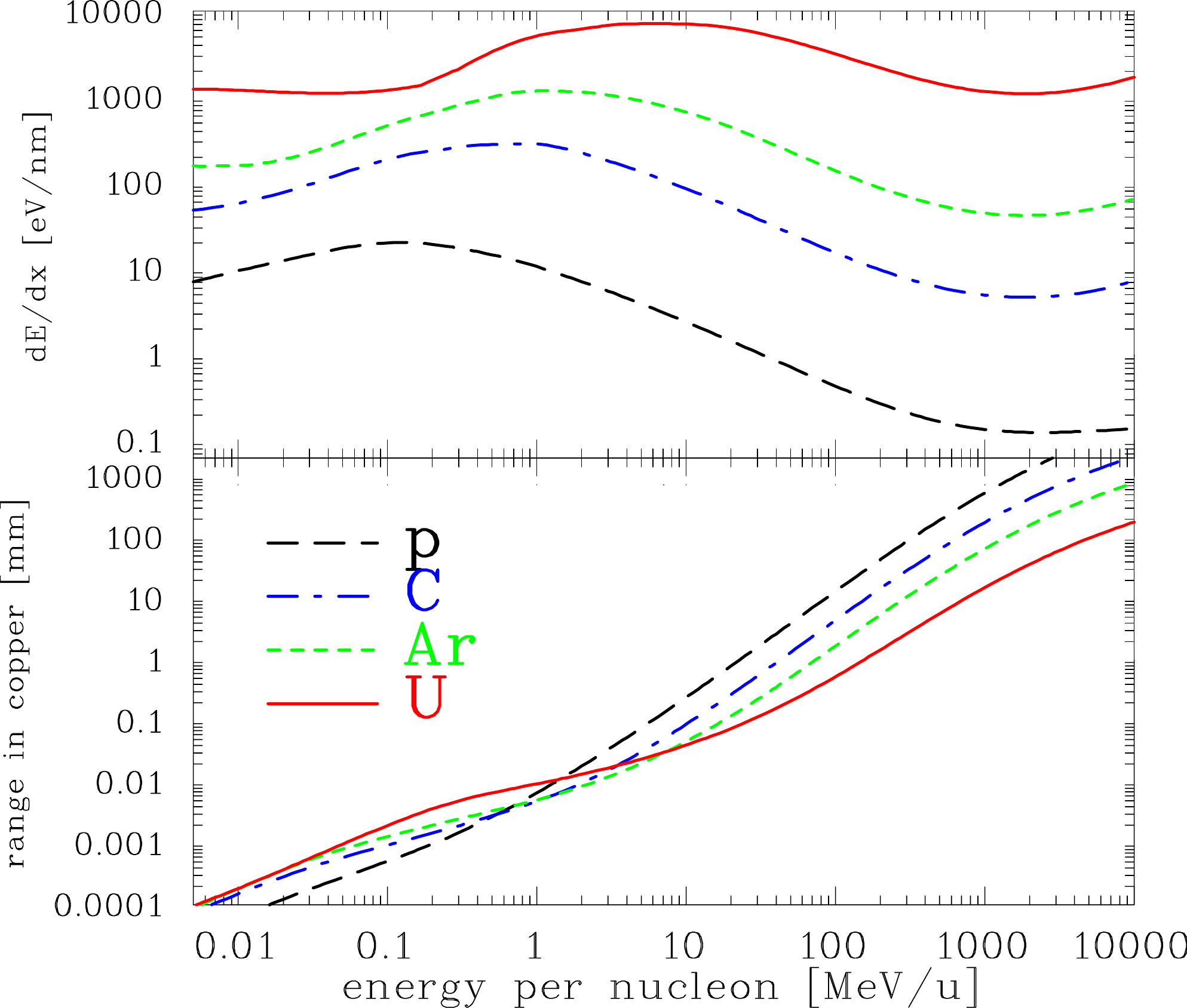}
\caption{The energy loss at the surface 
and the range in copper as a function of the kinetic energy for several
ions. The energy range is plotted from 5 keV/u to 10 GeV/u and the range
from 100 nm to 1 m. The calculation of the~electronic and nuclear stopping
uses the semi-empirical code SRIM \cite{ziegler-www}.}
\label{copper_range}
\end{figure}

For the discussion for all intercepting diagnostics like Faraday
cups, wire scanners, SEM-grids etc., 
the energy loss and the related range of the particles penetrating 
matter serves as the basic physical mechanism. 
The energy loss of a proton or an ion is mainly due to the collision
of the projectile with the electrons within solid target, the so-called
electronic stopping power. Due to the different masses of the~ion and the electron,
the energy transfer to the electron per collision is, in most cases, 
below 100~eV. The~electronic stopping power $\frac{\mbox{d}E}{\mbox{d}x}$ can be approximated by the semi-classical Bethe-Bloch formula 
which is written in its simplest from as \cite{leo-buch,knoll-buch,jackson-buch,ziegler-www,part-data-PhysRevD2018}
\begin{equation}
-\frac{\mbox{d}E}{\mbox{d}x} = 4 \pi N_{A} r_{e}^{2} m_{e} c^{2} \cdot 
\frac{Z_{t}}{A_{t}} \rho \cdot
\frac{Z_{p}^{2}}{\beta^{2}} 
\left[ 
\ln \frac{2 m_{e} c^{2} \gamma^{2} \beta^{2} }{I} - \beta^{2} 
\right]
\label{eq-bethe-bloch}
\end{equation}
with the constants: $N_{A}$ the Avogadro number, $m_{e}$ and $r_{e}$ the mass
and classical radius of an electron and $c$ the velocity of light. The target
parameters are: $\rho$ density of the target with nuclear mass $A_{t}$ and
nuclear charge $Z_{t}$; the quantity $ \frac{Z_{t}}{A_{t}} \rho $ correspond
to the electron density. $I$ is the mean ionization potential for
removing one electron from the target atoms; a rough approximation for
a target with nuclear charge $Z$ is $I \simeq Z \cdot 10$ eV, more
precise values are given e.g. in \cite{ziegler-www}.
The projectile parameters are: $Z_{p}$ nuclear charge of the ion with velocity
$\beta$ and Lorentz factor $\gamma =(1-\beta^{2})^{-1/2}$. This formula has to be modified since ions travelling through matter are not bare nuclei, but have
some inner electrons. An~effective charge is used instead of $Z_{p}$
calculated by e.g. semi-empirical methods described for the codes SRIM
\cite{ziegler-www} or LISE++ \cite{lise++www}. The
result of such semi-empirical calculation for the energy loss is shown in 
Fig.~\ref{copper_range} for different
ions into copper. The energy loss is maximal for ions with kinetic energy
around 100 keV/u to 7~MeV/u (corresponding to velocities $\beta \sim 1.5$ \%
to $ 12$ \%) depending on the ion species. These are typical energies of a
proton/heavy ion LINAC. Below 100 keV/u the energy loss decreases and nuclear
stopping becomes significant. Energies below 10 keV/u are typical for
proton/heavy ion sources mounted on a~high voltage platform. For relativistic
energies above 1 GeV/u, the energy loss is nearly constant; these are typical
energies of particles extracted from a synchrotron.

For the consideration of a Faraday cup (see section \ref{chapter-faraday-cup}), 
the range in the material is important.
Copper is often used for cups, due to its high heat conductivity. For a
particle accelerated to $E_{kin}$, the~range $R$ is calculated numerically
from the stopping power via
\begin{equation}
R =\int_{0}^{E_{kin}} \left( \frac{\mbox{d}E}{\mbox{d}x} \right)^{-1} \mbox{d}E
\end{equation}
and has an approximately scaling for ions 
above $\simeq 10 $ MeV/u \cite{leo-buch}
\begin{equation}
R \propto E_{kin}^{1.75} ~.
\end{equation}
The results are shown in Fig.~\ref{copper_range}. This range should be 
shorter than the mechanical dimension for a~practical Faraday cup
design; for energies
below 100 MeV/u the range is less than 1 cm. Or, in other words, cups are only
useful at LINACs or cyclotrons. For higher energies, the range is too large for the typical beam instrumentation sizes
and other techniques, e.g. particle detectors are applied. Moreover, at those particles energies are above the nuclear
Coulomb barrier where nuclear reactions are possible and charged secondary particles
might leave the material resulting in a wrong reading.

\subsubsection{Material interaction of electrons}
\begin{figure}
\centering \includegraphics*[width=50mm,angle=90]
{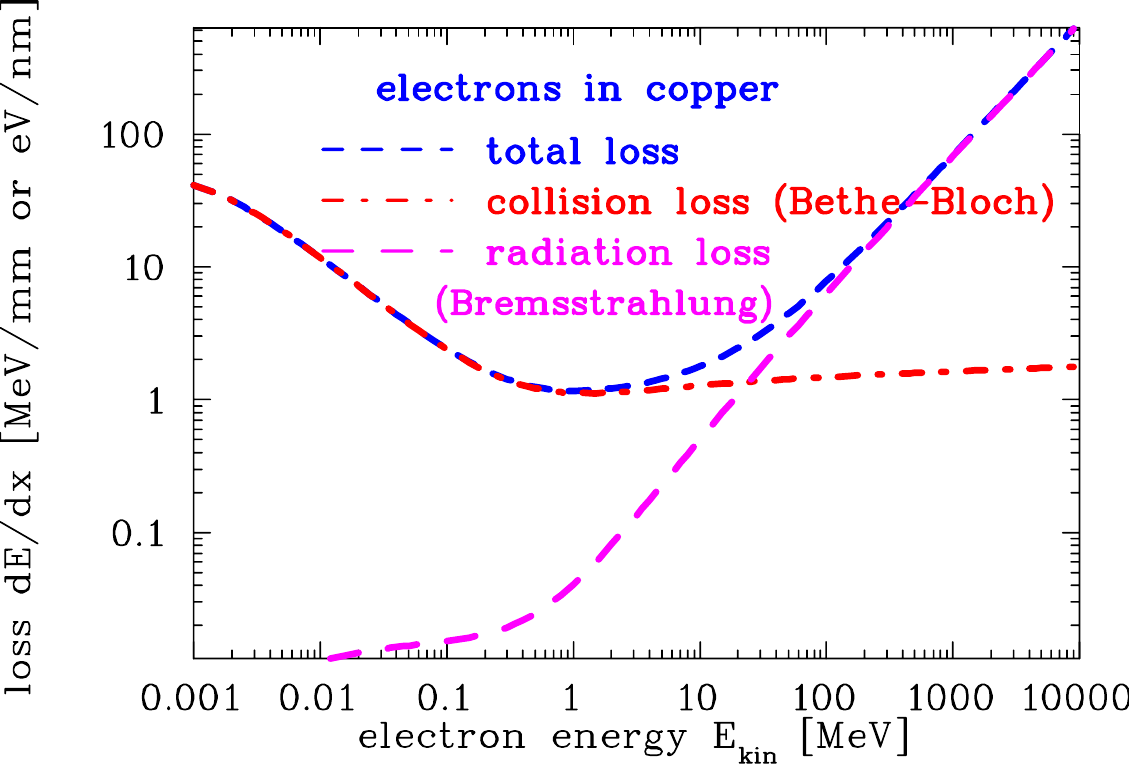}
\caption{The energy loss of electrons in copper for collision loss as
given by the Bethe-Bloch equation and radiation
losses as dominated by the emission of Bremsstrahlungs-photons, data from \cite{berg_NIST}.}
\label{energyloss-electrons}
\end{figure}

The stopping of electrons in matter differs from protons and ions, see
Fig.~\ref{energyloss-electrons} where the sum of the two relevant
processes of collisional loss $\mbox{d}E/\mbox{d}x|_{col}$ and radiation loss
$\mbox{d}E/\mbox{d}x|_{rad}$ are added. The collisional loss for electrons 
due to electronic stopping $\mbox{d}E/\mbox{d}x|_{col}$ 
is also described by a modified Bethe-Bloch formula;
the modification is caused by the equal mass of the projectile and the
target electrons and their indistinguishably, see e.g. \cite{part-data-PhysRevD2018}. 
This regime dominates only for energies below below 1 MeV as
corresponding to the a velocity of $\beta = v/c= 94$ \% and a
Lorentz-factor of $\gamma= \left( 1-\beta^2\right)^{-1/2}=2.96$. 
Even a small electron
accelerator reaches higher final energies. For energies
above a few 10~MeV the~radiation loss by Bremsstrahlung, i.e. the
emission of photons by acceleration or deceleration 
of an~electron in the vicinity of
the target nucleus, dominant, see e.g. \cite{leo-buch}. This radiation
loss scales roughly linear to the electron energy and quadratically to
the target charge $Z_t$ as $\mbox{d}E/\mbox{d}x |_{rad} \propto E \cdot Z_t^2$.
The~trajectories of the primary electrons in
the target are more curved than for ions, due to the possible high
energy- and momentum transfer in a single collision, therefore
electrons have much larger lateral straggling than ions. 
Moreover, the longitudinal straggling is more substantial than for ions resulting 
in a broader range distribution. 

At electron accelerators, Faraday cups are mainly used behind the electron gun for electrons with typically 100 keV energy. In this case, the construction is comparable to the Faraday cups used for proton beams, see Section~\ref{chapter-faraday-cup}. At higher energies, the required amount of material to stop the electrons and to absorb secondary charged particles (e.g $e^+ - e^-$ pair production from the Bremsstrahlungs-photons) increases significantly. A Faraday cup used for high energies up to some GeV is desribed in \cite{morgan_dipac05}. 

\begin{figure}
\parbox{0.45\linewidth}{
\centering \includegraphics*[width=55mm,angle=0]
{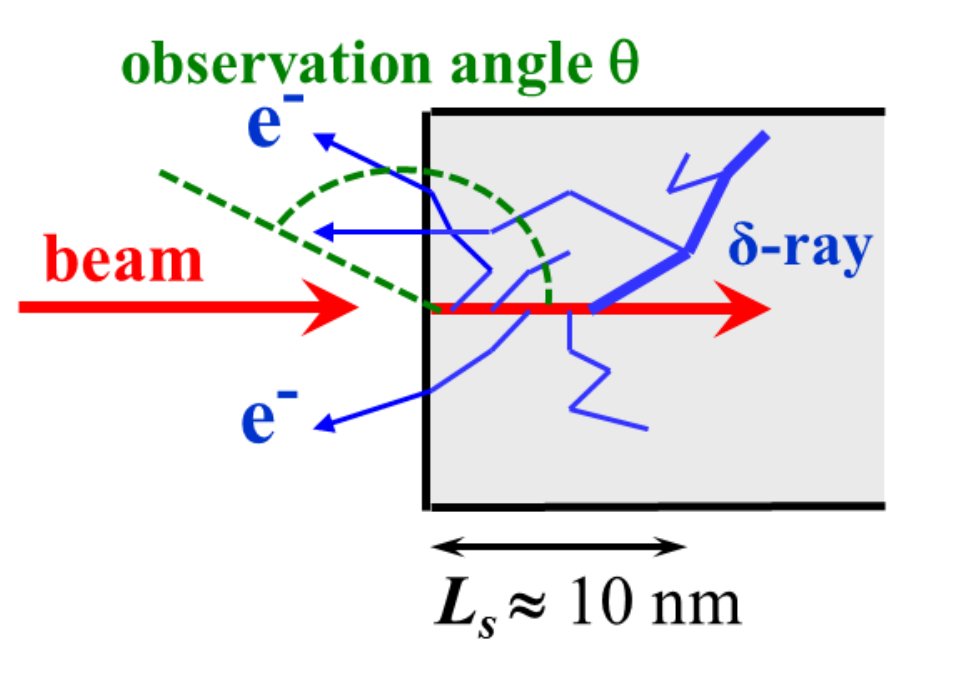}
}
\parbox{0.55\linewidth}{
\centering \includegraphics*[width=75mm,angle=0]
{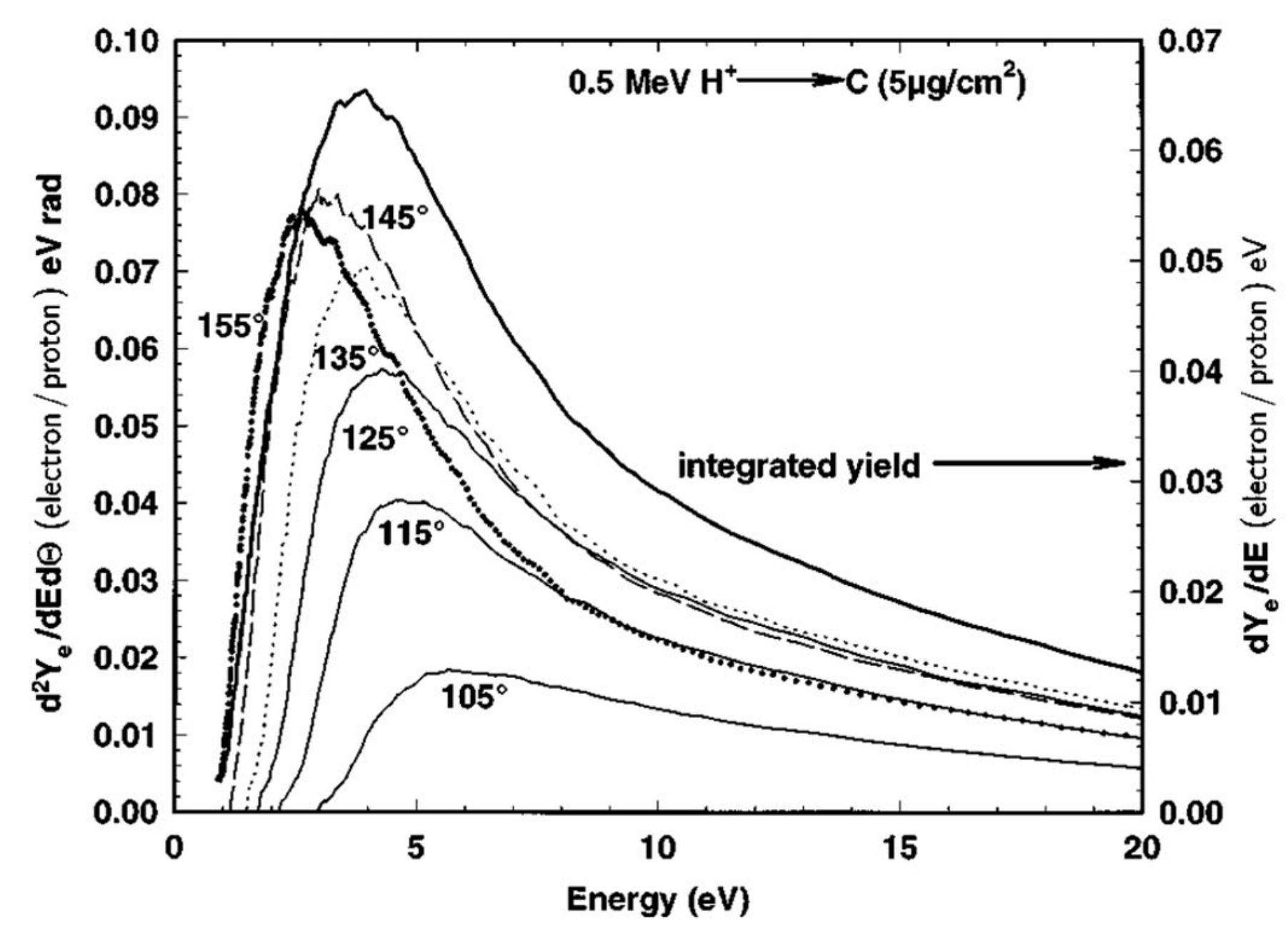}
}
\vspace*{-0.4cm}
\caption{Left: Schematic drawing of the secondary electron generation from a surface by charged particle impact; $L_s$ is the scattering length from
which electrons are able to reach the surface. Right: A typical measurement
of the electron energy spectrum for different emission angles and the
total yield in backward direction by 0.5 MeV proton impact on a carbon
foil \cite{drex-physreva96}.}
\label{scheme_secondary_electrons}
\end{figure}

\subsubsection{Secondary electron generation}
\label{chapter-sternglass}
When a charged particle, either an ion or an electron, travels through
matter it interacts with the target electrons as schematically depicted 
in Fig.~\ref{scheme_secondary_electrons} left. In a close, primary collision the
target electron can be accelerated to high energies much larger than
100 eV, sometimes these electrons are called $\delta$-rays. The energy 
distribution depends on the impact parameter and the
energy of the incoming particle. In a metallic solid-state material
the fast $\delta$-electrons collide with the surrounding electrons;
because both collision partners 
have equal masses the energy is transferred efficiently and more
electrons are liberated, see the schematics in 
Fig.~\ref{scheme_secondary_electrons}. This so-called
thermalization of the electrons inside the conduction band is
performed in a typical scattering length of $L_s \simeq 10$ nm, i.e. 
within the distance of
some 100 lattice planes. Due to this many collisions, there is only a
weak correlation between the direction of the~incoming particle and the
scattered electrons. If these
electrons reach the surface and have still an energy higher than the
work-function (typically 1 to 5 eV) they have a high
probability to escape from the metal. 
The current of the secondary electrons
is given by the so-called Sternglass formula
\cite{Stern-PhyRev57}
\begin{equation} 
I_{sec} = Y \cdot \frac{\mbox{d}E}{\rho \mbox{d}x} \cdot I_{beam}
\label{sem-yield_general}
\end{equation} 
with $ I_{beam}$ being the current of incoming particles and 
$Y$ being the yield factor describing the amount of secondary emission
per unit of energy loss $ \frac{dE}{\rho dx}$ at the surface of the
metal. The yield $Y$ depends on the metal (via the work-function) and
the surface quality, e.g. modified by possible adsorbed gas contamination. 
The mean kinetic energy of the escaping secondary electrons is in the order of
10 eV and has some dependence on the emission angle;
an example for the energy spectrum is shown in
Fig.~\ref{scheme_secondary_electrons} and discussed in more details in 
\cite{drex-physreva96}. 
To first order the angular distribution $P(\theta)$ of the electron emission
can be approximated by a $P(\theta) \propto \cos \theta$ law, where
$\theta$ is the angle of the trajectory with respect to the~surface
($\theta = 180^o$ means backscattering).

\subsection{Faraday cups}
\label{chapter-faraday-cup}
\begin{figure}
\parbox{0.67\linewidth}{
\centering \includegraphics*[width=90mm,angle=00]
{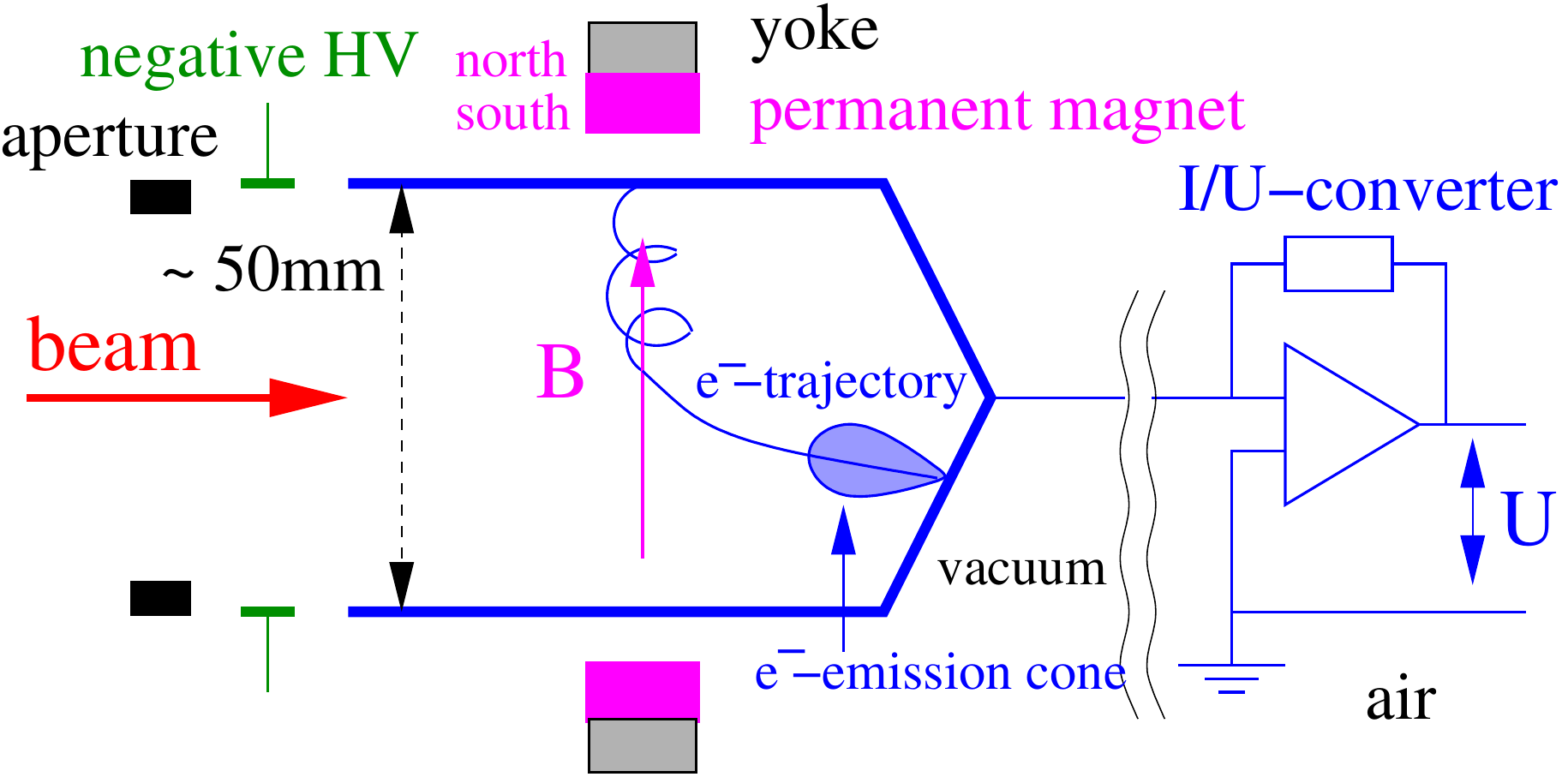}
}
\parbox{0.32\linewidth}{
\centering \includegraphics*[width=45mm,angle=00]
{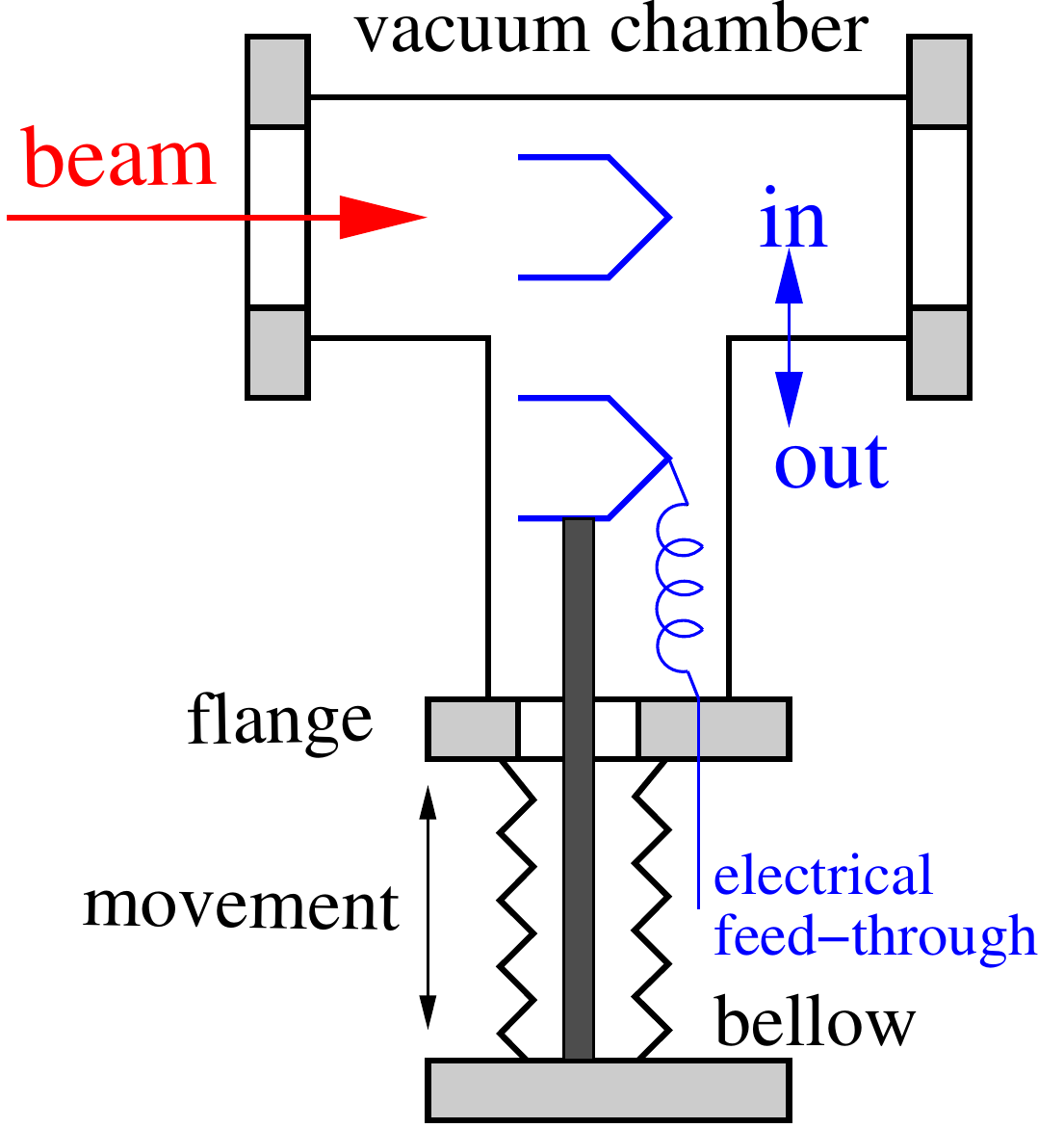}
}
\caption{Left: Scheme of an un-cooled
Faraday cup with magnetic and electric secondary electron
suppression. Right: Scheme for inside and outside position of destructive
diagnostics devices.}
\label{cup-low-int}
\end{figure}

\begin{figure}
\parbox{0.5\linewidth}{ 
\centering \includegraphics*[width=40mm,angle=0]
{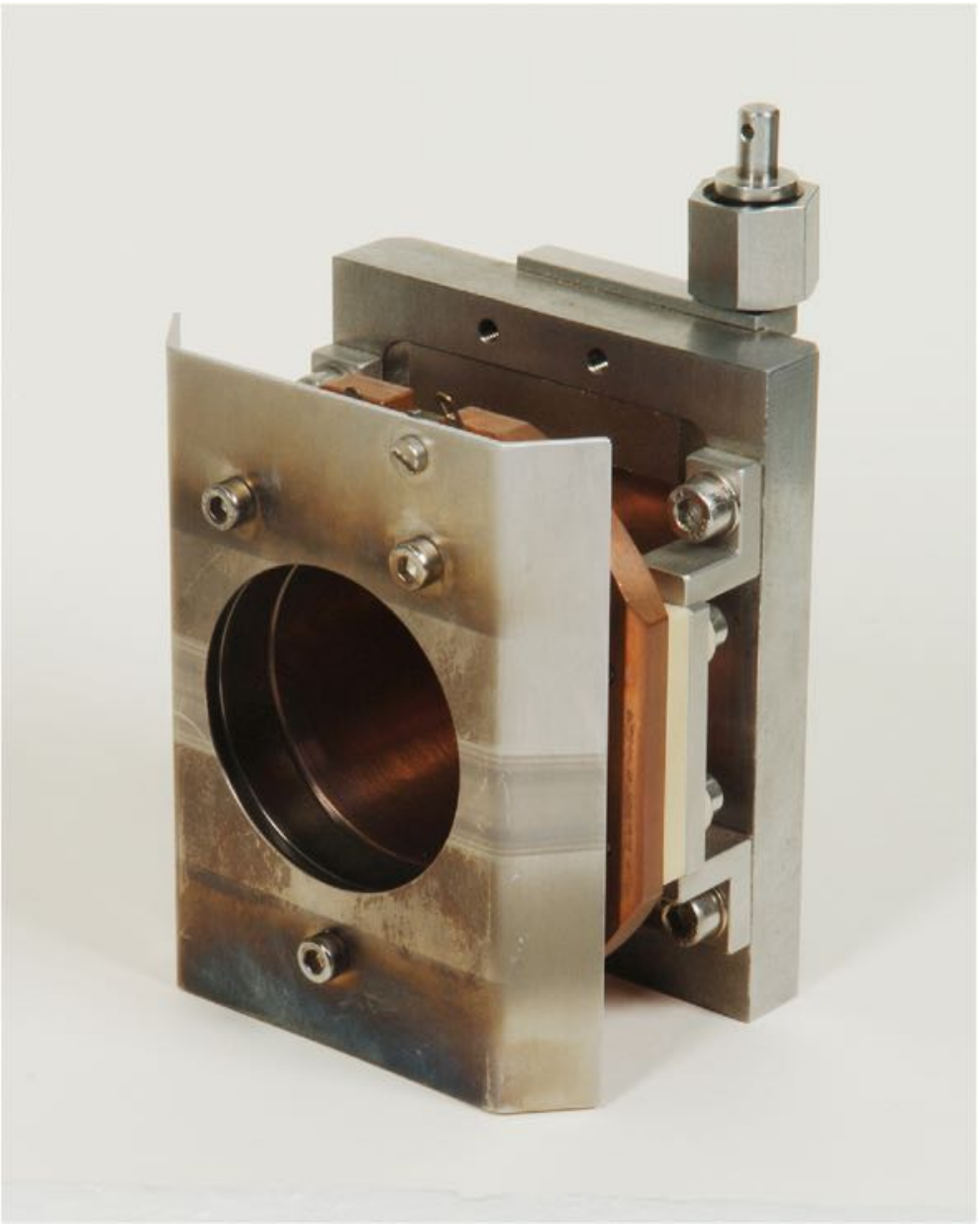}
}
\parbox{0.5\linewidth}{ 
\centering \includegraphics*[width=40mm,angle=00]
{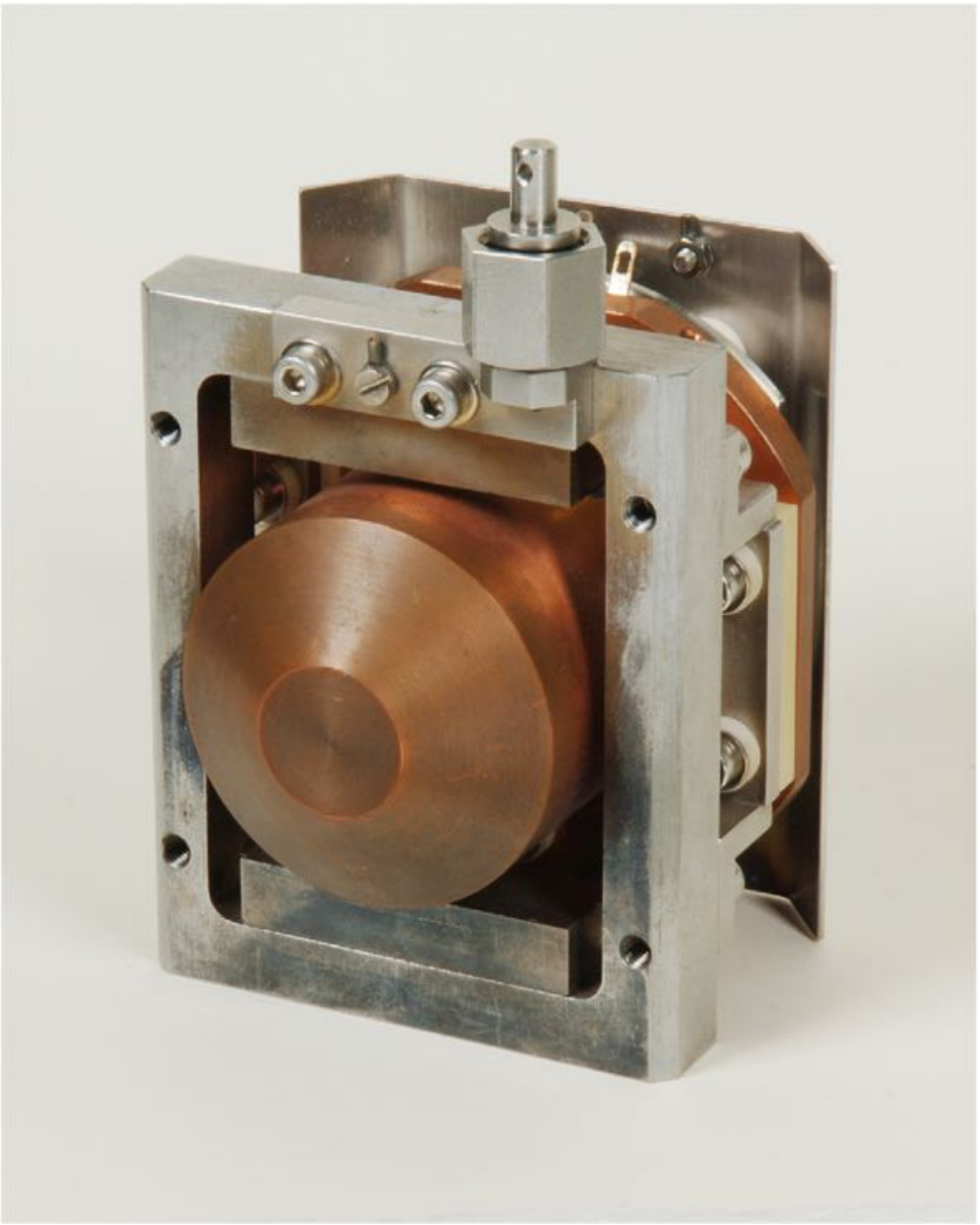}
}
\caption{Photo of an \O 50 mm uncooled
Faraday cup with magnetic and electric secondary electron suppression.}
\label{foto-faraday-front}
\end{figure}

A Faraday cup is a beam stopper supposed to measure the electrical current of
the beam. The basic cup design is shown in Fig.~\ref{cup-low-int} and
a photo in Fig.~\ref{foto-faraday-front}: An isolated
metal cup is connected to a current sensitive pre-amplifier. As shown for an
active beam transformer, the pre-amplifier consist of a low impedance input
and a conversion to a voltage. Range switching is achieved by using different
feedback resistors for the operational-amplifier. With a Faraday cup, much
lower currents can be measured as compared to a transformer: A measurement of
10 pA for a dc-beam is possible with a low noise current-to-voltage
amplifier and careful mechanical design; this is 5 orders of magnitude more
sensitive than a~dc-transformer. Low current measurement is important, e.g.,
for the acceleration of radioactive beams.

\begin{figure}
\parbox{0.60\linewidth}{ 
\centering \includegraphics*[width=60mm,angle=0]
{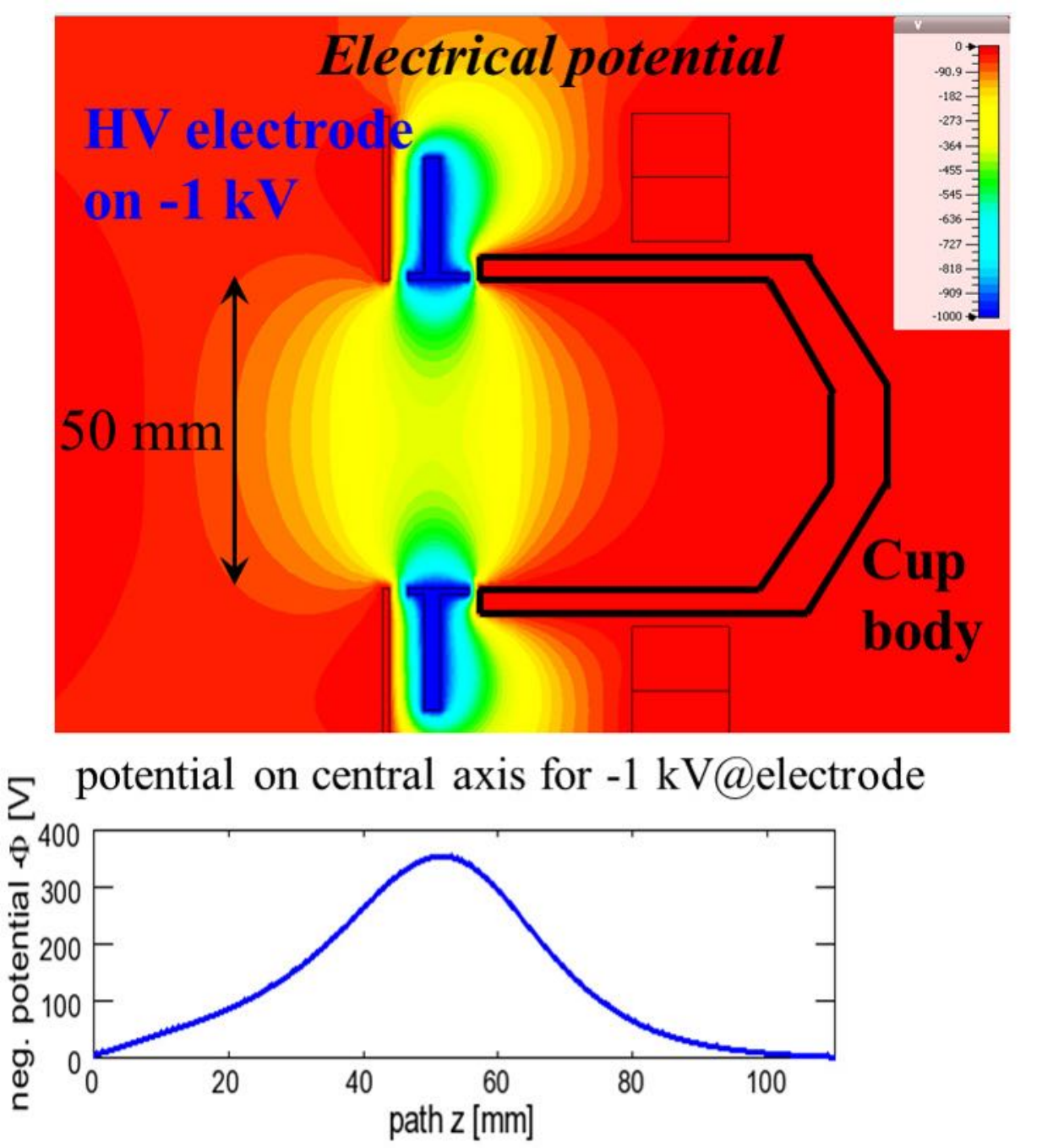}
}
\parbox{0.35\linewidth}{ 
\hspace*{-2cm} \includegraphics*[width=55mm,angle=0]
{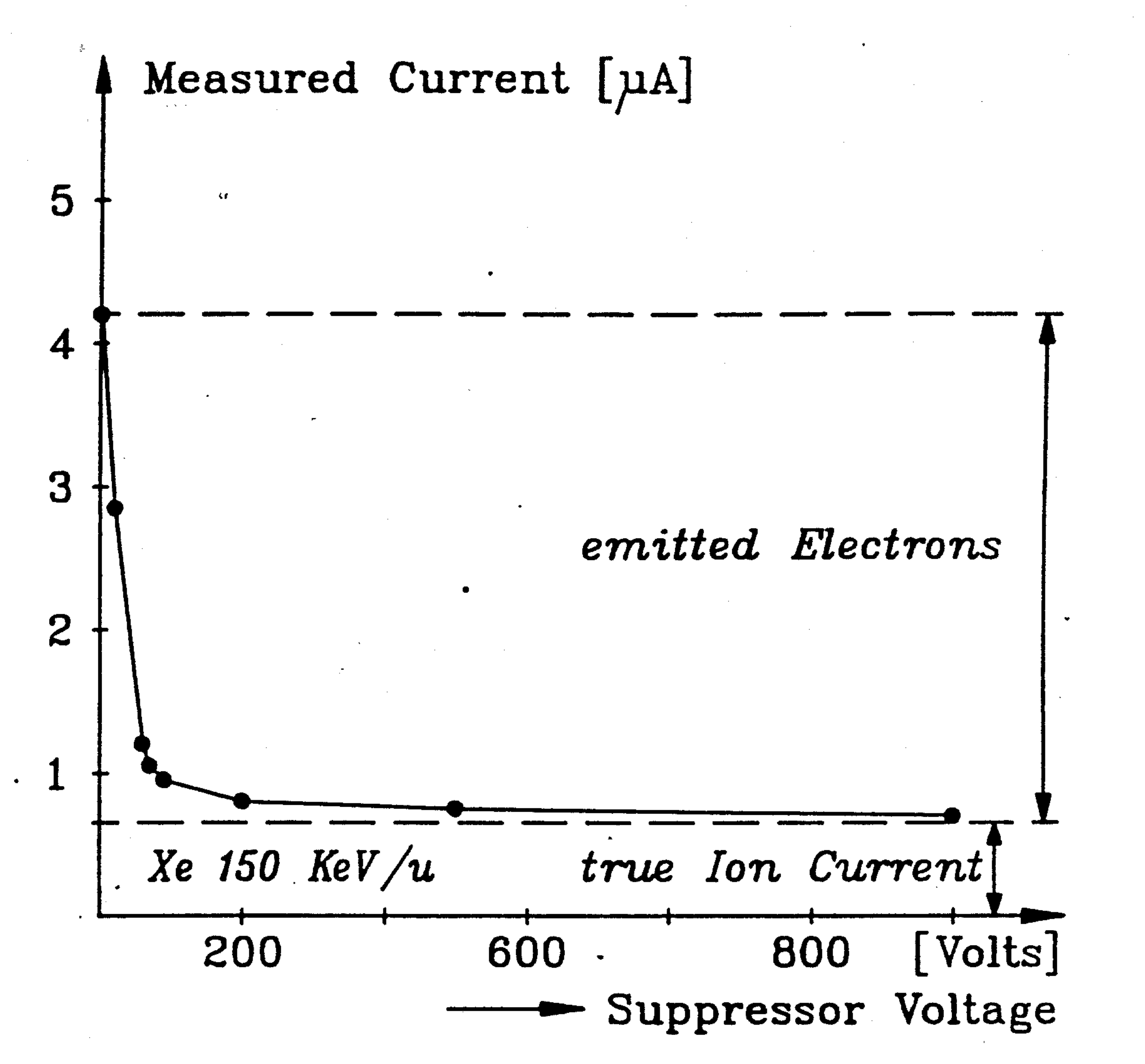}
}
\caption{Left: The electric potential of the Faraday cup as shown in 
Fig.~\ref{foto-faraday-front} with the high voltage electrode biased
by -1 kV is shown at the cylindrical symmetry plane. The beam enters
from left. The negative potential along the~central axis is depicted at
the bottom. 
Right: The effect of secondary electrons suppression inside a Faraday cup as
determined by the beam current determination a function of the applied voltage.}
\label{cup-suppresion-voltage}
\end{figure}

\begin{figure}
\centering \includegraphics*[width=110mm,angle=0]
{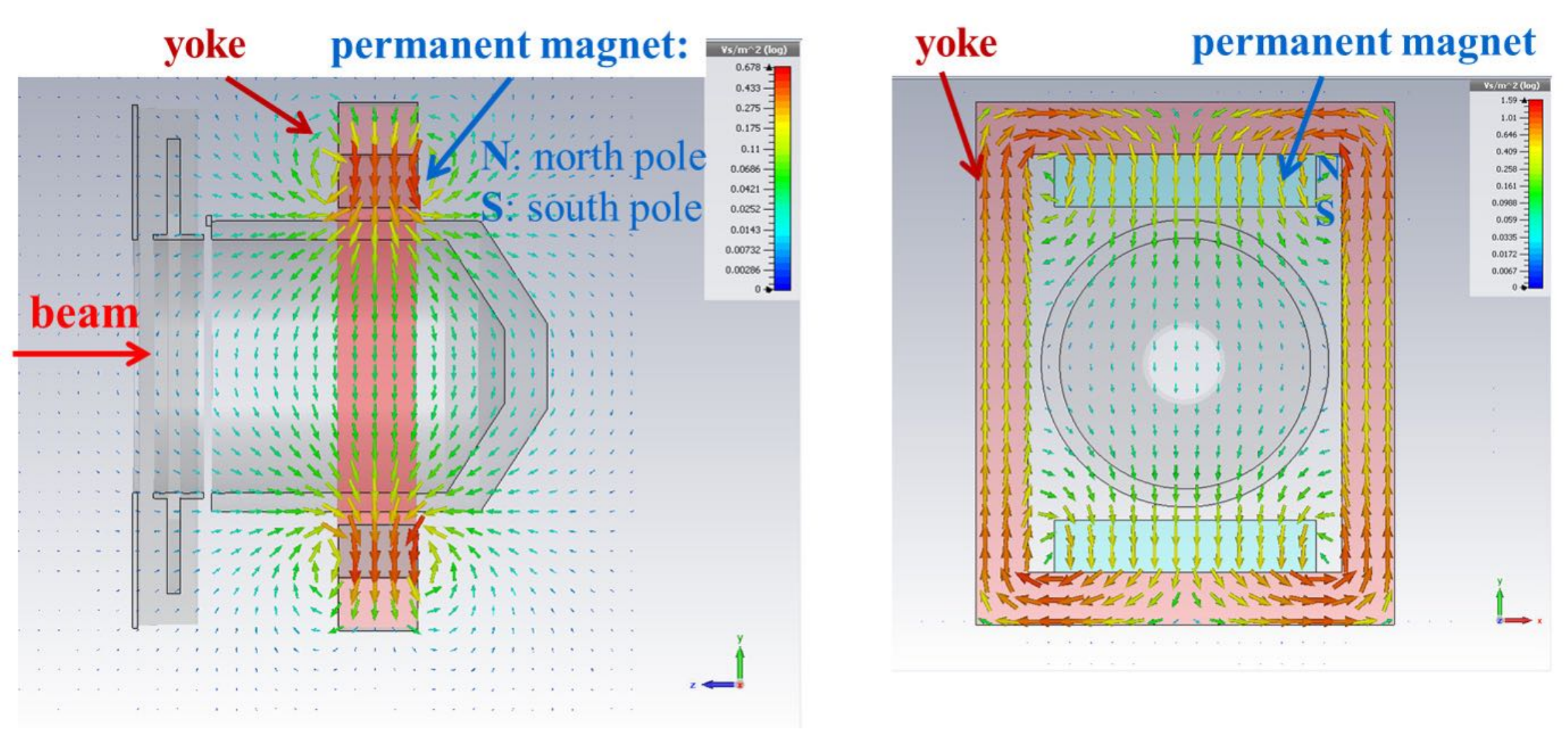}
\caption{Magnetic field lines of the arrangement of Co-Sm permanent
magnets within the yoke for the Faraday cup of
Fig.~\ref{foto-faraday-front}. The
homogeneous field strength is $B\sim 60$ mT.}
\label{cup-magnetic-field}
\vspace*{-0.1cm}
\end{figure}

When an accelerated particle hits a surface, secondary electrons are
liberated, see Section~\ref{chapter-sternglass}. The flux of
these electrons are
proportional to $\cos \theta$, where $\theta$ is the angle of the electron
trajectory with respect to the surface; their average energy is below
$\simeq 10$ eV.
If these electrons leave the insulated cup, the reading of the beam
current is wrong by this amount. A secondary electron suppression has to be
foreseen. It can be realized by:
\begin{itemize}
\item Using a high voltage suppression close to the entrance of the cup. By applying a voltage well above the mean energy of the secondary electrons,
they are pushed back to the cup surface, see Fig.~\ref{cup-suppresion-voltage}, left. 
The disadvantage of this method is related to the fact that the electrical field on the~beam axis is lower than on the edges; in the depicted case the maximum potential is about 35 \% of the potential applied to the electrode. The emission of the
maximal energetic electrons occurred opposite to the~beam direction, and the potential has to be chosen to be higher than the~kinetic
energy of at least 99 \%
of the electrons. A measurement with varying 
suppression voltage is shown in Fig.~\ref{cup-suppresion-voltage}, right. 
\item By using a magnetic field created by permanent magnets. In this field
$B$, the secondary electrons spiral around the magnetic field lines with the
cyclotron radius $r_{c}$
\begin{equation}
r_{c}=\frac{\sqrt{2m_{e}E_{kin,\perp}}}{eB} 
\end{equation}
with $m_{e}$ the electron mass, $e$ its charge and $E_{kin,\perp}$ the kinetic
energy of the velocity component perpendicular to the field lines. For the example of $E_{kin,\perp}=10$ eV and a
field of 10 mT the radius is $r_{c}\sim 1$ mm. With permanent magnets, field 
lines perpendicular to
the beam axis can be created by permanent magnets, see
Fig.~\ref{cup-magnetic-field}.
\end{itemize}

\section{Beam position monitors}
\label{kap-pickup}

\begin{figure}
\centering \includegraphics*[width=80mm,angle=0]
{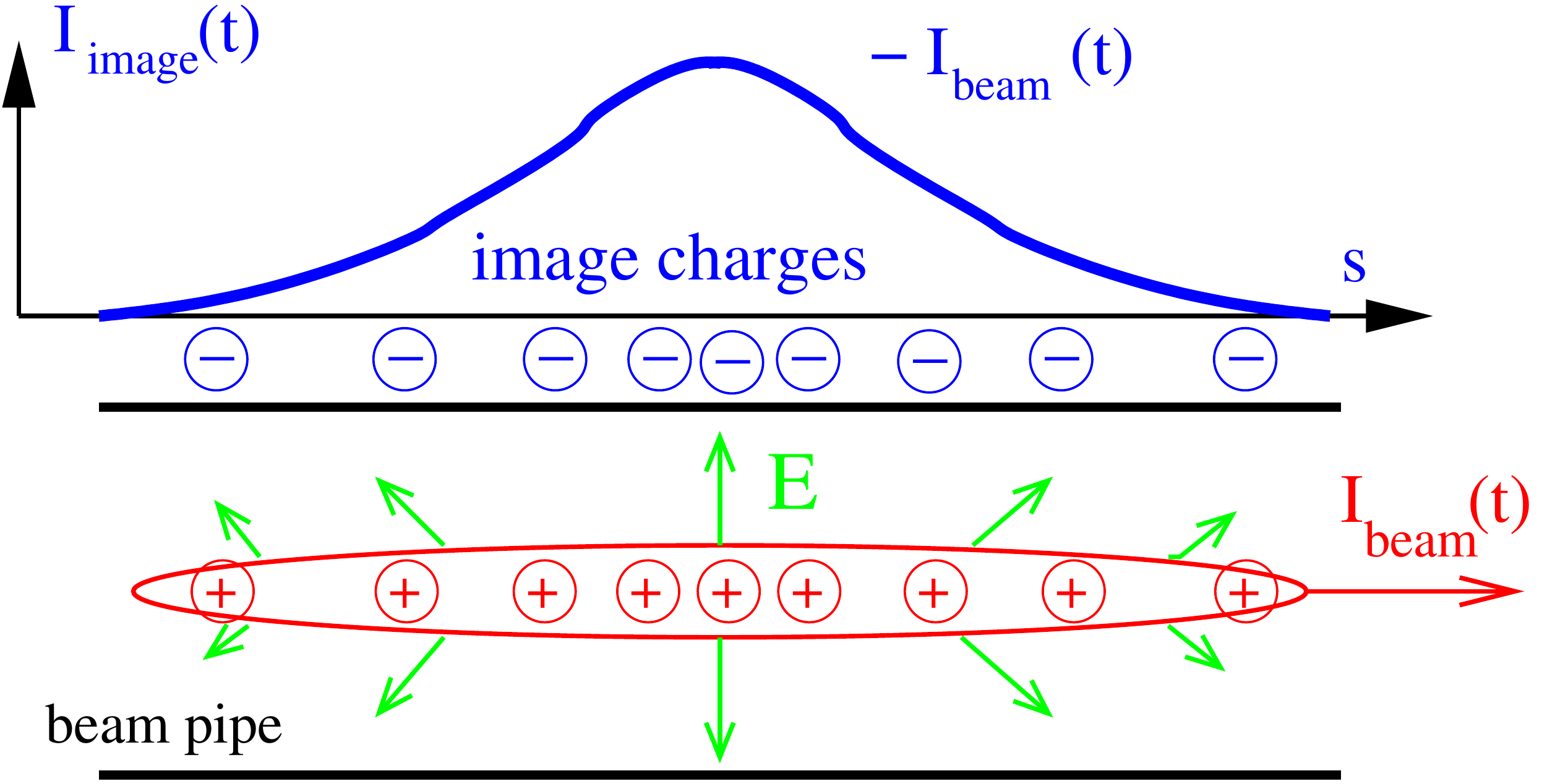}
\caption{The beam current induces a wall current of the same magnitude and time structure 
but reversed polarity.}
\label{scheme-wall-current}
\vspace*{-0.3cm}
\end{figure}

The position of a beam in terms of its transverse centre is very
frequently displayed as 
determined by pick-up plates in a non-invasive manner. It is measure by insulated metal
plates where an image current is induced by the electro-magnetic field of the beam particles, 
see Fig.~\ref{scheme-wall-current}. Because the electric field of a~bunched beam
is time-dependent, an ac signal is seen on the plate, and the coupling is done
using rf technologies. Only time-varying signals can be detected by
this principle as generated by a bunched beam.
The principle signal shape, as well as the most often
used types of pick-ups, are described. The~application is the determination of
the beam position, i.e. the transverse centre-of-mass of the bunches. 
To this end, four pick-up plates are
installed, and the difference of the signals yields the~centre-of-mass in both
transverse axes. The device of the plates is called \textbf{P}ick-{\bf
U}p \textbf{PU} the corresponding 
installation for the~centre-of-mass determination is referred as \textbf{B}eam 
\textbf{P}osition \textbf{M}onitor \textbf{BPM}. Various measurements based on this position information are possible but are only discussed briefly at the end of this section.

\subsection{Signal treatment of capacitive pick-ups}
\label{kap-pickup-treatment}

\begin{figure}
\centering \includegraphics*[width=110mm,angle=0]
{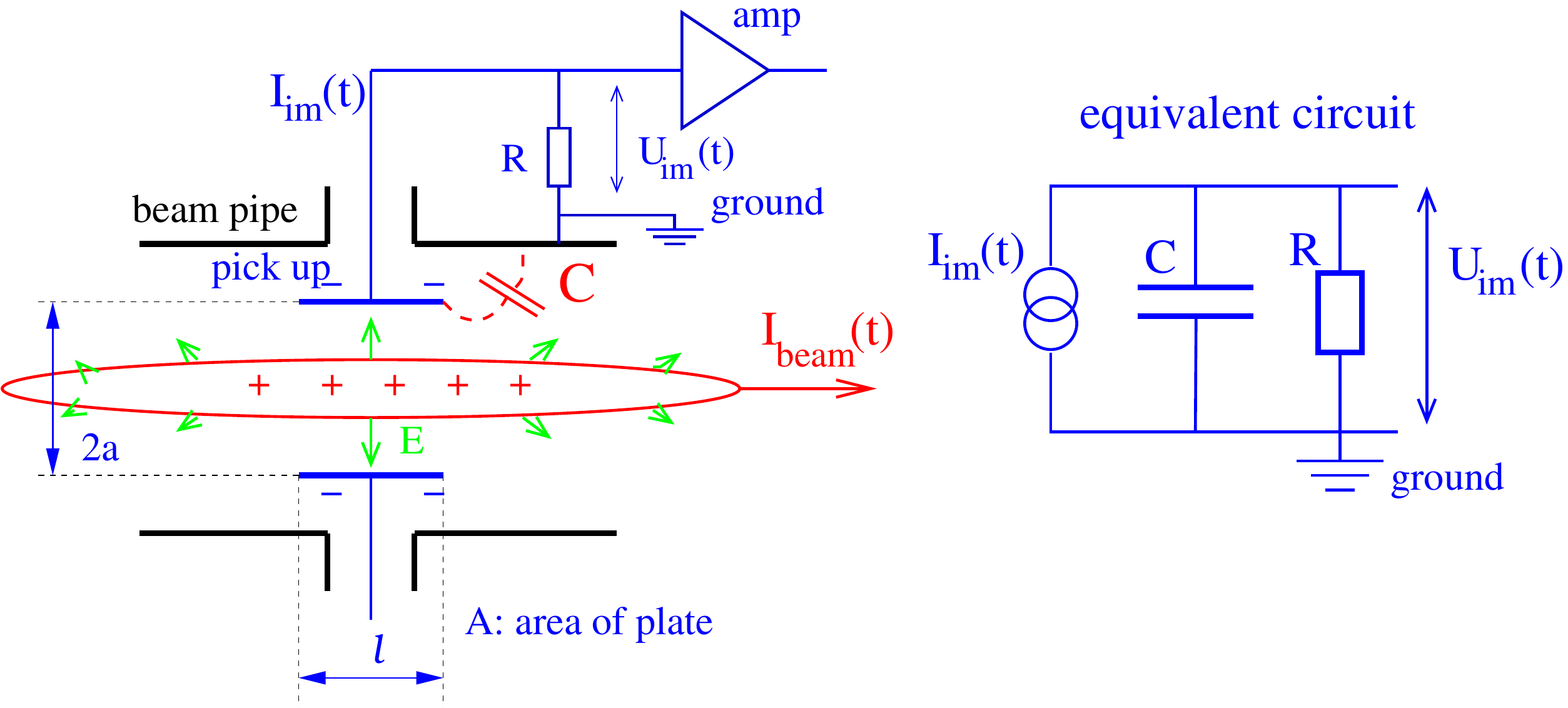}
\caption{Scheme of a pick-up electrode and its equivalent circuit.}
\label{cap-pickup-principle}
\end{figure}

As shown in Fig.~\ref{cap-pickup-principle}, a capacitive pick-up consists of
an electrode inserted in the beam pipe. Here the~induced image charge of
the beam is coupled via an amplifier for further processing. The electrode at a~distance $a$ from the beam centre has an area of $A$ and a length in
longitudinal direction of $l$. 
As the~signal we measure the voltage $U_{im} (\omega)$ at a resistor $R$ in the equivalent circuit of Fig.~\ref{cap-pickup-principle} as realized by the input matching of an amplifier or the ADC input. We use the formula in frequency domain 
\begin{equation}
U_{im} (\omega) = R \cdot I_{im} (\omega)
= Z_{t}(\omega, \beta) \cdot I_{beam} (\omega) ~~~.
\label{eq-transfer-impedance}
\end{equation}
For all types of pick-up electrodes, the general quantity of transfer
impedance $Z_{t} (\omega, \beta)$ is defined by Eq.~(\ref{eq-transfer-impedance}) in the frequency domain according
to Ohm's law. The transfer impedance describes
the effect of the beam on the pick-up voltage and it is dependent on
frequency, on the velocity of the beam particles $\beta$ and on geometrical
factors. It is very helpful to make the description in the frequency domain,
where the~independent variable is the angular frequency $\omega$, related to
the time domain by Fourier transformation. The exact treatment of the signal is discussed in the literature, in particular in Refs.~\cite{schmick_cas2018,smaluk_book,brandt_cas2007,strehl_book,forck_juas,shafer-biw90, wendt-dipac11}; here we discuss only the results.

\begin{figure}
\centering \includegraphics*[width=55mm,angle=90]
{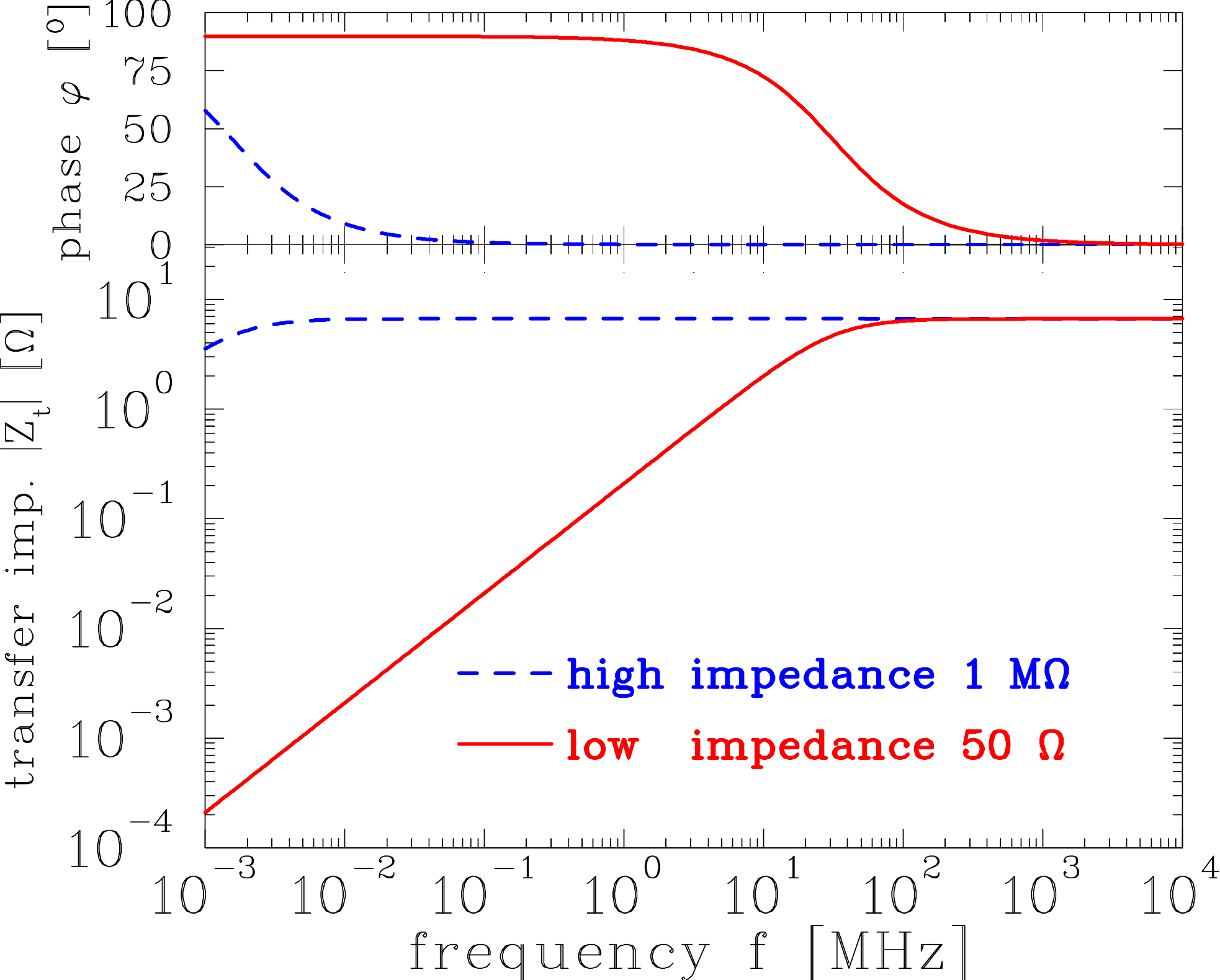}
\caption{Absolute value and phase of the transfer impedance for a $l=10$ cm
long cylindrical pick-up 
with a~capacitance of $C=100 $ pF and an ion velocity of
$\beta = 50 \%$ for high (1 M$\Omega$) and low (50 $\Omega$) input impedance
of the amplifier.}
\label{high-pass-pickup}
\vspace*{-0.3cm}
\end{figure}

\begin{figure}
\centering \includegraphics*[width=33mm,angle=90]
{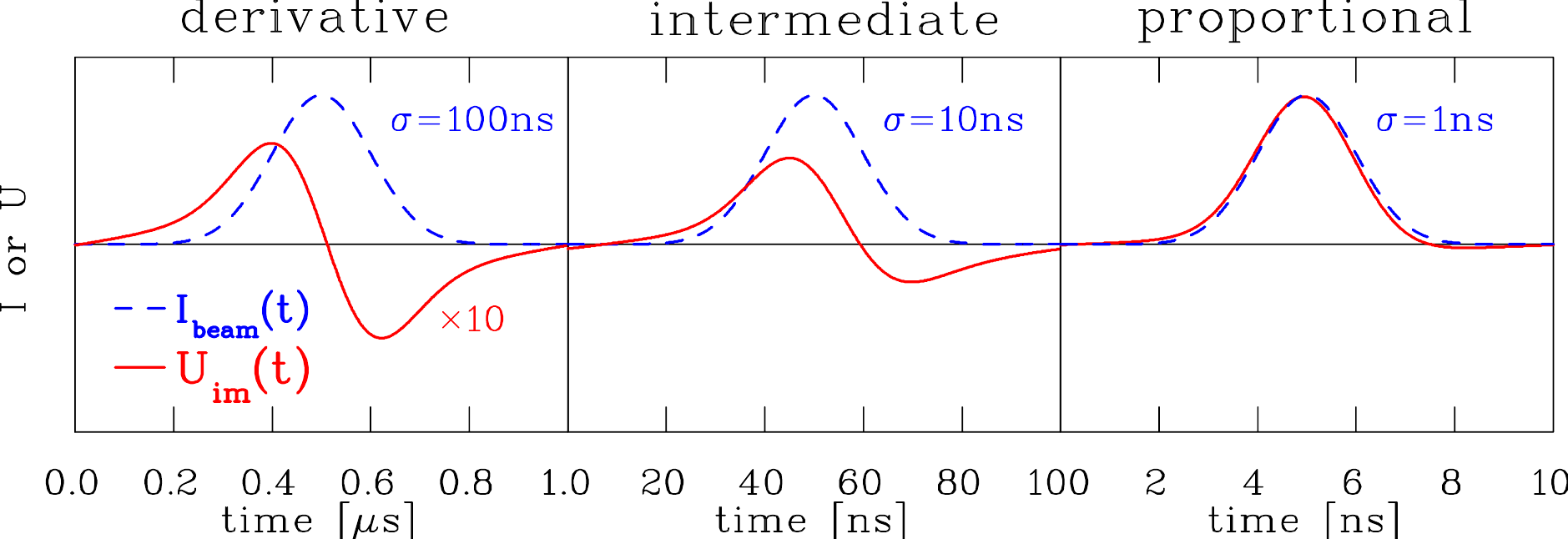}
\vspace*{0.2cm}

\centering \includegraphics*[width=30mm,angle=90]
{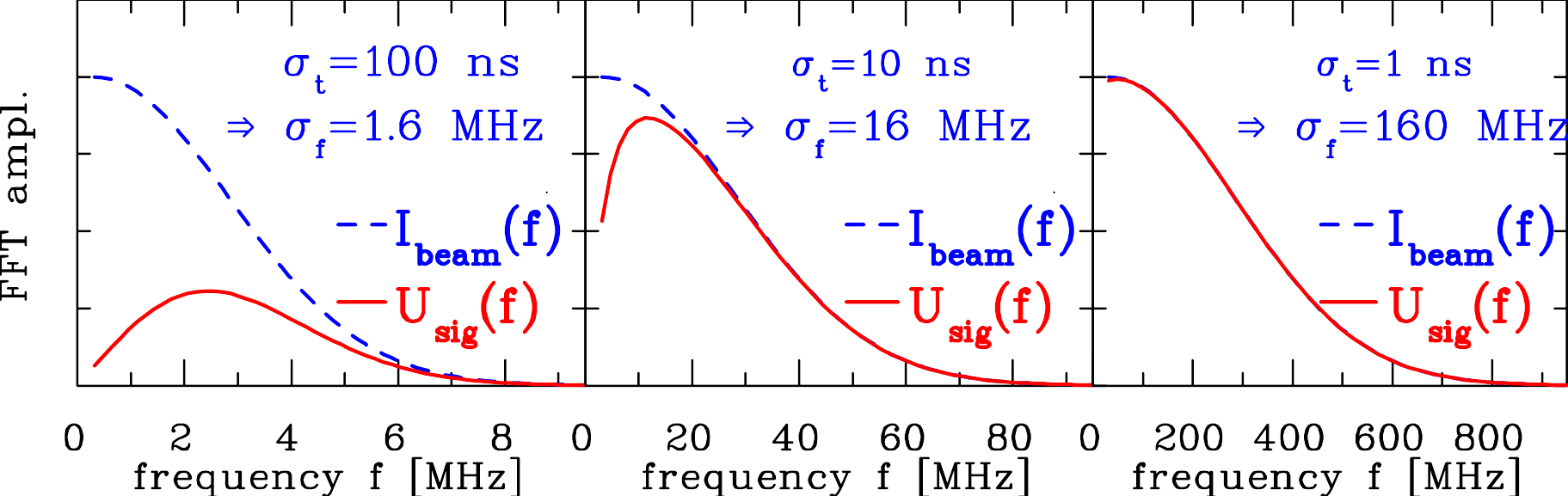}
\caption{Simulation of the image voltage $U_{im} (t)$ for the values of
the pick-up used in Fig.~\ref{high-pass-pickup} terminated with $R=50$
$\Omega$ for three different bunch lengths of Gaussian distribution with
$\sigma$ of 100 ns, 10 ns and 1 ns, respectively. The cut-off frequency
$f_{cut}$ is 32 MHz. Note the different time scales. (The bunch length in
the last case is artificially short for a proton synchrotron.) 
A Gaussian function in time domain of width of $\sigma_{t}$ has a 
Fourier transformation described by a Gaussian function of width 
$\sigma_{f}=1/(2\pi \sigma_{t})$ centred at $f=0$.}
\label{pickup_signal_simu}
\vspace*{-0.3cm}
\end{figure}

The transfer impedance is a complex (in the mathematical sense) function for a electrode of the~given geometry defined in Fig.~\ref{cap-pickup-principle} with a capacitance $C$ for the signal recording with an input impedance of $R$ as 
\begin{equation}
Z_{t} (\omega, \beta) = \frac{1}{\beta c} \cdot 
\frac{1}{C} \cdot \frac{A}{2 \pi a } \cdot \frac{i\omega RC}{1+i\omega RC}~~~.
\label{eq_zt_general}
\end{equation}

This is a description of a first-order 
high-pass filter for the transfer impedance $Z_{t} (\omega, \beta) $ 
with a~cut-off frequency $f_{cut}= \omega_{cut}/2\pi= (2\pi RC)^{-1}$. 
For the case of the so called linear-cut
electrodes used at proton synchrotrons (see Section~\ref{chapter-linearcutBPM}), a typical value of the
capacitance is $C=100 \mbox{~pF}$ with a length of $l=10 \mbox{~cm}$. The
high-pass characteristic is shown in Fig.~\ref{high-pass-pickup} for a 50
$\Omega$ and a high impedance 1 M$\Omega$ amplifier input resistor. In the
figure the absolute value
\begin{equation}
\left| Z_{t} ( \omega) \right| = \frac{1}{\beta c} \cdot \frac{1}{C} \cdot \frac{A}{2\pi
a} \cdot
\frac{\omega/\omega_{cut}}{\sqrt{1+\omega^{2}/\omega_{cut}^{2}}} \mbox{~~~~and
the phase relation~~~~} \varphi (\omega)= \arctan \left(\omega_{cut}/ \omega
\right)
\end{equation} 
is shown. A pick-up has to match the interesting frequency range, which is
given by the accelerating frequency and the bunch length. In a proton
synchrotron typical values of the accelerating frequency are in the range from
1-10 MHz, while for LINACs and electron synchrotrons typically 100 MHz to 3
GHz are used. 

We can distinguish two extreme cases for the transfer
impedance of Eq.~(\ref{eq_zt_general}): 
\begin{itemize} 
\item \textit{High frequency range $ \omega \gg \omega_{cut}$:} Here the
transfer impedance converges to
\begin{equation}
Z_{t} (\omega) \propto 
\frac{i\omega/\omega_{cut}}{1+i\omega/\omega_{cut}} \longrightarrow 1.
\end{equation}
The resulting voltage drop at $R$ is for this case
\begin{equation}
U_{im} (t) = \frac{1}{\beta c C} \cdot \frac{A}{2\pi a} \cdot I_{beam} (t).
\end{equation}
Therefore the pick-up signal is a direct image of the bunch time structure
with no phase shift, i.e. $\varphi=0$. To get a low cut-off frequency
$\omega_{cut}=1/RC$, high impedance input resistors are used to monitor long
bunches, e.g. in a proton synchrotron. The calculated signal shape is
shown in Fig.~\ref{pickup_signal_simu} (right). Note that in the figure a 50
$\Omega$ termination is considered, leading to a large value of the cut-off
frequency $f_{cut}= 32 $ MHz. In the application of a proton synchrotron,
high impedance ($\sim 1$ M$\Omega$) termination yielding a much lower value of
the cut-off frequency $f_{cut} = 10 $ kHz in this case as shown in 
Fig.~\ref{high-pass-pickup} where the condition corresponds to the enlarged 
flat part of the depicted transfer impedance. 
\item \textit{Low frequency range $\omega \ll \omega_{cut}$:} 
The transfer impedance is here
\begin{equation} 
Z_{t} (\omega) \propto
\frac{i\omega/\omega_{cut}}{1+i\omega/\omega_{cut}} \longrightarrow i
\frac{\omega}{\omega _{cut}}.
\end{equation} 
Therefore the voltage across $R$ is in this case
\begin{equation}
U_{im} (t) = \frac{R}{\beta c} \cdot \frac{A}{2 \pi a} \cdot i \omega
I_{beam}= - \frac{R}{\beta c} \cdot \frac{A}{2 \pi a} \cdot \frac{\mbox{d}I_{beam}}{\mbox{d}t} 
\end{equation}
using the frequency domain relation $I_{beam} = I_{0} e^{-i\omega t}$.
We see that the measured voltage is proportional to the derivative of the beam
current. This can also be seen from the phase relation of the high-pass filter
in Fig.~\ref{high-pass-pickup}, where a phase shift of 90$^{o}$ corresponds to
a derivative. The signal is bipolar, as shown in
Fig.~\ref{pickup_signal_simu} (left). 
\end{itemize}

The fact that 
a pick-up has a high-pass characteristic leads to the fact that dc-beam cannot be detected (dc-beams are available, e.g. behind the ion source or for a transfer line during slow extraction).

\subsection{Characteristics for position measurement by BPMs}

\begin{figure}
\centering \includegraphics*[width=65mm,angle=0]
{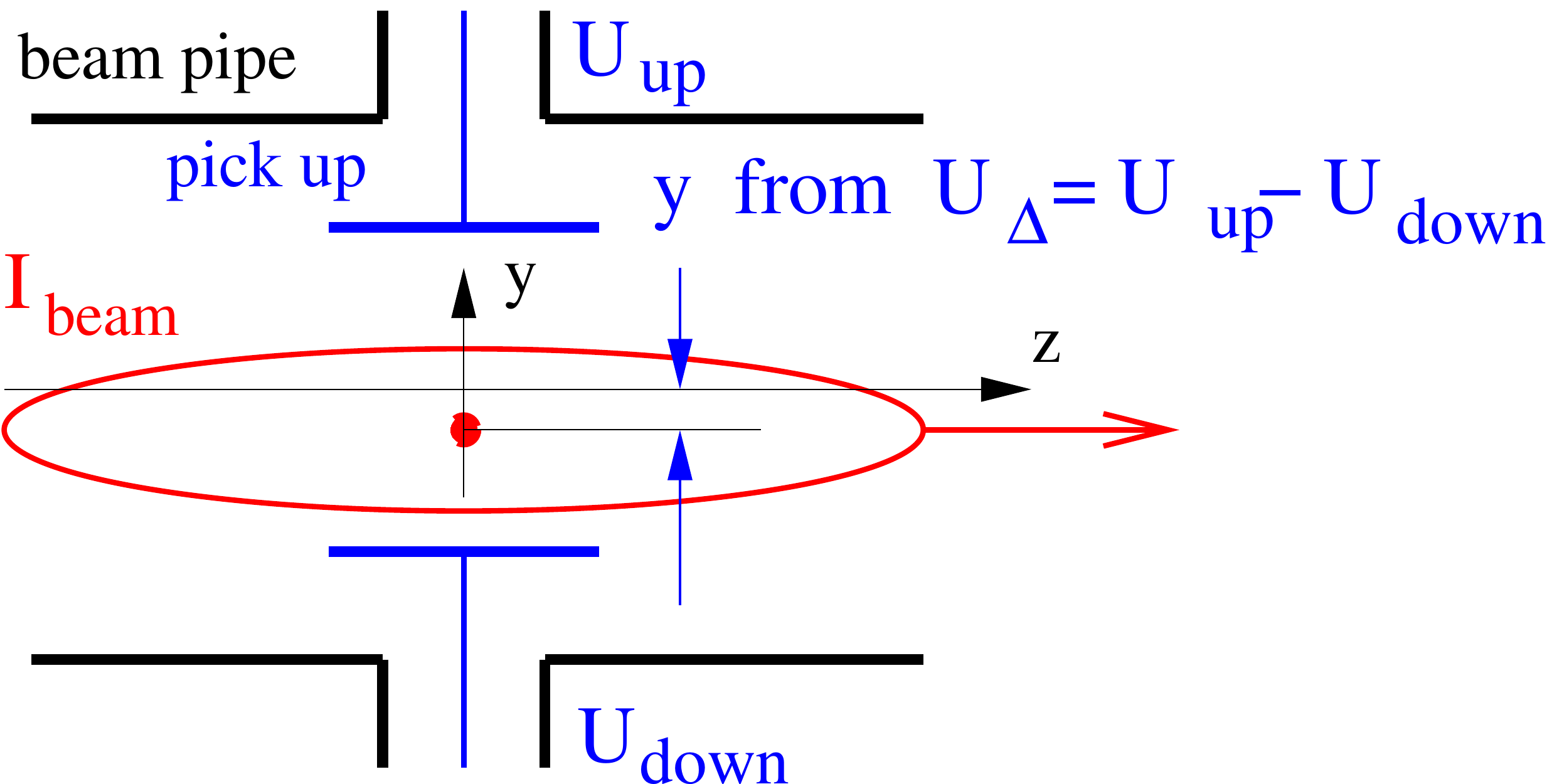}
\vspace*{-0.1cm}
\caption{Schematics of a BPM for the vertical position reading based on the
proximity effect.}
\label{scheme-prox-effect}
\end{figure}

The deviation of the beam centre with respect to the centre of the vacuum
chamber is frequently monitored using four isolated plates or buttons by determining
the voltage difference $\Delta U_{x} = U_{right} - U_{left}$ or $\Delta U_{y} =
U_{up} - U_{down}$ of opposite plates. The closer distance to one of the
plates leads to a higher induced voltage. This is called the 'proximity
effect' and schematically shown in Fig.~\ref{scheme-prox-effect}. Normalizing
to the total signal $\Sigma U_{x} = U_{right} + U_{left}$, the horizontal
displacement $x$ can be obtained via
\begin{equation}
x = \frac{1}{S_{x}} \cdot \frac{U_{right} - U_{left}}{U_{right} + U_{left}} = 
\frac{1}{S_{x}} \cdot 
\frac{\Delta U_{x}}{\Sigma U_{x}}~~~~ (\mbox{horizontal}) 
\label{eq-position-sensi-hor}
\end{equation} 
which is independent of the beam intensity. For the vertical plane the
position $y$ is given by
\begin{equation}
y = \frac{1}{S_{y}} \cdot \frac{U_{up} - U_{down}}{U_{up} + U_{down}} 
= \frac{1}{S_{y}} \cdot \frac{\Delta U_{y}}{\Sigma U_{y}}~~~~ 
(\mbox{vertical}). 
\label{eq-position-sensi-vert}
\end{equation} 
This position measurement is the most frequent application of pick-ups, hence
they are 
called \textbf{B}eam \textbf{P}osition \textbf{M}onitor \textbf{BPM}. The proportional 
constant $S_x $ respectively $ S_y $ between the measured normalized voltage
difference and the beam displacement is called position sensitivity and 
its unit is $S=[\%/\mbox{mm}]$. Sometimes the inverse is used $k = 1/S$ 
given in units of $k=\mbox{[mm]}$ 
and is called BPM position constant or position sensitivity as well. 
It is possible that the position sensitivity 
depends on the beam position itself, corresponding to a non-linear voltage 
response for a large beam displacement and, additionally on the
evaluation frequency, hence it is a function $S=S(x,y,\omega)$.

For typical beam displacements less than 1/10 of the beam pipe aperture, the difference $\Delta U $ is lower by about this factor compared to the sum voltage $\Sigma U$, i.e.,
\begin{equation}
\mbox{typically} ~~~~~\Delta U \ll \frac{\Sigma U }{10} ~~~.
\end{equation}
Sensitive signal processing is required for the difference voltage to
achieve a sufficient signal-to-noise ratio. 

\subsection{Position measurement using button BPMs}
\begin{figure}
\parbox{0.40\linewidth}{
\centering
\includegraphics*[width=40mm,angle=0]
{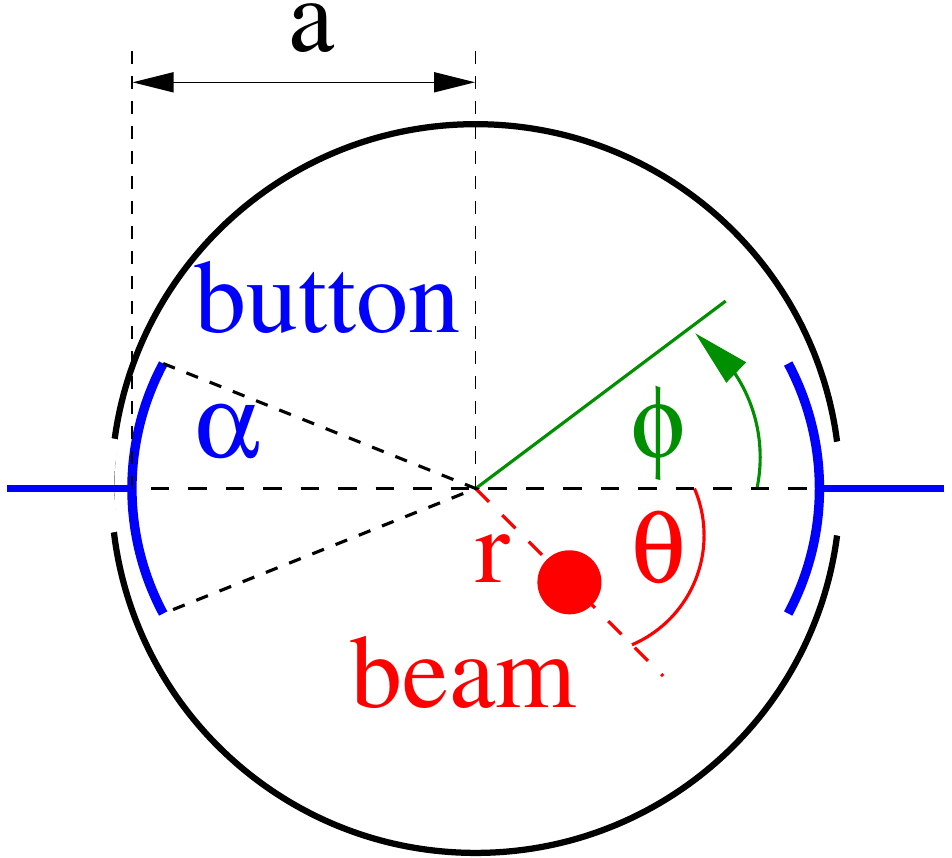} 
} 
\parbox{0.58\linewidth}{
\centering
\includegraphics*[width=43mm,angle=90]
{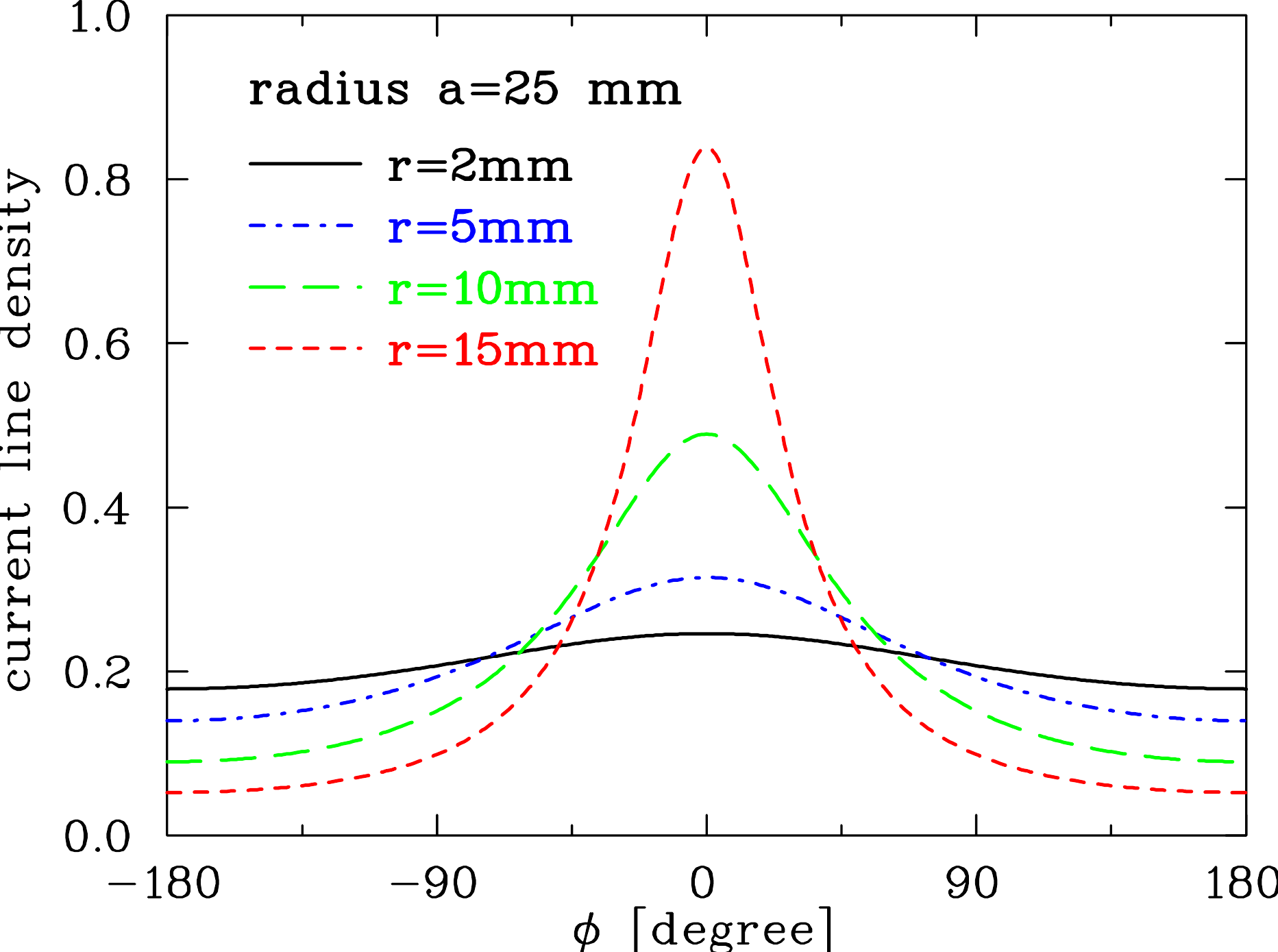} 
}
\caption{Schematics for a button BPM and the image current density generated by a 'pencil' beam at different displacements $r$ for an azimuth $\theta =0$.}
\label{scheme-button-bpm}
\end{figure}

For a round arrangement of buttons, a simple 2-dimensional electro-static model 
can be used to calculate the voltage difference as a function of beam displacement. 
According to Fig.~\ref{scheme-button-bpm} we assume a thin, 'pencil' beam having a transverse 
extension much lower the beam pipe radius of 
current $I_{beam}$ which is located off-centre by the amount $r$ at an angle $\theta $. 
The wall current density $j_{im}$ at the beam pipe of radius $a$ is given as 
a function of the azimuthal angle $\phi $ as 
\begin{equation} 
j_{im} (\phi ) = \frac{I_{beam}}{2\pi a} \cdot 
\left( \frac{a^2 - r^2}{a^2 + r^2 -2ar\cdot \cos (\phi -\theta )} \right) 
\label{eq-button-current-density} 
\end{equation}
and is depicted in Fig.~\ref{scheme-button-bpm}; see \cite{pine-luen06} for a
derivation of this formula.

\begin{figure}
\parbox{0.49\linewidth}{
\centering \includegraphics*[width=45mm,angle=90]
{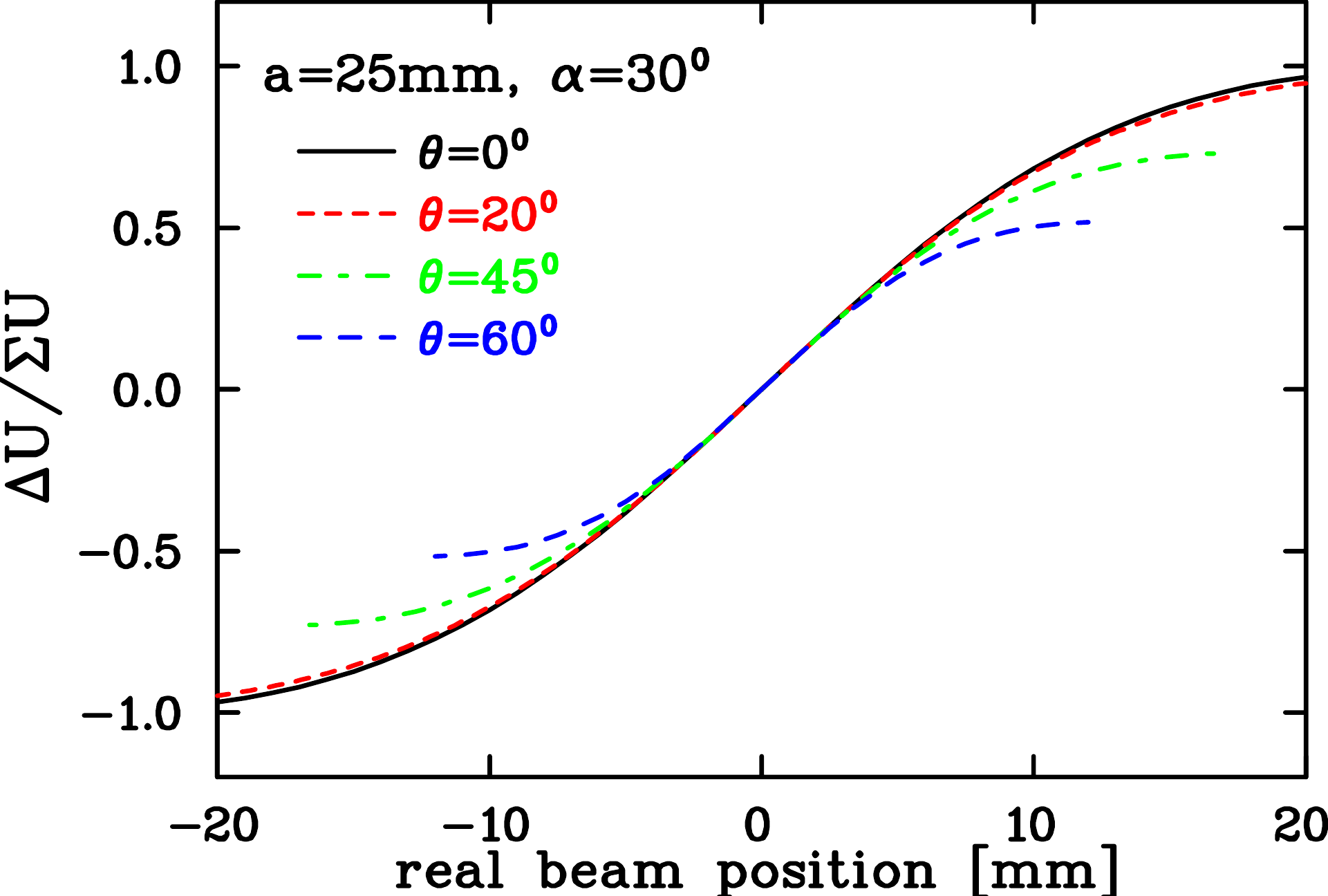}
} 
\parbox{0.49\linewidth}{
\centering \includegraphics*[width=58mm,angle=0]
{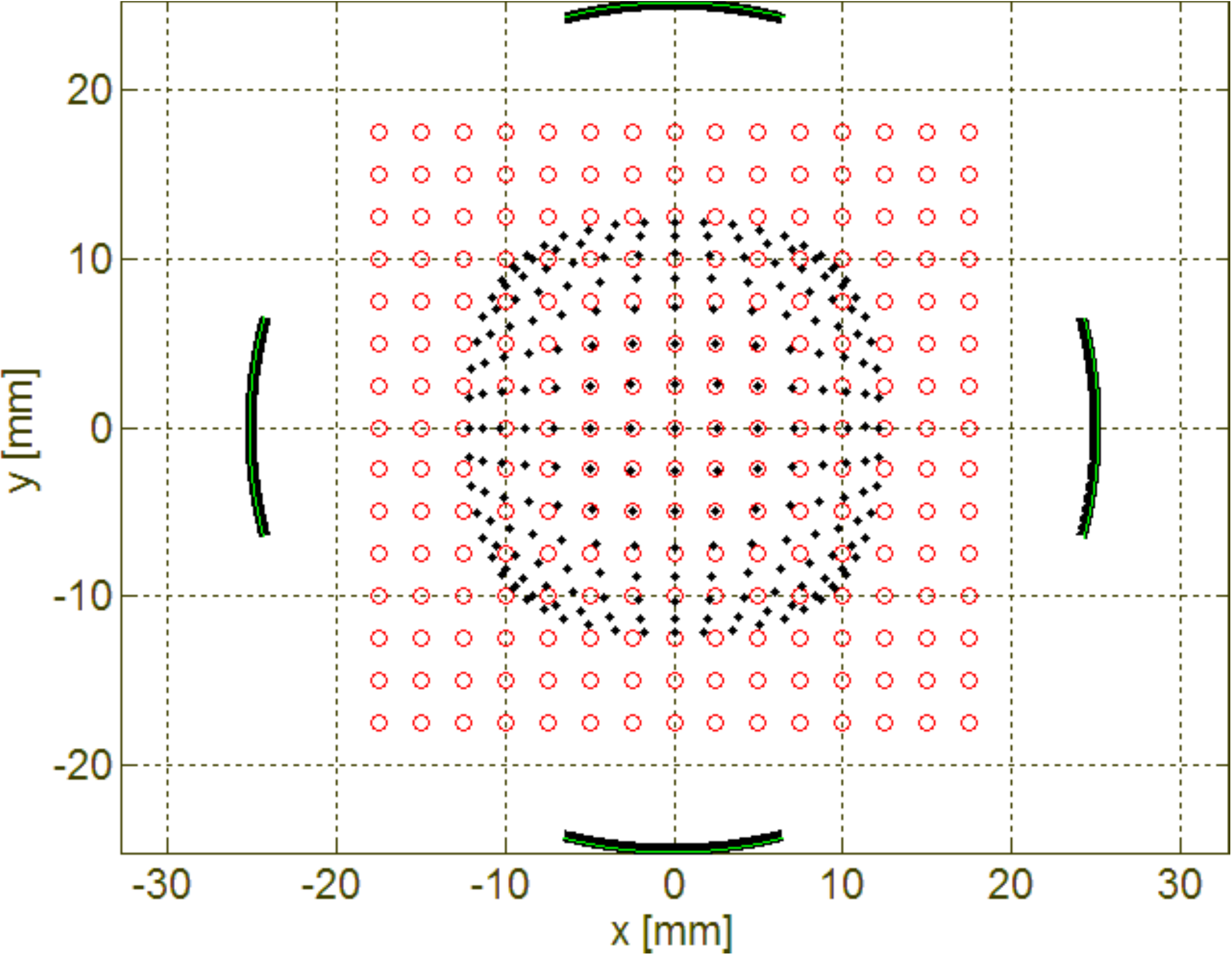}
}
\caption{Left: Horizontal position calculation for different azimuthal
beam orientation $\theta $ for an BPM electrode arrangement of angular coverage $\alpha = 30^o$ 
and a beam pipe radius $a=25$~mm as a function of horizontal beam displacement Right: In the position map the open
circles represent the real beam position and the dots are the~results of the $1/S \cdot \Delta U/ \Sigma U$ algorithm with
$S=7.4$~\%/mm for the central part.} 
\label{buttonBPM_xycoubling_cylinder}
\end{figure}

As discussed above, this 
represents the proximity effect, where the current density depends on 
the~distance with respect to the beam centre. 
The button-type BPM electrodes covering an angle $\alpha$ and the~image current 
$I_{im}$ is recorded as given by:
\begin{equation} 
I_{im} = a \int\limits_{-\alpha/2} ^{+\alpha/2} j_{im} (\phi ) d\phi ~~~.
\label{eq-button-signal}
\end{equation} 
The resulting signal difference for opposite plates as a function of
horizontal beam displacement, corresponding to $\theta =0^o$, shows a
significant non-linear behaviour as displayed in
Fig.~\ref{buttonBPM_xycoubling_cylinder}. The signal voltage is calculated
according to Eq.~(\ref{eq-transfer-impedance}) as $U=R \cdot I_{im}$. It can be seen that for the normalised
difference $\Delta U/ \Sigma U$ the linear range continues to a larger beam
offsets. The~non-linearity increases if
the~beam centre moves outside the horizontal axis, which is depicted in
Fig.~\ref{buttonBPM_xycoubling_cylinder} for different values of the~azimuthal
orientation $\theta $ as a function of horizontal displacement $x= r
\cdot \cos\theta$ according to
Eqs.~(\ref{eq-button-current-density}) and (\ref{eq-button-signal}). The~non-linearity can be influenced by the button size as represented in
this 2-dimensional approach by the~angle $\alpha$. However, in
the central part, the reading is nearly independent of the orientation leading
to a universal position sensitivity $S$. The dependence between the
horizontal and vertical plane is better depicted in the~position map of
Fig.~\ref{buttonBPM_xycoubling_cylinder}, right: Here the real beam positions
with equidistant steps are plotted as well as the~results using $1/S \cdot
\Delta U/ \Sigma U$ calculations with $S$ fitted at the central part.
The~preceding 2-dimensional electro-static model delivers satisfying
result for typical electron beams of relativistic beam velocities
i.e. TEM-like field pattern; only minor corrections are necessary for the 
case of a circular pick-up.

\begin{figure}
\parbox{0.55\linewidth}{
\centering \includegraphics*[width=80mm,angle=0]
{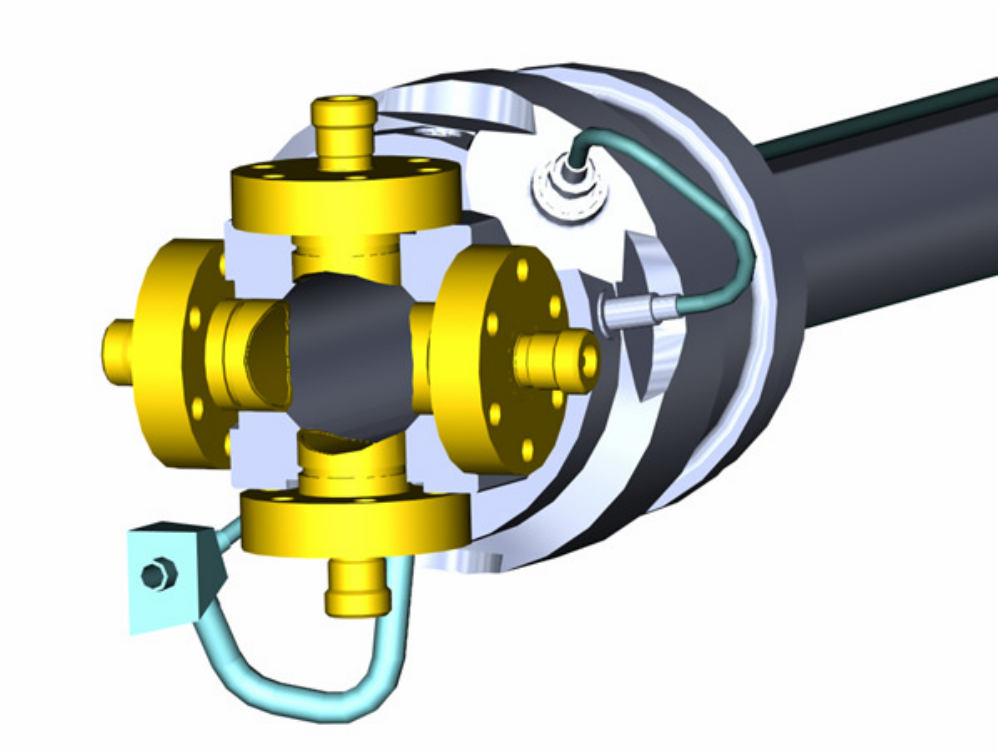}
} 
\parbox{0.44\linewidth}{
\centering \includegraphics*[width=55mm,angle=0]
{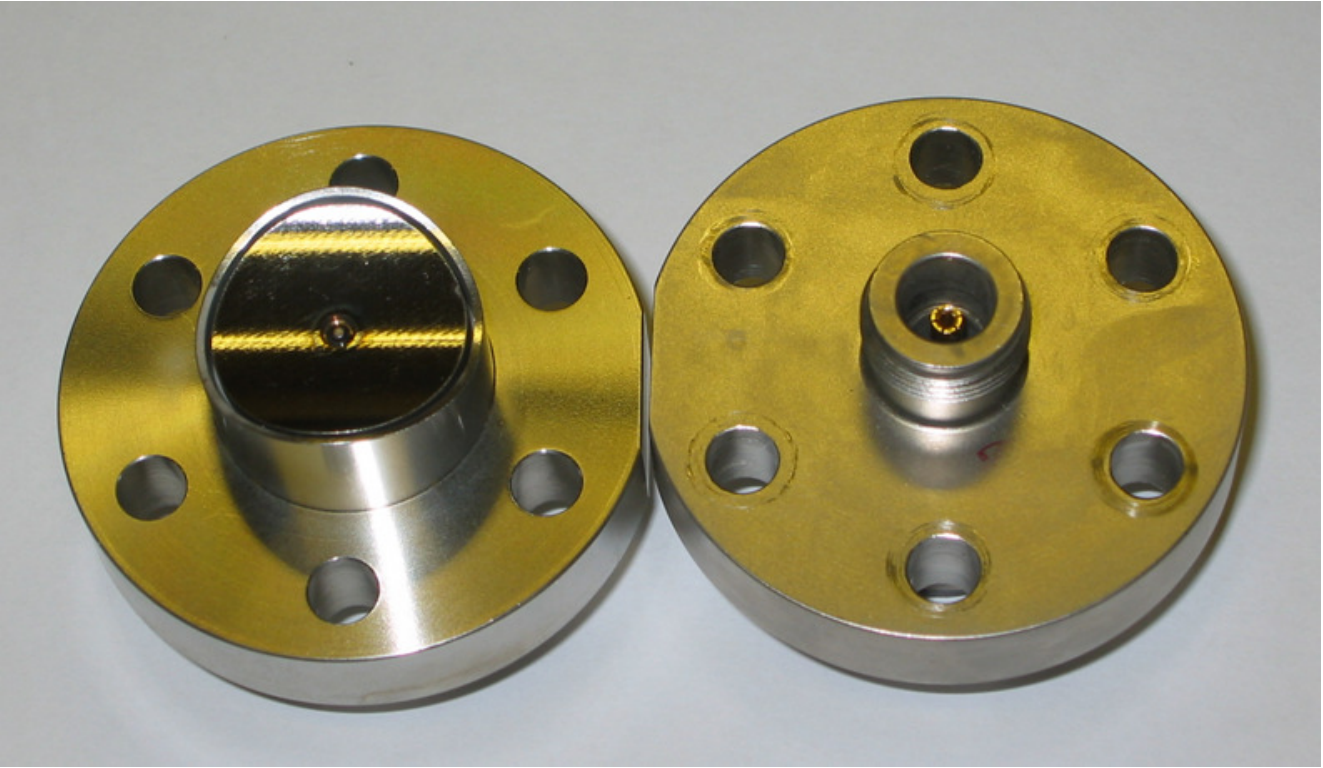}
\centering \vspace*{-0.2cm} \hspace*{-4cm}
\includegraphics*[width=60mm,angle=0]
{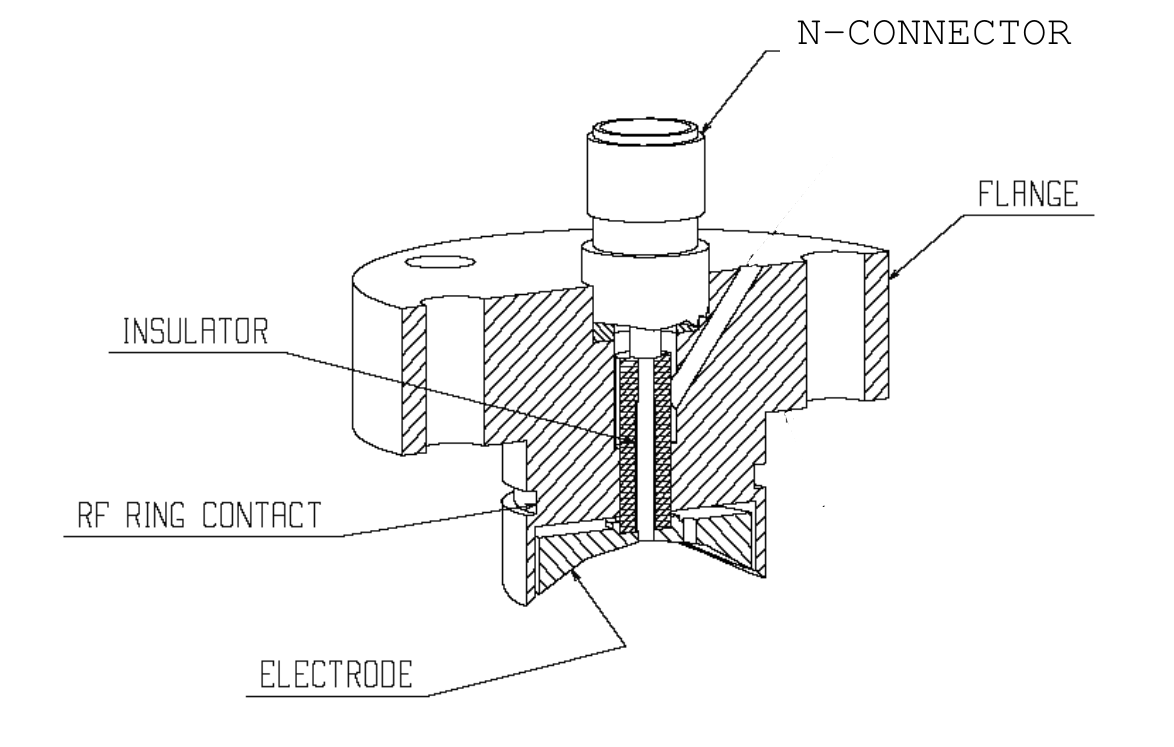}
}
\caption{Left: The installation of the curved \O~24 mm button BPMs at
the LHC beam pipe of \O~50 mm \cite{bocc-abi06}. Right: Photo
of a BPM used at LHC, the air side is equipped with an N-connector as
well as a technical drawing for this type.} 
\label{drawing-LHC-button}
\end{figure}

A BPM arrangement obeying the circular beam pipe
geometry is installed at CERN LHC and depicted in
Fig.~\ref{drawing-LHC-button}. Due to the size and the round
electrodes, this arrangement is commonly called button BPM. 
The reason for such installation is to
ensure a smooth transition from the regular beam pipe to the BPM region
and to prevent for excitation of an electro-magnetic field by the beam,
the so-called wake-field. In most other cases, planar buttons are
used due to their simpler mechanical realisation. 

Button pick-ups are the most popular devices for electron accelerators. They
consist of a circular plate of typically 10 mm diameter mounted flush with the
vacuum chamber. The cross-section of the~chamber is not changed by this insertion, to avoid excitation of wake-fields
by the beam. The button itself should have a short vacuum feed-through with a
smooth transition to the 50 $\Omega$ cable, to avoid excitation of standing
waves and to reach a bandwidth up to 10 GHz.

\begin{figure}
\centering \includegraphics*[width=115mm,angle=0]
{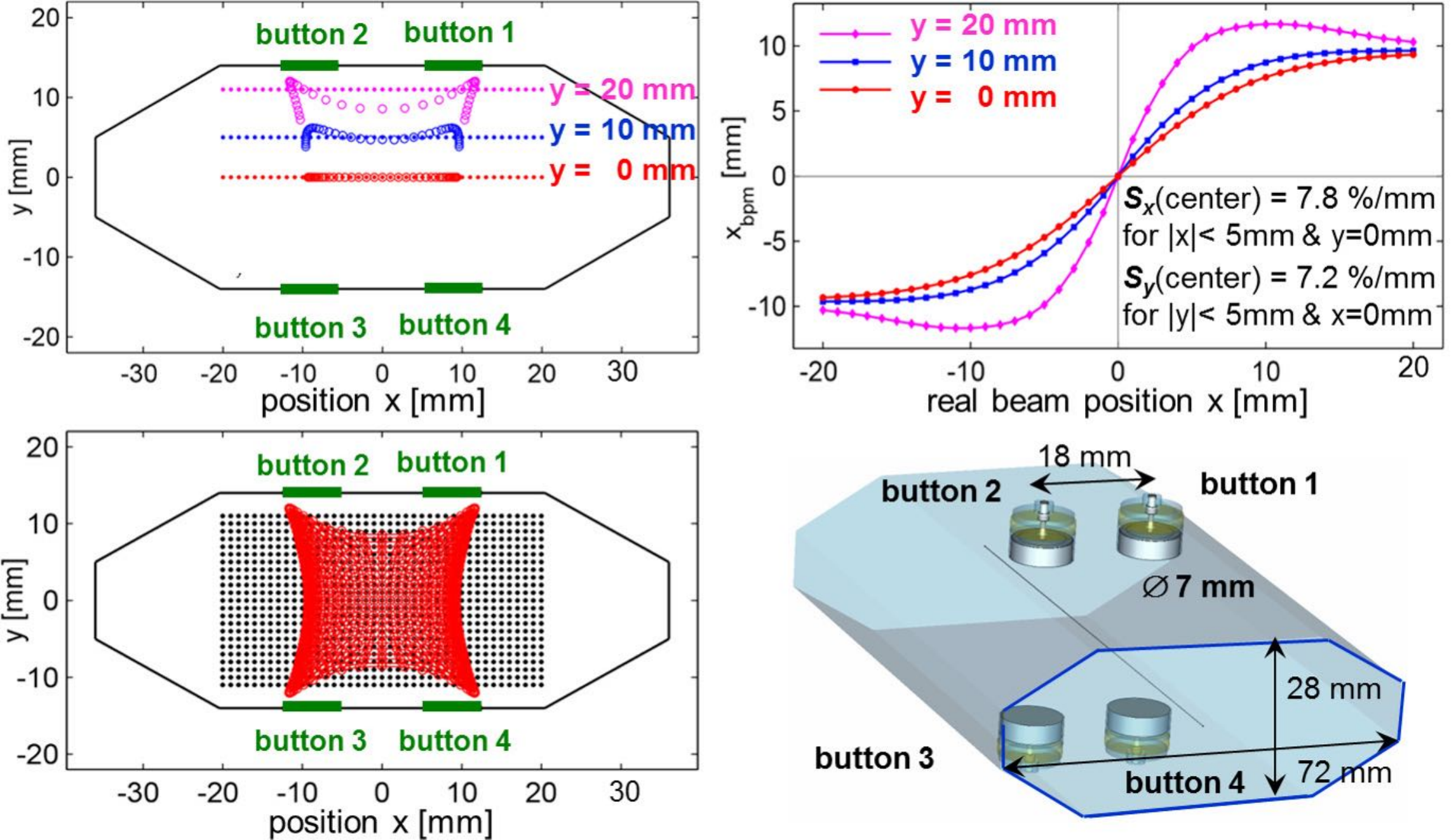}
\caption{Numerical calculation of the position map for the arrangement
at the synchrotron light source ALBA; from \cite{nos-ibic14}. Top left: Results of position
calculation for a horizontal scan from $-20~\mbox{mm} < x <
20~\mbox{mm}$ for three vertical offsets using constant horizontal
$S_x (center)$ and vertical $S_y (center)$ position
sensitivities. Top right: The calculated position for the horizontal
scan for these three vertical offsets; the horizontal position
sensitivity $S_x (center)$ at the central part is determined from
the slope of the curve with zero vertical offset. Bottom left:
Position map for the entire horizontal and vertical range. Bottom
right: The geometry of the \O~7 mm button pick-ups of 18 mm
horizontal distance within the chamber.} 
\label{calc-button-ALBA}
\end{figure}

Fig.~\ref{calc-button-ALBA} shows a
typical setup used at the storage ring of a synchrotron light source within a planar arrangement, where the buttons are not mounted
in the horizontal plane only to avoid synchrotron light hitting the feed-through.
The position is evaluated via
\begin{equation}
\mbox{horizontal:~~} x= \frac{1}{S_{x}} \cdot 
\frac{(U_{1} + U_{4}) - (U_{2} + U_{3})}{U_{1} + U_{2} + U_{3} + U_{4}}
\mbox{~~~~vertical:~~} y= \frac{1}{S_{y}} \cdot 
\frac{(U_{1} + U_{2}) - (U_{3} + U_{4})}{U_{1} + U_{2} + U_{3} + U_{4}}
\end{equation}
using all four electrode voltages for the position determination. An appropriate size and location of the~buttons for a linear sensitivity within a maximal range at the central
part and comparable values of the~related horizontal and vertical
position sensitivity has to be calculated numerically, see
e.g. \cite{nos-ibic14}. 
The~optimised location depends on the size and
distances of the electrodes, as well as the chamber cross-section. The inherent geometry of the vacuum chamber 
leads to a non-linear position sensitivity and a strong coupling
between horizontal and vertical displacements (i.e. non-linear
position sensitivity 
$S(x,y)$ in 
Eq.~(\ref{eq-position-sensi-hor})). In
contrast to the circular arrangement, even at the central part the
horizontal position sensitivity $S_x (x,y)$ depends significantly on
the vertical displacement $y$ as well and a corresponding dependence
for the vertical direction $S_y (x,y)$.

\subsection{Linear-cut BPMs for proton synchrotron}
\label{chapter-linearcutBPM}
\begin{figure}
\parbox{0.48\linewidth}{ \centering 
\includegraphics*[width=55mm,angle=0]
{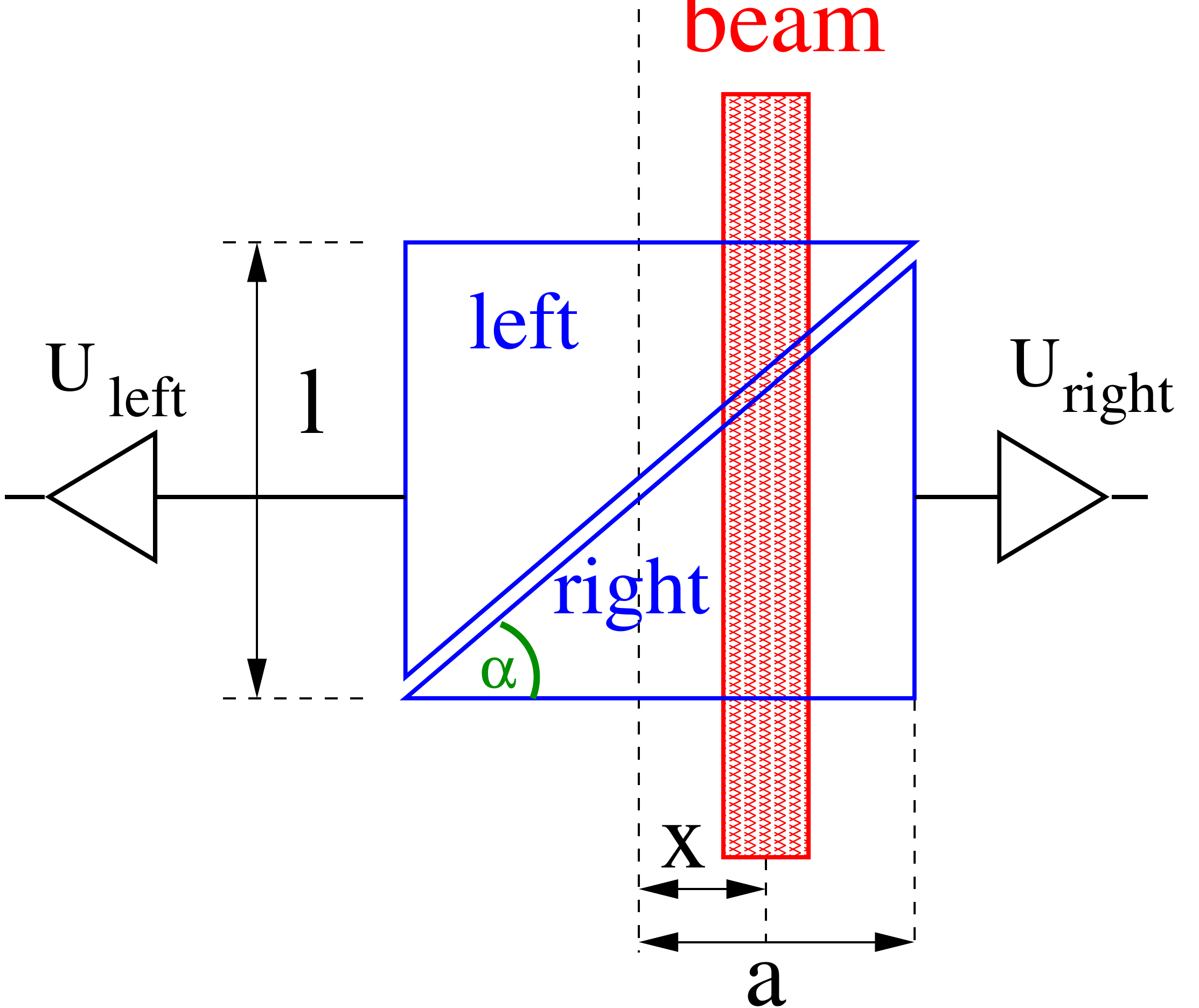}
\vspace*{-0.0cm} } \parbox{6cm}{ \centering
\includegraphics*[width=55mm,angle=0]
{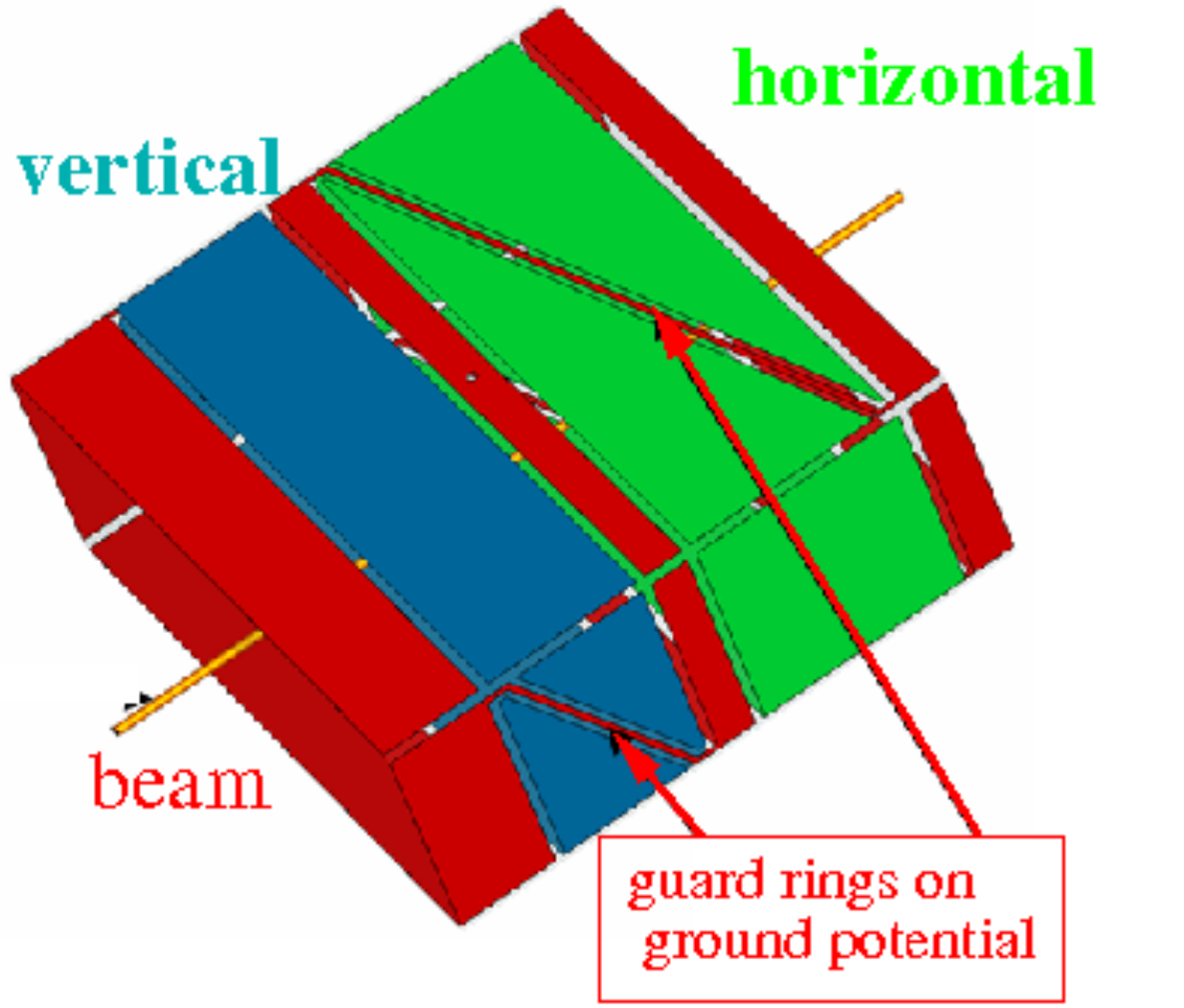}
\vspace*{-0.0cm} }
\caption{Left: Scheme of the position measurement using the
so-called linear-cut and an example of an electrode arrangement for the
horizontal plane. Right: Schematic drawing for a linear-cut BPM for both planes.}
\label{scheme-shoebox}
\end{figure}

Related to the long bunches at proton or ion synchrotrons of typically several meters, 
long electrodes of typically 20 cm are installed to enhance the signal strength. 
A box-like device is used normally, to get a precise linear dependence with 
respect to the beam
displacement, see Fig.~\ref{scheme-shoebox}. For the previously discussed
button pick-up geometries, the signal of the plate closer to the beam's
centre-of-mass is larger than that of the more distant plate; this is called
the proximity effect. By contrast, 
the linear-cut BPMs have another principle: The influenced signal
is proportional to the actual plate length at the beam centre position.
For a given beam displacement $x$ the electrode's image voltage
$U_{im}$ is proportional to the length 
$l_{left}$ and $l_{right}$ of the beam projected on the electrode
surface as shown for the horizontal direction in
Fig.~\ref{scheme-shoebox}, left. For triangle electrodes with
half-aperture $a$ one can write: 
\begin{equation}
l_{right}= \left( a+x \right) \cdot \tan \alpha \mbox{~~~~ and ~~~~}
l_{left}= \left( a-x \right) \cdot \tan \alpha ~~~ \Longrightarrow
~~~ x = a \cdot \frac{l_{right} -l_{left}}{l_{right}+ l_{left}} ~~~. 
\end{equation}
The position reading is linear and can be expressed by the image voltages as
\begin{equation}
x = a \cdot \frac {U_{right}-U_{left}} {U_{right}+U_{left}}
\equiv \frac{1}{S_x} \cdot \frac {\Delta U_x} {\Sigma U_x} ~~~
\Longrightarrow ~~~ S_x=\frac{1}{a} 
\label{diag:cut:eq}
\end{equation}
which shows that the position sensitivity for this ideal case is
simply given by the inverse of the half-aperture. 
Compared to other types of pick-ups, the position sensitivity is
constant for nearly the full range of displacements. Moreover, 
the position reading in the horizontal plane is independent on the~beam position in the vertical plane. Due to the linearity the position
reading is independent on the beam size, which is of importance for
the relatively large beam size compared to the chamber aperture for
low energy proton or ion synchrotrons. This position linearity 
is the main advantage of the linear-cut type as compared to the button
BPM type. As a summary the basic parameters for linear-cut and button BPMs are compared in 
Table~\ref{table-shoebox-button-comp}.

\begin{table}
\centering
\caption{Simplified comparison between linear-cut and button BPM.}
\begin{tabular}{lll} \hline \hline
& \textbf{Linear-cut BPM} & \textbf{Button BPM} \\ \hline 
Precaution & bunches longer than BPM & bunches comparable to BPM \\ 
BPM length (typ.) & 10 to 20~cm per plane & \O ~0.5 to 5~cm \\ 
Shape & rectangular or cut cylinder & orthogonal or planar orientation\\ 
Mechanical realization & complex & simple\\ 
Coupling & often 1~M$\Omega$, sometimes 50~$\Omega$ & 50~$\Omega$ \\ 
Capacitance (typ.) & 30 - 100 pF & 3 - 10 pF \\ 
Cut-off frequency (typ.) & 1 kHz for $R=1$M$\Omega$ & 0.3 to 3 GHz for $R=50$~$\Omega$\\ 
Usable bandwidth (typ.) & 0.1 to 100 MHz & 0.3 to 5 GHz \\ 
Linearity & very linear, no $x$-$y$ coupling & non-linear, $x$-$y$ coupling\\ 
xPosition sensitivity & good & good \\ 
Usage & at proton synchrotron, & proton LINAC, all electron acc.\\
& $f_{acc}<10$~MHz & $f_{acc}>100$~MHz \\ \hline \hline
\end{tabular}
\label{table-shoebox-button-comp}
\end{table}

For the position measurement, other techniques are used in addition: this comprises stripline BPMs working like a directional coupler to separate the signal of counter-propagating beams in a collider. Moreover, 
monitors are used based on the measurement of the magnetic field by so-called inductive pick-ups or the
excitation of cavity modes within a so-called cavity BPM. Further
descriptions and citations can be found in Refs.~\cite{schmick_cas2018,smaluk_book,brandt_cas2007,strehl_book,forck_juas,shafer-biw90, wendt-dipac11, lor-biw98}.

\subsection{Electronic treatment for position determination}
\label{kap-pickup-electronic}

To get the position of the beam, the signals from the electrodes have to be
compared. For this comparison, the signal shape (differentiation or
proportional behaviour) is of minor importance. The electronics used for this
purpose are described only briefly in this contribution; a detailed review is given in
\cite{schmick_cas2018,smaluk_book,brandt_cas2007,vis-dipac00,wendt-ibic14}. For the position resolution, the
signal-to-noise ratio is 
amplifier noise, as well as the electronic noise of the following devices
contribute. Therefore a minimum bunch current is needed for a reliable
position measurement. Two different principles are commonly used based on analogue electronics, the so
called broadband and narrow-band processing:

\begin{figure}
\centering \includegraphics*[width=100mm,angle=0]
{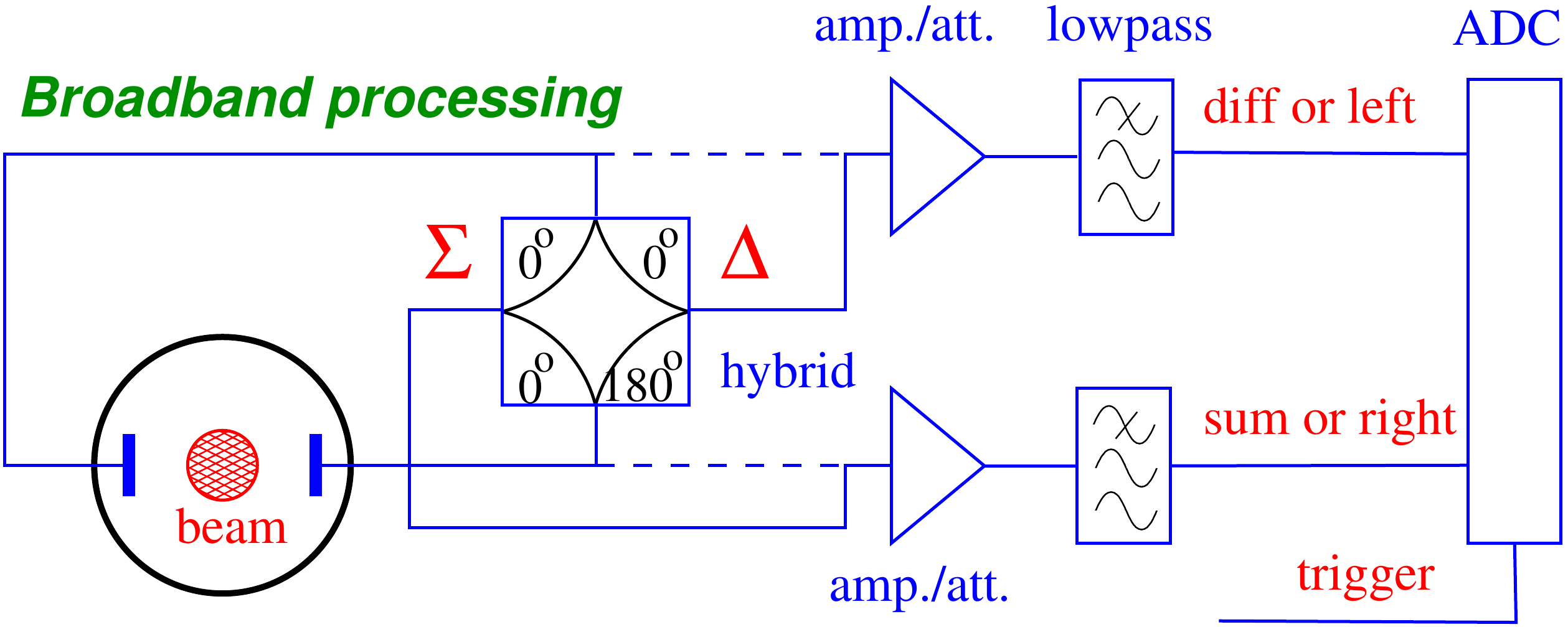}
\caption{Scheme of a broadband signal processing. The plate signals
are either fed directly to the amplifier (dashed line) or via a
hybrid the sum and difference is generated in an analogue manner
(solid line).} 
\label{pick-elec-broadband}
\vspace*{-0.1cm}
\end{figure}

\begin{figure}
\centering \includegraphics*[width=100mm,angle=0]
{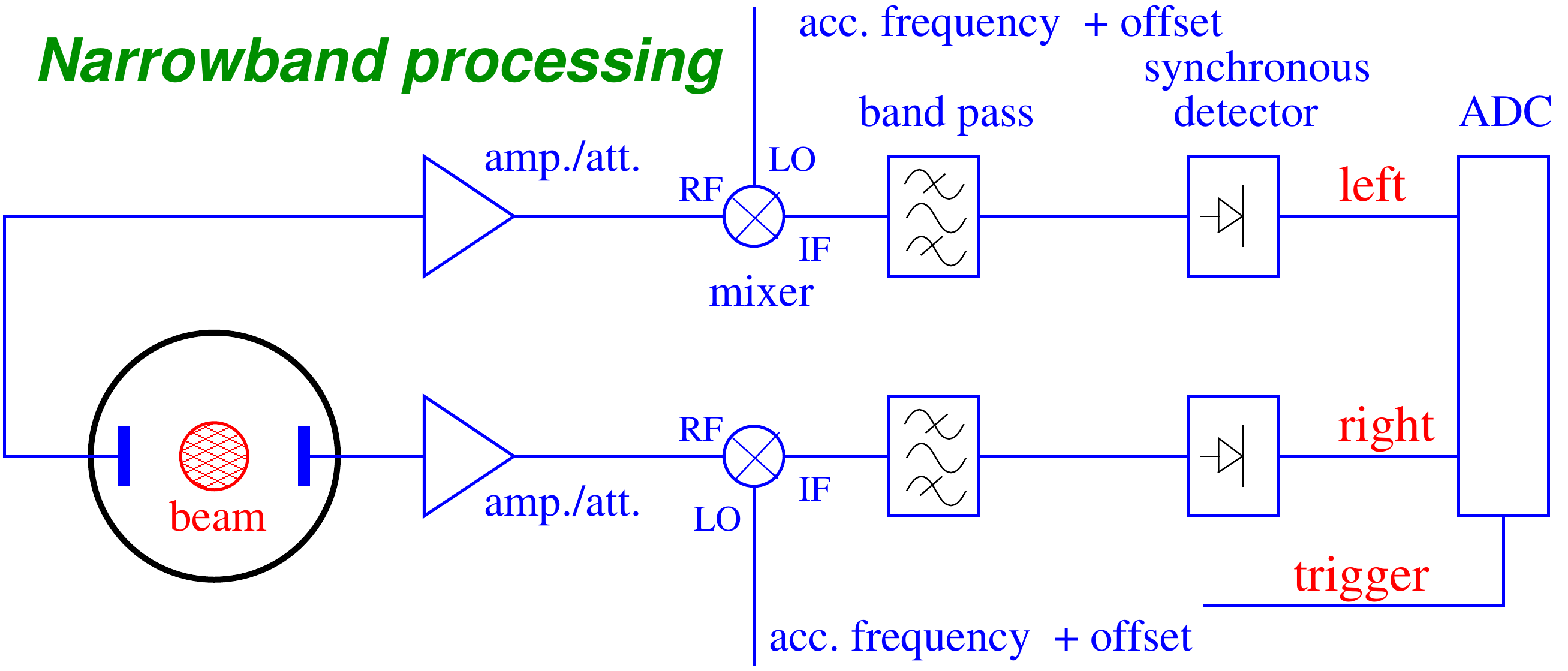}
\caption{Schematic signal analysis for a narrow-band treatment.}
\label{pick-elec-narrowband} 
\vspace*{-0.3cm}
\end{figure}

\textit{Broad-band processing:} In the broadband case, as shown in
Fig.~\ref{pick-elec-broadband}, the signals 
from the individual plates are amplified (or even attenuated) to adapt the
signal strength to the ADC input level. The sum and difference signal is then
calculated from the digitised values. For noise reduction and alias-product
suppression, a low-pass filter is used matched to the sample rate of the
ADC. In many applications, the~sum and difference voltages are
analogously generated by a 180$^o$ hybrid or a differential
transformer. Because they are purely passive devices, they can be mounted quite
close to the BPM plates even in case of high radiation. The
resulting sum and difference signals are then stored in the ADC. 
The difference signal,
which is usually lower by at least a factor of 10, can be amplified by a
higher amount than the~sum signal to exploit the full ADC range. 
The analogue electronics is required to match the signal shape to the
properties of the ADC. 
With the help of high-speed digital signal processing, the bunch signal is then
reduced to one value per bunch and the beam position is calculated from the
sum and difference value. 
The precision
of this method is lower as compared to the narrow-band processing described
below. For electron machines, with small beam size, a resolution of 100 $\mu$m
can be achieved by this broadband processing. The advantage is the
possibility to perform a bunch-by-bunch analysis (i.e. measuring the position of
always the same bunch rotating in a synchrotron) by using an appropriate
trigger setting or precise digital signal processing. The broadband
scheme is installed at transfer lines between synchrotrons where only
one or few bunches are transferred.

\begin{figure}
\centering \includegraphics*[width=53mm,angle=90]
{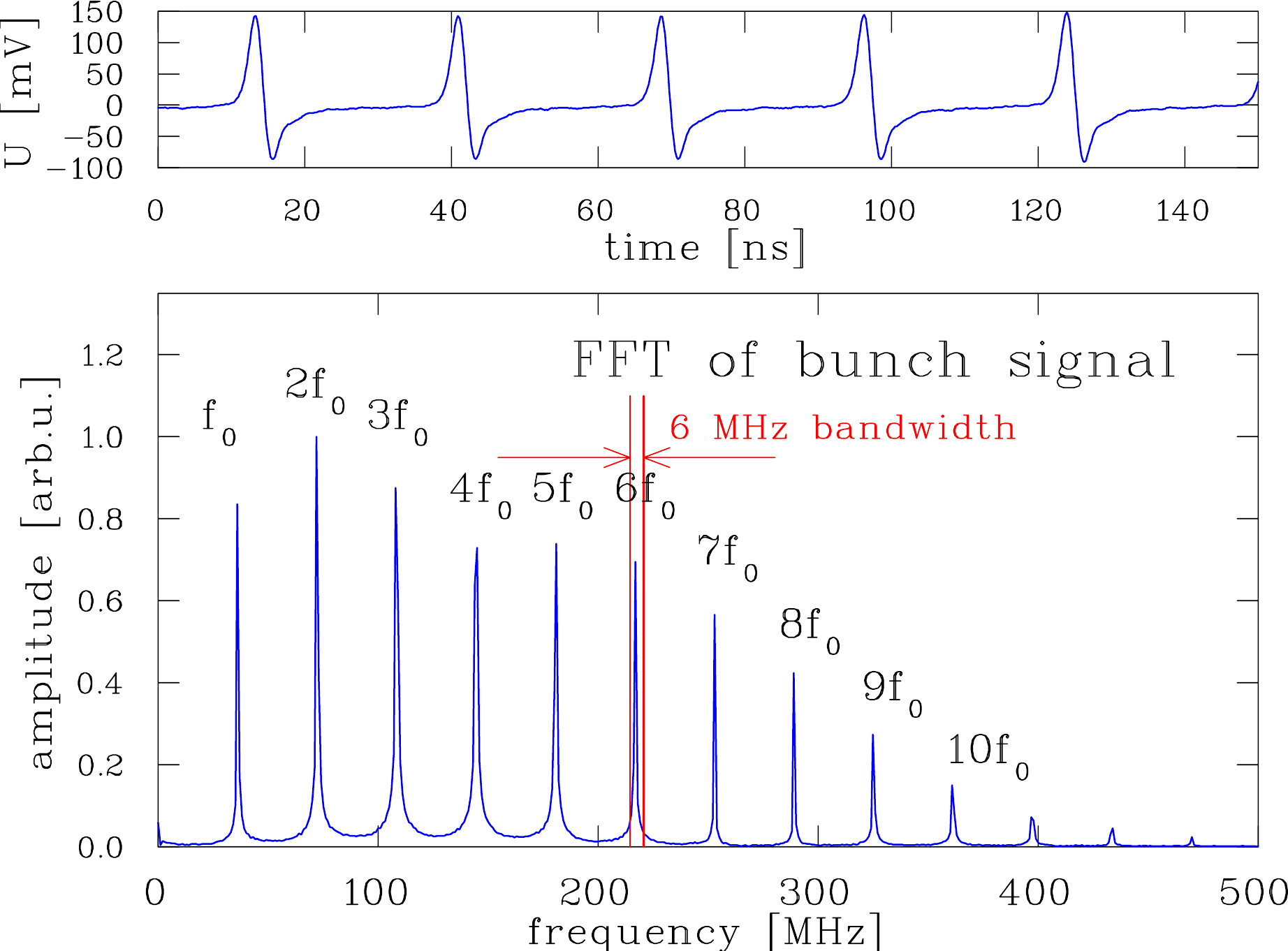}
\caption{Bunch signal (top) and its Fourier transformation (bottom) at the 
GSI-LINAC of a 1.4 MeV/u beam. In the
Fourier spectrum each line corresponds to the indicated harmonics of the 36
MHz accelerating frequency. The~position evaluation is done at $6
f_{0}= 216.8$ MHz by a 6 MHz narrow-band processing as indicated by the red
lines.}
\label{unilac_phasensonden-fourier}
\vspace*{-0.3cm}
\end{figure}

\textit{Narrow-band processing:} 
The narrow-band processing is used to get a higher precision of the~position
reading, attaining 1 $\mu$m in electron machines by this electronics
scheme. The better
signal-to-noise ratio is achieved by reducing the bandwidth of the signal
processing by several orders of magnitude. As an illustrative example, the
Fourier-Transformation of the signal from the GSI-LINAC pick-up is shown in
Fig.~\ref{unilac_phasensonden-fourier}. The spectral power of the signal is
mainly available at the bunch repetition harmonics $nf_{0}$. The~position
evaluation uses only this considerable signal power within the band of frequency span
$\Delta f$. By this an effective noise
reduction is achieved by a limitation of the signal processing
bandwidth, because the unavoidable thermal noise voltage $U_{eff}$ at a resistor
$R$ scales with the square root of the~bandwidth $\Delta f$ as
$U_{eff}=\sqrt{4 k_B T R \Delta f}$ with $k_B$ being the Boltzmann
constant and $T$ the temperature. Technically, the pick-up signal is mixed with the
accelerating frequency using standard rf-components. The resulting quasi-dc signal is then digitised by an ADC, and the position is calculated via software. In general, this
technique is called heterodyne mixing, and the same principle is used in a
spectrum analyser. The mixing corresponds to an average over many bunches
leading to much higher precision in the~position reading but does not allow a turn-by-turn observation. Besides for a synchrotron, such 
narrow-band processing is applied at LINACs or behind cyclotrons with a pulse length above 10 $\mu$s. In case only one or few bunches are transferred, e.g. between two synchrotrons, the narrow-band processing cannot be applied. 

Motivated by the intensive usage in telecommunication the sampling rate and a voltage resolution of ADCs have significantly improved in the last decades; an overview on ADC technologies can be found in Refs~\cite{schmick_cas2018,brandt_cas2007}. Modern technologies tend to
digitise the signal on an early stage of the signal path and replace
the analogue conditioning by digital signal processing, e.g. performing
the narrow-band treatment in the digital domain. The related methods compared
to the analogue treatment are reviewed in \cite{wendt-ibic14}.

\subsection{Trajectory measurement}
\begin{figure}[th]
\centering \includegraphics*[width=130mm,angle=0]
{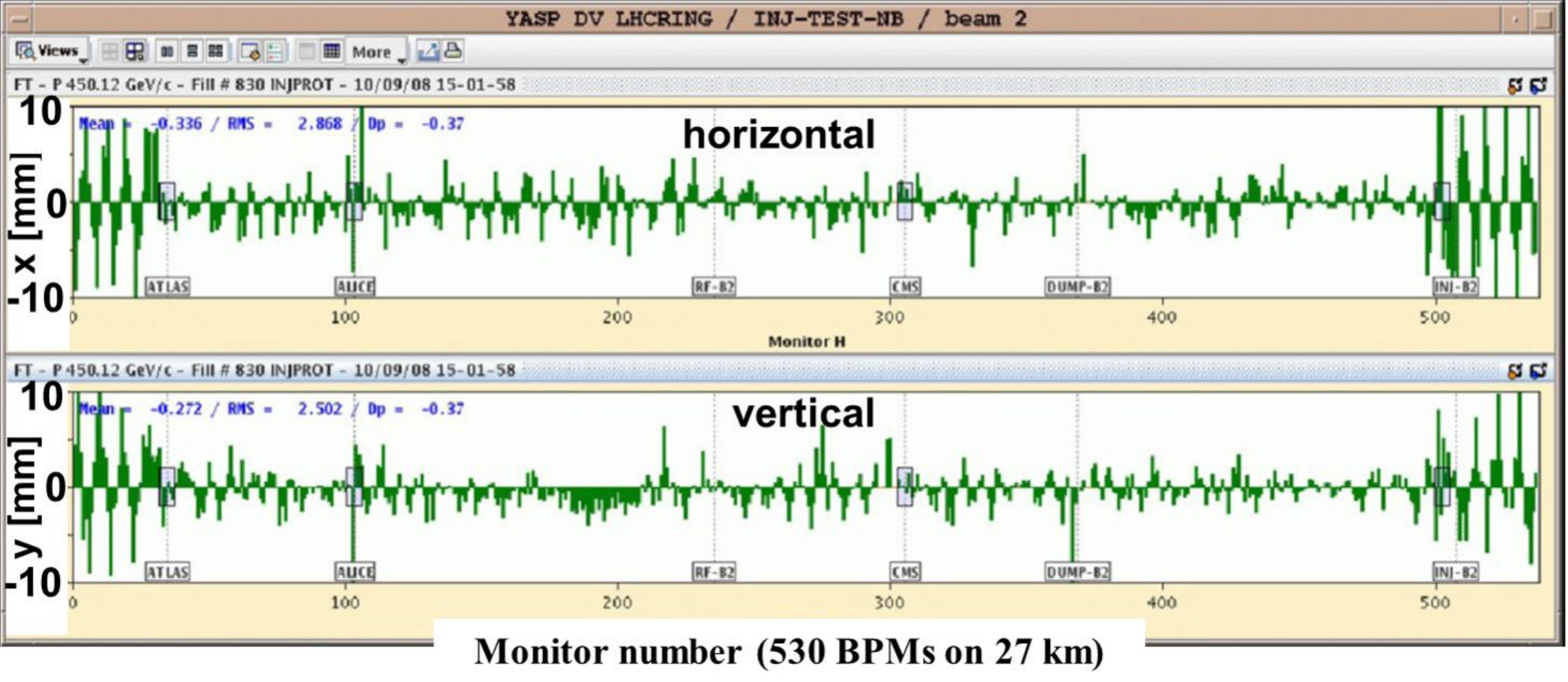}
\caption{Measurement of the trajectory of a single bunch injected in the LHC with a position display at the 530 BPMs around the synchrotron \cite{jon-biw10}.}
\label{LHC-trajectories-measurement}
\vspace*{-0.3cm}
\end{figure}

With the term trajectory the position of a single bunch during its
transport through a transport line or within a synchrotron is meant, and it is the basic
information of the initial alignment of the accelerator
setting. Because the single bunch position is monitored, a broadband
processing for the BPM electronics is used. As an example of a measurement
Fig.~\ref{LHC-trajectories-measurement} depicts the measurement of the
trajectory of a single bunch injected into the LHC on the first days
of its commissioning \cite{jon-biw10}. The position reading at each of
the 530 BPM on the 27 km long synchrotron is displayed showing some
oscillations caused by mismatched injection parameters, which could be
aligned later on with the help of this diagnostics. 
To this respect, the LHC is treated as a transfer line. 

\subsection{Closed orbit measurement at a synchrotron}
\begin{figure}[th]
\centering \includegraphics*[width=100mm,angle=0]
{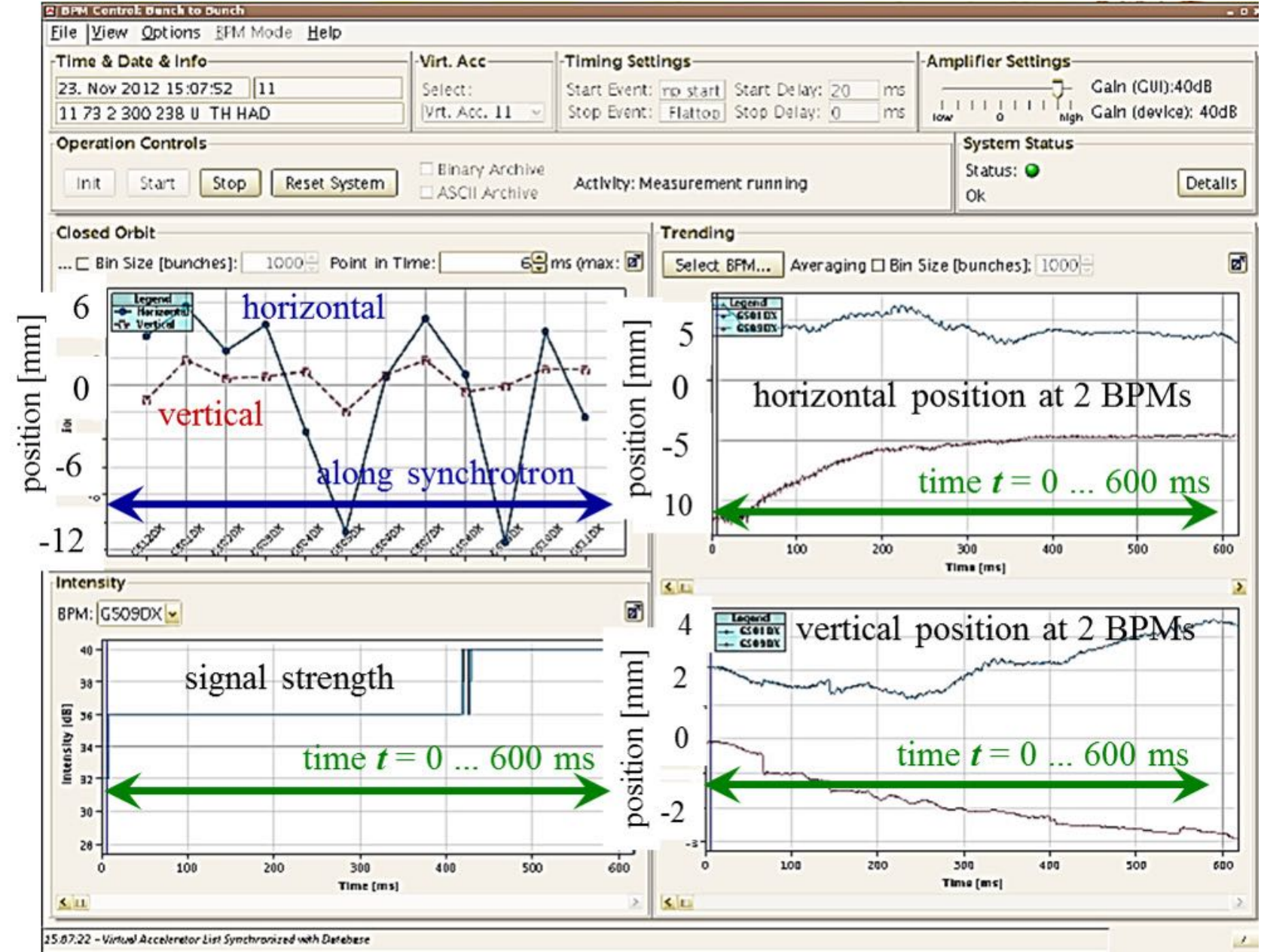}
\caption{Example of a position measurement done at the GSI synchrotron for a
partly misaligned beam during a~0.6 s long acceleration. The upper
left plot shows the 
position around the ring at 12 locations at the begin of the~ramp; 
the upper right plot shows the horizontal position during the
acceleration ramp for 
two BPM locations, the~lower right plot the vertical position
for these BPMs; the lower left plot is a signal power $\propto 
U_{\Sigma}^{2}$ during the~ramp for a~single BPM. The panel at the top is used 
for the control of pre-amplifiers and timing.}
\label{sis-position-hkr}
\vspace*{-0.3cm}
\end{figure}

The closed orbit within a synchrotron is measured via the position averaged over many turns; the required accuracy is in the order of 10 $\mu$m for proton accelerators and about 1$\mu$m for synchrotron light sources with typical averaging duration in the the order of 10 ms. 
In a synchrotron, several BPMs for the~determination of the closed orbit
are installed to enable sufficient correction possibilities. An example of the~use of a position
measurement is displayed in Fig.~\ref{sis-position-hkr} during the
acceleration in the GSI synchrotron.

The position reading of the BPMs around the ring can be used as the input
of a feedback loop to do an active beam correction to compensate systematic 
and temporal misalignments, such a feedback system is installed in
most synchrotrons, see e.g. \cite{rehm-ibic19}. To enable proper automatic corrections, four BPMs per tune 
value, separated approximately by about $\mu_{\beta} \simeq 90^{o}$ betatron phase advance, is a typical choice.

\subsection{Tune measurement}
\label{kap-btf}
One essential parameter of a synchrotron is its tune as given by the amount of betatron 
oscillations per turn. The principle of
a tune measurement is related to excite a coherent transverse 
betatron oscillation and the position determination by BPMs on a turn-by-turn basis. 

\begin{figure}
\centering \includegraphics*[width=30mm,angle=90]
{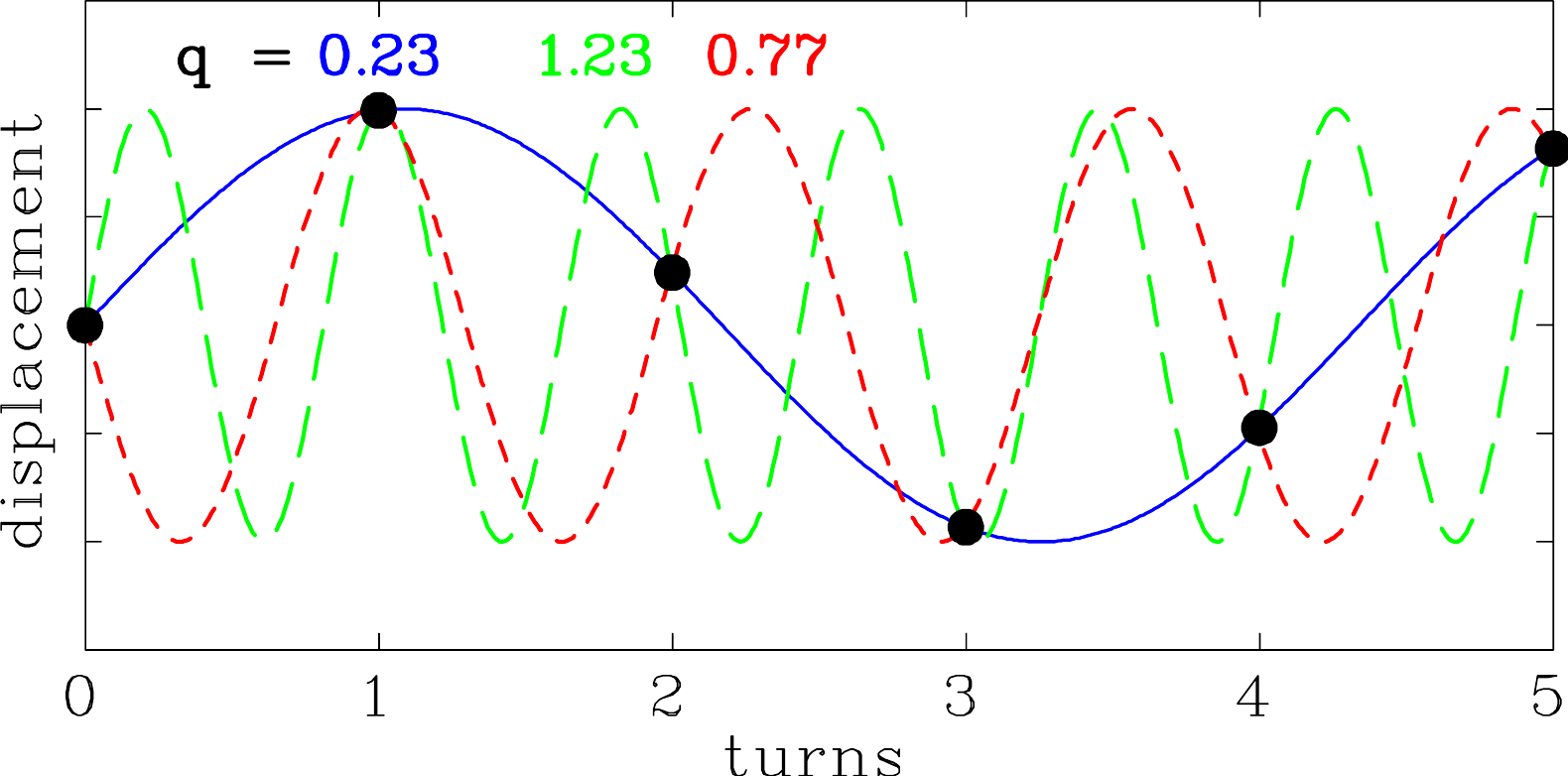}
\caption{Schematic of the tune measurement for six subsequent turns determined
with one pick-up, together with the three lowest frequency fits.}
\label{tune-multi-calc}
\vspace*{-0.1cm}
\end{figure}

The tune value can be split in two parts as $Q= Q_{n} + q$ with $Q_{n}$ is
the integer part and $q$ fractional part; the measurement determines only the fractional part $q$. The integer number of oscillations $Q_{n}$
cannot be seen, but this is normally of no interest, as it is known from
calculations. This behaviour is schematically shown in
Fig.~\ref{tune-multi-calc} for the case of a turn-by-turn reading of one
BPM:
A sine function can be fitted through the position data,
but one cannot distinguish between the lowest frequency (here $q=0.23$) and
higher values, leading to the fact that the fractional part is only uniquely
defined for $0<q<0.5$. To distinguish between a value below or above 0.5, the
focusing by the quadruples (i.e. the tune) can be slightly changed and the
direction of the shift of $q$ can be observed.

\subsubsection{The kick method, time-domain method} 
\label{kap-btf-kick-method}
\begin{figure}
\parbox{0.34\linewidth}{ \centering \includegraphics*[width=55mm,angle=0]
{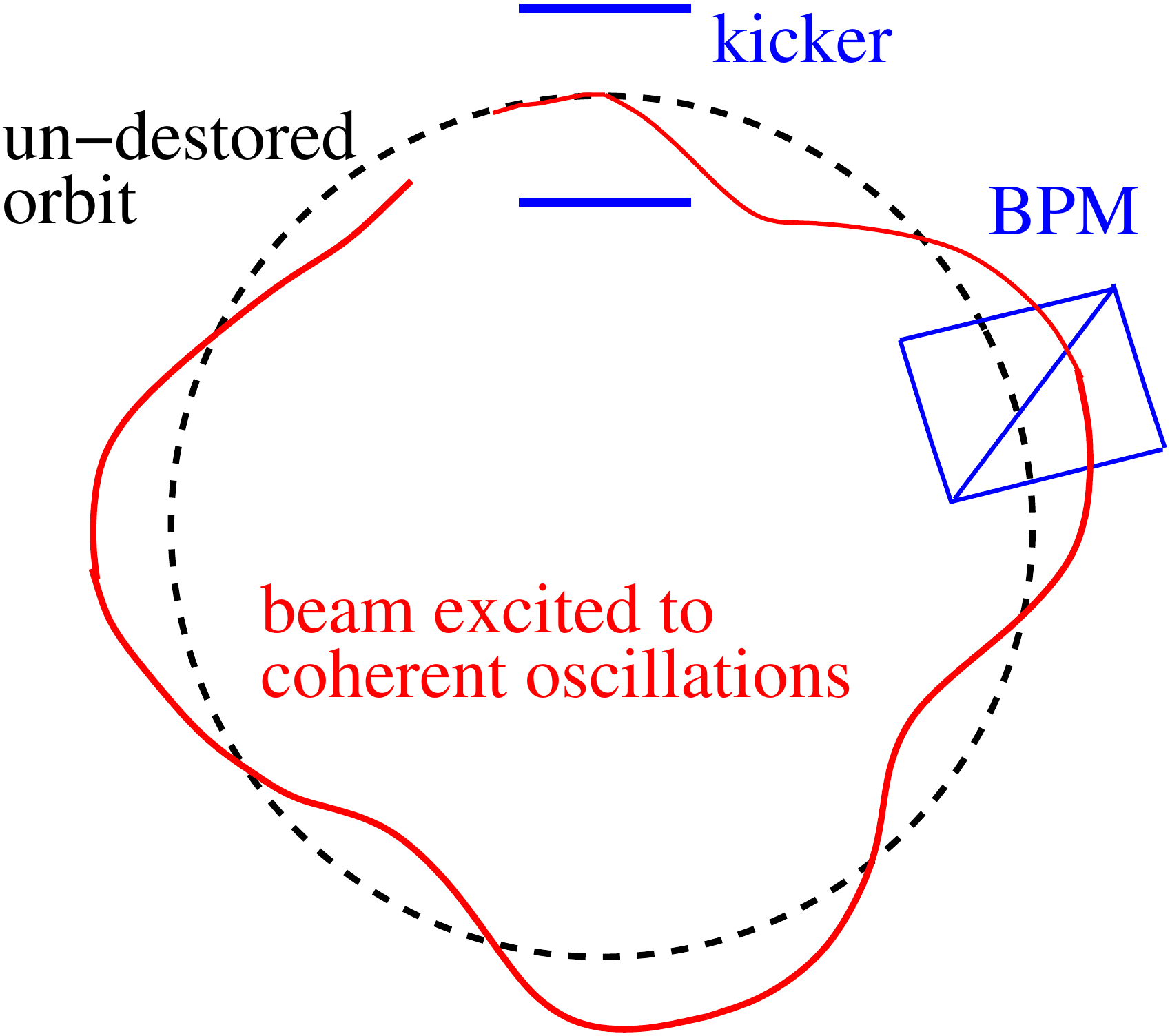} }
\parbox{0.65\linewidth}{ \centering
\includegraphics*[width=45mm,angle=0]
{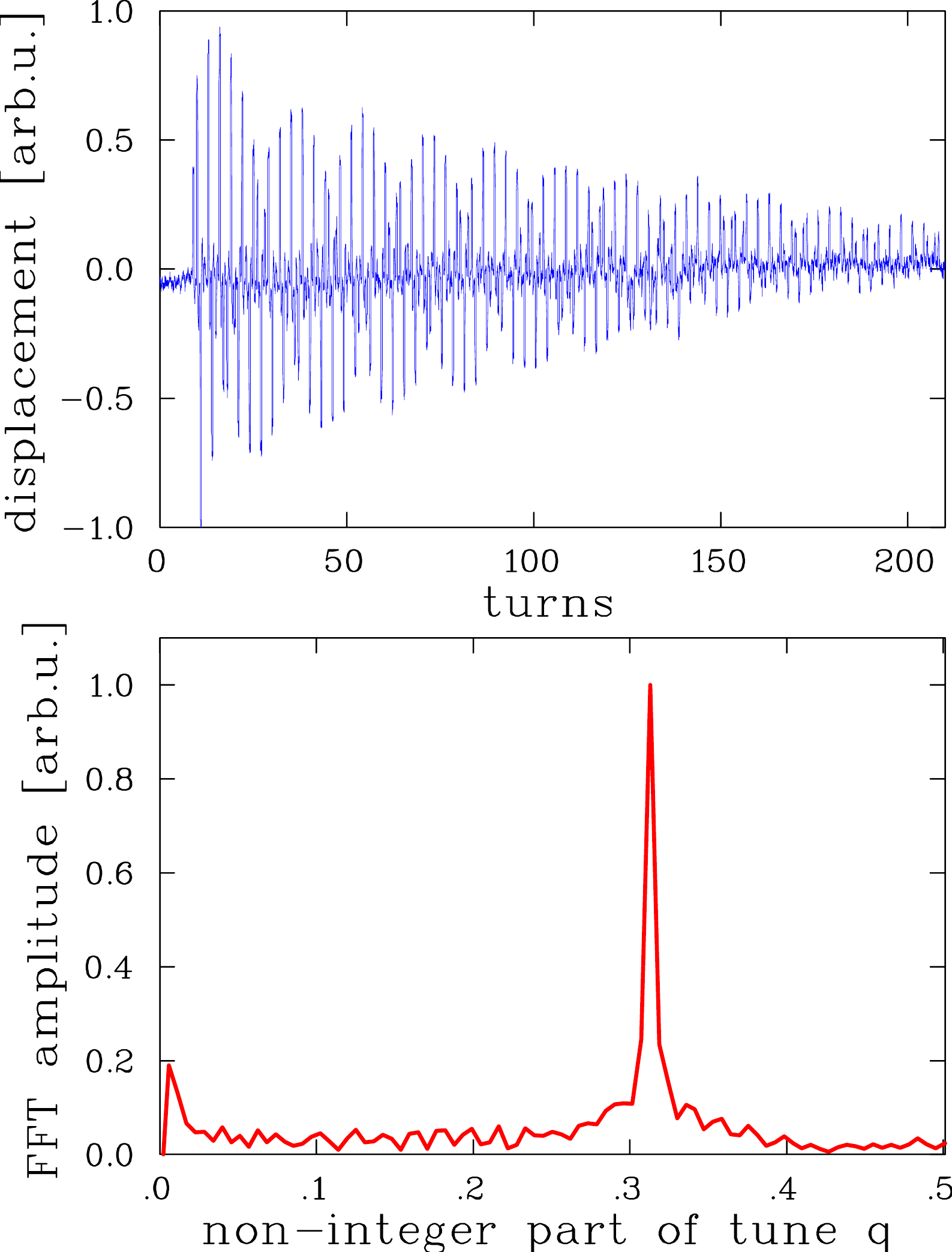}
\vspace*{-0.1cm} }
\caption{Left: Scheme of the beam excitation and detection of coherent
bunch oscillations by a BPM. Right: Beam oscillations after a kick
excitation recorded in the time domain 
for 200 turns (top) and its Fourier transformation for $q$ determination
(bottom) at the GSI synchrotron.}
\label{sis-tune-kick}
\end{figure}

For this method, coherent betatron oscillations are excited by a fast
kick. This kick has to be much shorter than the revolution time $1/f_{0}$ and is provided in most cases
by a so-called kicker magnet. The~strength of the kick has to be chosen to prevent for beam losses. 
The beam position is monitored turn-by-turn
(broadband processing only), and it is stored as a function of time. The
Fourier transformation of the~displacements gives the fractional part $q$ of
the tune. An example from the GSI
synchrotron is shown in Fig.~\ref{sis-tune-kick}.

\begin{figure}
\centering \includegraphics*[width=32mm,angle=90]
{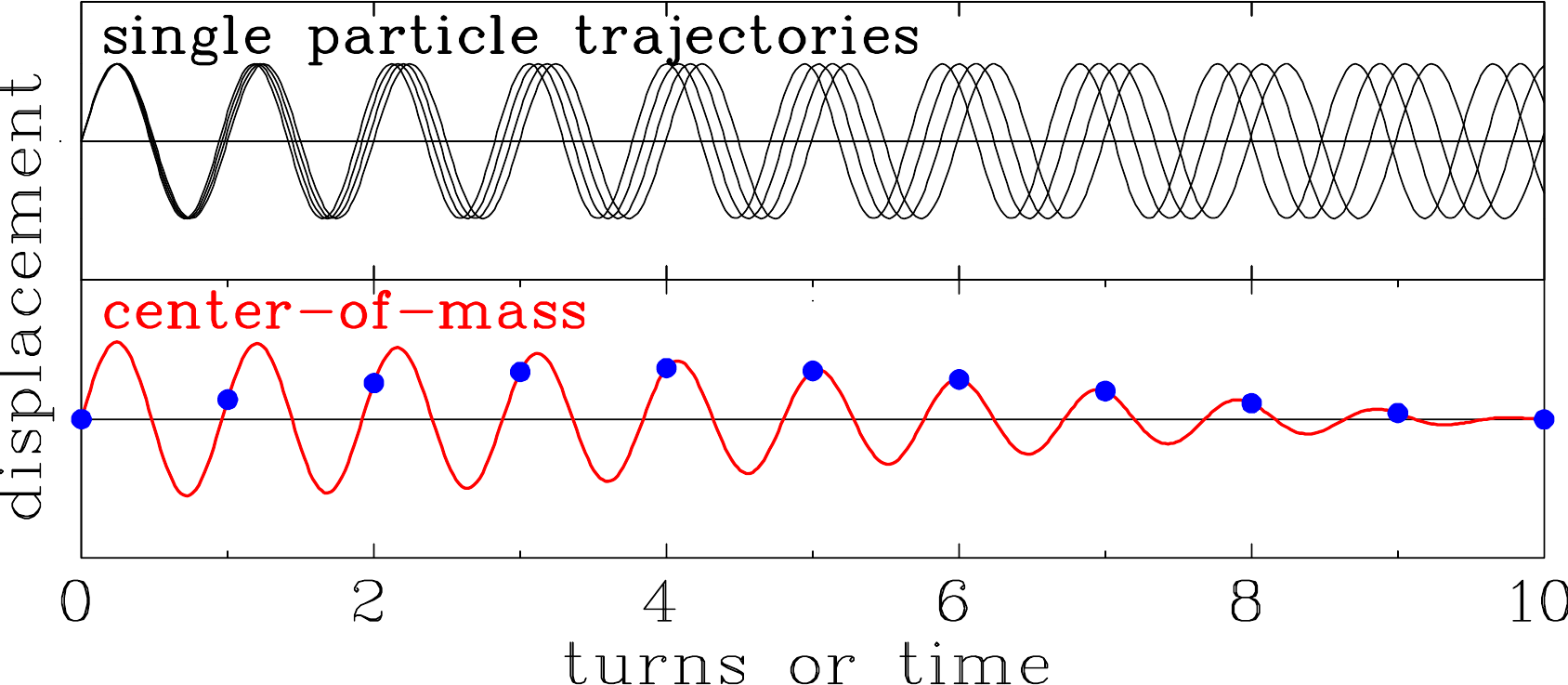}
\caption{Plot of the individual trajectories of four particles 
after a kick (top) and the resulting coherent sum signal as measured by a
pick-up (bottom). (The tune spread is much too large for a real machine.)}
\label{landau-prinz}
\vspace*{-0.1cm}
\end{figure}

To understand the limitation of this method, one has to look at the single
particle behaviour. At the~time of the kick, the particles start a free
betatron oscillation with the same phase, but each particle has a~slightly
different betatron frequency due to the tune spread. Even if the individual
oscillations last for a~long time, the coherent motion detected by a pick-up
is damped as a consequence of the de-coherence of the individual motion given
by the time of the inverse betatron frequency spread, see
Fig.~\ref{landau-prinz}. To achieve an adequate
resolution of typically $\Delta q \simeq 10^{-3}$, at least $N=500$ turns of
non-vanishing displacements are needed. In several practical cases, the
de-coherence time is shorter than this border as demonstrated in
Fig.~\ref{sis-tune-kick}. Hence, this straightforward method is not very
precise.

An additional drawback of such a measurement is a blow-up of the transverse
emittance of the~stored beam, because the excitation of the individual
particles are damped much slower than the coherent motion as depicted in
Fig.~\ref{landau-prinz}. For proton synchrotrons this can be of relevance, in contrary to electron synchrotrons where the trnasverse particle motion is
damped relatively fast by synchrotron radiation.

\subsubsection{Transverse excitation by a frequency chirp}

For this method, an rf method for tune determination uses a beam excitation
driven by a continuous sine wave stimulus. This wave is typically swept slowly
within some 100 ms over a certain band, and the~response of the beam is measured in
coincidence. The displayed spectrum represents the {\bf B}eam {\bf T}ransfer
{\bf F}unction {\bf BTF}, which can be defined as the transverse velocity
response to the given acceleration by the kicker. Besides the name BTF
measurement, this method is sometimes called 'beam excitation by a~frequency
chirp'.

\begin{figure}
\parbox{0.49\linewidth}{ \centering \includegraphics*[width=45mm,angle=90]
{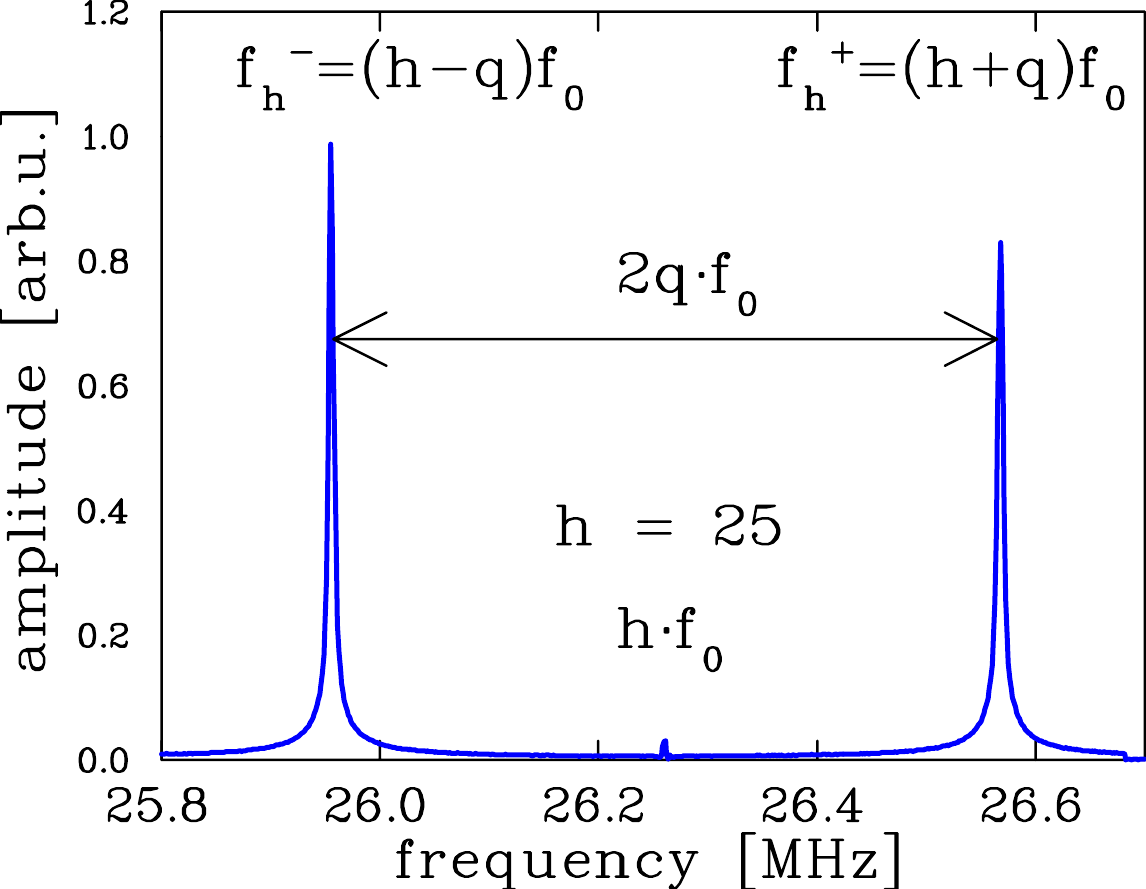} }
\parbox{0.49\linewidth}{ \centering \includegraphics*[width=46mm,angle=90]
{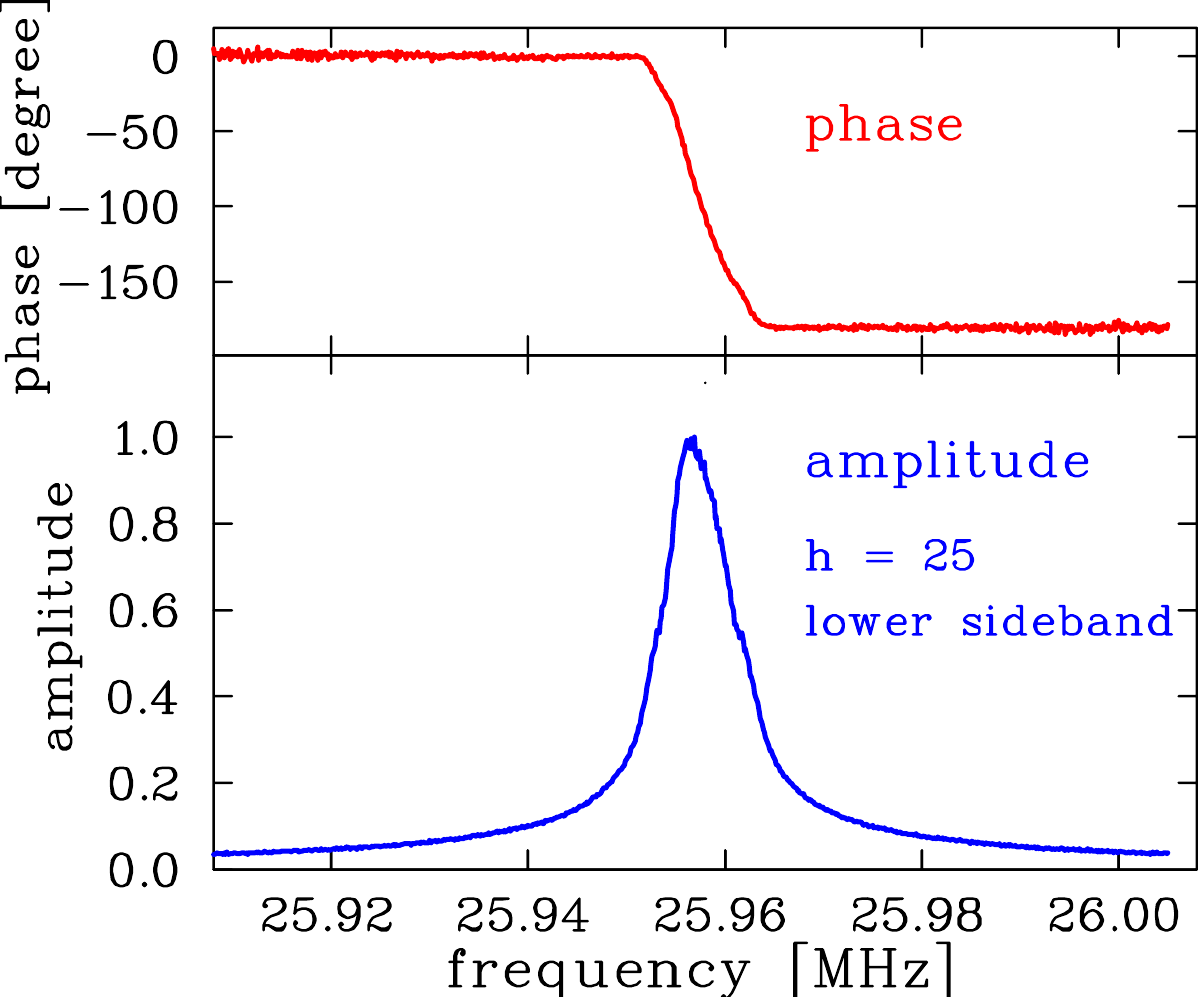} }
\caption{BTF measurement at the GSI synchrotron with a 500 MeV/u U$^{73+}$
beam having a revolution frequency of $f_{0}=1.048$ MHz. The wide scan
around the $h=25^{th}$-harmonics is shown left and a scan around a lower
sideband centred at 
$f_{25}^{-}$ on the right for amplitude and phase. 
The fractional tune is $q=0.306$ for these data.}
\label{btf-sis-2sidebands}
\vspace*{-0.3cm}
\end{figure}

Due to practical reasons, a harmonics of the revolution frequency $f_{0}$ is
used for the excitation, and the resulting resonance line is expressed in terms
of these harmonics. An example of such a measurement with an excitation around
the $h=25^{th}$ revolution harmonics is shown in Fig.~\ref{btf-sis-2sidebands}.
The beam reacts as a driven oscillator with a resonance as given by the tune
value. Because only the fractional part can be measured, this resonance is
located at $f^{\pm}_{h} = (h\pm q) f_{0}$ on both sides of the revolution
frequency and are called lower and upper sideband. From the distance of the
sidebands, the fractional part of the tune $q$ is determined. The resonance
nature of the BTF measurement can also be seen from the phase shift of $180^{o}$
during resonance crossing, as shown in on the right side of
Fig.~\ref{btf-sis-2sidebands}. With this method, a high precision of the tune value $q$ up to $10^{-4}$ can be
reached.

\subsubsection{Noise excitation of the beam}
\begin{figure}
\centering
\includegraphics*[width=110mm,angle=0]
{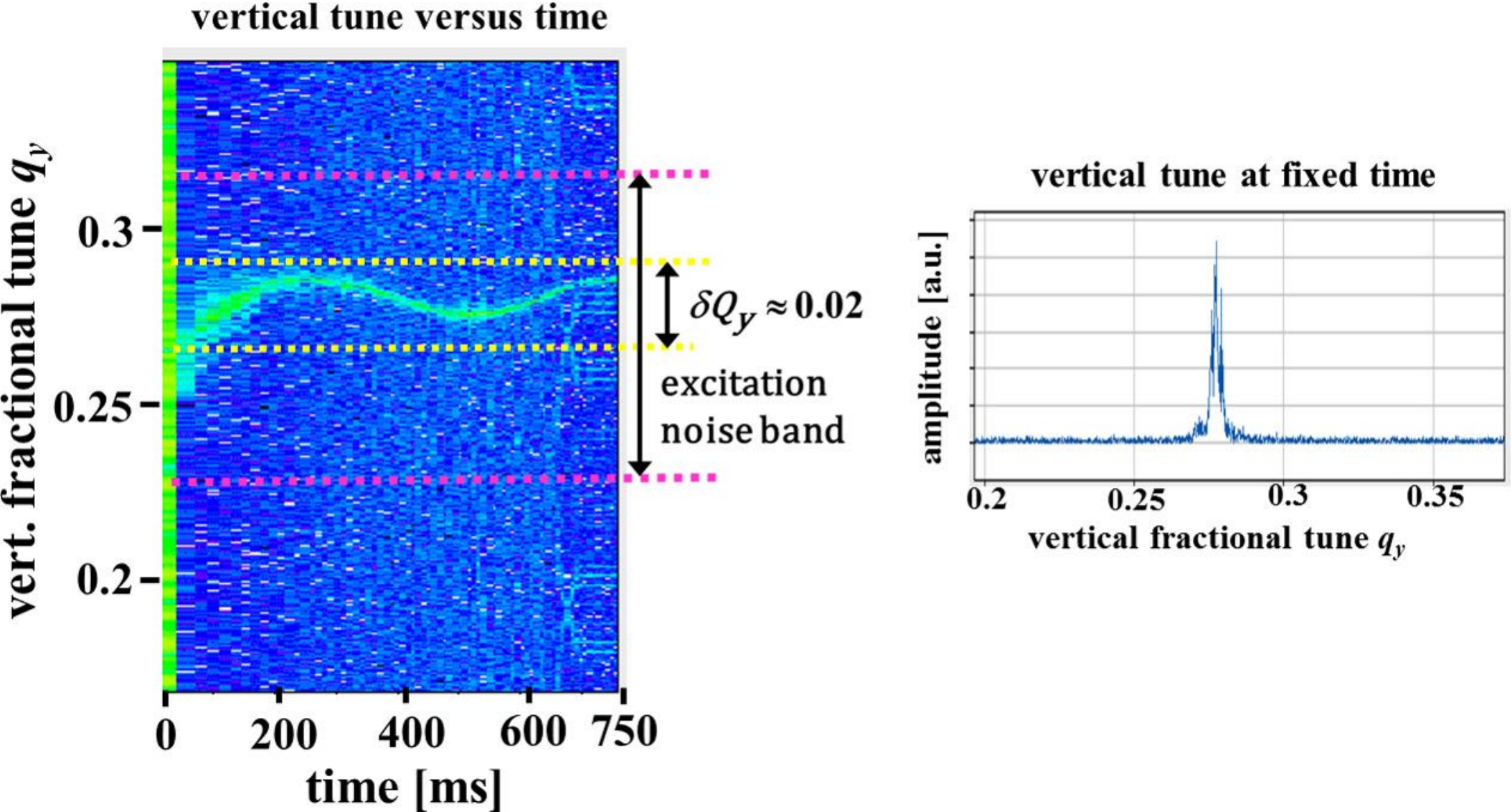}
\caption{Left: Measurement of the tune at the GSI synchrotron for an
Ar$^{10+}$ beam accelerated from 11 to 300~MeV/u within 0.75 s. Displayed is the~Fourier transformation of
displacements of successive individual bunches over 4096 turns
excited by the depicted noise band during acceleration as a function
of time. The maximum of the Fourier transformation is colour-coded in
a linear manner. Right: Single tune spectrum 110 ms after the ramp start
averaged over 4096 turns corresponding to $\sim 15$ ms.} 
\label{tune_noise_sis}
\vspace*{-0.1cm}
\end{figure}

Instead of a single frequency excitation, a wide-band rf signal can be applied:
The excitation takes place for all frequencies at the same
time, using white noise of adequate bandwidth being larger than the tune variation. 
The~position reading is done
in the time domain using a fast ADC recording each bunch. The~position
data for each bunch are 
Fourier analysed, and the spectrum is drawn corresponding to the~tune
value. An example is shown in 
Fig.~\ref{tune_noise_sis} from the GSI synchrotron during acceleration. 
The~resolution for the tune is given by the inverse of
the averaging time for one Fourier transformation calculation (for the~data in
Fig.~\ref{tune_noise_sis} 
in the order of some
ms, corresponding to 4096 turns), leading to about $\Delta Q 
\sim 3 \cdot 10^{-3}$ for one single scan. Due to the band excitation,
a time-resolved observation of the~tune is possible during the entire
acceleration ramp.

\subsection{Further measurements using BPMs}

Using the results of position measurements at a synchrotron, various beam 
parameters can be determined; some of them are briefly summarised here:

\textit{Chromaticity measurement:} The chromaticity $\xi$ is the coupling between the relative momentum offset $\Delta p$ with respect to the reference particle with momentum $p_0$ and the tune offset $\Delta Q$ according to $\frac{\Delta Q}{Q_0} = \xi \cdot \frac{p}{p_0}$. By a slight momentum change using a detuned rf-frequency, the particles can be slightly accelerated or decelerated. The tune is then measured and plotted as a function of the momentum deviation; the slope gives the value of $\xi$. More detail related to this advanced beam diagnostics is discussed in the corresponding articles in Refs.~\cite{schmick_cas2018,brandt_cas2007}.

\textit{$\beta$-function measurement:} For the tune measurement, the bunch position at one BPM is read and analysed. As there are many BPMs installed in a synchrotron (as needed for the closed orbit feedback), a~measurement can be done following one single bunch position from one BPM to the next while exciting the beam to coherent betatron oscillations. The bunch movement corresponds to the betatron phase advance $\Delta \mu$ from one BPM location at $s_0$ to the next location $s_1$. For this the values of the $\beta$-function at this location can be calculated with the help of the formula $\Delta \mu = \int_{s0} ^{s1} \frac{\mbox{d}s}{\beta}$ and compared to the expected value from beam optics calculations. The general method, advanced technologies and analysing methods are discussed in \cite{schmick_cas2018,brandt_cas2007,tomas-prst17} 

\textit{Dispersion measurement:} The dispersion $D(s)$ along the accelerator location $s$ is coupling between the relative momentum offset $\Delta p$ with respect to the reference particle with momentum $p_0$ and the~position offset $\Delta x(s)$ according to $\Delta x(s) = D(s) \cdot \frac{\Delta p}{p_0}$. The value of $D(s_{BPM})$ at the location $s_{BPM}$ of a BPM can be determined by introducing a slight momentum offset (as described for the chromaticity measurement) and a position measurement. A plot of the position offset versus the momentum deviation delivers the dispersion at this location as discussed in Refs.~\cite{schmick_cas2018,brandt_cas2007}.

\section{Measurement of beam profile} 
\label{kap-profile} 
The transverse beam profile 
can be influenced by quadrupole magnets installed transfer lines. A measurement is important to control the 
beam width, as well as the transverse matching between different part of an 
accelerating facility. In a transport line, a large number of bending, focusing, and correction 
magnets give rise to the need for many profile measurement stations. 
Usually, the size of an electron beam is less than 1 mm, while proton beams have typically large diameters, up to some cm resulting in different technical requirements for the beam instrument. 
Depending on the beam particles, current, and energy, a large variety of devices exists; we can only briefly discuss the most popular ones. The beam spot can be directly observed by intercepting it with scintillator screens and viewing the emitted fluorescence with a camera. When charged particles with relativistic 
velocities pass a metallic foil, radiation is emitted as described by 
classical electrodynamics; the process is called optical transition radiation 
OTR. A camera records this light in an arrangement comparable to the one for scintillation screens. 
Secondary electron emission grids SEM-grid are widely used as an electronic alternative to achieve a large dynamic range; a grid of wires with typically 1 mm spacing is inserted. Instead of a fixed grid, one single wire can be scanned through the beam to get high spatial resolution; this is called Wire Scanner. A third approach is a so-called ionization profile monitor IPM, where the ionization products of the interaction of the beam with the residual gas atoms or molecules inside the non-perfect vacuum of the tube are detected. In these devices, the resulting electrons or ions are accelerated by an external electrical field to a detector having a spatial resolution. The ionization profile monitor is well suited as non-destructive 
methods for proton synchrotrons. 
High energetic electrons emit synchrotron radiation if the trajectory is curved. 
Monitoring this light by synchrotron radiation monitors SRM yields direct 
information about the beam spot.

\subsection{Scintillation screen} \label{kap-scint-screen} 
\begin{figure}
\parbox{0.5\linewidth}{ \centering \vspace*{-3.0cm} 
\includegraphics*[width=50mm,angle=0] 
{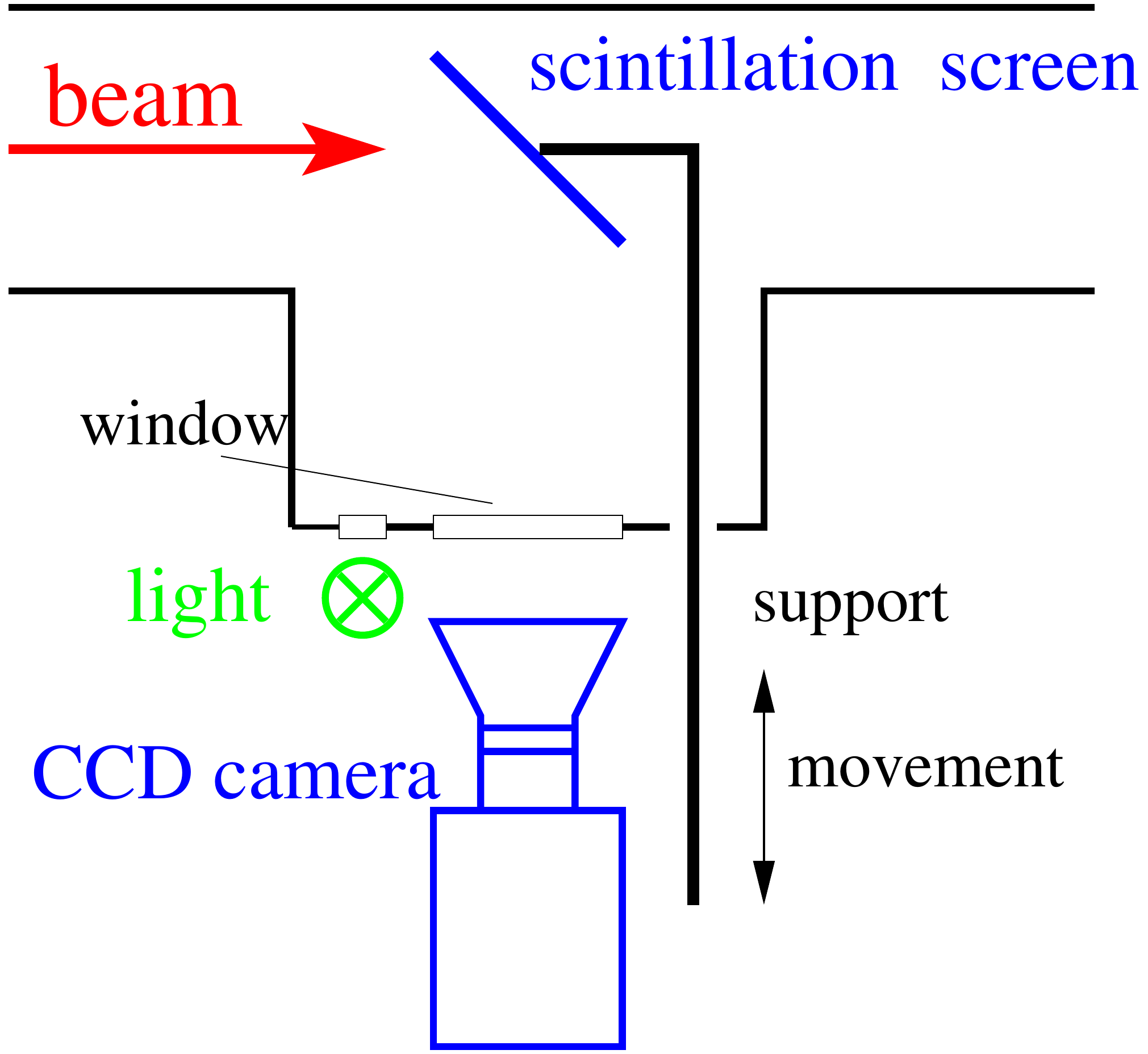} \vspace*{-0.5cm} } 
\parbox{0.49\linewidth}{ \includegraphics*[width=50mm,angle=0] 
{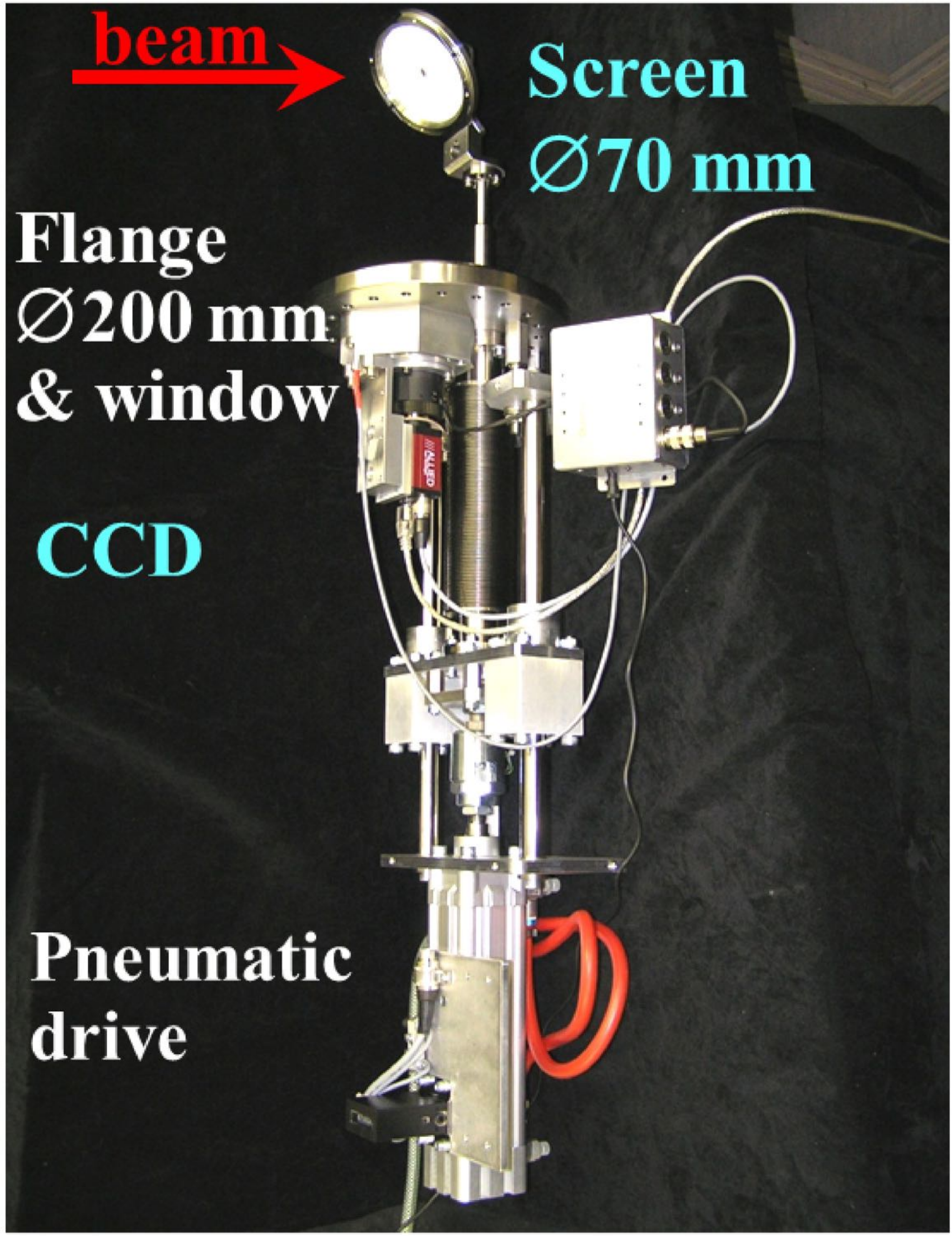} }\caption{Left: The 
scheme of a intercepting scintillation screen is shown. Right: A photo from a 
P43 phosphor scintillation screen of \O 70 mm and the CCD camera are mounted 
on a \O 200 mm flange with pneumatic drive is depicted.} 
\label{scheme-scintillator} \end{figure}

The most direct 
way of beam observation is recording the light emitted from a scintillation 
screen intersecting the beam, monitored by a commercial video, CMOS (complementary metal-oxide-semiconductor)
or CCD (charge-coupled device) 
camera as reviewed in Refs.~\cite{wala-cas18,wala-ibic18,wala-dipac11,ische-prab14}. These 
devices are installed in nearly all accelerators from the source up to the 
target and is schematically shown in Fig.~\ref{scheme-scintillator} together 
with a realization where the pneumatic drive mounted on a \O 200 mm flange.

When a charged particle penetrates a material the energy loss (as discussed in Section~\ref{kap-current-energyloss}) can be 
transformed to fluorescence light. In Table \ref{table-screens} some properties for frequently used scintillator materials are 
given; for more details see \cite{knoll-buch,wala-dipac11,Lecoq-buch}. 
The important properties of such a scintillator are: 
\begin{itemize} 
\item High light output matched to the optical system of the 
camera in the optical wavelength range of $400 \mbox{~nm~} < \lambda < 700 
\mbox{~nm}$. 
\item High dynamic range, i.e., good linearity between the incident particle flux and the light output. In particular, a possible saturation of the light gives rise to a deformation of the recorded profile. 
\item No absorption of the emitted light to prevent artificial broadening by the stray light inside the~material. 
\item Fast decay time, to enable the observation of possible variations of the beam size. 
\item Good mechanical properties for producing up to \O 100 mm large screens. 
\item Radiation hardness to prevent permanent damage. 
\end{itemize} 

\begin{table}
\centering
\caption{Chemical composition and some basic optical properties of inorganic
scintillation screens. The last four materials are so called phosphor screens, 
where powder is deposited on glass or metal plates.}
\begin{tabular}{llllrr}
\hline \hline
\textbf{Abbreviation} & \textbf{Type} & \textbf{Material} & \textbf{Activator} & \textbf{max. emission} & \textbf{decay time} \\ \hline
Quartz & glass & SiO$_{2}$ & non & 470 nm & $<$ 10 ns\\
Alumina & ceramics & Al$_{2}$O$_{3}$ & non & 380 nm & $\sim$ 10 ns\\
Chromox & ceramics & Al$_{2}$O$_{3}$ & Cr & 700 nm & $\sim$ 10 ms\\ 
YAG &crystal & Y$_{3}$Al$_{5}$O$_{12}$ & Ce & 550 nm & 200 ns\\ 
LuAG &crystal & Lu$_{3}$Al$_{5}$O$_{12}$ & Ce & 535 nm & 70 ns\\ 
Cesium-Iodide & crystal & CsI & Tl & 550 nm & 1 $\mu$s\\ 
P11 & powder & ZnS & Ag & 450 nm & 3 ms\\ 
P43 & powder & Gd$_{2}$O$_{2}$S & Tb & 545 nm & 1 ms\\ 
P46 & powder & Y$_{3}$Al$_{5}$O$_{12}$ & Ce & 530 nm & 300 ns\\ 
P47 & powder & Y$_{2}$Si$_{5}$O$_{5}$ & Ce \& Tb & 400 nm & 100 ns\\ 
\hline \hline
\end{tabular}
\label{table-screens}
\end{table}

\begin{figure}
	\centering
	\includegraphics*[width=90mm,angle=0]
	{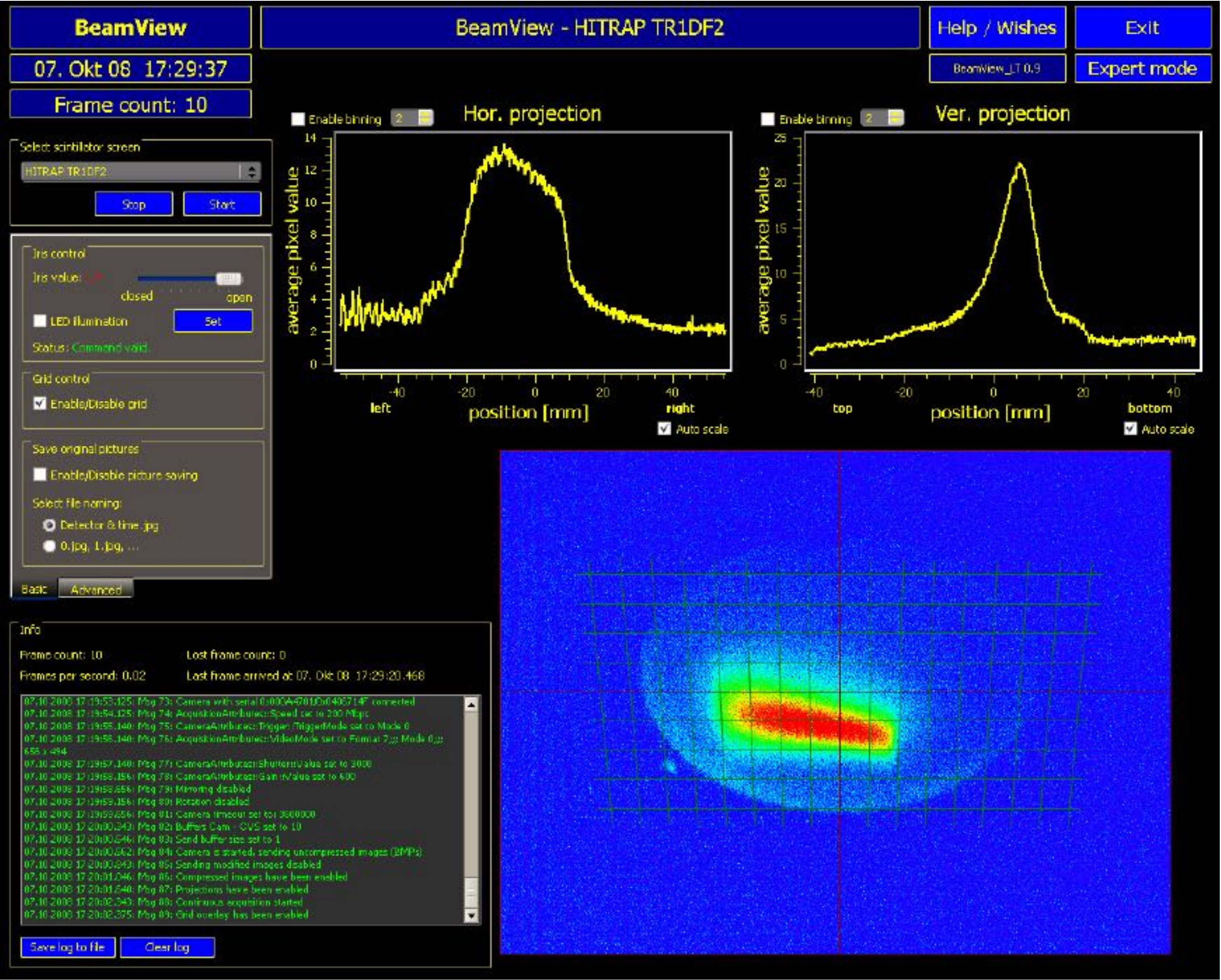} 
	\caption{Beam spot of a low current 4 MeV/u beam at GSI
		on a YAG:Ce screen recorded 
		with a digital CCD camera. The original black-white image is converted to
		false-colour for better visibility. The projection on the~horizontal
		and vertical axes are shown in the upper part \cite{hase-pcapac08}.} 
	\label{hitrap-yag}
\end{figure}

The material Chromox is a ceramic, which can be fabricated like pure
Al$_{2}$O$_{3}$, used widely as an~insulating material. The scintillation is
due to the Cr activators; chemically it is comparable to ruby. The~disadvantages are its long decay time of $\sim$10 ms and the large absorption
inside the material. Nevertheless, due to its robustness, it is quite often
used. The well-known Ce activated crystal materials like YAG
(Y$_{3}$Al$_{5}$O$_{12}$:Ce) have better optical properties and are widely used
for particle counting applications \cite{knoll-buch}. However, it is difficult to
produce crystalline disks of about 10 cm diameter out of this material. Instead of
the expensive single-crystal disk, one can use small grains of $\sim 10$
$\mu$m sizes, deposited on several mm thick glass or metal disks. These
phosphor screens are also used very frequently in analogue oscilloscopes,
electron microscopes and image intensifiers. P46 is an example for a~phosphor
screen powder offering a much lower production cost compared to the chemically
equivalent single-crystal YAG:Ce. The properties of doped materials are
strongly dependent on the activator concentration, therefore 
Table~\ref{table-screens} gives only approximated values.
The sensitivities of different materials span several orders of
magnitude and depend on the particle species as well, see
e.g. \cite{wala-ibic18,wala-dipac11}. As scintillation screens are used in 
all types of accelerators, a proper material choice is required. 
The beam image from a YAG:Ce screen is shown in Fig.~\ref{hitrap-yag} for a
low current beam of only $\sim 10^{6}$ ions at 4 MeV/u, proving the~high
sensitivity of that material. 

For high intensity beams, it must be ensured that the material is not
destroyed by the absorbed beam power. In particular, for 
slow heavy ions with a range comparable to the screen thickness 
this restricts the use, see also the discussion in Section~\ref{kap-current-energyloss}. 
A disadvantage of the screen is related to the~interception. The used material
is so thick (several mm) that it causes a significant energy loss, 
so it cannot be
used for the diagnostics of a circulating beam inside a synchrotron.

\subsection{Optical transition radiation screens}
At electron accelerators with relativistic particles 
the profile is determined from the electromagnetic radiation
at an intercepting thin metallic foil by the so-called \textbf{O}ptical
\textbf{T}ransition \textbf{R}adiation \textbf{OTR}. Generally, light is generated by a classical
electromagnetic process including special relativity, 
as caused by a~charged particle passing from one
medium into another. While passing to the 
vacuum in front of the~foil, the particle
has a certain electromagnetic field configuration, which is different
from the field inside the medium, because the foil has a dielectric constant $\epsilon_r \neq 1$, i.e. different from the vacuum value. 
By approaching the foil, the
particle's electromagnetic field leads to a time-dependent
polarization at the~foil boundary. When the charged particle transverses the foil, this field configuration is changed suddenly. The change of this polarization
at the foil surface generates
the radiation with a characteristic intensity and angular distribution.

\begin{figure}
\parbox{0.52\linewidth}{ 
\centering \includegraphics*[width=80mm,angle=0]
{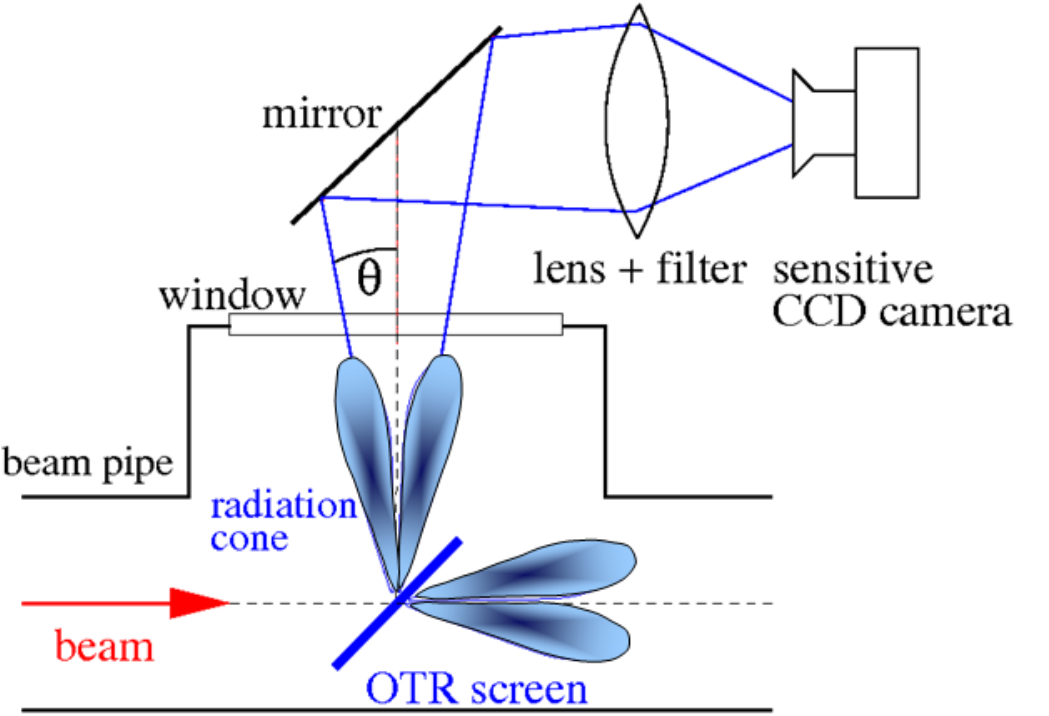} }
\parbox{0.47\linewidth}{
\centering \includegraphics*[width=63mm,angle=0]
{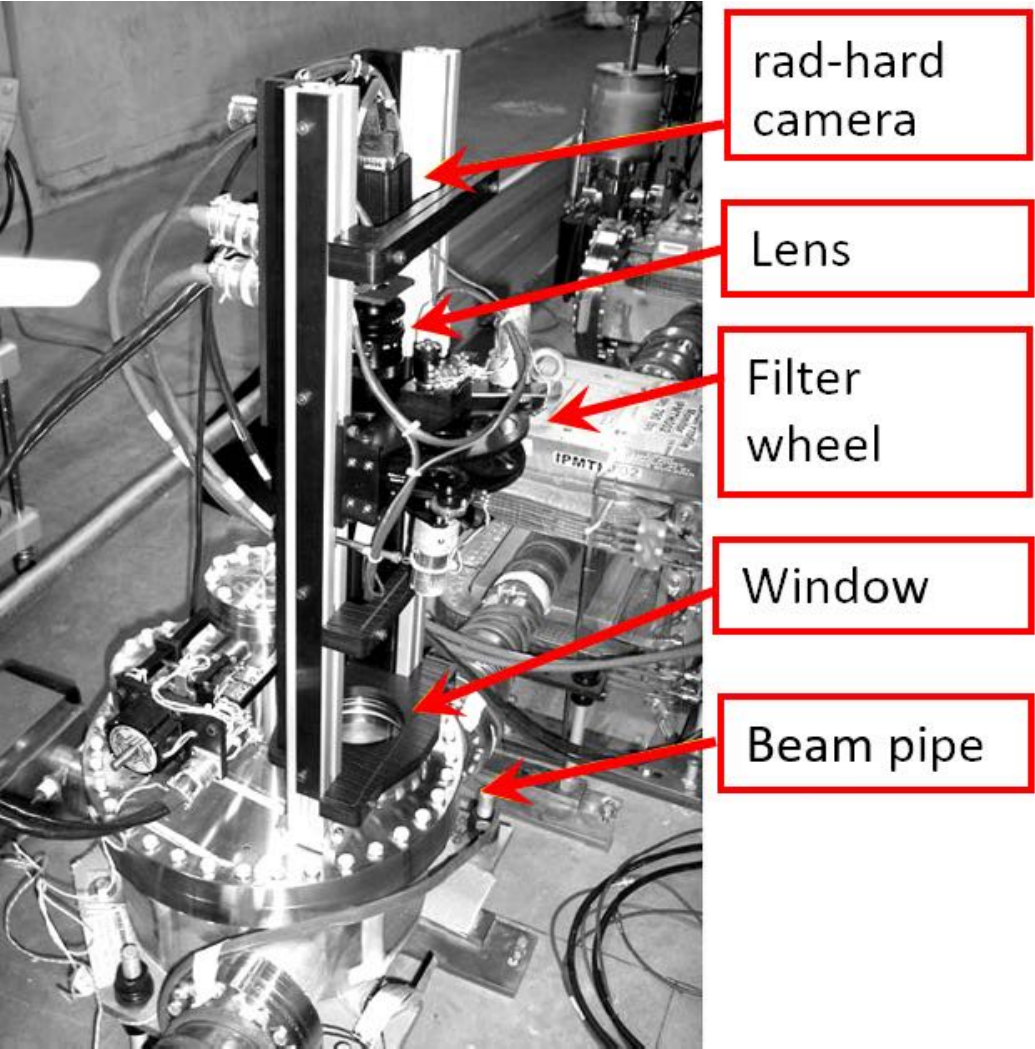} }
\caption{Left: The scheme of an OTR screen measurement. Right: Photo for an OTR monitor at Fermilab \cite{scar-pac07}}
\label{scheme-otr}
\end{figure}

A typical setup of an OTR measurement is shown in Fig.~\ref{scheme-otr}.
The foil is inserted under an angle of 45$^o$ with respect to the beam path in most
cases. The foil is made of Aluminum or Aluminum coated on Mylar with
a thickness 1 $\mu$m or less. The light is emitted in the forward
direction as well as at 90$^o$, because the metallic surface acts as
a mirror. Depending on the particle energy, the angular distribution
is peaked at the angle $\theta = 1/\gamma $ with $\gamma$ the
relativistic Lorentz factor. For typical values, 100 to 1000 beam particles yield one photon in the optical
wavelength range. With appropriate optics, an image of the foil is recorded with a camera. For a technical perspective, the installation is comparable to the one of a scintillation screen but with a different intersecting target. 

\begin{figure}
\centering
\includegraphics*[width=43mm,angle=90]
{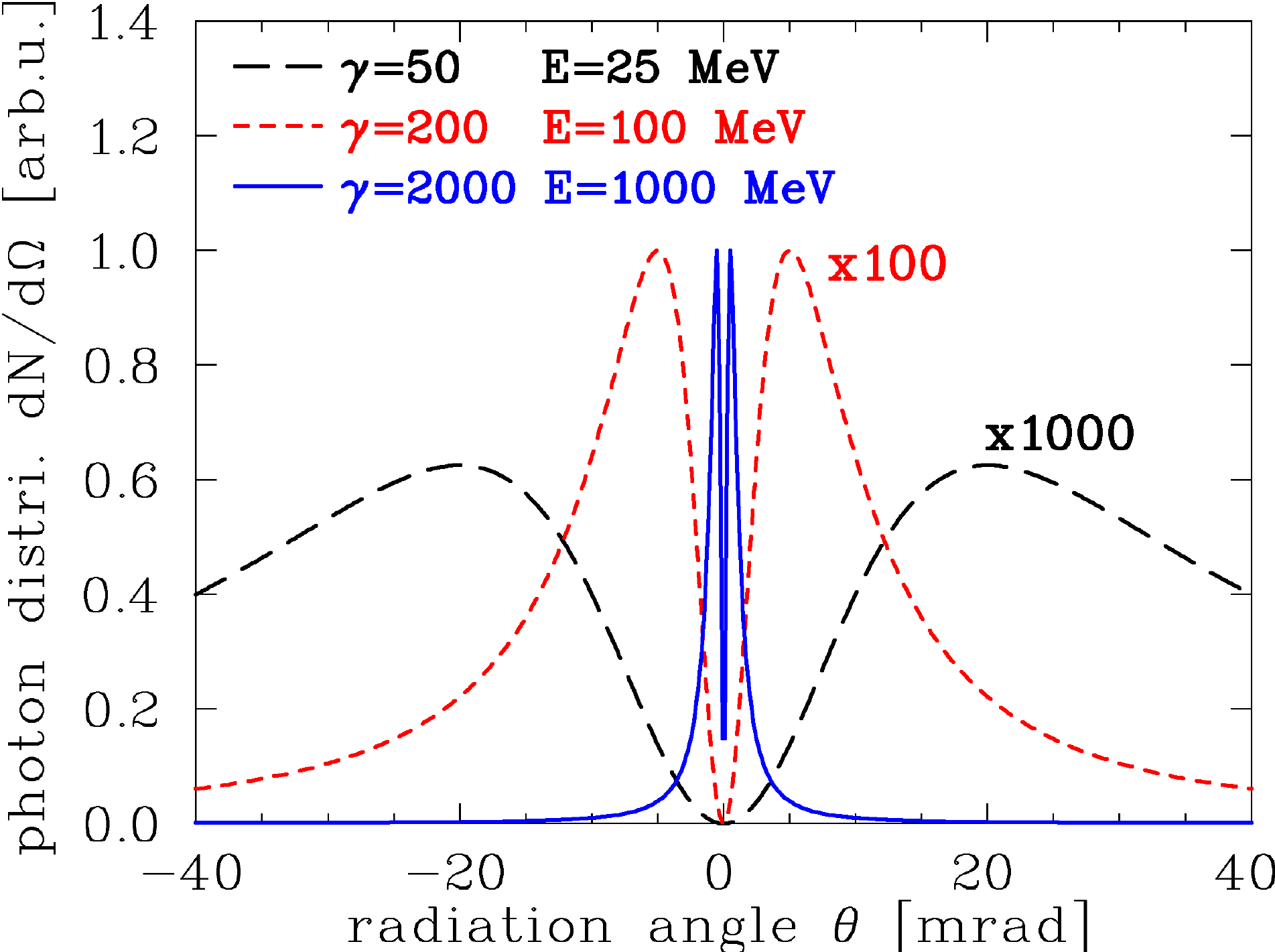} 
\caption{The intensity distribution of OTR as a function of the
observation angle for three different value of the~Lorentz-factor 
$\gamma$. Additionally, the corresponding
beam energies for electrons are given. The photon intensity is enhanced by
a factor of 1000 for $\gamma=50$ and 100 for $\gamma=200$ in the plot.}
\label{otr-angle-distri}
\end{figure}

The general process is treated in e.g.
\cite{jackson-buch,war-jap75,bos-nim85}
leading to closed but extensive formula. For relativistic beam
particles and some other valid assumptions, the radiated energy $\mbox{d}W$ emitted into
a solid angle $\mbox{d}\Omega$ per frequency interval $\mbox{d}\omega$ can be
approximated by
\begin{equation}
\frac{\mbox{d}^2 W}{\mbox{d}\Omega \mbox{d} \omega} =
\frac{2e^2 \beta^2}{\pi c} \cdot
\frac{\theta^2}{\left( \gamma^{-2} + \theta ^2 \right)^2 }
\label{eq-OTR-simple}
\end{equation}
with $c$, $e$, $\gamma$ are 
the velocity of light, the elementary charge and the relativistic Lorentz-factor,
respectively. The observation is oriented at an angle $\theta$ perpendicular
to the beam path, the so-called specular angle,
see Fig.~\ref{scheme-otr}.
There is no difference concerning the radiation for electrons
or protons moving with the same Lorentz-factor, 
reflecting the fact that only the electromagnetic fields of
the beam particles are concerned. 
Note that the radiated energy does not depend
on the frequency $\omega$, i.e., the whole spectrum is covered.
This is valid up to the plasma frequency of the metal, which is for
most metals in the deep UV at about 10~eV. 
The radiated energy of Eq.~\ref{eq-OTR-simple} is converted to the
number of photons by $W= N_{photon} \cdot \hbar \omega$ observed
within a wavelength interval from $\lambda_{begin}$ to $\lambda_{end}$
in the optical region by the camera and integration over the
interval $\lambda_{begin} ... \lambda_{end}$. This 
yields the number of photons per solid angle
\begin{equation}
\frac{\mbox{d} N_{photon}}{\mbox{d} \Omega } = N_{beam} \cdot
\frac{2e^2 \beta^2}{\pi \hbar c} \cdot
\log \left( \frac{\lambda_{begin} }{\lambda_{end} } \right) \cdot
\frac{\theta^2}{\left( \gamma^{-2} + \theta ^2 \right)^2 }
\label{eq-otr-angle}
\end{equation}
with $N_{beam}$ is the number of beam particles. This function is plotted
in Fig.~\ref{otr-angle-distri} for three different values of $\gamma$.
The radiation is more focused for
higher energies, 
having the advantage that a larger fraction of the
photon reaches the camera. At electron accelerators, OTR is used
even at moderate energies above $\simeq 100$ MeV, corresponding to a
Lorentz-factor $\gamma \simeq 197$. 
For protons this Lorentz-factor, or
equivalent a beam energy 
for proton of 184 GeV, is only reached at 
some high energy facilities like at CERN SPS and LHC where OTR screens
are installed \cite{bovet-dipac99,scar-pac07}.

At modern, LINAC-based free electron laser facilities, the usage of OTR was hindered by te occurrence of so-called coherent OTR emission \cite{wesch-dipac11}. For intense, very short bunches (below typically 100 fs duration) with a longitudinal substructure (typically on the fs-scale), photons are emitted during a short time and from a small area. As these are electro-magnetic waves emitted with a phase and angular correlation, interference might occur leading to an interference pattern at the camera location. The detectable emission characteristic in terms of spatial distribution and wavelength spectrum depends significantly on longitudinal substructure of the bunch. For those beam parameters, the OTR-foil image does not represent the transverse beam distribution and exclude the usage of OTR for profile measurement; as an~alternative scintillation screens were reverted at those facilities.         

The OTR profile determination has some advantages and disadvantages compared 
to a scintillating screen measurement \cite{iriso-dipac09,jung-dipac03}:
\begin{itemize}
\item
OTR is mainly used for electron beams with energies above 100 MeV.
For proton beams, the~needed Lorentz-factor $\gamma$ is only reached at the very high energy accelerators. Scintillation screens can be used for any beam and energy.
\item
The usable light emitted from an OTR screen depends significantly on the Lorentz-factor $\gamma$; for lower beam energies several orders of magnitude less light is emitted as compared to scintillation screens. Hence, more sensitive and expensive cameras have to be used for an OTR monitor.  
\item
OTR is based on a classical electromagnetic process leading to linearity between the number of photons and the beam intensity (disregarding the aforementioned coherent photon emission for very short bunches). Scintillation is a more complex process and depends significantly on the~material; saturation and radiation damage might occur, falsifying the profile reading. 
\item
For the OTR process, the number of photons and their distribution is independent of the thickness of the foil. Hence, very thin foils of pure Al or
Al-coated on Mylar foils down to 0.25 $\mu$m are used. Scintillation screens are generally thicker to guaranty their mechanical stability. For scintillation screens the amount of emitted photons scales with the thickness of the material. Hence, they are better suited for low intensity beams.   
\item
OTR and scintillation screen are intercepting monitors. However, for OTR the thin foil minimizes the influence on the beam due to the low scattering probability. It can also be applied at high beam power, because of the low energy loss in the thin, low-Z material leading to a lower influence on the~beam. Scintillation screens are thicker and the heating by the beam might modify the~scintillation process. Moreover, most of the materials contain high-Z material leading to a larger energy loss.    
\end{itemize}

\newpage
\subsection{Secondary electron emission grid SEM-grid}
\label{kap-semgrid}
\begin{figure}
\parbox{0.56\linewidth}{
\centering
\includegraphics*[width=75mm,angle=0]
{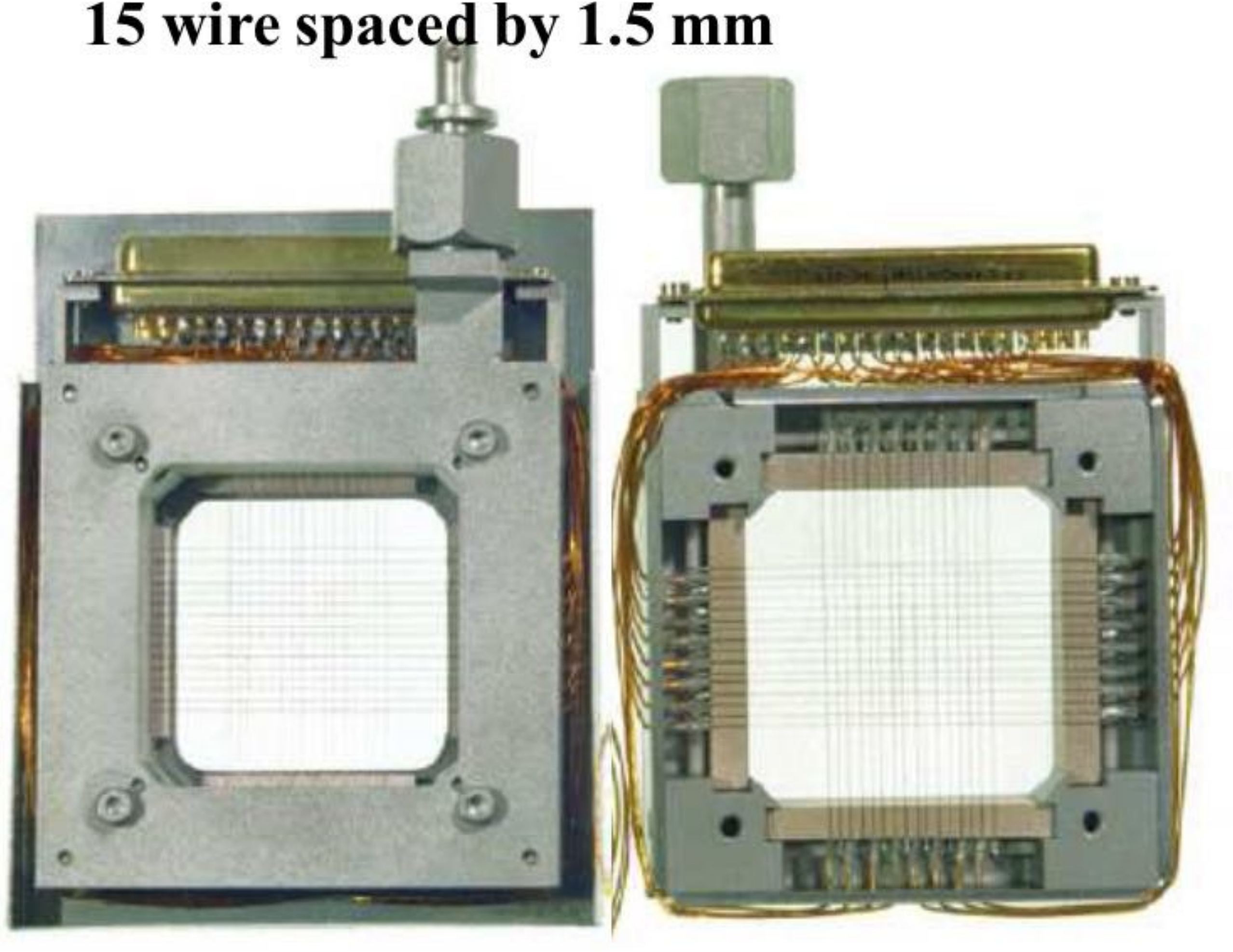} }
\parbox{0.42\linewidth}{
\centering
\includegraphics*[width=45mm,angle=0]
{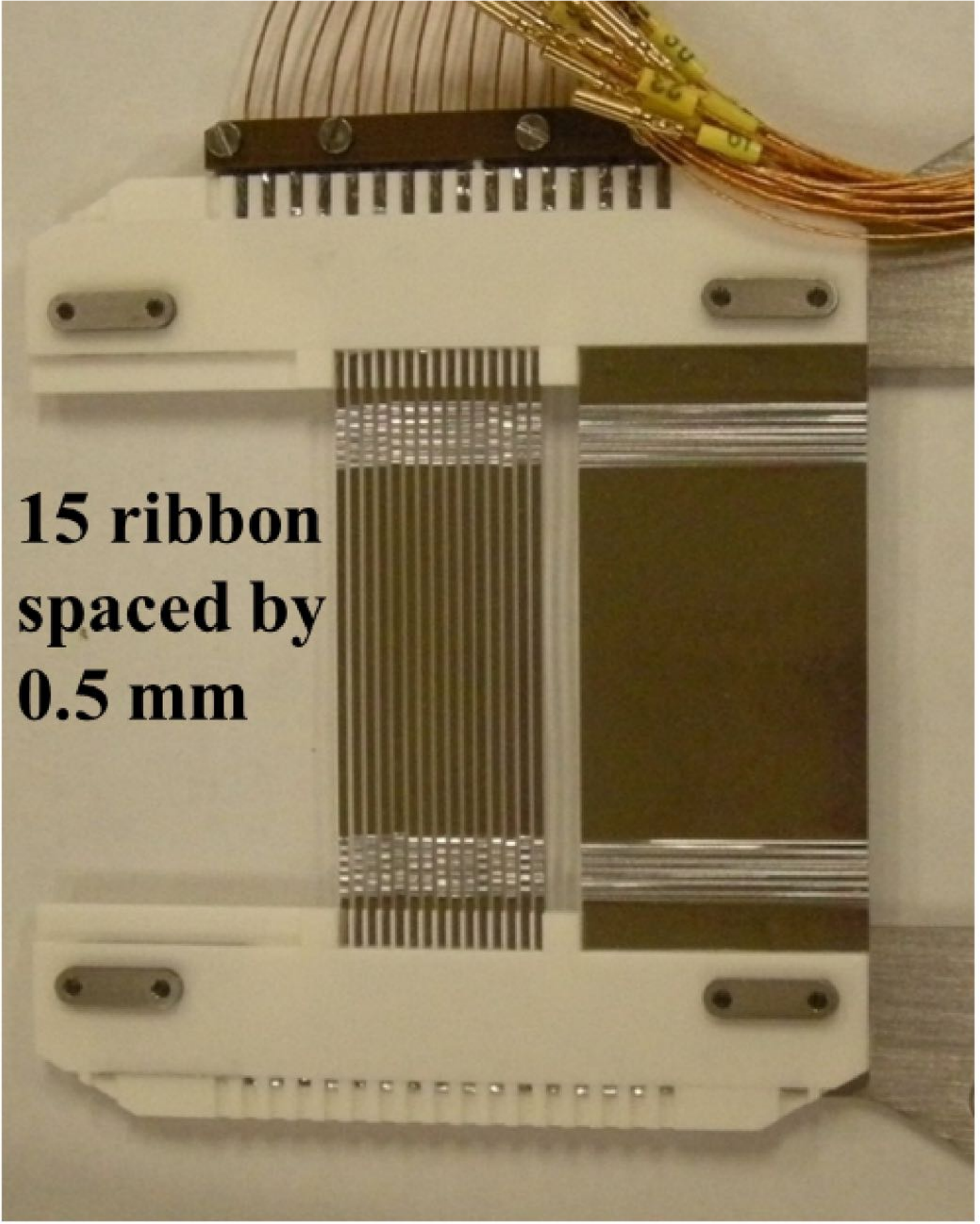} }
\caption{Left: SEM-grid for both planes with 15 tungsten wires spaced by 1.5 mm
as typically used at a proton LINAC. 
The individual wires are insulated with glass-ceramics.Middle: The
same SEM-grid with removed tin cover. 
Right:
SEM-grid based on ribbon made from 25 $\mu$m thick aluminium foil by laser
cutting; the spacing is 0.5~mm. This type is used for high energetic
proton above typically 1 GeV. The
devices are mounted on a~pneumatic drive to move it in and out of the
beam (not shown).}
\label{unilac-grid-foto}
\end{figure}

When particles hit a surface, secondary electrons are knocked-out form 
the surface, as described in
Section~\ref{chapter-sternglass}. For the profile determination, 
individual wires or ribbons interact with the beam; this is called a 
\textbf{ S}econdary \textbf{E}lectron E\textbf{M}ission \textbf{SEM-grid} or a harp as reviewed
in \cite{plum-biw04}. 
Each of the wires
has an individual current-to-voltage amplifier. This is an electronic
alternative to a scintillation screen with a much higher dynamic range,
i.e., the ratio of minimal to maximal detectable current is orders of magnitude
larger. For the metallic wires of about 100 $\mu$m diameter and typically 1 mm spacing tungsten is often used owing their excellent refractory properties related to the very high melting temperature.
In particular, at low
energy LINACs this is important because no cooling can be applied due to the
geometry. A photo of such a device is shown in Fig.~\ref{unilac-grid-foto} and the specifications are given in Table \ref{tab-unilac-grid}. For applications at higher beam energies, ribbons might be used as an alternative.

\begin{table}[htb!]
\centering
\caption{Typical specification for a SEM-grid used at proton and heavy ion
LINACs.} 
\begin{tabular}{ll}
\hline \hline
\textbf{Parameter} & \textbf{Value}\\ \hline
Diameter of the wires & 0.05 to 0.5 mm\\
Spacing & 0.5 to 2 mm\\
Length & 50 to 100 mm\\
Material & W \\
Insulation of the frame & glass or Al$_{2}$O$_{3}$\\
number of wires & 10 to 100\\
Max. power rating in vacuum & 1 W/mm\\
Min. sensitivity of I/U-conv. & 1 nA/V\\
Dynamic range of the electronics & 1:10$^{6}$\\
Integration time & 1 $\mu$s to 1 s\\
\hline \hline
\end{tabular}
\label{tab-unilac-grid}
\end{table}

Generally, the SEM-Grid is mounted on a drive to be inserted into the beam path for the measurement. 
For low energies, at proton LINACs the particles are stopped in
the material or undergo a~significant energy loss. 
The ratio diameter-to-spacing 
of the wires determines the attenuation of the~beam current (and of cause
also the signal strength on the individual wires). Typically only 10 \% of the~beam area is covered by the wires, and in this sense, the profile measurement
is nearly non-destructive. For energies above 1 GeV/u, 
the relative energy loss is negligible, and ribbons are used to increase the signal strength.

\subsection{Wire scanner}
Instead of using several wires with individual, expensive electronics, a
single wire can be swept through the beam \cite{plum-biw04}, which is called wire scanner. 
The advantage is that the resolution 
is not limited by the wire spacing and therefore this technique is often used
at electron LINACs with beam sizes in the sub-mm range. It can also be
applied in proton synchrotrons owing the small amount of intercepting
matter. 

\begin{figure}
\parbox{0.5\linewidth}{
\centering
\includegraphics*[width=60mm,angle=0]
{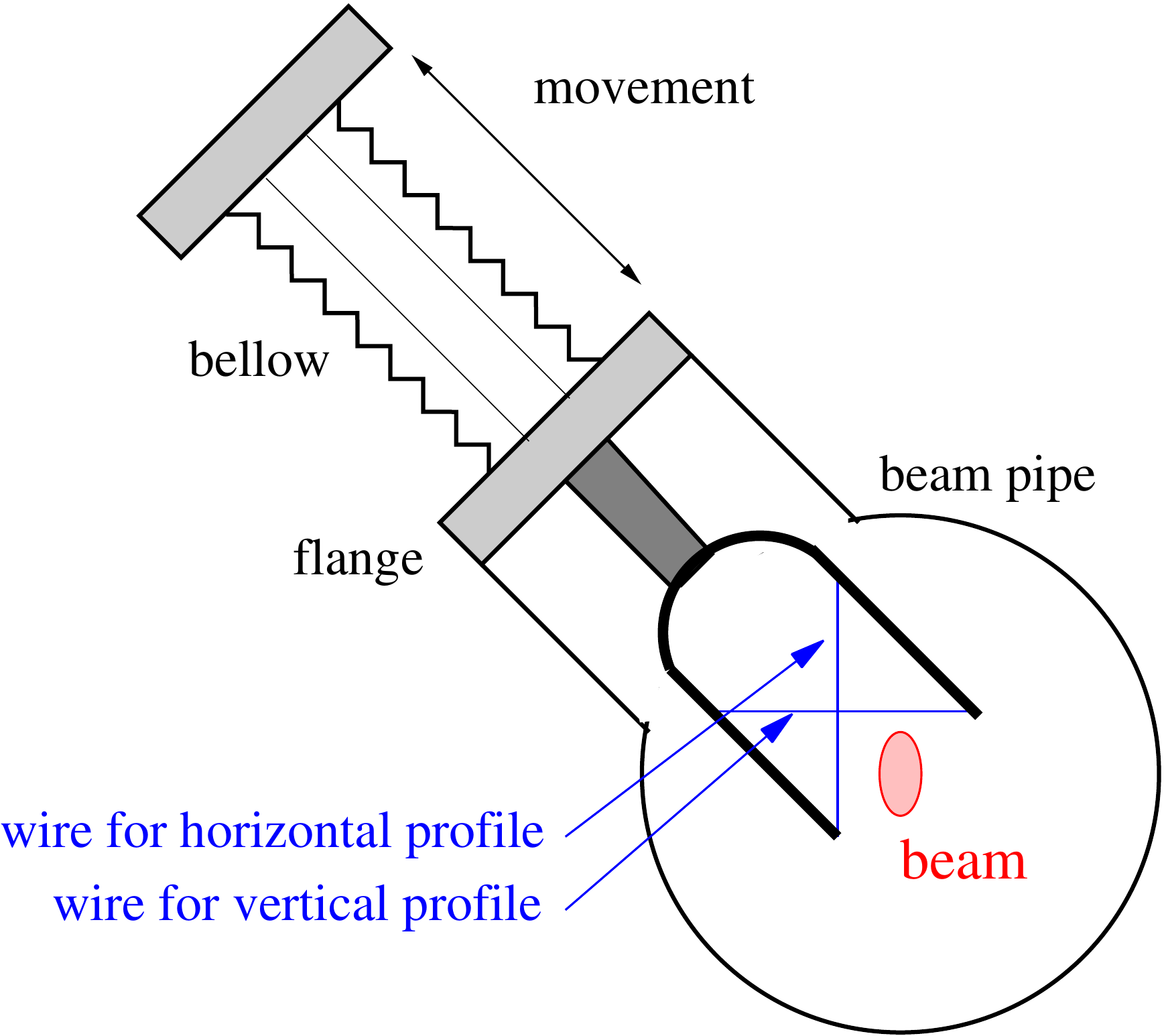} 
}
\parbox{0.49\linewidth}{
\centering
\includegraphics*[width=47mm,angle=180]
{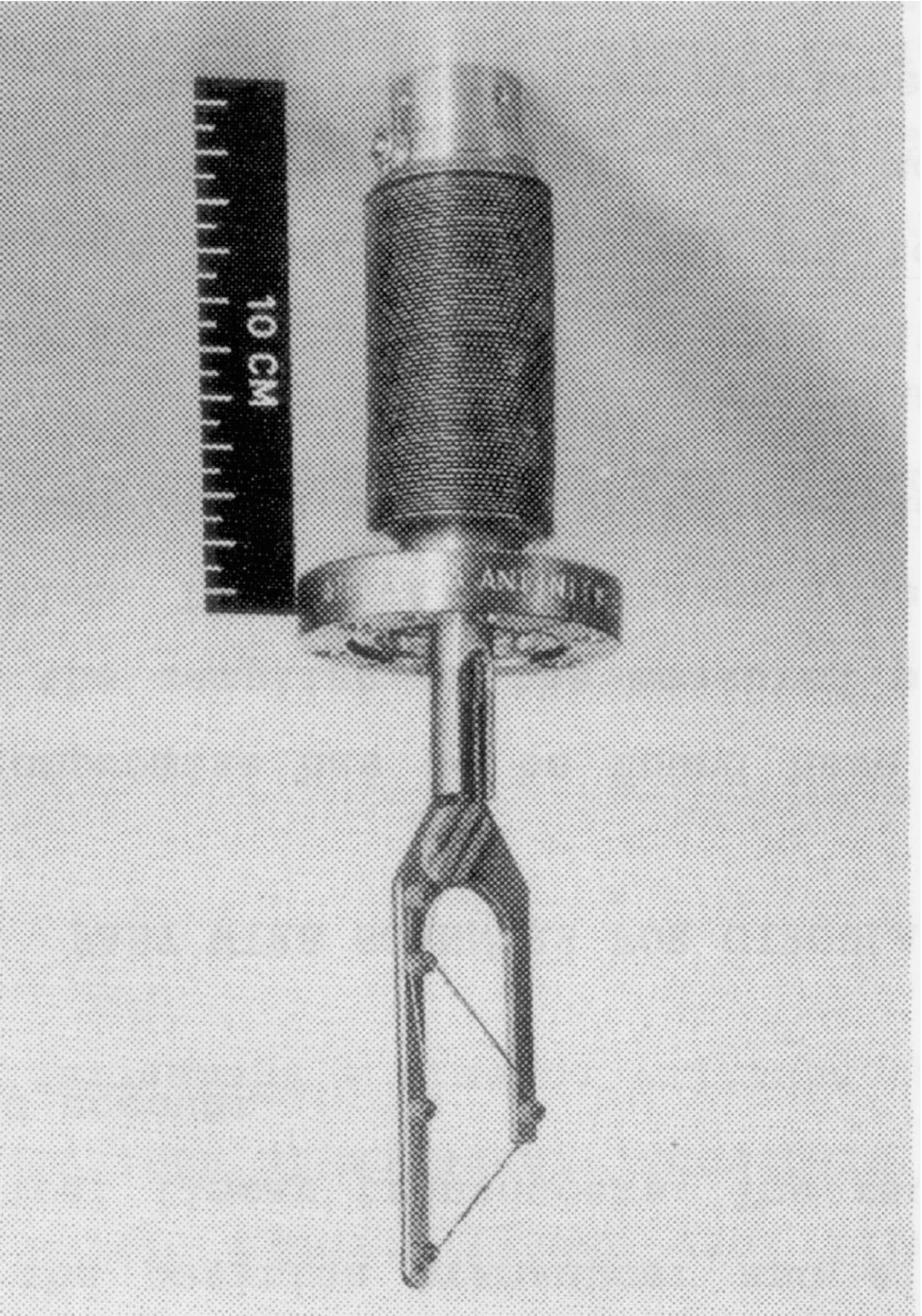} 
}
\caption{Left: Schematics of the movement. Right: Linear scanner used at CERN \cite{col-ieee} }
\label{linear-scanner}
\end{figure}

There are two types of installations: We first discuss the \textit{slow linear wire scanner}. The wire is mounted on a linear 
stepping motor drive for slow movement operated with a typical velocity of less than 
1 cm/s with as depicted in Fig.~\ref{linear-scanner}. The slow wire scanner is used at many transfer lines at LINAC facilities for all types of beams. 
A resolution of 1 $\mu$m 
is reached in electron accelerators \cite{field-nim95}. 
To get this low beam width reading $\sigma_{beam}$ for a Gaussian beam, 
deconvolution of the contribution of the~wire
(4$\mu$m thick carbon wire in this case) from the measured width $\sigma_{meas}$ 
has to be done according to $\sigma_{beam}^{2} = \sigma_{meas}^{2} -4 \cdot r_{wire}^{2}$.
In most cases, the wire is mounted on a fork, which is inserted into the~beam 
pass by a drive mounted on 45$^{o}$. Then only one drive is 
sufficient for measuring both transverse planes by mounting the wires
in a crossed orientation, as shown in Fig.~\ref{linear-scanner}. If the
signal generation is performed by reading the secondary emission current then
two isolated wires can be mounted compactly on a type of fork. If the signal is
generated by beam loss monitors outside the beam pipe, 
the two wires have to be separated (as shown in the Fig.~\ref{linear-scanner}
left), so the wires cross the beam one after the other.

\begin{figure}[tb]
\parbox{0.57\linewidth}{
\centering
\includegraphics*[width=95mm,angle=0]
{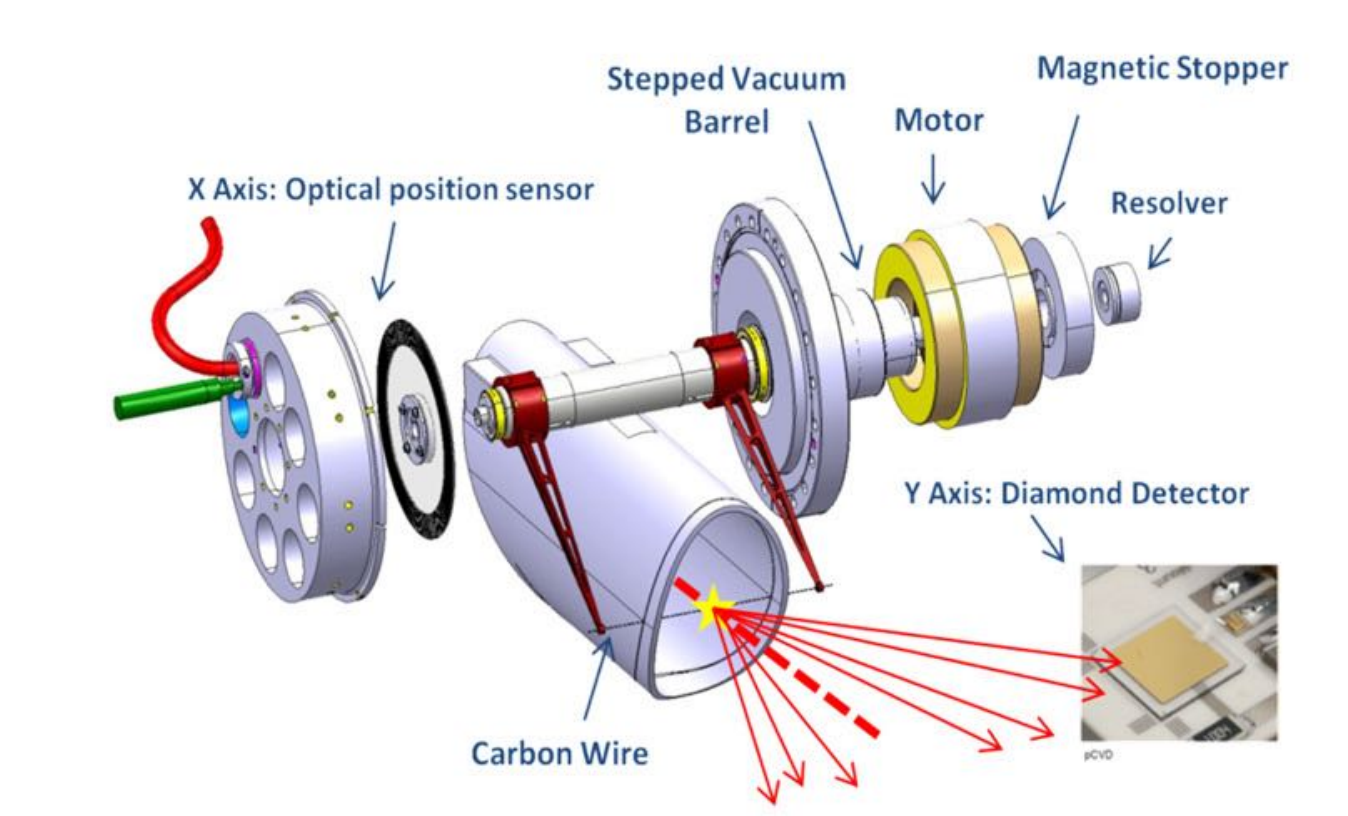} }
\parbox{0.42\linewidth}{
\centering \includegraphics*[width=70mm,angle=0]
{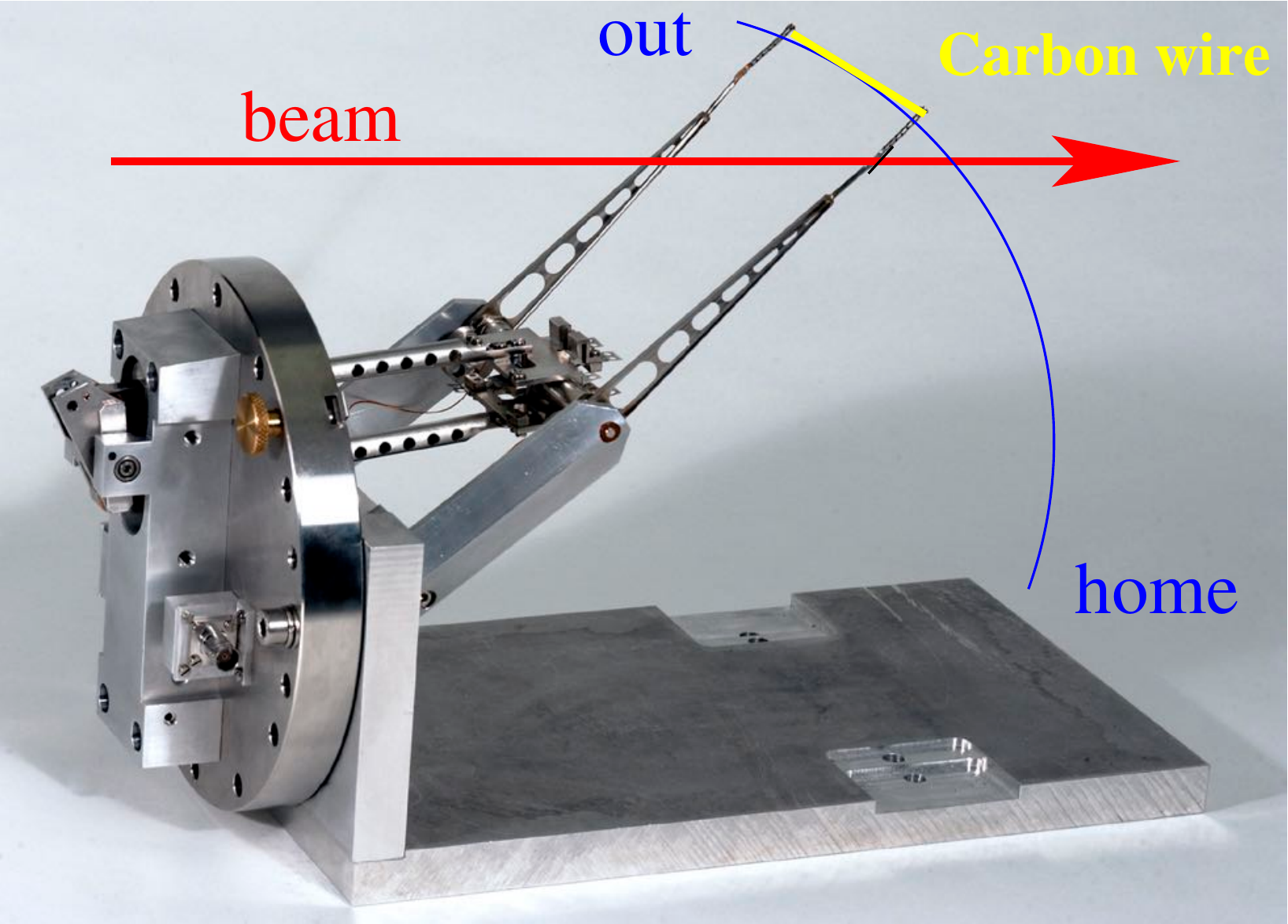} }
\caption{Left: Scheme of the entire detection system for a flying wire. Right: Pendulum scanner fork and wire used at CERN
synchrotron \cite{burg-dipac03}.}
\label{pendulum-scanner}
\end{figure}

An arrangement with a straight wire on a fast pendulum mechanics is a second type of so-called \textit{fast flying wire scanner} often used at a proton synchrotron is depicted 
in Fig.~\ref{pendulum-scanner} together with a schematic drawing of the entire arrangement. 
Scanning velocities up to 10 m/s can be achieved with a unique motor-driven mechanism. As the wire material, carbon or SiC is used due to its
low weight and low nuclear charge $Z$, resulting in a low energy deposition in
the wire (Bethe-Bloch Eq.~\ref{eq-bethe-bloch} gives $dE/dx \propto \rho \cdot
Z/A$). In addition, these materials can withstand high
temperatures without melting. The thickness can be down to 10
$\mu$m. However, due to the single scanned wire, the profile is not taken at a single instant, even with high scanning
velocity. Therefore only the steady-state distribution can be probed.

\begin{figure}
\parbox{0.52\linewidth}{
\centering
\includegraphics*[width=33mm,angle=270]
{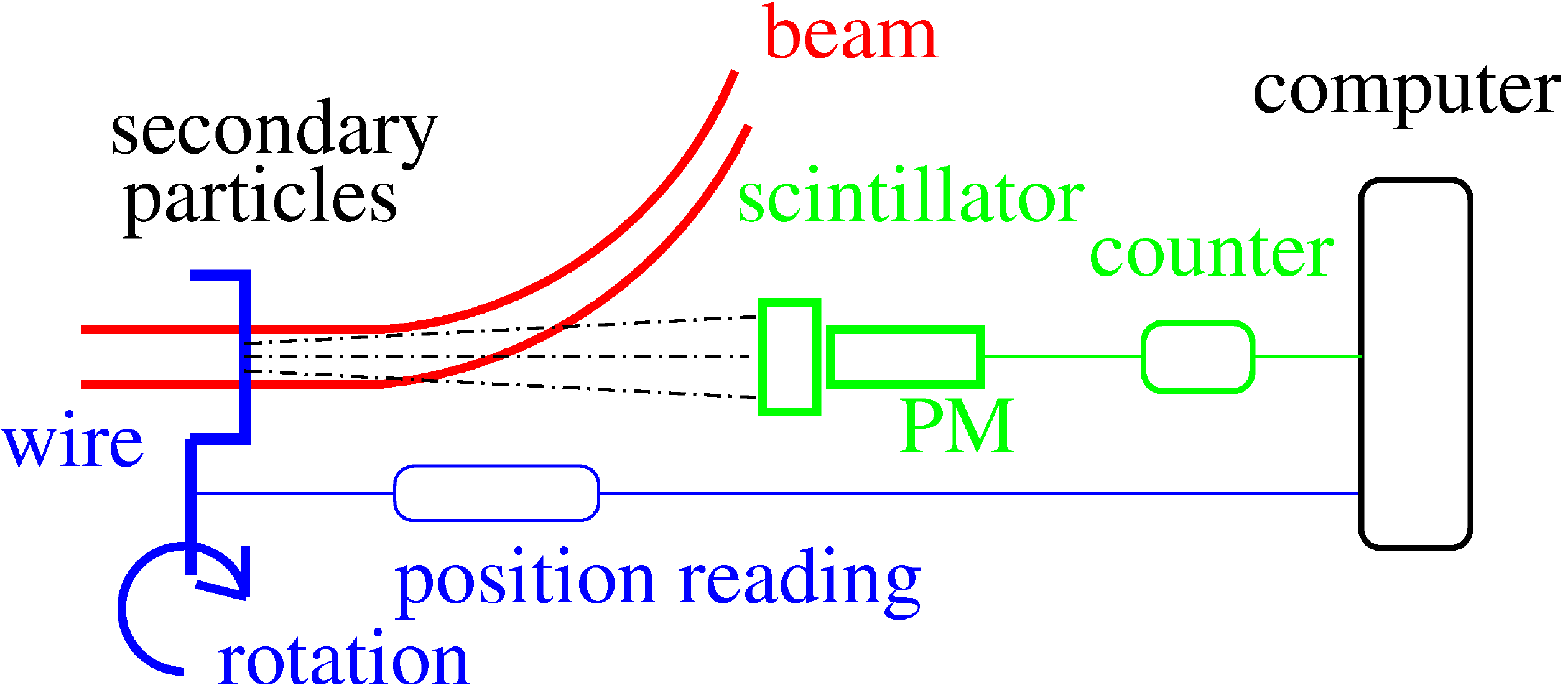} 
}
\parbox{0.47\linewidth}{
\includegraphics*[width=70mm,angle=0]
{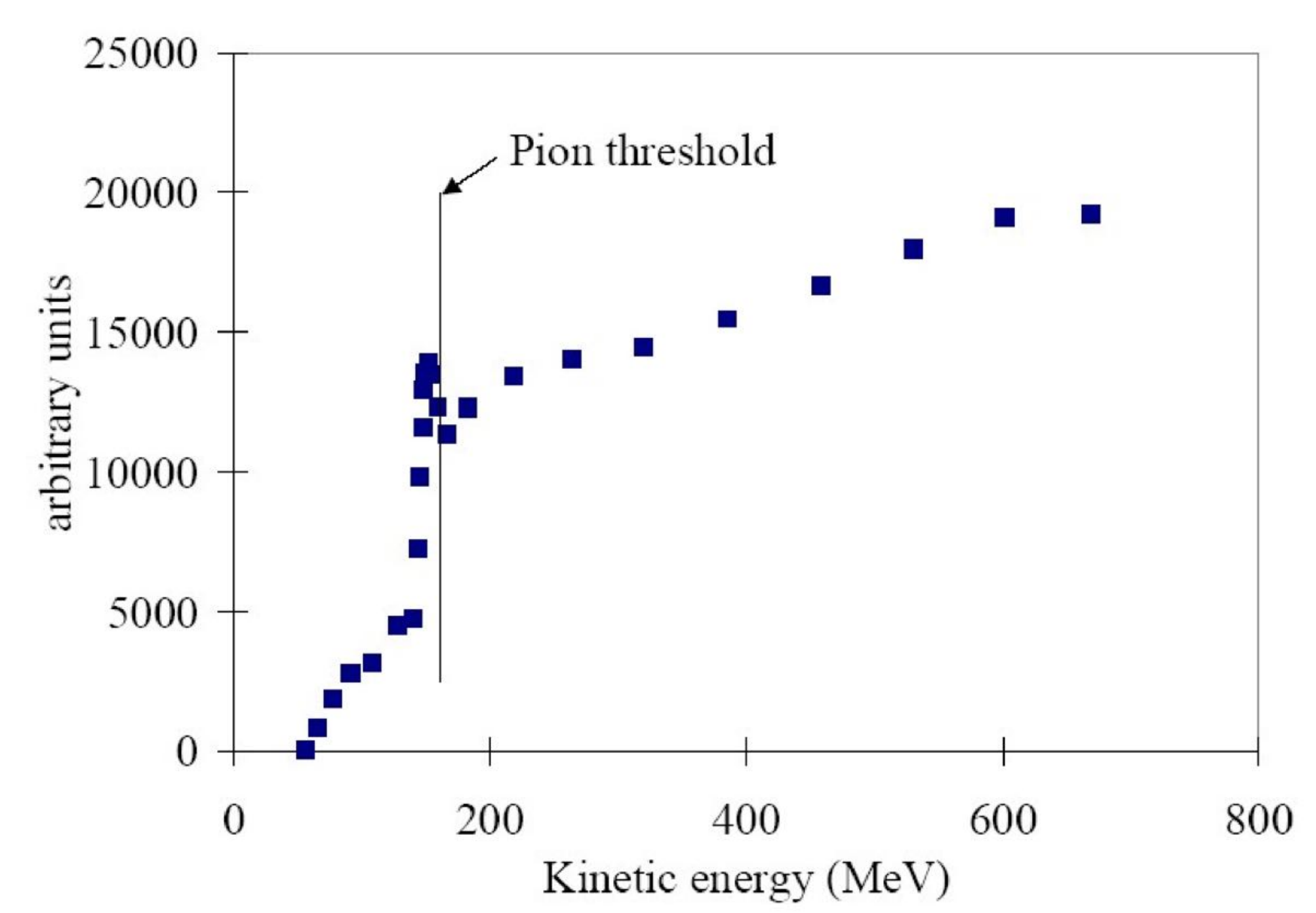} 
}
\caption{Left: 
Scheme of a wire scanner using the production of secondary particles
as the signal source. Right: The total rate from a wire scanner under proton
impact at CERN PS-Booster, 
showing the rate increase above the~$\pi$-threshold at $\sim 150$
MeV \cite{raich-dipac05}.}
\label{scheme-second-wire}
\end{figure}

For the display of profiles, the position of the wire, determined
by the position-encoder, is plotted on the horizontal axis. The beam signal
for the vertical axis can be deduced from the current given by the emitted
secondary electrons, like for a SEM-grid. This is done in particular for low
energy protons and heavy ions. In most cases for beam energies larger than 
150 MeV/u for ions (the threshold for $\pi$-meson production) 
or 10 MeV for electrons the signal is deduced by monitoring the secondary particles outside of the beam
pipe, see Fig.~\ref{scheme-second-wire}. These secondary particles might be hadrons created by the nuclear interaction of the proton or ion projectiles and
the wire, having enough kinetic energy to leave the vacuum chamber. For the
case of electron accelerators, the secondary particles are mainly 
Bremsstrahlung-photons. The detector is just a beam loss
monitor, e.g. a 
scintillator installed several meters away (see also Section~\ref{kap-beamloss}). The count-rate is plotted as a
function of the wire position as a precise representation of the beam
profile. 

The fast flying wire is mainly used for at synchrotrons where the fast scanning velocity of typically $v_{scan} = 10$ m/s results in only a few passages of the individual particle through the wire. For high energetic particles, the resulting energy loss is sufficiently small so that the particles can still be stored (more precisely: stay within the acceptance of the synchrotron). This enables the usage of flying wires as transverse profile diagnostics for the stored beam.

A comparison of the slow linear wire scanner and the SEM-grid shows advantages and
disadvantages of both types \cite{plum-biw04}:
\begin{itemize}
\item 
With an SEM-grid the beam intensity is sampled concurrently,
whereas a moving wire samples the parts of the profile at different locations at
different times. Therefore variations of the beam intensity in time will be
mixed with transverse intensity variations using a scanning device.
\item 
In case of pulsed beams a synchronization between the readout of the electronics and the beam pulse is required as well as a movement within the beam pause. Synchronization between readout and beam pulse is easier to achieve for a SEM-grid.
\item
The resolution of a SEM-grid is fixed by the wire spacing (typically 1 mm), 
while a wire scanner can have much higher resolution, down to 10 $\mu$m, due to its
constant movement. 
\item 
The electronics for data acquisition is cheaper for a scanning
system. An SEM-grid requires one channel per wire. 
\item
For the cost of the mechanics, it is vice versa: The precise vacuum actuator
for the scanner is more expensive than the pneumatic drive needed for a
SEM-grid. 
\end{itemize}

\subsection{Ionization profile monitor IPM}

\begin{figure}
\parbox{0.49\linewidth}{
\centering
\includegraphics*[width=60mm,angle=0]
{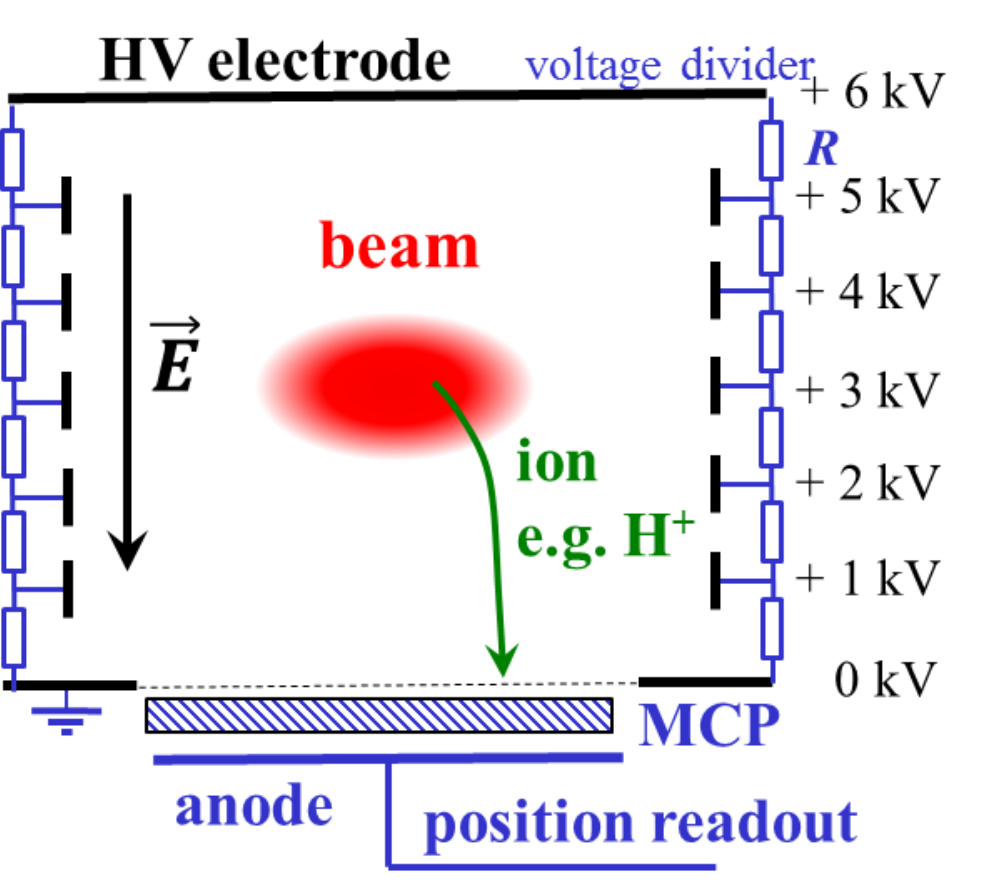} 
}
\parbox{0.49\linewidth}{
\centering
\includegraphics*[width=50mm,angle=0] 
{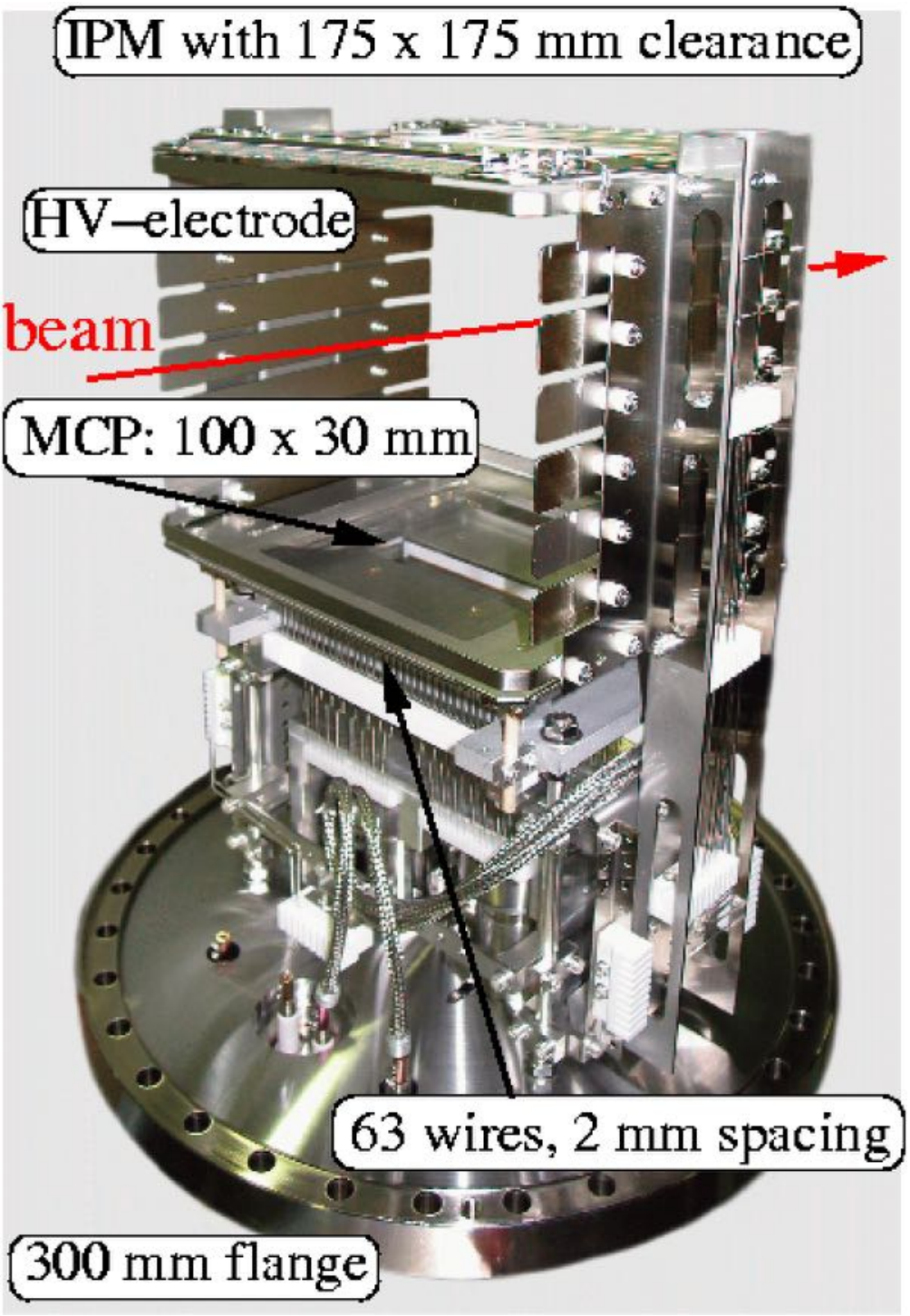}
}
\caption{Left: Scheme of an ionization profile monitor for the horizontal
profile determination. Right: 
The large aperture ionization profile monitor installed at the GSI
synchrotron for the horizontal direction. The readout behind the MCP (bottom) is done with an array of 63
wires with 2 mm spacing.}
\label{scheme-rgm}
\end{figure}

A frequently used non-destructive method for the profile determination 
is the \textbf{I}onization 
\textbf{ P}rofile \textbf{M}onitor \textbf{IPM} sometimes also called \textbf{R}esidual
\textbf{G}as \textbf{M}onitor \textbf{RGM}. 
These monitors are installed in nearly every proton
synchrotron for the detection of beam sizes between some mm and several cm; a review is given in \cite{forck-ipac10,forck-dipac05}. The principle is based on the detection of ionized products from a collision of the beam particles with the residual gas atoms or molecules present in the vacuum
pipe. A scheme for such a~monitor is shown in Fig.~\ref{scheme-rgm}. Due
to electronic stopping process, 
electrons are liberated, and electron-ion pairs are generated. An
estimation of the signal strength can be obtained by Bethe-Bloch 
formula Eq.~(\ref{eq-bethe-bloch}) discussed in Section~\ref{kap-current-energyloss}. Due to the single collision behaviour, the process can be approximated by considering some
100 eV for the creation of one electron-ion pair. To separate the electron-ion 
pairs, an~external electric field of typically 1 kV/cm is applied by metallic plates installed in the vacuum pipe, but outside the beam path. To achieve higher field homogenity, 
guiding strips are installed at the side to get a 
smooth transition from the HV side to the ground plane. For most
applications, the residual
gas ions are detected in this setup. A realization from the GSI synchrotron is 
shown in Fig.~\ref{scheme-rgm} having a large opening of 17 cm
\cite{giac-dipac05}. For electron synchrotrons, IPMs are seldom used, due to the smaller electron
beam dimensions. 

Typical pressures for LINACs and transfer lines are in the range
$10^{-8} - 10^{-6}$ mbar and for synchrotrons $10^{-11} - 10^{-9}$ mbar. For the LINAC case, the vacuum pressure
is high, and the energy loss is larger due to the lower beam energy. Enough
electron-ion pairs 
are generated to give a measurable current (down to 0.1 nA per strip) 
of secondary ions 
for direct detection by a sensitive SEM-grid like wire array, e.g. \cite{egberts-dipac11}. 
In a synchrotron, the pressure and the stopping power is lower. A \textbf{M}ulti
\textbf{C}hannel \textbf{P}late \textbf{MCP} particle detector acting as a 'pre-amplifier' to enable spatial resolved single particle detection \cite{gys-nima15,wiza-nima79}.

\begin{figure}
\parbox{0.45\linewidth}{
\centering
\includegraphics*[width=60mm,angle=0]
{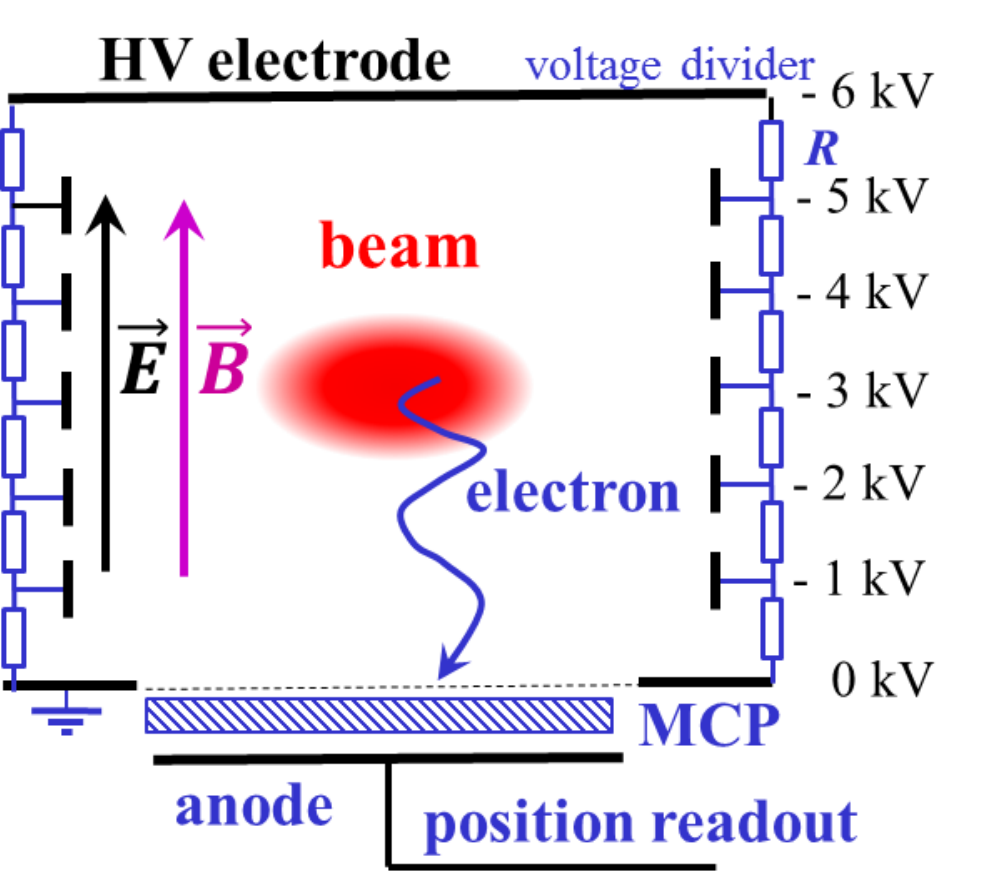} 
}
\parbox{0.54\linewidth}{
\hspace*{0.6cm}
\includegraphics*[width=47mm,angle=90]
{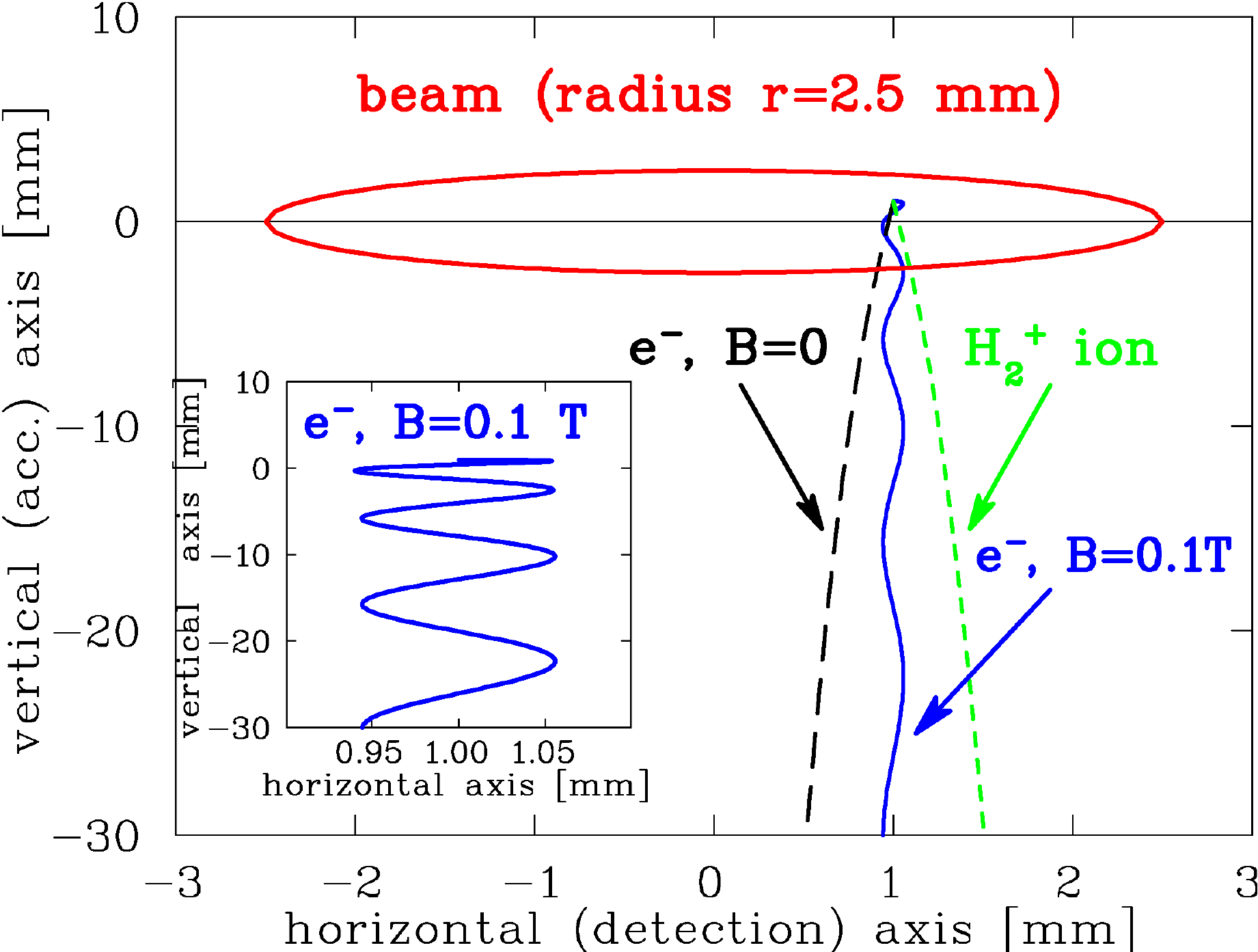} 
}
\caption{Left: Schematic trajectory of an electron in an ionization profile monitor in the presence of a homogeneous 
electric $E$ and magnetic field $B$. Right: The comparison of the 
trajectories of an
H$_{2}^{+}$-ion and an electron with and without a magnetic field 
visualizes the broadening due to the beam space-charge field. 
By applying a magnetic field the electron is fixed to the field line, as
shown in the insert. 
The beam parameters are: $10^{9}$ U$^{73+}$ within a~10~m long bunch
with a radius of 2.5 mm.}
\label{scheme-rgm-mag}
\end{figure}

For the measurement of intense beams, detection of electrons is more suited
to overcome the~space-charge induced broadening as it causes a modification of the residual gas ion's trajectory, see
Fig.~\ref{scheme-rgm-mag}. The~HV is reversed, to accelerate now the
negative electrons created by the beam interaction toward the~detector.
If an additional magnetic field $B$ 
is applied parallel to the external electrical
field, the~electrons spiral around the magnetic field lines with the
cyclotron radius $r_{c}=\sqrt{2m_{e}E_{kin, \perp}}/{eB}$ 
with $E_{kin, \perp}$ being the~kinetic energy fraction perpendicular to the magnetic field. This
transverse kinetic energy is
determined by the initial transverse velocity given by 
the kinematics of the atomic collision. The cross-section depends
on the beam and residual gas molecule properties \cite{icru-rep95},
resulting in kinetic energies for the emitted electron up to keV for
the typical case. 
The cyclotron radius, and therefore the~detector resolution, 
is mainly determined by this initial energy and is nearly independent
of the beam space-charge field.
The~projection of the 
trajectory in the plane perpendicular to the field is a circle with less than
0.1 mm radius (corresponding to the spatial resolution of the MCP) 
if a magnetic field of $B\simeq 0.1$ T is applied. 
The~movement along the field is a linear acceleration with a typical time of
flight of $\sim 2 $ ns. The~necessary magnetic field of about 0.1 T 
is generated in most cases with a dipole
magnet.

\subsection{Synchrotron radiation monitor}
For electron accelerators, the effect of synchrotron radiation, 
emitted by accelerated
electrons, can be used for a profile determination. As known from classical
electrodynamics \cite{jackson-buch} the radiation power $P_{synch}$ 
for a momentum change $\mbox{d}p/\mbox{d}t$ of a particle with mass $m_{0}$ is given by
\begin{equation}
P_{synch}=
\frac{e^2c}{6\pi \varepsilon _0(m_0c^2)^2}\left[ \frac{\mbox{d}p}{\mbox{d}t}\right] ^2 = \frac{e^2c}{12\pi \varepsilon _0} \frac{\gamma ^4}{\rho ^2}.
\label{PS1}
\end{equation}
For a circular accelerator, the direction of the momentum is changed in the bending dipole magnets of radius $\rho$ leading to the emission of synchrotron radiation, which is significant only for relativistic energies with a Lorentz-factor above at least $\gamma > 100$. Therefore, the process is of relevance for typical electron synchrotrons (for electrons $\gamma = 100 $ corresponds to $E_{kin}=50.6$ MeV). However, for proton accelerators is can be only applied for energy above some 100 GeV (for protons $\gamma = 100 $ corresponds to $E_{kin}=92.9 $ GeV).

\begin{figure}
\centering
\includegraphics*[width=100mm,angle=0]
{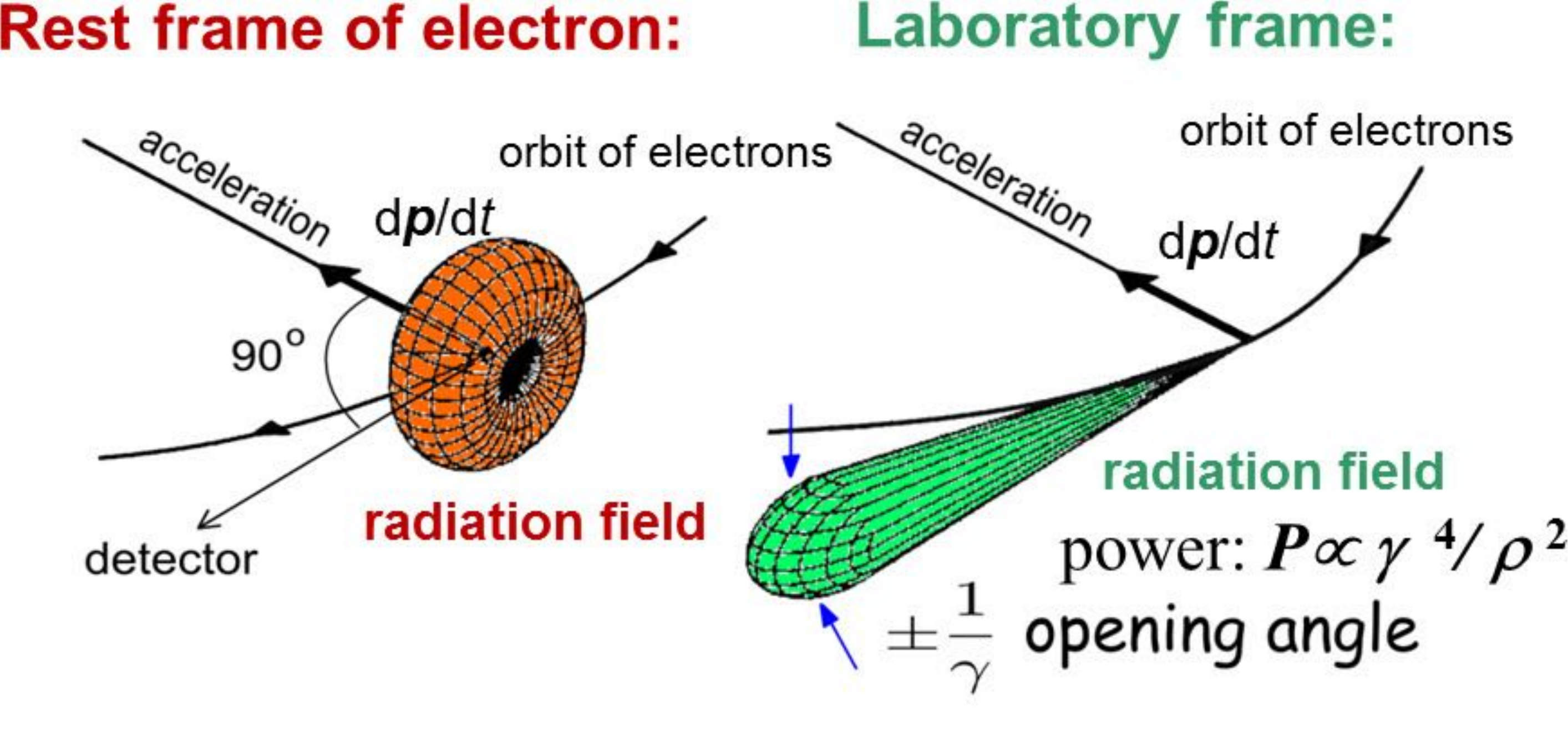}
\vspace*{-0.6cm}
\caption{Forward peaked synchrotron radiation for relativistic particles from
a bending magnet.}
\label{scheme-sr-allgemein}
\end{figure}

In the centre-of-mass system, the radiation is emitted perpendicular to the momentum
change. The~Lorentz transformation to the laboratory frame gives a factor of
$\gamma$, 
yielding a forward peak distribution with an opening cone of half-angle $\gamma$ as
demonstrated in Fig.~\ref{scheme-sr-allgemein}. 

\begin{figure}
\centering
\includegraphics*[width=90mm,angle=0]
{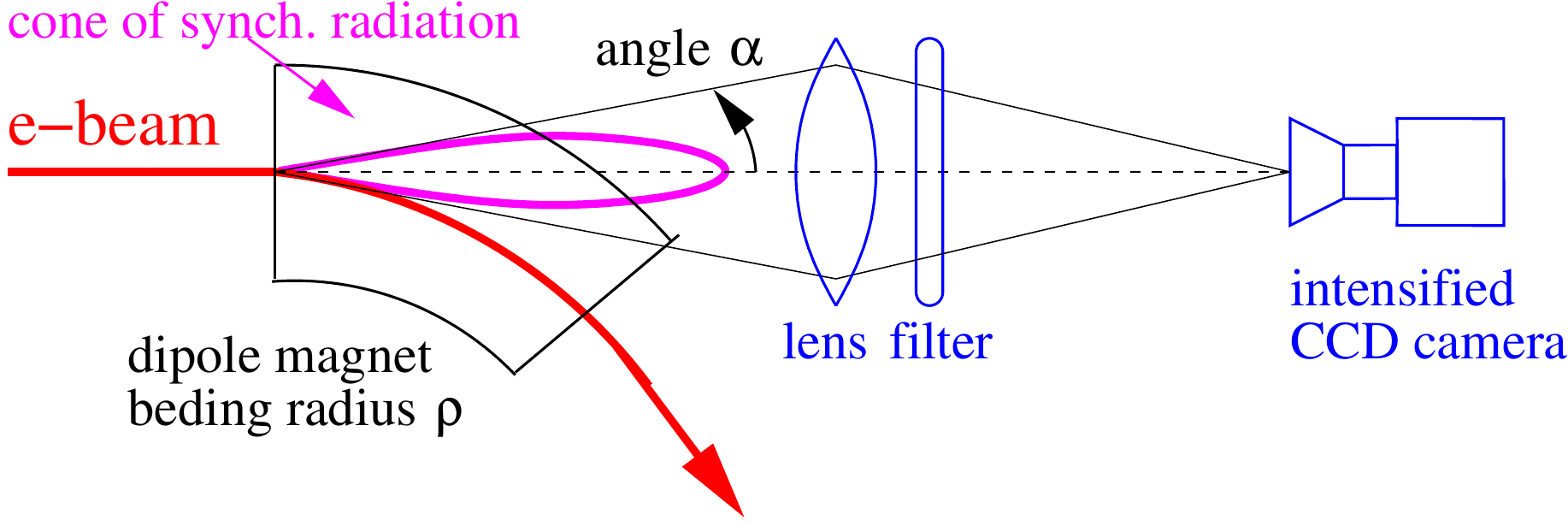} 
\caption{Scheme for a synchrotron radiation profile
monitor observing the radiation from a dipole.} 
\label{scheme-sr-general}
\end{figure}

The light emitted by the electron's bend in a dipole magnet can be used
to monitor its horizontal and vertical beam profile, as schematically 
shown in Fig.~\ref{scheme-sr-general} and reviewed in \cite{kube-dipac07}. . 
For diagnostic purposes, 
the optical part of the emitted spectrum is observed in most cases by using
optical band-pass filters. For this
wavelength, high-quality optics are available, and standard CCD or CMOS cameras can be
used. A typical resolution limit for the related optics is in the order of $\sim 100$ $\mu$m, as related to the~diffraction pattern created at the finite size lenses and filters \cite{kube-dipac07, wilke-biw94}. A higher resolution
can be achieved by observing the light from wigglers or undulators if
they are installed in the synchrotron.

\begin{figure}
\parbox{0.57\linewidth}{
\includegraphics*[width=95mm,angle=0]
{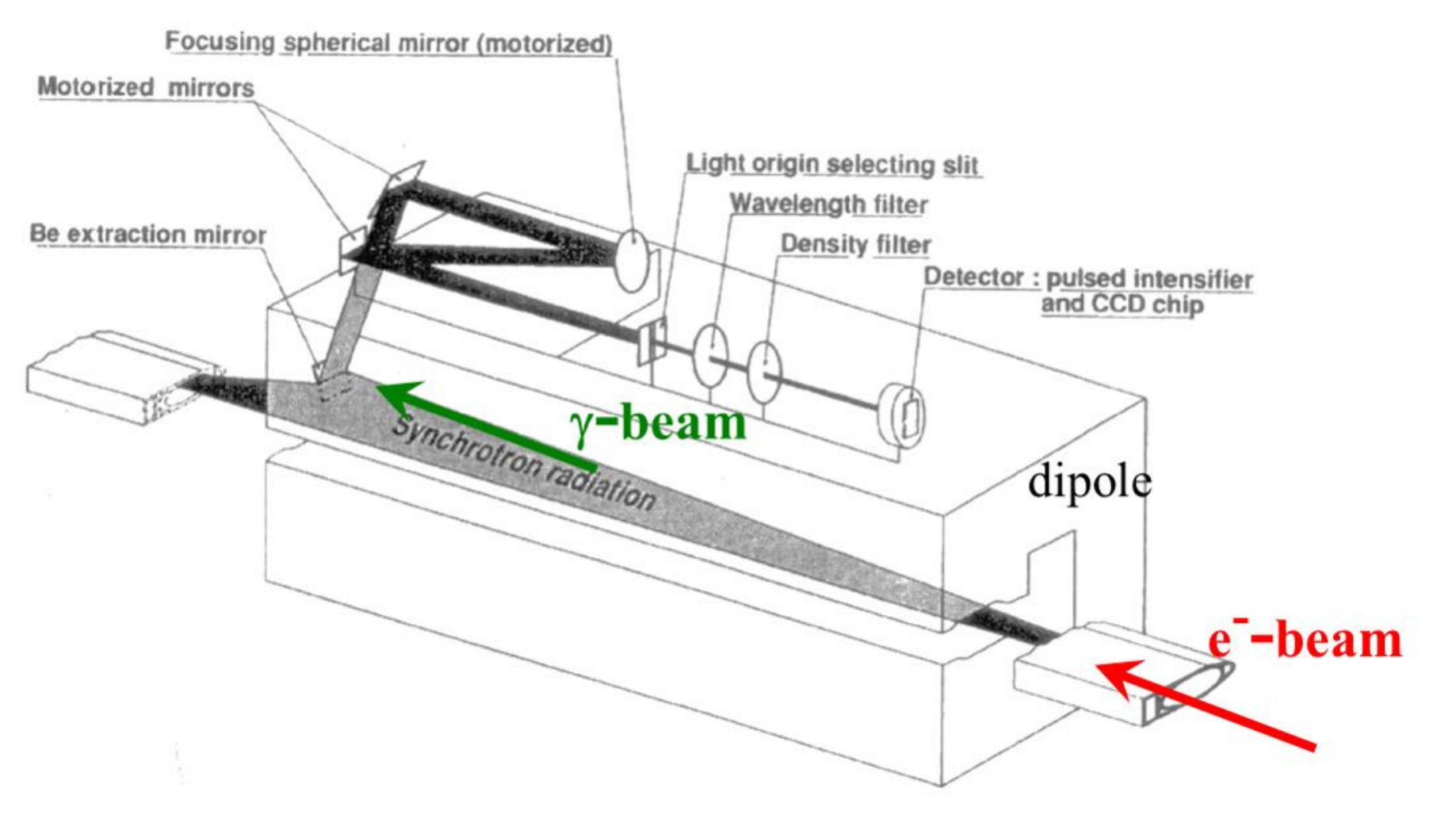} }
\parbox{0.42\linewidth}{
\centering
\includegraphics*[width=60mm,angle=0]
{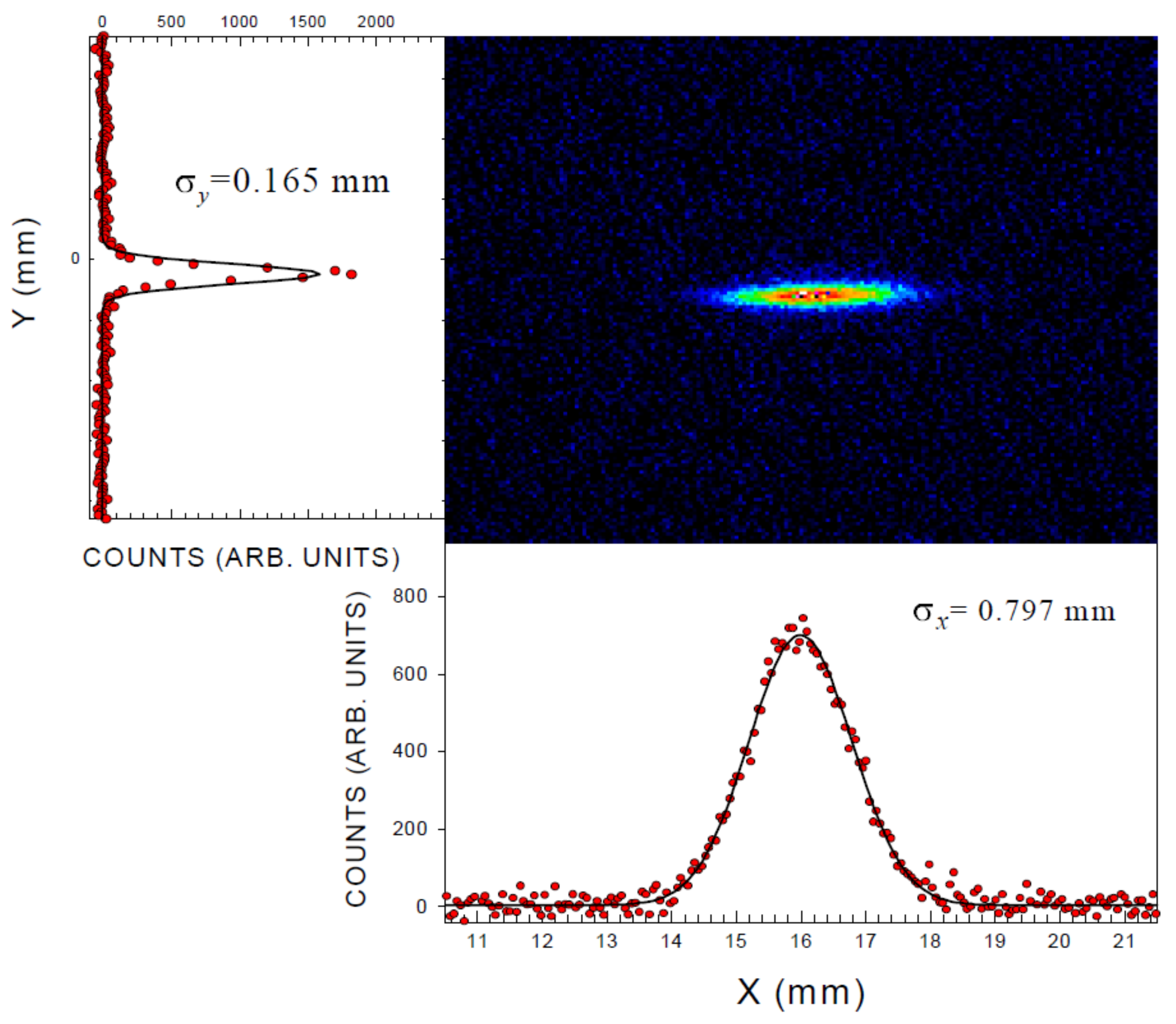} }
\caption{Left: The synchrotron radiation profile monitor at LEP. The optical system
is installed close to dipole magnet with bending radius of
3100 m \cite{bovet-pac91}. Right: Image of the electron beam in the APS accumulator ring by a~synchrotron radiation monitor \cite{yang-pac97}.}
\label{scheme-sr-lep}
\end{figure}

A realization of a synchrotron light profile monitor is shown in
Fig.~\ref{scheme-sr-lep} at CERN LEP \cite{bovet-pac91}. 
The~bending radius is here
3.1 km and the diffraction gives the most significant contribution to the~resolution by about $\sigma \sim 300$ $\mu$m, which is comparable to the real
electron beam size close to the final energy. The~setup consists of a metallic
mirror to deflect the light out of the plane of the beam. Owing to the~high power
of the synchrotron radiation, the mirror has to be cooled, and possible deformations can spoil the~optical quality. With the help of some curved mirrors, the light is
focused, and an optical filter selects the wavelength range of blue or UV
light and detected with a camera. Even though the
resolution is not too high,
this relatively straight forward system can be quite helpful due to
its non-destructiveness, as demonstrated in Fig.~\ref{scheme-sr-lep} 
from the synchrotron light source APS \cite{yang-pac97}. 

\begin{figure}
\centering
\includegraphics*[width=145mm,angle=0]
{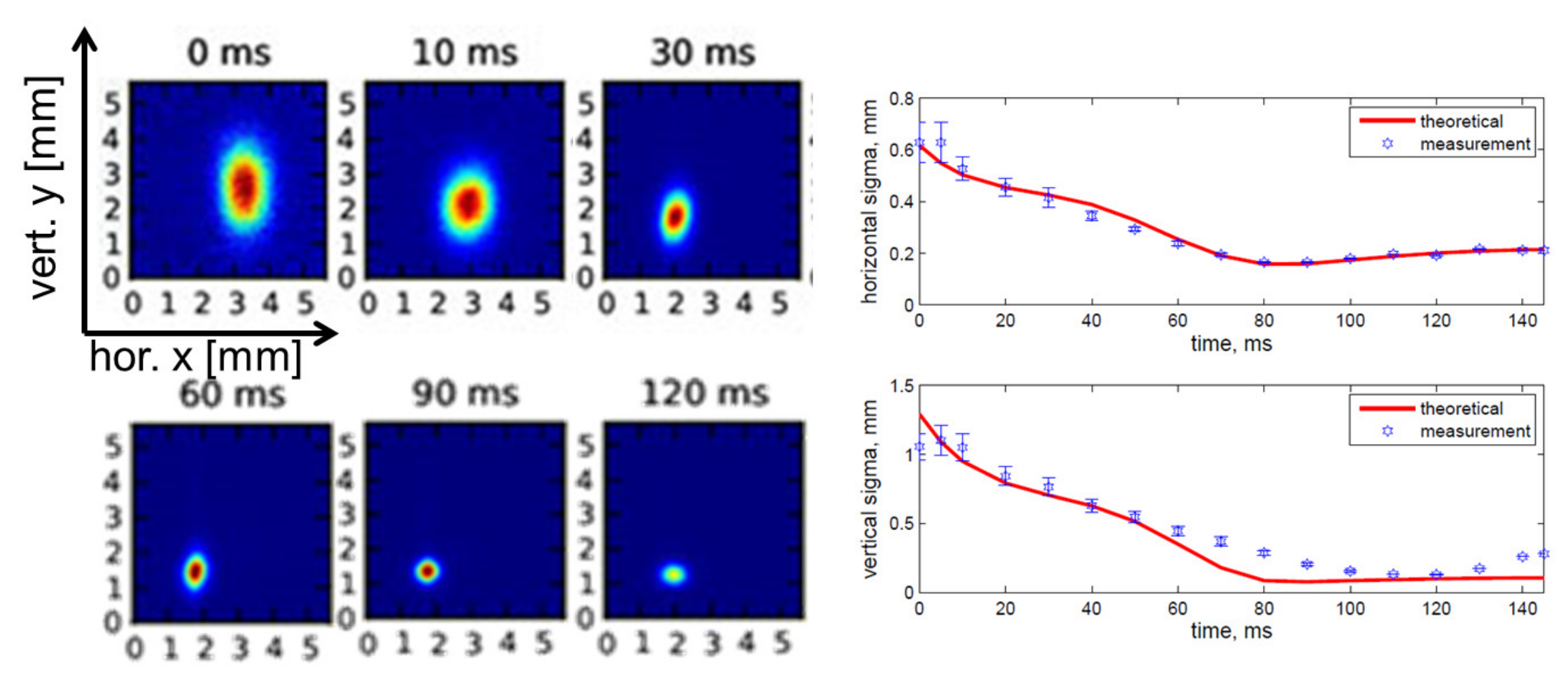} 
\caption{Left: Examples of two dimensional beam image by a synchrotron light monitor at ALBA booster synchrotron. Right: The fitted horizontal and vertical beam width during the acceleration 0.1 to 3 GeV within 130 ms and the comparison to simulations \cite{iriso-ipac11}.}
\label{ALBA-booster-adia-damping}
\end{figure}

An example for the application of a synchrotron radiation monitor is shown in Fig.~\ref{ALBA-booster-adia-damping} during the~acceleration at the ALBA booster synchrotron\cite{iriso-ipac11}. For the two-dimensional beam image, the width in the horizontal $\sigma_{x}$ and vertical $\sigma_{y}$ direction can be calculated and compared to simulations. A decrease of beam size is visible due to adiabatic damping. However, there an additional contribution to the beam width by the relative momentum spread $\Delta p /p_0$ of the beam particles and the non-zero dispersion $D$ at the measurement location as $sigma_{tot} = \sqrt{ \varepsilon \beta + (D\cdot \Delta p/p_0)^{2}} $ and the contribution by the quantum fluctuations related to the emission of photons \cite{iriso-ipac11}.

The diffraction limit of the standard monitor can be compensated by using an interference technique by
a double-slit as
known from astronomy. From the distance of the minima of the interference
pattern, the beam width can be calculated. Due to the non-coherent emission
by an ensemble of electrons, the fringes fade-out for large
beam 
sizes. A resolution down to the 1 $\mu$m range has been realized with this
method in combination with an elaborated analysis tool, see e.g. \cite{kube-dipac07, mitsh-cas98}.

Another method to achieve a higher resolution is related to the observation of much shorter wavelengths. X-ray pin-hole cameras are used for this purpose. 
Only the emitted x-rays, scraped
by a typically 20 $\mu$m aperture of high Z-material are recorded with a
detector. An example is described in \cite{tord-dipac07} A typical resolution concerning the beam width of 10 $\mu$m can be reached using such X-ray observation scheme \cite{kube-dipac07}; but this requires a more sophisticated technical realization than for the observation in the~optical range. 

At dipoles in transfer lines, such synchrotron light monitors could be installed as well, which is realized at some facilities, e.g. \cite{scheidt-dipac05}. However, there might be some technical problems as the amount of light emitted by the single pass of the electrons might be insufficient. Instead, OTR screens due to their more straightforward technical realization serve as a frequently used profile diagnostics in transfer lines.

\section{Measurement of transverse emittance}
\label{kap-emittance} 

The emittance describes the quality of a beam. Its determination is based on
profile measurements; it is unimportant what method of profile measurement is used, as long as it has an adequate resolution. In the
following, we start with a slit-grid device, where the spatial
coordinate is fixed by an aperture and the angle distribution is
measured. This is suited for particles having a range in matter 
below $\sim 1$ cm i.e. proton
or ions with $E_{kin}< 100$ MeV/u. The emittance can also be determined by
fitting the beam envelope, measured at one location with different focusing
conditions or at different locations. This can be applied to transfer lines
for all particle conditions. 
However, here it is problematic to include emittance
growth due to space-charge forces, i.e., a blow-up of the transverse size due to
the forces between the~charged particles. A detailed review on the involved
physics, technologies and data acquisition is given in \cite{stock-biw06}. 

In a synchrotron it is sufficient to
measure the beam profile at one location only. For the stationary state of 
stable storage the orientation of the ellipse is fixed for a given location$s$.
This is equivalent to the knowledge of the lattice functions dispersion 
$D(s)$ and $\beta$-function $\beta (s)$, is fixed. Therefore, only the~absolute value $\epsilon $ is of interest. The emittance is calculated from the beam width $\sigma$ via
\begin{equation} 
\epsilon = \frac{1}{\beta(s)} \left[
\sigma^{2} - \left( D(s) \frac{\Delta
p}{p}\right)^{2} \right] 
\end{equation} 
which is valid for the horizontal and vertical plane, respectively. If
only horizontal bends occur in a~facility, the dispersion in vertical
direction is in most cases $D_y=0$ and equation 
simplifies to $ \epsilon_y = \frac{ \sigma_y^{2}}{\beta_y(s)} $. Hence, the following discussion concerns transfer lines only.

\subsection{Definition of the emittance}
\label{kap-emi-general}
\begin{figure}
\centering
\includegraphics*[width=65mm,angle=0] 
{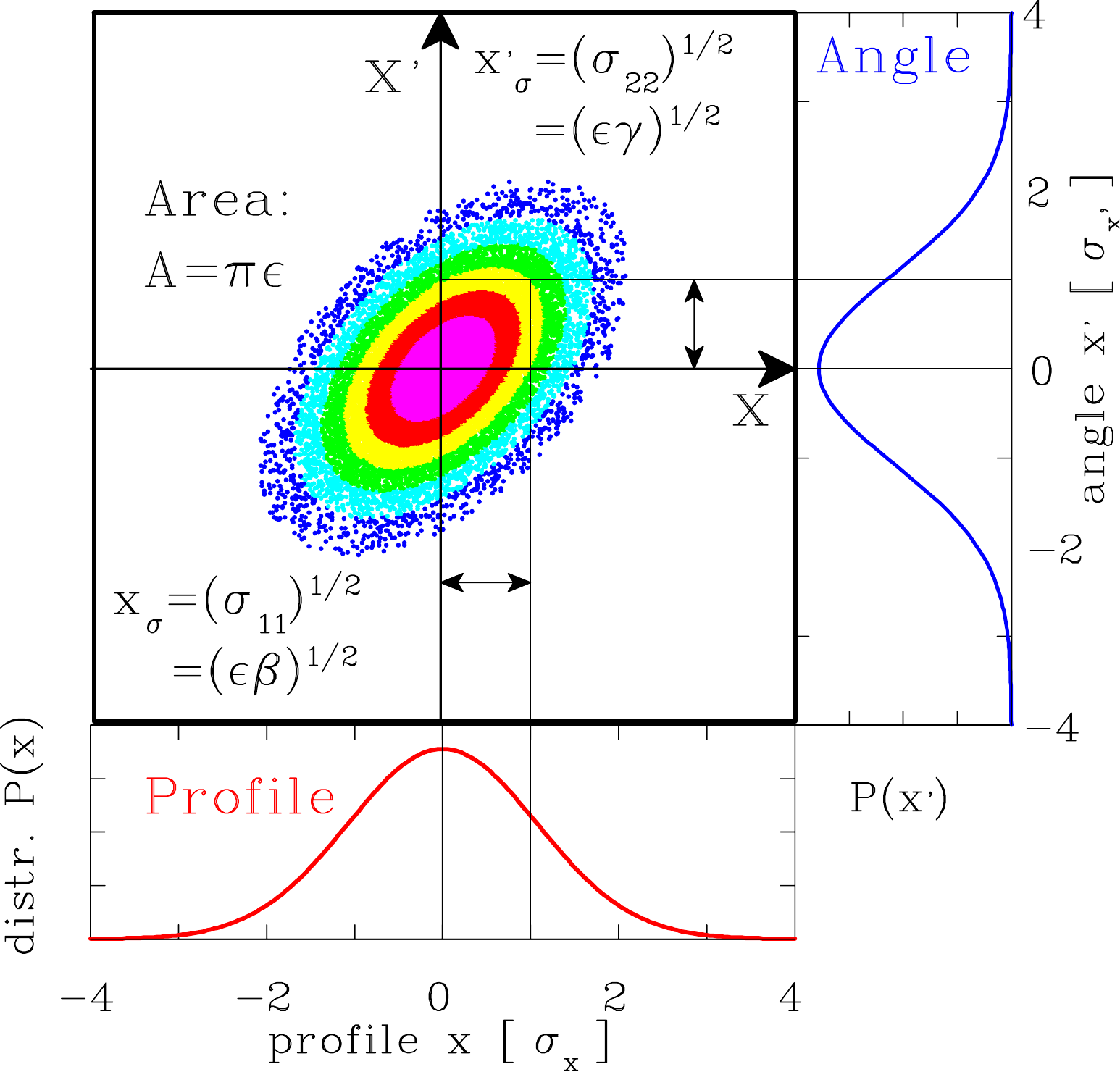}
\vspace*{-0.4cm}
\caption{The emittance ellipse and the projection to space- and angle 
coordinates for a Gaussian density distribution. The values of the 
independent variables are given in units of the standard deviation.}
\label{scheme-emittance}
\end{figure}

The motion of the beam particle can be described by a linear 
second-order differential
equation. This assumes the absence of any non-linear coupling, like space
charge 
forces or beam-beam effects as well as coupling between the two transverse and the 
longitudinal planes. 
The beam quality is given by the phase space volume, which is in this
case a constant
of the motion. The emittance for one plane is defined by 
\begin{equation}
\epsilon_{x}= \frac{1}{\pi} \int_{A} \mbox{d}x \mbox{d}x',
\end{equation}
where $A=\pi \epsilon$ is the area of the phase space occupied by the
beam, see Fig.~\ref{scheme-emittance}. 
A determination of the~emittance is equivalent to the determination of
the distribution of the spatial coordinate $x$ (i.e. the~beam profile), the
distribution in angle $x'$ and the correlation between $x$ and $x'$.

The interpretation of an area assumes a hard-edge, homogeneous 
distribution. A more
realistic case uses a density 
distribution $\rho({x,x'})$ defined at the position of the vector $\vec{x}=
(x,x')$ for each location $s$ along the beam path. 
The profile distribution $P(x)$ is obtained by integrating the density $\rho$
over $x'$ as $P(x) = \int \rho (x,x') dx'$; the parameter
$\sqrt{\sigma_{11}}$ is the standard deviation of this distribution.
$\sqrt{\sigma_{22}}$ is the corresponding value for the angular distribution
$P(x')$ obtained by integrating $\rho(x,x')$ over $x$ as $P(x') = \int \rho
(x,x') dx$. $\sigma_{12}$ is called covariance as it 
describes the correlation between $x$ and $x'$ and
is related to the orientation of the ellipse in the phase space. 
With the help of these three parameters
$\sigma_{ij}$ the~so-called beam matrix {\boldmath $\mathbf{\sigma}$} can be
defined in the following way:
\begin{equation}
\mbox{\boldmath $\mathbf{\sigma}$}(s) = \left( 
\begin{array}{ll}
\sigma_{11}(s) & \sigma_{12}(s) \\ 
\sigma_{12}(s) & \sigma_{22}(s)
\end{array}
\right)
\end{equation}
at one location $s$ in the beam line. The beam matrix {\boldmath $\sigma$} is
a representation of the beam ellipse at this location $s$ and varies along the
beam path. It is used in connection with the transfer matrix $\mathbf{R}$,
which describes the action of the optical elements along the beam path in a 
linear approximation (i.e. for optical elements like drifts,
dipoles, quadrupoles or solenoids) by
\begin{equation}
\mbox{\boldmath $ \sigma $}(s_{1}) = \mathbf{R} \cdot \mbox{\boldmath $\sigma
$} (s_{0}) \cdot \mathbf{R^{T}}
\label{eq-sig-general}
\end{equation}
from a location $s_{0}$ to $s_{1}$. 

The absolute value of the emittance at each location can be defined using this
notation as
\begin{equation}
\epsilon_{x} = \sqrt{\det \mbox{\boldmath $ \sigma $}} =
\sqrt{\sigma_{11}\sigma_{22}- \sigma_{12} ^{2}}
\label{emi-und-sigma}
\end{equation} 
and corresponds to an occupation of 15 \% of the full phase space area. The
unit of this value is m$\cdot$rad or more frequently mm$\cdot$mrad. To be
consistent with the geometrical interpretation of an area surrounded by an
ellipse, the number is multiplied by $\pi$; in this case 39 \% of the beam is
inside the area of $\pi \cdot \sqrt{\sigma_{11}\sigma_{22}- \sigma_{12}
^{2}}$. In other words, the Gaussian beam quality at a given location $s$
is fully described by the beam matrix $\mbox{\boldmath $\sigma$} (s)$.

Frequently the Twiss parameters are used, which are the beam matrix
elements normalized by the~emittance as
\begin{equation}
\alpha = -\sigma_{12}/\epsilon ~~~~~~~~~~
\beta = \sigma_{11}/\epsilon ~~~~~~~~~~
\gamma = \sigma_{22}/\epsilon.
\end{equation}
The beam matrix is then
\begin{equation}
\mbox{\boldmath $ \sigma $} = \epsilon \cdot \left( 
\begin{array}{cc}
\beta & -\alpha \\ 
-\alpha & \gamma
\end{array}
\right) ~~~.
\end{equation}
The width of profile- and angular distribution is given by 
\begin{equation}
x_{\sigma} = 
\sqrt{\sigma_{11}} = \sqrt{\epsilon \beta} ~~~~~~~\mbox{and}~~~~~~~
x'_{\sigma} = \sqrt{\sigma_{22}} = \sqrt{\epsilon \gamma} .
\end{equation}
Their geometric meaning is one standard deviation of the transverse profile 
and angular distribution of the beam. The geometrical size of the phase
space ellipse is changed along the beam pass $s$; therefore the parameters
$\alpha (s)$, $\beta(s)$ and $\gamma (s)$ are functions on the position
$s$. In particular, for a synchrotron $\beta(s)$ is called the beta-function,
describing the beam size via 
$x_{\sigma} (s) =\sqrt{\epsilon \cdot \beta(s)}$. 
For any arbitrary phase-space
distribution the beam emittance can be calculated via the statistical moments of a 2-dimensional distribution $\rho (x,x')$.

To describe the beam quality via the emittance
the $rms$ value ({\em r}oot {\em m}ean {\em s}quare) can be calculated as
\begin{equation}
\epsilon_{rms} = \sqrt{\det \left(
\begin{array}{cc}
\langle x^{2}\rangle & \langle xx'\rangle\\
\langle xx'\rangle & \langle {x'}^{2}\rangle\\
\end{array}
\right)} =
\sqrt{\langle x^{2}\rangle \langle {x'}^{2}\rangle -
\langle xx'\rangle^{2}} 
\label{eq-emi-statdev}
\end{equation}
using the statistical moments $\langle ...\rangle$; this is equivalent to Eq.~(\ref{emi-und-sigma}).

The emittance is a quantity defined in the laboratory frame due to the
definition of the divergence. When a beam is accelerated, the divergence
shrinks ('adiabatic damping') due to the changing ratio of longitudinal
velocity $v_{s}$ to transverse velocity $v_{x}$: 
$x'=v_{x}/v_{s}$. To compare emittance for
different longitudinal momenta $p_{s}=m_{0}\cdot \gamma_{rel} \cdot v_{s}$, 
the normalized emittance $\epsilon _{norm}$ 
\begin{equation}
\epsilon _{norm} = \frac{v_{s}}{c} \gamma_{rel} \cdot \epsilon
\label{eq-emitt-norm}
\end{equation}
is referred to a value $v_{s}/c \cdot \gamma_{rel}=1$ with $c$ is the velocity of
light and $\gamma_{rel}= 1/\sqrt{1-(v_{s}/c)^{2}}$ is the~relativistic Lorentz factor. The normalized emittance is constant
under ideal accelerating conditions. 

A measurement of emittance means a determination of the numerical value of
$\epsilon$ as well as the~orientation and shape of the phase space 
distribution represented by the matrix elements $\sigma_{{ij}}$ or the Twiss parameters $\alpha, \beta, \gamma$ correspondingly. 

\subsection{Slit-grid method}
\begin{figure}
\centering
\includegraphics*[width=110mm,angle=0]
{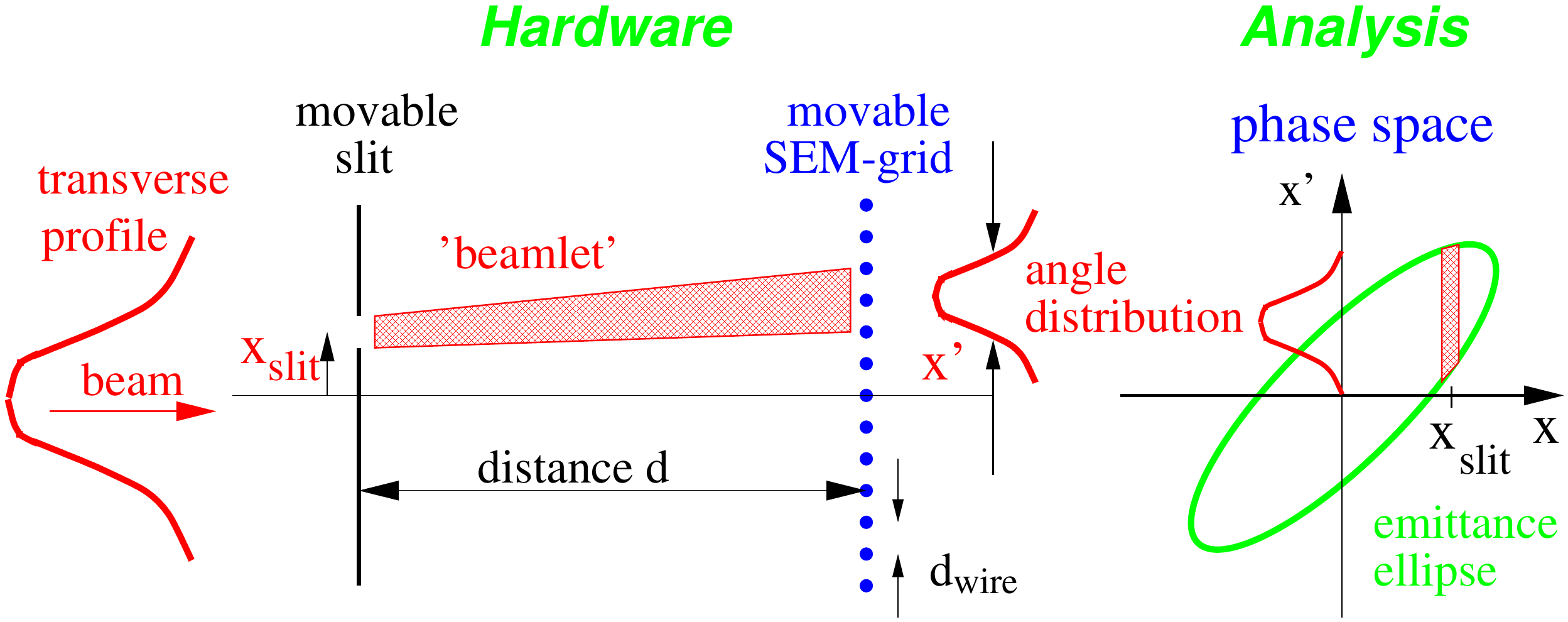} 
\caption{Scheme of a slit-grid emittance measurement device.}
\label{scheme-slit-grid}
\end{figure}

\begin{figure}
\centering
\includegraphics*[width=65mm,angle=0]
{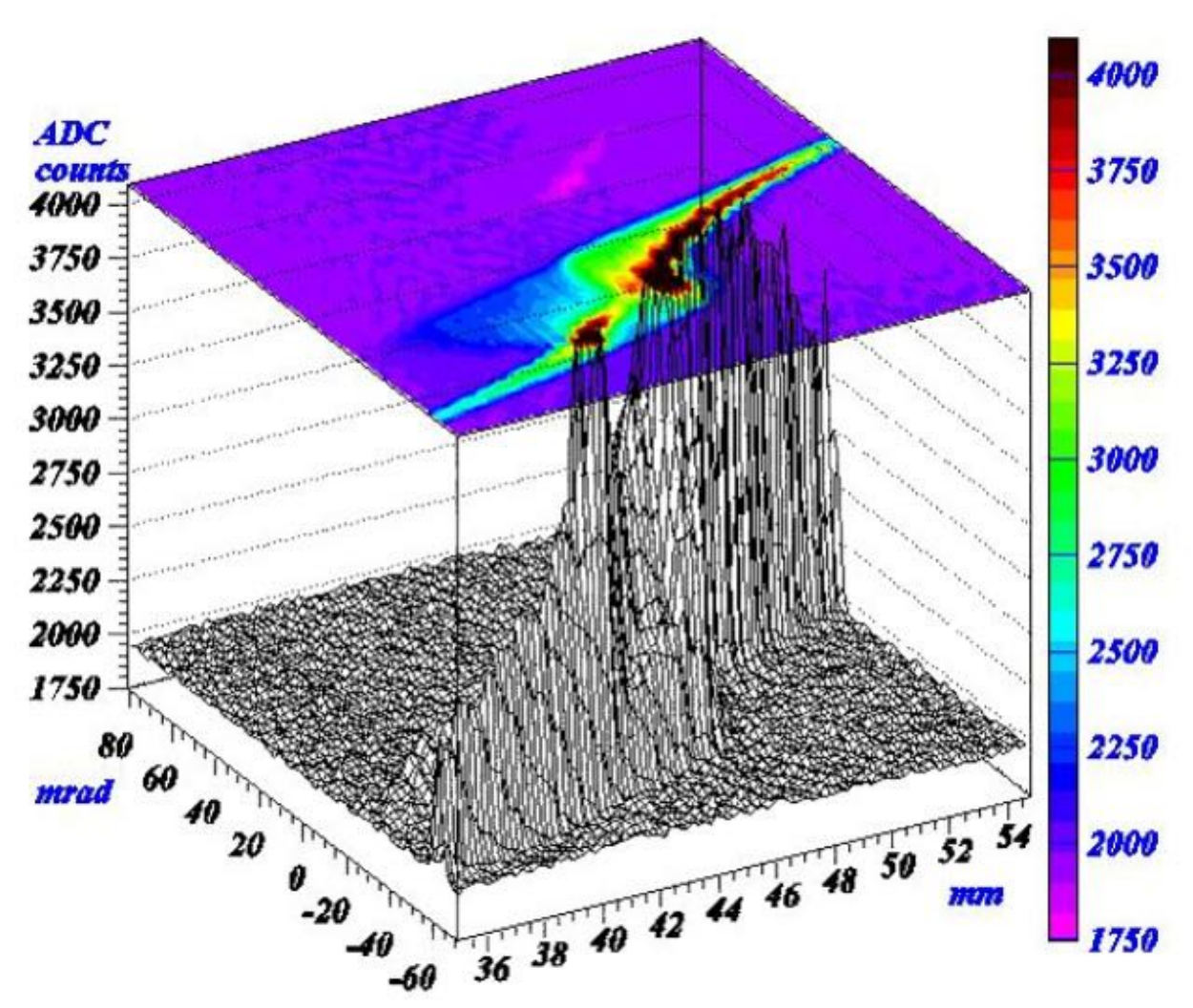} 
\caption{Emittance measurement using a slit-grid device with a low energy Ar$^{4+}$ ion of 60 keV at the Sprial2 facility depicting an mountain and contour plot \cite{ausset-dipca09}.}
\label{slit-grid-3dim}
\end{figure}

A popular method at proton or ion LINACs is the slit-grid device depicted in 
Fig~\ref{scheme-slit-grid}, where the beam particles have a penetration depth below 
1 cm. Here the position $x$ is fixed for one direction with a~thin slit having an opening of typically 0.1 to 0.5 mm to filter out only
a small fraction of the beam at a~known location. In the
perpendicular direction, the full beam is transmitted to get a large
signal. The~angle $x'$ is determined with an SEM-grid having a distance from the
slit of 10 cm to 1 m depending on the~ion velocity. In the field-free drift
space the trajectories of the particles, combined in a 'beamlet', 
are straight lines.
The angle distribution gives the contribution to the~emittance plot in the phase space at the
slit location. The slit is then scanned through the beam to get all
positions. After making the full scan, the~emittance is plotted, and the
$rms$-value of the emittance $\epsilon_{rms}$ is calculated using the statistical moments as given in Eq.~(\ref{eq-emi-statdev}) from the measured data. 
A fit with an elliptical shape is
drawn to the~data, and the Twiss parameters can be calculated. An example is shown in Fig.~\ref{slit-grid-3dim} for a low energy ion beam as a
contour or three-dimensional plot. This method can also determine more pathological phase-space
distributions, not only Gaussian distributions. Close to an ion source, this
happens quite often due to the~large space-charge forces or the large profile width, where aberrations of the magnets could be significant.

According to Fig.~\ref{scheme-slit-grid}, the resolution for the space coordinate $\Delta x$
is limited by the slit width $\Delta x = d_{slit}$. The angle resolution
$\Delta x'$, measured at the distance $d$, 
is given by the radius of the wire $r_{wire}$ and the width of the slit
resulting in $\Delta x' = (d_{wire} +
2r_{wire})/d$. The size of discrete elements in the phase space is given by 
$\Delta x \cdot \Delta x'$. This leads to a discretization error, in
particular in the case of small beam sizes (focused beam) 
or small angle
distributions (parallel beam). The resolution is improved by scanning the SEM-grid in steps lower than the distance of the wires $d_{wire}$, increasing the density of the~discrete elements in the~phase space analysis. This can lead
to overlapping elements because their size $\Delta x \cdot \Delta x'$ 
stays constant. The same holds for a movement of the slit with a step size
lower than the slit width.

\subsection{Quadrupole variation}
\label{kap-emi-quad}
\begin{figure}
\centering
\includegraphics*[width=95mm,angle=0]
{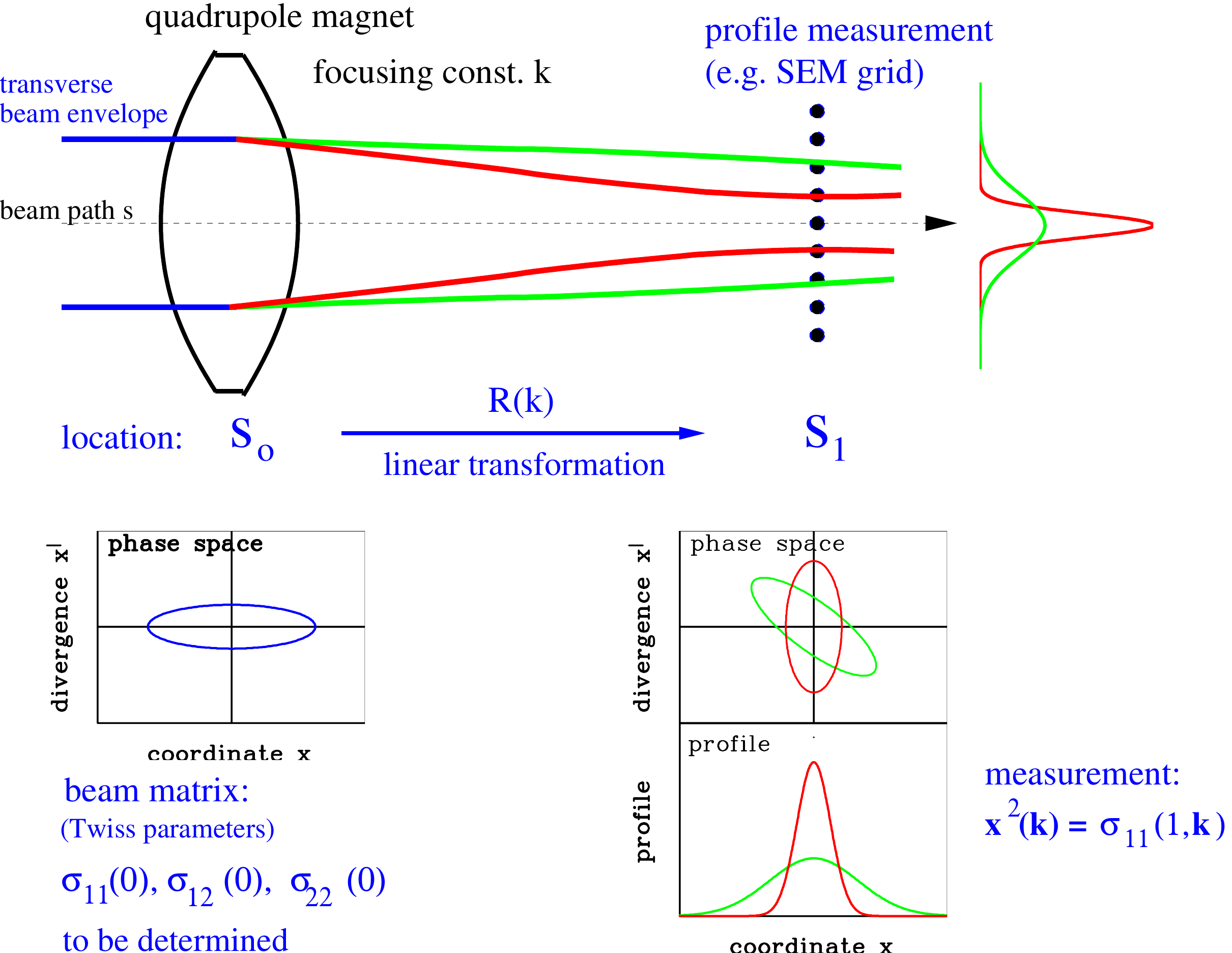} 
\caption{Variation of a quadrupole strength 
for the determination of the emittance at the location $s_{0}$.}
\label{scheme-quad-variation}
\end{figure}

At a transfer line, the emittance can be determined from a series of profile measurements
by changing the focusing strength of a quadrupole as schematically shown in
Fig.~\ref{scheme-quad-variation}. 
To derive the emittance from such a measurement, 
linear transformations are used. 

For a straight, non-dispersive 
transfer line, the
transformation from a location $s_{0}$ to $s_{1}$ is given by the $2\times 2$
transfer matrix $\mathbf{R}$. 
A determination of the emittance at the position $s_{0}$ is equivalent to the~evaluation of the beam matrix {\boldmath $\sigma $} 
at this position. The beam
matrix is transformed to a second location $s_{1}$ with the help of the
product of 
transfer matrices for the individual elements $\mathbf{R}= \prod
\mathbf{R_{elements}}$ from the~quadrupole to the profile measurement
location via
\begin{equation}
\mbox{\boldmath $ \sigma $} (1) = \mathbf{R} \cdot 
\mbox{\boldmath $ \sigma $} (0) \cdot \mathbf{R^{T}}
\label{eq-sig-allgemein}
\end{equation}
The beam width $x_{rms} (1)$ is measured at $s_{1}$ and the equation for the
element 
$\sigma_{11} (1) $ is given by
\begin{equation}
x_{rms}^{2} (1) \equiv 
\sigma_{11}(1) = R_{11}^{2} \sigma_{11} (0) + 2 R_{11}R_{12}\sigma_{12} (0) +
R_{12}^{2} \sigma_{22} (0).
\label{eq-sigma-quad-vari}
\end{equation}
This is a linear equation for the unknown three beam matrix elements 
$\sigma_{ij} (0)$ at
location $s_{0}$, in Fig.~\ref{scheme-quad-variation} 
in front of the focusing quadrupole magnet.
To get a solution, we need at least three different settings
of the~quadrupole strength $k_{i}$, and therefore different transfer matrices 
$\mathbf{R}(k_{i})$ leading to three different readings of the profile
width as depicted in Fig.~\ref{scheme-quad-variation}. 
Assuming $i=1,2 ... n$ different settings of the quadrupole strength $k_{1},
k_{2} ... k_{n}$ and $n$ measurements of the beam width 
$x_{rms}^{2} (1, k_{i}) \equiv \sigma_{11}(1, k_{i})$ 
a redundant system of linear equations is obtained in the form 
\begin{eqnarray}
\hspace*{-0.8cm} \nonumber
\sigma_{11}(1, k_{1}) &=& R_{11}^{2}(k_{1}) \cdot \sigma_{11} (0) + 2
R_{11}(k_{1})R_{12}(k_{1})\cdot \sigma_{12} (0) + R_{12}^{2}(k_{1}) \cdot
\sigma_{22} (0) \mbox{~~focusing~}k_{1} \\ \nonumber
\hspace*{-0.8cm} 
\sigma_{11}(1, k_{2}) &=& R_{11}^{2}(k_{2}) \cdot \sigma_{11} (0) + 2
R_{11}(k_{2})R_{12}(k_{2})\cdot \sigma_{12} (0) + R_{12}^{2}(k_{2}) \cdot
\sigma_{22} (0) \mbox{~~focusing~}k_{2} \\ \nonumber
& :&\\ 
\hspace*{-0.8cm} 
\sigma_{11}(1, k_{n}) &=& R_{11}^{2}(k_{n}) \cdot \sigma_{11} (0) + 2
R_{11}(k_{n})R_{12}(k_{n})\cdot \sigma_{12} (0) + R_{12}^{2}(k_{n}) \cdot
\sigma_{22} (0) \mbox{~~focusing~}k_{n} 
\label{system-eq-quad-vari}
\end{eqnarray}
The solution of this system are the values of the beam matrix
$\sigma_{{ij}}(0)$ (or equivalently the Twiss parameters) 
at the location $s_{0}$, the
entrance of the quadrupole magnet. With theses values, the size and
orientation of the phase space ellipse is fixed.
For three measurements ($k_{1}, k_{2}, k_{3}$) 
we can have a unique solution, but nothing is
learned about the errors. Therefore, more than three measurements have to be performed, leading
to a redundant system of linear equations. The solution is performed by a least-square fit to
the~best parameters of $\sigma_{ij} (0)$ or by solving the linear regression
problem via so-called normal equations. Both algorithms are described in textbooks of Linear Algebra or Numerical Mathematics. 

\begin{figure}
\centering
\includegraphics*[width=70mm,angle=0]
{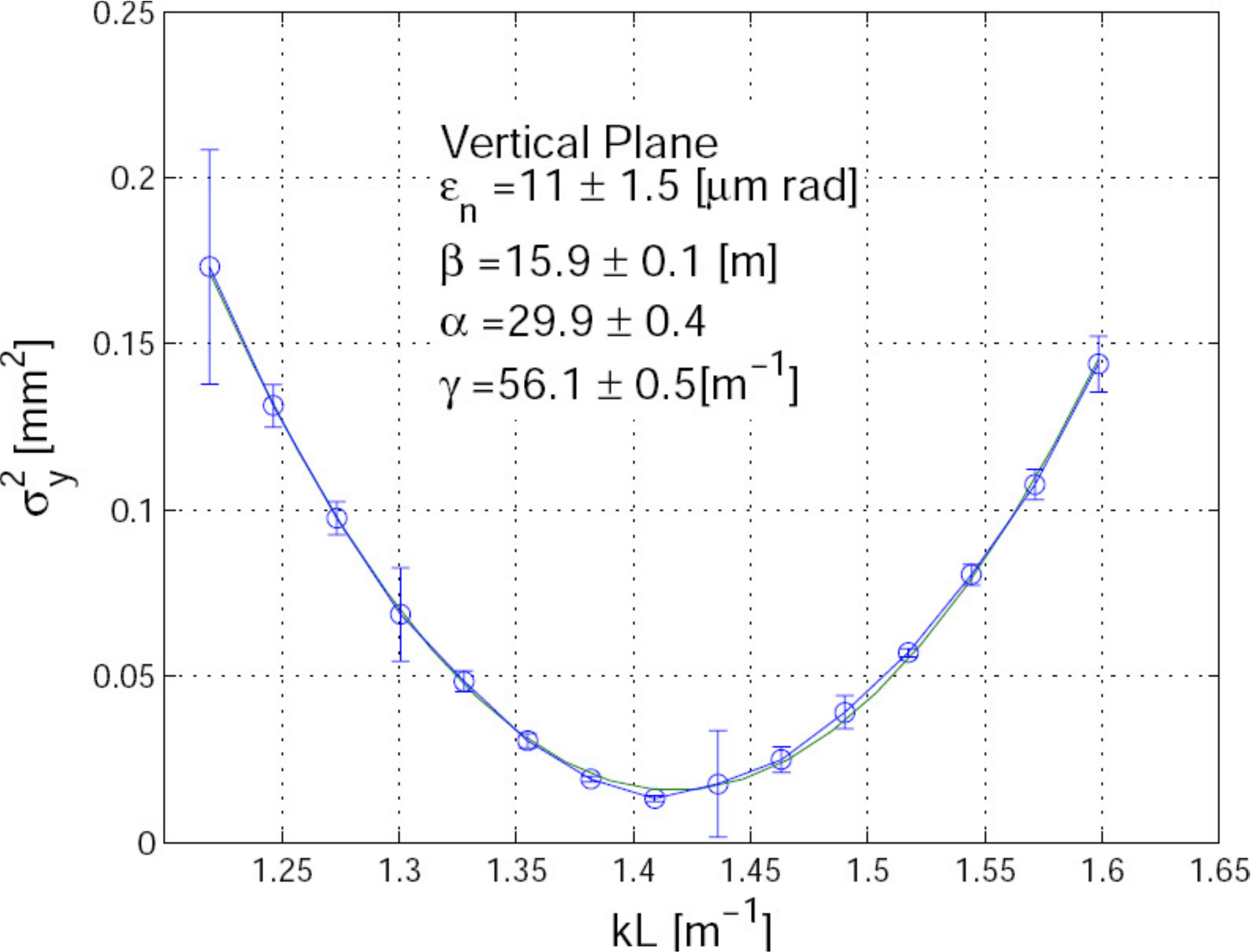} 
\caption{Profile width determined with a YAG:Ce screen at Elettra electron
LINAC for a
quadrupole variation and the parabolic fit \cite{penco-epac08}.}
\label{result-quad-variation}
\end{figure}

An example is shown in
Fig.~\ref{result-quad-variation} performed at the electron LINAC at Elettra for a beam with 107~MeV using a YAG:Ce scintillation screen for profile
determination. To get a small
error for the~emittance determination, it is recommended to pass a beam waist
by the quadrupole variation. Only in this case, a~parabola can be fitted
through the profile data with sufficient accuracy.

Instead of solving the redundant system of linear equations depicted in Eq.~(\ref{system-eq-quad-vari}), one can start
from the parabola fit of the beam size squared as a function of the quadrupole
strength as shown in Fig.~\ref{result-quad-variation}. Most frequently
the transfer line comprises of a quadrupole followed by a drift toward
the~profile measurement 
location. A quadratic dependence is observed for the following reason: 
Assuming a~thin lens approximation with a focal length $f$ of the quadrupole
action one can write the transfer matrix as:
\begin{equation} 
\mathbf{R_{focus}}=\left( 
\begin{array}{cc}
1 & 0 \\ 
-1/f & 1
\end{array}
\right)
\equiv
\left(
\begin{array}{cc}
1 & 0 \\ 
K & 1
\end{array}
\right).
\end{equation} 
After a drift of length $L$ one gets the transfer matrix of the
transfer line 
\begin{equation} 
\mathbf{R} = \mathbf{R_{drift}} \cdot \mathbf{R_{focus}}
=\left( 
\begin{array}{cc}
1 + LK & L \\ 
K & 1
\end{array}
\right).
\end{equation} 
Inserting this matrix into Eq.~(\ref{eq-sig-allgemein}) for the transformation 
of
the beam matrix $\mbox{\boldmath $ \sigma $} (1) = \mathbf{R} \cdot 
\mbox{\boldmath $ \sigma $} (0) \cdot \mathbf{R}^{T}$ 
one gets for the measured beam matrix element 
$\sigma_{11} (1) = x^{2}_{rms}(1)$:
\begin{equation}
\sigma_{11} (1) = L^{2} \sigma_{11}(0) \cdot K^{2} 
+ 2 \left( L \sigma_{11}(0) + L^{2} \sigma_{12}(0) \right) \cdot K 
+ \sigma_{11}(0) + 2 L \sigma_{12} (0) + L^{2} \sigma_{22}(0) ~~~.
\label{eq-emi-fit-messung}
\end{equation}
This is the expected quadratic function in the quadrupole gradient $K$. 
From the beam
width measurement for various quadrupole settings a parabola fit is performed
as shown in Fig.~\ref{result-quad-variation}. This yield the beam matrix elements $\sigma_{ij}$ or the Twiss parameters $\alpha, \beta, \gamma$ as well as the albolute value of the beam emittance $\epsilon$.

In the above discussion, it is assumed that the method is
performed at a location where no dispersion $D$ is
present. The lattice function dispersion $D$ gives rise to an
additional contribution
to the~beam width $\Delta x$ by the momentum deviation $\Delta p$ of the
particles with respect to the mean momentum $p_0$ as
$\Delta x= D \cdot \Delta p/p_0$. The dispersion effect 
is added in a quadratic manner to the measured beam width as $x^2_{rms}=
\sigma_{11}(0) + (D\cdot \Delta p/p_0)^2$. To avoid the measurement of
the lattice function $D$ and the~beam's longitudinal momentum spread, the location for 
emittance determination should be dispersion free.
The~described method is based on linear transformations and conservation of the
emittance during the manipulation by the quadrupole. Moreover, an elliptical shape of the
emittance is assumed. Depending on the~general beam properties, this
is a reasonable assumption as long as non-linear effects, like an aberration of
magnets by higher-order field contributions or space-charge
forces are low. If the~investigations are done with intense beams, an emittance blow-up may occur, depending on the particle
density, i.e., on the~beam size. However, the size is to be changed, and even a waist
has to be created. For high intensity beams, self-consistent,
iterative algorithms
have to be applied to get an estimation of the emittance at the~quadrupole
entrance, see, e.g. \cite{schulz-linac00,dimov-hb16}.

\section{Measurement of bunch structure}
\label{kap-longi}

\begin{figure}
\parbox{0.45\linewidth}{ \centering \includegraphics*[width=31mm,angle=0]
{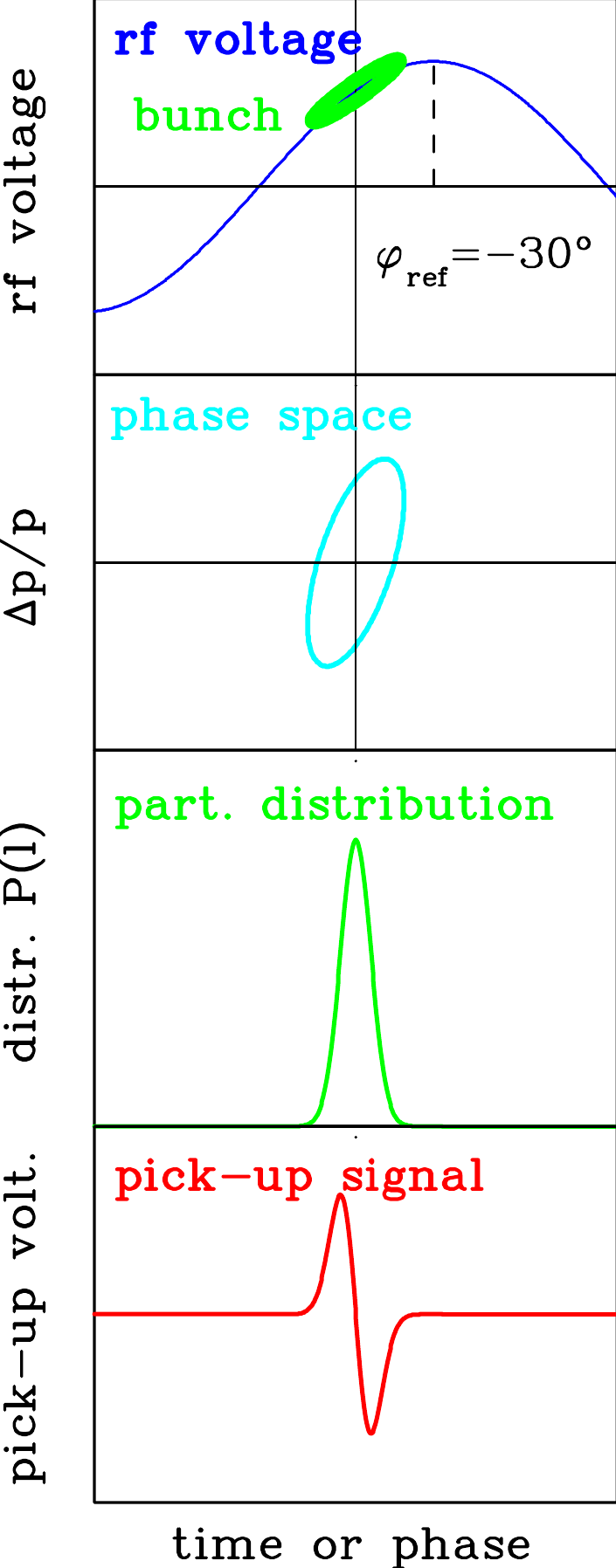}
} \parbox{0.45\linewidth}{ \centering \includegraphics*[width=65mm,angle=0]
{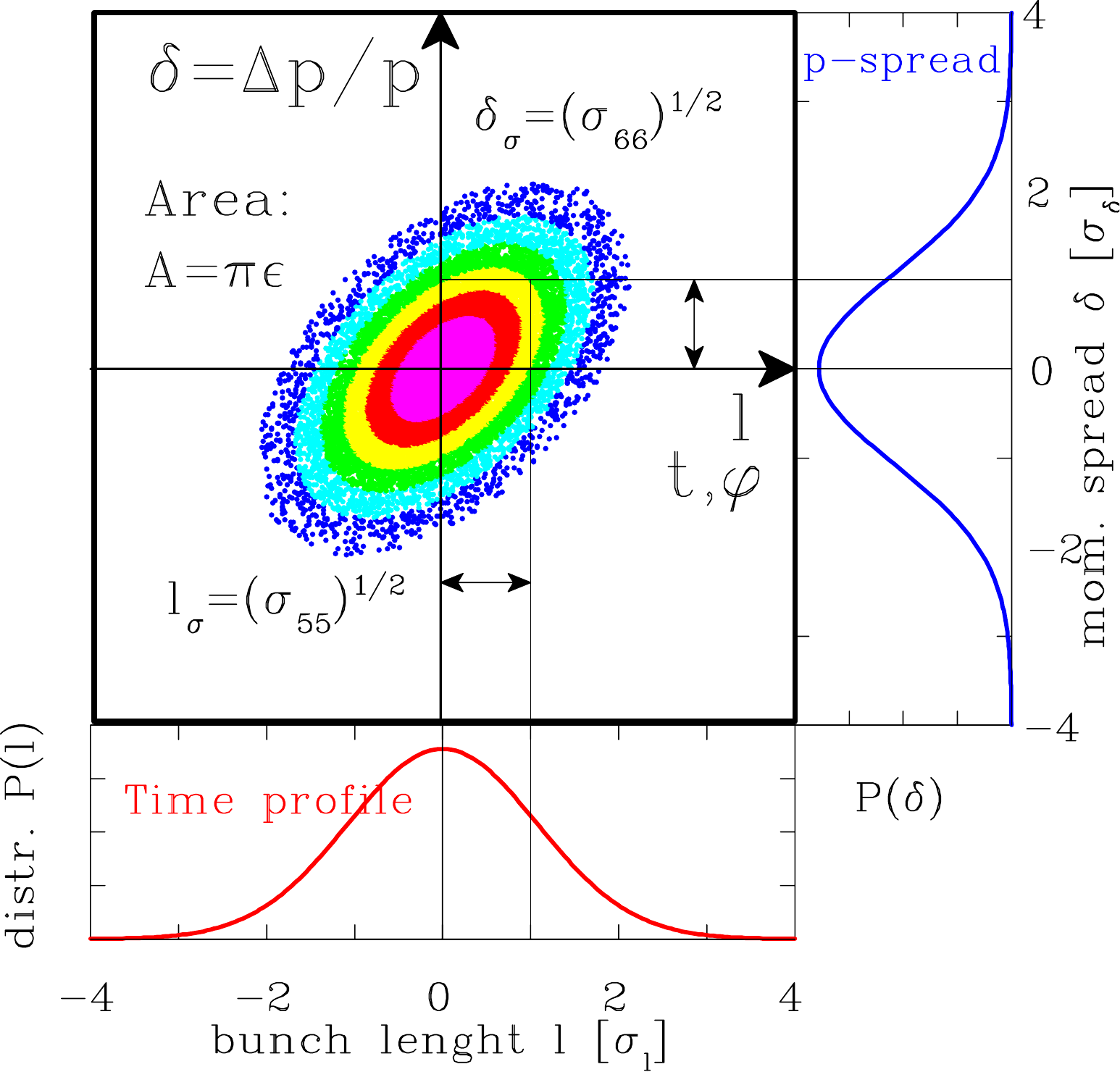} }
\caption{Left: 
The relation between the accelerating voltage and the longitudinal
emittance as well 
as the time distribution of the bunches as measured with the differentiated
pick-up signal. Right: The longitudinal phase space plot of a Gaussian
distribution.}
\label{scheme-longi-pickup}
\vspace*{-0.1cm}
\end{figure}

Longitudinal parameters are as important as the transverse ones. The
longitudinal phase space is spanned by:
\begin{itemize} 
\item The longitudinal spread of the bunch $l$ is given in units of 
length (mm), time (ns)
or phase ($^{o}$~degrees with respect to the accelerating frequency), 
see Fig.~\ref{scheme-longi-pickup}. The mean
value is the centre of the bunch relative to the rf or relative to the ideal
reference particle. The corresponding transverse value is the transverse
beam profile. For an un-bunched dc beam, e.g. in a proton 
storage ring, the~quantity is of no meaning.
\item The momentum spread $\delta = \Delta p/p_0$ is the deviation relative to
the momentum $p_0$ of the ideal reference particle. Instead of the momentum, it is sometimes common for proton LINACs and cyclotrons 
to relate the quantity beam energy $\Delta
E_{kin}/E_{kin}$ or even only $\Delta E_{kin}$ is given. The~corresponding transverse value is the beam divergence distribution.
\end{itemize} 
The value of emittance $\epsilon_{long}$ is given by the product of the
two quantities
\begin{equation}
\epsilon_{long} = \frac{1}{\pi} \int_{A} \mbox{d}l \mbox{d}\delta.
\end{equation}
where $A$ is the area of the phase space occupied by the beam particles, 
see Fig.~\ref{scheme-longi-pickup}.
Linear transformations can be applied in the same way as for the transverse
case, see Section~\ref{kap-emi-general}. The normalized longitudinal
emittance
\begin{equation}
\epsilon_{long}^{norm} = \frac{v_{s}}{c} \gamma_{rel} \cdot \epsilon_{long}
\end{equation}
is preserved under ideal conditions with $v_{s}$ is the longitudinal
velocity and $\gamma_{rel}$ is the relativistic Lorentz factor.

Using a pick-up or a fast current transformer, the projection of the phase space on the 
time axis could be 
determined for some beam parameters, resulting in the bunch position and bunch structure. This is the case for most proton synchrotrons, and with the help of tomographic reconstruction, the~full longitudinal phase space can be determined from a series of turn-by-turn bunch measurement, see, e.g. \cite{hancock-prab,hancock-www-note}. However, some conditions have to be
fulfilled for a pick-up or transformer to guarantee an adequate bunch shape measurement:
Firstly, the signal from the pick-up or transformer must reflect the bunch shape, i.e. the frequency spectrum of the bunch must fit in the bandwidth of the recording electronics; for short bunches, e.g. at high energy electron accelerators this might not be the case, see Section~\ref{kap-streak-camera-meas}. Secondly, the bunch has to be much longer in the longitudinal direction than the
pick-up. For bunches with a length comparable to the pick-up, a further
condition is that the beam must be sufficiently relativistic so that the
beam's electric field is essentially transverse. For the transverse electric
field component $\vec{E}_{\perp}$, Lorentz transformation \cite{jackson-buch} results an enhancement in the
lab-frame compared to the rest-frame of the moving charge by 
\begin{equation}
E_{\perp, lab} (t_{lab}) = \gamma_{rel} \cdot E_{\perp, rest} (t_{rest})
\label{eq-e-field-transimple}
\end{equation}
tincluding the
transformation of the time $t_{rest}$ to $t_{lab}$. For bunches shorter than the
pick-up, the signal does not reflect the longitudinal bunch shape.
For electron beams, the bunch length is
so short that the~bunch structure is smeared out by integration on the pick-up
capacitance; expressed in terms of electrical parameters by the
limited bandwidth. Here the monitoring of
synchrotron radiation is used in connection with the fast optical method using
streak cameras. At LINAC-based light sources, the~time resolution of a streak camera is insufficient and the method of electro-optical modulation is briefly discussed at the~end of the section. 

A measurement of the energy- or momentum spread $\delta=\Delta p/p$ is not
discussed in detail here. A magnetic spectrometer can be used for this
purpose, having a point-to-point focus in connection with small slits located at
appropriate locations. This topic is described in textbooks on beam optics, e.g. in
\cite{woll-buch}.

\subsection{Bunch structure measurement at circular light sources}
\label{kap-streak-camera-meas}
In most cases, longitudinal diagnostics at electron accelerators is not done
using pick-ups, although, due to the relativistic velocities, it would be
possible. However, the bunch length, e.g. at a synchrotron light source
is typically only of the order of several
ten ps, so that even a pick-up bandwidth of several GHz is insufficient to
reproduce the detailed bunch structure. Here we profit from the emission of
synchrotron light at a dipole, or preferably from an insertion device. A review
of this technique is given in \cite{scheidt-epac00}.

\begin{figure}
\centering 
\includegraphics*[width=110mm,angle=0]
{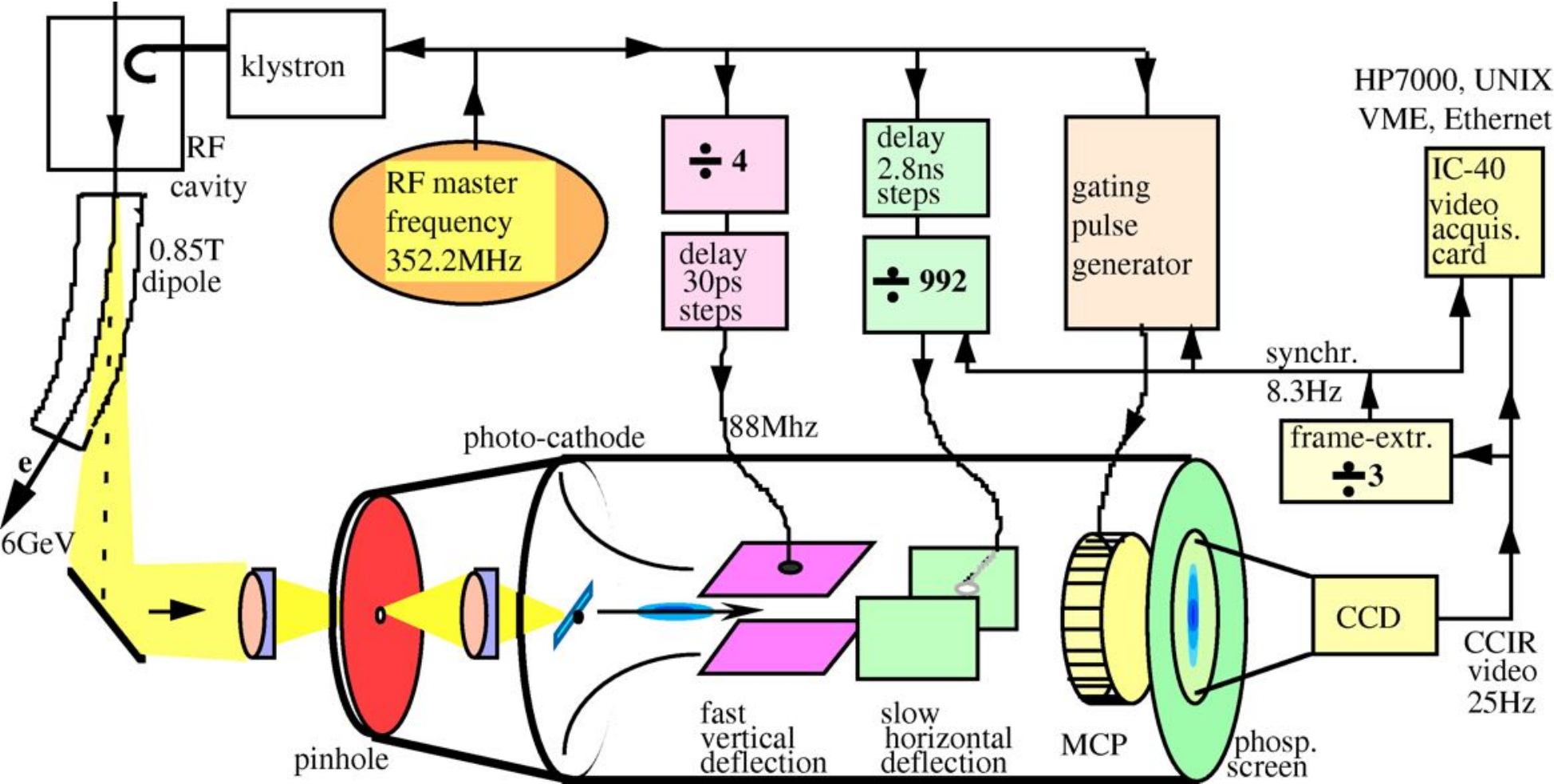}
\caption{Principle of a streak camera measurement for bunch width determination
at electron accelerators ESRF \cite{scheidt-epac96}.}
\label{scheme-streak-scheidt}
\vspace*{-0.1cm}
\end{figure}

\begin{figure}
\centering 
\includegraphics*[width=100mm,angle=0]
{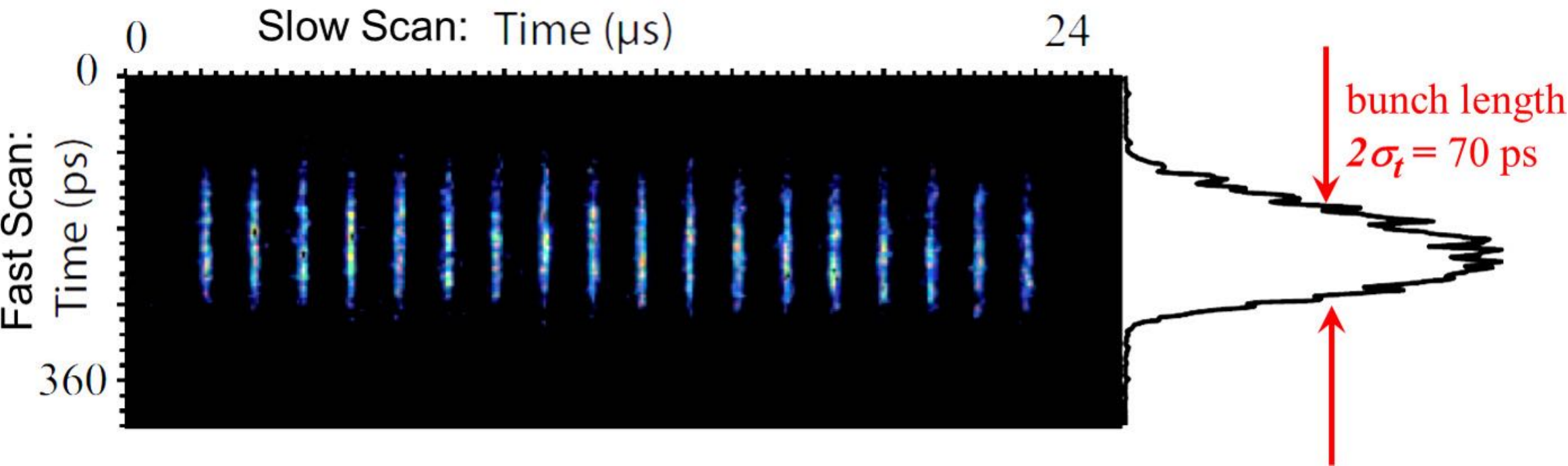}
\caption{Bunch length measurement with a streak camera, using 
synchrotron light from individual bunches emitted by the passage
through a dipole at SOLEIL \cite{labat-dipac07}. The horizontal axis
scaling is 24 $\mu$s full scale for the bunch repetition (slow scan
direction) and the vertical axis is 360 ps full scale for the bunch
structure (fast scan direction).}
\label{streak-results-2}
\vspace*{-0.1cm}
\end{figure}

The principle of such a measurement is shown in 
Fig.~\ref{scheme-streak-scheidt}. It uses a streak camera as a commercially available device, which allows visible light observations with a time resolution down to typically 1 ps. The synchrotron light emitted by the bent electrons beam in the optical to UV wavelength range is 
used. It is focused and scraped by a pin-hole onto the photo-cathode of a 
streak camera. The photo-electrons from the photo-cathode are 
accelerated and pass a fast deflector driven by a frequency which is locked to the accelerating frequency for synchronization. After a certain drift space, the electrons are amplified by an MCP and converted to photons by a phosphor screen. With a standard camera, the phosphor screen image is stored. The different arrival time of the synchrotron light, and therefore the bunch structure, is converted to a difference in space with a full-scale 
of typically several 100 ps and a resolution in the~ps range. A second, 
perpendicular pair of plates is driven with a much lower frequency to separate the~images from the individual bunches.

An example of a bunch length 
measurement is depicted in Fig.~\ref{streak-results-2} as performed with synchrotron radiation from a dipole 
\cite{labat-dipac07}. The slow scan for the separation of the individual bunches is displayed in the horizontal axis. The vertical axis has a full scale of 360 
ps and the bunch width is only $\sigma \simeq 40$ ps, demonstrating the high resolution of such a system. Those short bunches are needed by some synchrotron radiation users for time-resolved spectroscopy. Effects like head-tail oscillations or longitudinal bunch oscillation due to coupled bunch instabilities are visible with the help of the streak camera measurements as well, see, e.g. \cite{scheidt-epac00}.

\subsection{Electro-optical bunch structure measurement at linear light sources}
At modern LINAC-based light sources, the process of free-electron lasing depends non-linearly on the~temporal structure of the bunches. A typical bunch width is in the order of 100 fs with a non-Gaussian bunch structure and significant shot-to-shot variations might occur. The time resolution of $\sim 1$~ps concerning the light observation by the aforementioned streak camera is insufficient. 

\begin{figure}
	\parbox{0.49\linewidth}{\centering 
		\includegraphics*[width=85mm,angle=0]
		{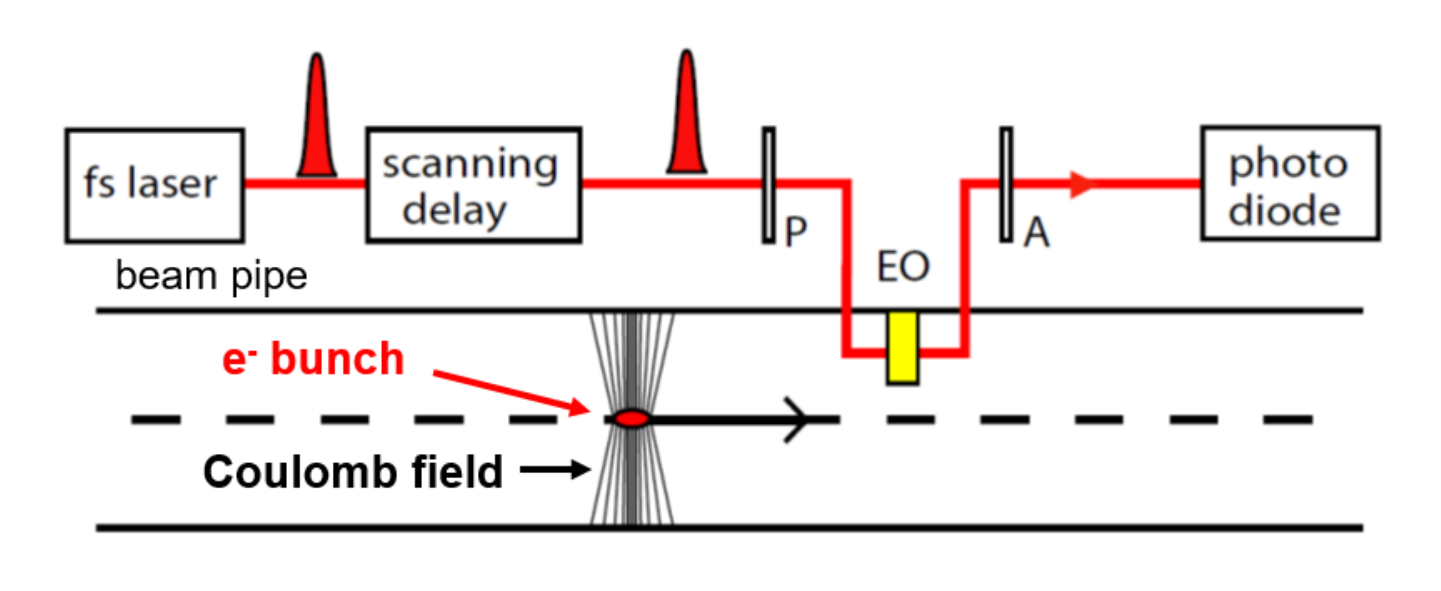}}
	\parbox{0.49\linewidth}{\centering 
		\includegraphics*[width=52mm,angle=0]
		{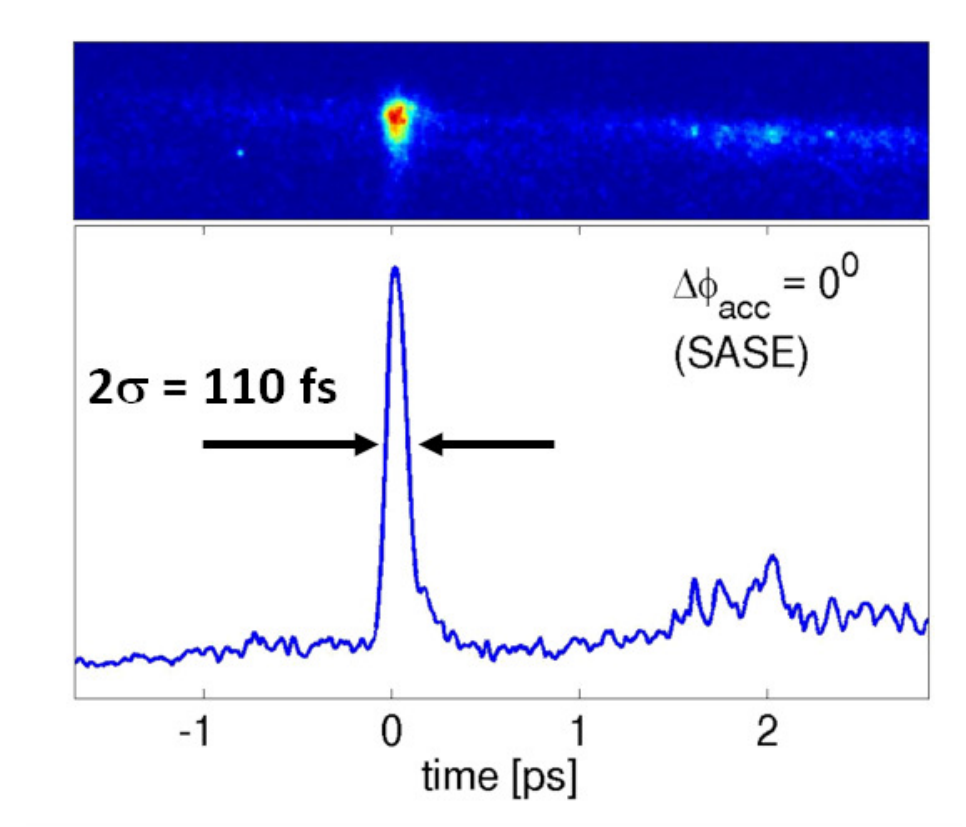}} 
	\caption{Left: Electro-optical scanning by a short laser pulse for relativistic electron beams (P: polariser, A: analyser for the polarization, EO: electro-optical crystal), courtesy B. Steffen. Right: Single shot bunch structure measurement (camera image and projection) using EOTD at FLASH \cite{steff-prab09}.}
	\label{scheme-EOS-scan-steffen}
	\vspace*{-0.1cm}
\end{figure}

\begin{figure}
	\centering 
	\includegraphics*[width=140mm,angle=0]
	{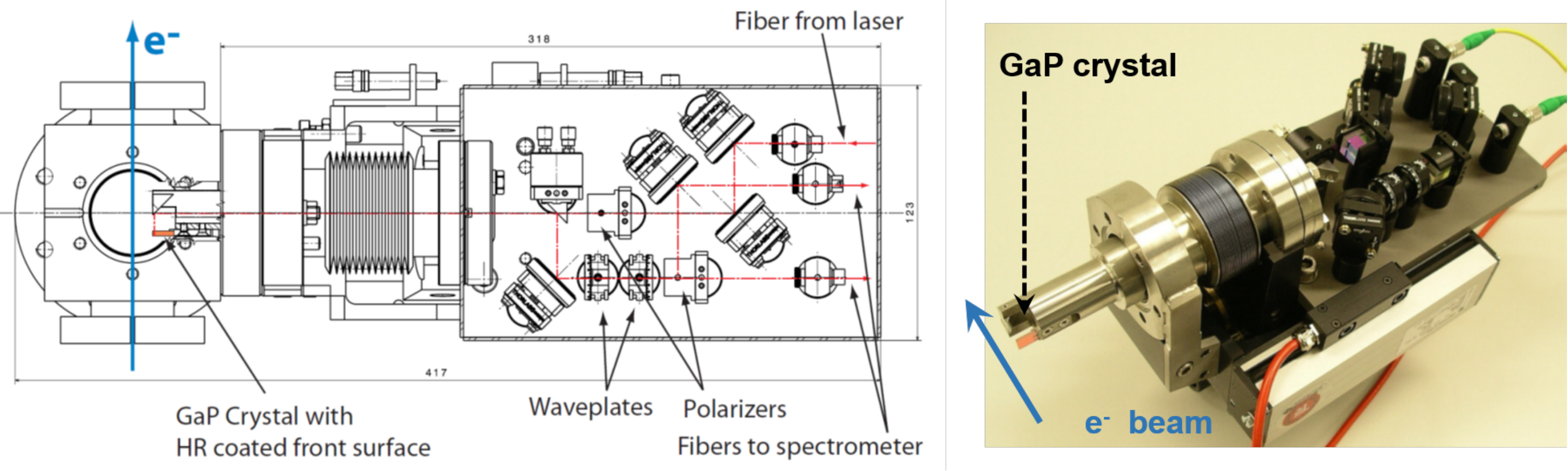}
	\caption{Birefringent crystal and optics for an electro-optical bunch shape monitor at the PSI FEL \cite{steff-dipac09}.}
	\label{photo-EOS-steffen-dipac09}
	\vspace*{-0.1cm}
\end{figure}

The technique of electro-optical sampling is suited for a time resolution down to some 10 fs. Electrons at that LINAC-based light sources are accelerated to highly relativistic velocities leading to an~enlarged and temporally compressed transversal electric field as described by Eq.~(\ref{eq-e-field-transimple}). This electrical field is detected by a birefringent crystal mounted typically 5 mm outside of the beam path, see Fig.~\ref{scheme-EOS-scan-steffen} left. In most cases, a GaP or ZnTe crystal of thickness less than 100 $\mu$m is installed. The underlying effect of electro-optical modulation is a change of the index of refractivity between two orthogonal crystal planes depending on the external electric field; this leads to a rotation of the polarisation of transmitted light \cite{hecht-buch}. The rotation of polarisation, as caused by the time-dependent electric field of the electron bunch, can be probed by a co-propagating laser which is linear-polarised before the crystal. With the~help of a subsequent analysing polariser, this is transferred to a time-dependent intensity modulation. A short laser pulse of typically 10 fs duration is used for the probing; at most installations a Titanium-Sapphire (Ti:Sa) laser or Ytterbium fibre laser crates the short pulse. In the basic installation, the arrival time between this laser pulse and the electron bunch is varied by an optical delay (i.e. varying path length of the laser light) to probe the temporal bunch structure as schematically depicted in Fig.~\ref{scheme-EOS-scan-steffen}. A~photo of the~beamline installation is presented in Fig.~\ref{photo-EOS-steffen-dipac09}. This configuration is a scanning method (the~controllable laser delay is varied each shot and acts as the time axis) and not suited in case of shot-to-shot variations of the electron bunches; therefore this method is seldom installed.

\begin{figure}
	\centering 
	\includegraphics*[width=160mm,angle=0]
	{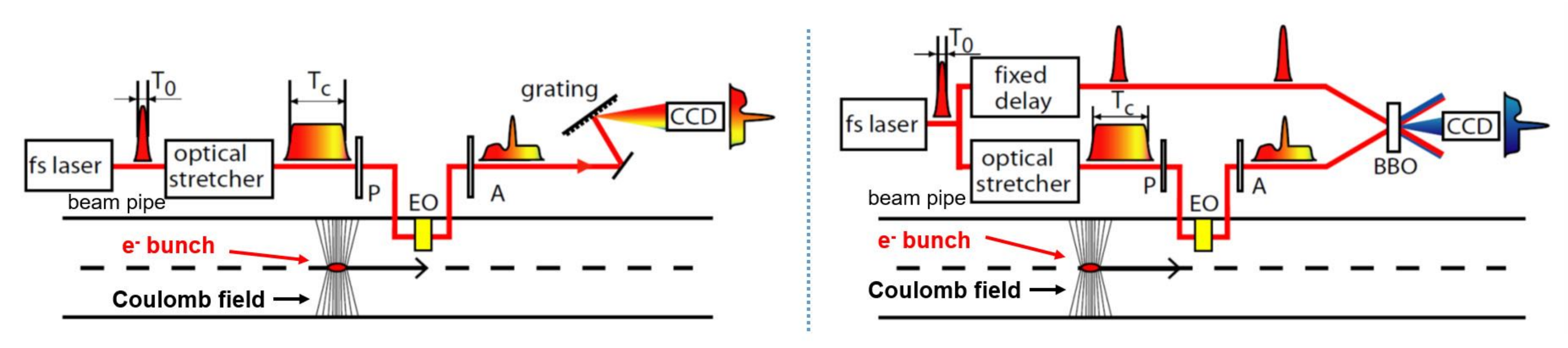}
	\caption{Left: Scheme of the electro-optical spectral decoding EOSD method. Right: Scheme of the electro-optical temporal decoding EOTD method, courtesy B. Steffen.}
	\label{scheme-EOSD-EOTD-steffen}
	\vspace*{-0.1cm}
\end{figure}

To enable a single shot observation, the setup is modified by two different methods as depicted in Fig.~\ref{scheme-EOSD-EOTD-steffen}. The first method is called \textbf{E}lectro-\textbf{O}ptical \textbf{S}pectral \textbf{D}ecoding \textbf{EOSD}: Generally, a laser pulse of finite duration is composed by light within a range of wavelengths. An optical stretcher \cite{hecht-buch} separates the different wavelengths in time by a colour-dependent optical path length in a dispersive element like prisms, providing a correlation between the wavelength and the arrival time at the crystal. The electron bunch field temporally modifies the wavelength-dependent light in the electro-optical crystal. After passing the analysing polariser, the change in beam-induced polarisation is reflected in a~wavelength-dependent intensity variation. By an optical spectrometer, this colour modulation is transferred to a~spatial profile, which is finally monitored by a camera. The second method is called \textbf{E}lectro-\textbf{O}ptical \textbf{T}emporal \textbf{D}ecoding \textbf{EOTD} and is based on two laser beams: The same method of beam-induced modulation is installed. However, the original laser beam is spit before the optical stretching. Both laser beams are combined with an angle inside a non-linear crystal \cite{hecht-buch}. (Barium Boron Oxide BBO is often used for this purpose, providing a $2^{nd}$-harmonics generation as often used for frequency doubling of a~laser beam.) The modulation detection is performed by frequency mixing of both laser beams as the~non-linear crystal  emitting the sum frequency of both laser beam, i.e. acts as a single-shot correlation measurement. Owing to the non-coaxial propagation inside the crystal, the light of the sum frequency is emitted from different locations, leading to a spatial separation of the beam-induced intensity modulation. Finally, a camera can monitor the related image. A detailed comparison of both methods is described in \cite{steff-prab09}. The achievable time resolution of the EOSD method is better than 50 fs. In contrast, the achievable time resolution by the EOTD method is about 25 fs. However, EOTD has the disadvantage of a larger laser power requirement related to the frequency mixing. A measurement of a single pulse of 100 fs duration at a FEL facility obtained by the EOTD method is shown in Fig.~\ref{scheme-EOS-scan-steffen} right.

\section{Beam loss detection}
\label{kap-beamloss}
\begin{figure}
\centering
\includegraphics*[width=100mm,angle=0]
{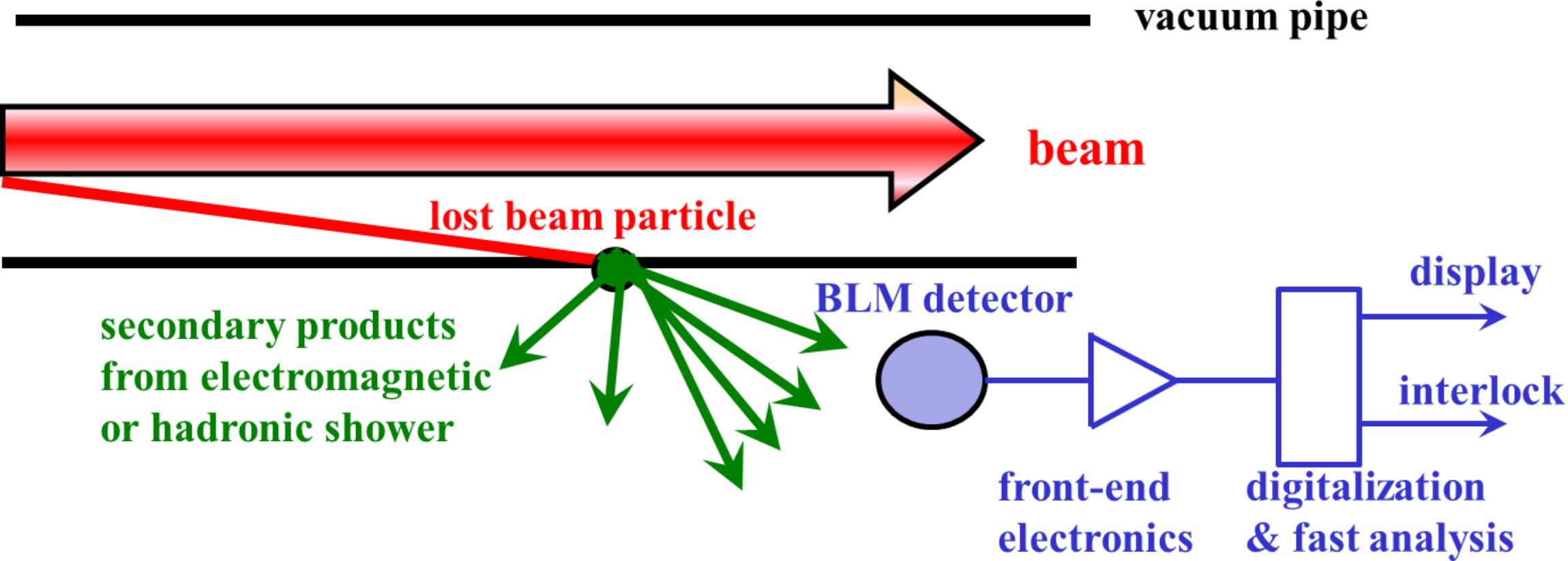} 
\caption{Scheme for the installation of Beam loss monitors outside of the vacuum pipe.}
\vspace*{-0.3cm}
\label{scheme_BLM_installation}
\end{figure}

In a real accelerator, the transmission from the source to the target is never
100 \%. The fraction of beam particles lost has to be controlled carefully to
achieve optimal
transmission. The lost beam particles cause some activation of the
accelerator components by nuclear reactions. Moreover, the surrounding material
can be destroyed by the radiation, as well as by the heating caused by the
particles' energy loss. To detect the shower of secondary particle, as originated 
by nuclear interaction processes, 
a large variety of \textbf{B}eam \textbf{L}oss \textbf{M}onitors \textbf{BLM} exists. 
At nearly every high current accelerator facility, these monitors are
installed for the protection of the accelerator components. The relatively cheap 
BLM instruments are mounted outside
of the vacuum pipe at crucial locations as schematically depicted in 
Fig.~\ref{scheme_BLM_installation}. Their signals are a relevant
information to prevent unwanted loss during the operation, e.g. caused by
malfunctions of components. A careful analysis of the
location and time structure of possible losses has to be performed before 
the choice of the suitable types of beam loss monitors can be made, see
Refs.~\cite{schmick_cas2018,brandt_cas2007,witt-epac02,Zhuk-biw10,shafer-biw02} for reviews. 

\subsection{Secondary particle production}
\label{kap-interaction}
\begin{figure}
\centering
\vspace*{-0.5cm}
\includegraphics*[width=140mm,angle=0]
{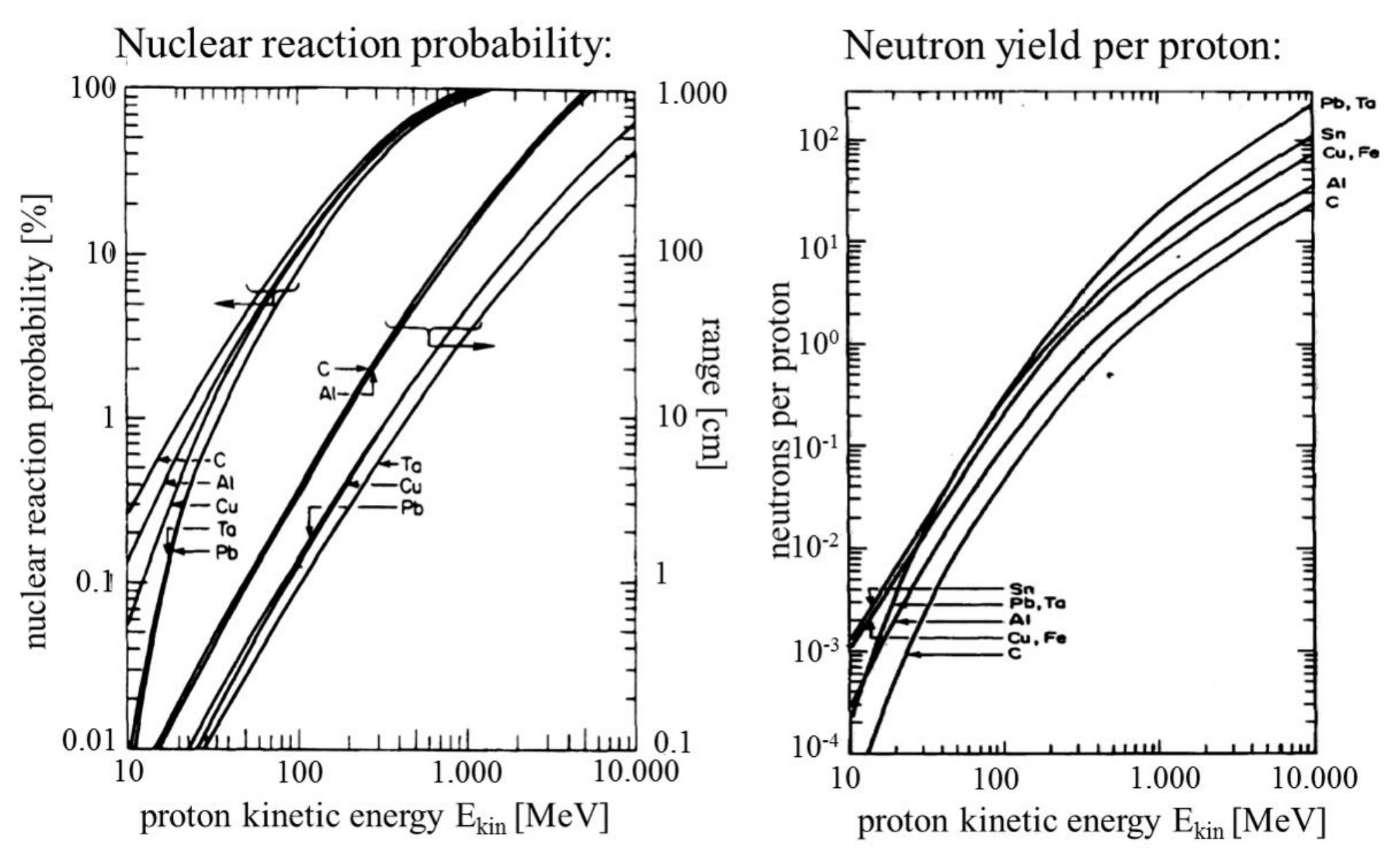} 
\vspace*{-0.4cm}
\caption{Left: The ranges of protons in various metals and the probability of
an inelastic nuclear reaction within the range 
as a function of the proton energy, i.e. within a thick target. 
Right: The total number of neutrons per proton incident on a thick target of
various metals as a function of the proton energy \cite{tho-handbook}.}
\label{fig-nuclear-crossection}
\end{figure}

When a high energy particle hits the vacuum pipe or any
other material, secondary particles are generated. The relevant
processes are described, e.g. in \cite{sull-buch,tho-handbook}.
Here only a brief overview is given:
\begin{itemize}
\item
\textit{Interaction of electrons:}
For electron energies above $\sim 100$ MeV, Bremsstrahlung dominates the~slow-down process in materials, as shown in Fig.~\ref{energyloss-electrons}.
The created high energetic $\gamma$ photons give rise to further particles
via $e^+ - e^-$ pair production. If the energy of the
$\gamma$ photons are high enough, other particles, like $\mu^{\pm},
\pi^{\pm}...$ can also be produced, 
an electro-magnetic shower is generated. Moreover, the nucleus can be
excited to so-called giant resonances: This is a collective nuclear excitation,
where the neutrons oscillate against the protons. The dipole mode of
these giant resonances has a threshold of about 6 MeV for typical materials.
The de-excitation proceeds with high probability via neutron emission as a
$(\gamma , n)$ 
reaction. For higher energies also $(\gamma , p)$ and $(\gamma , np)$
channels are open.\\ \hfill 
When the electron is slowed down below $\sim 10 $ MeV,
ionization loss by electronic stopping dominates, i.e. the creation of an electron vacancy at the atomic shell which is then filled by an outer electron leading to a soft
x-ray emission. Those x-ray photons are absorbed within a short distance of mm to cm in typical metals.
\item
\textit{Interaction of protons:}
Beside electronic stopping, as shown in 
Fig.~\ref{copper_range}, nuclear
interactions are possible. As a first step, we define the term 'thick target' for the case the interaction length is comparable to the range as given by
the electronic stopping. In these
thick targets, the probability of a nuclear reaction rises to
nearly 100 \% for energies above 1 GeV, see
Fig.~\ref{fig-nuclear-crossection}. Most of the final channels of these
nuclear reactions include neutron emission. As an example, 
for the stopping of 1 GeV proton in copper or iron, $\sim 10 $ 
fast neutrons are liberated. The neutron
yield scales approximately with $E_{kin}$ for energies above 1 GeV.
In addition, hadron showers (by the strong interaction)
are possible, resulting in various species of 'elementary' particles.
\end{itemize}

Common to all interactions is the production of radioactive nuclei leading
to activation of the~accelerator components. The surrounding material relatively quickly stops the emitted charged
particles. However,
the neutrons, produced by most primary interactions, can travel long
distances. Some of the~beam loss monitors are therefore sensitive to these
neutrons. Except for the production of radioactive nuclei, all
processes are fast, compared to the timescale of interest in accelerator
physics, i.e. faster than $\sim 10 $ ns. In this sense, a beam loss monitor
reacts promptly to the particle loss. Owing to the kinematics of the primary
interaction, the secondaries are emitted into a (more or less) forwardly peaked
angular distribution. This leads to a spatial resolution of the~loss detection
by the monitor position close to the~loss point.

\subsection{Types of beam loss monitors}
\begin{figure}
\centering
\includegraphics*[width=50mm,angle=270]
{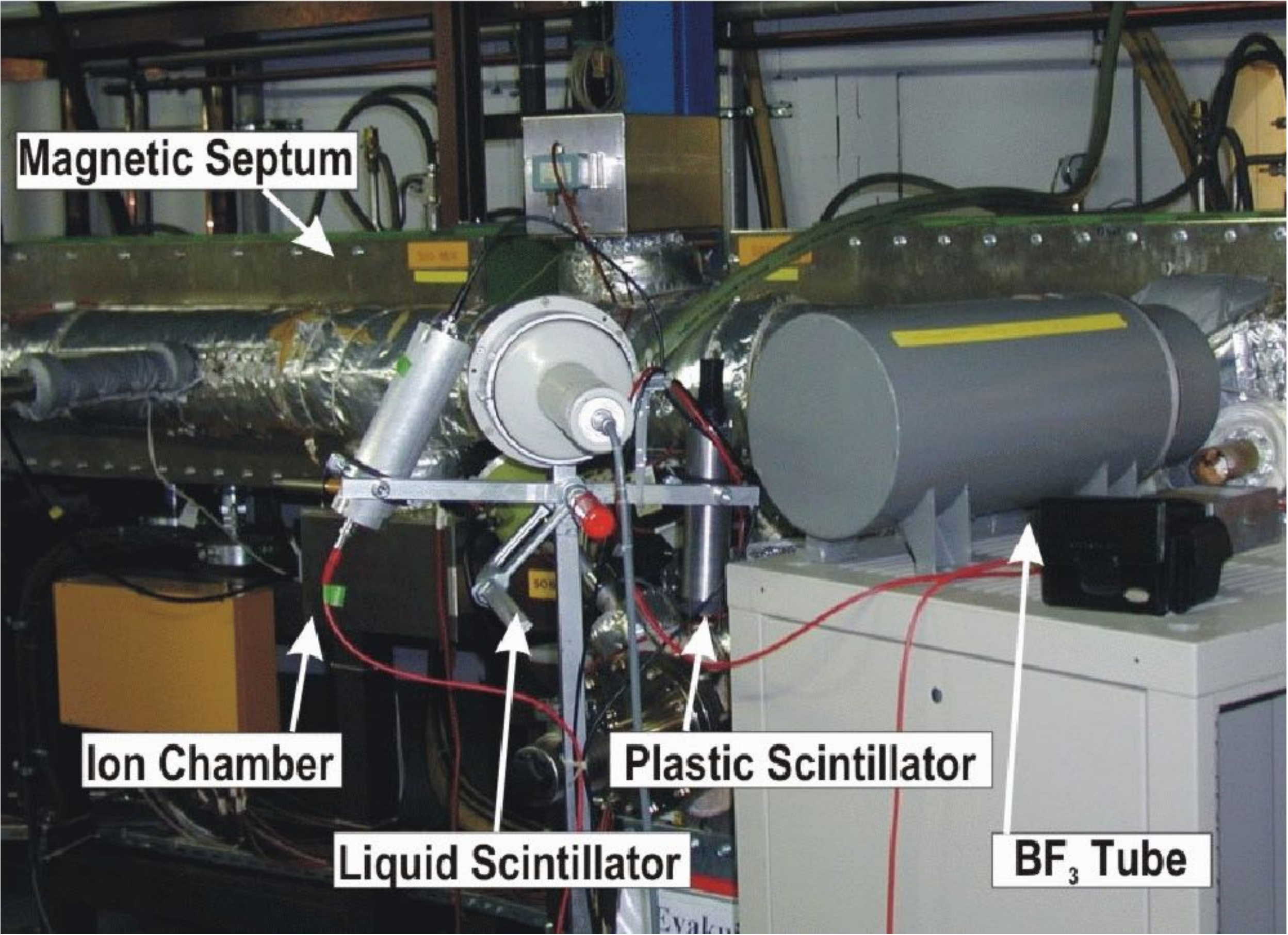} 
\caption{The tested beam loss detectors installed at the extraction from the
GSI heavy ion synchrotron \cite{forck-dipac01}.}
\vspace*{-0.3cm}
\label{foto-beamloss-detector}
\end{figure}

It is the task for beam loss monitors to localize the position and time
of the loss. Their signals should be proportional to the amount of
loss at this position. High sensitivity is needed to measure
low losses. A~high dynamic range is required to deal with a sudden
loss of a sizable fraction of the beam.
Depending on the application,
a bunch-to-bunch resolution on a $\sim
10 $ ns time scale is needed, as well as up to 100~ms for a slow detection.
All loss monitors are installed outside of the
vacuum pipe, detecting mostly secondary particles.
These can be neutrons, which
are not much absorbed by the surrounding material, charged particles
like protons, $e^-$ and $e^+$ or $\gamma$-rays.
Fig.~\ref{foto-beamloss-detector} shows a photo of some frequently
used types as tested at the GSI synchrotron. An overview of the different
types is given in \cite{schmick_cas2018,brandt_cas2007,witt-epac02,Zhuk-biw10,shafer-biw02}.

\subsubsection{Plastic scintillators}
\begin{figure}
\parbox{0.58\linewidth}{\centering
\includegraphics*[width=73mm,angle=0]
{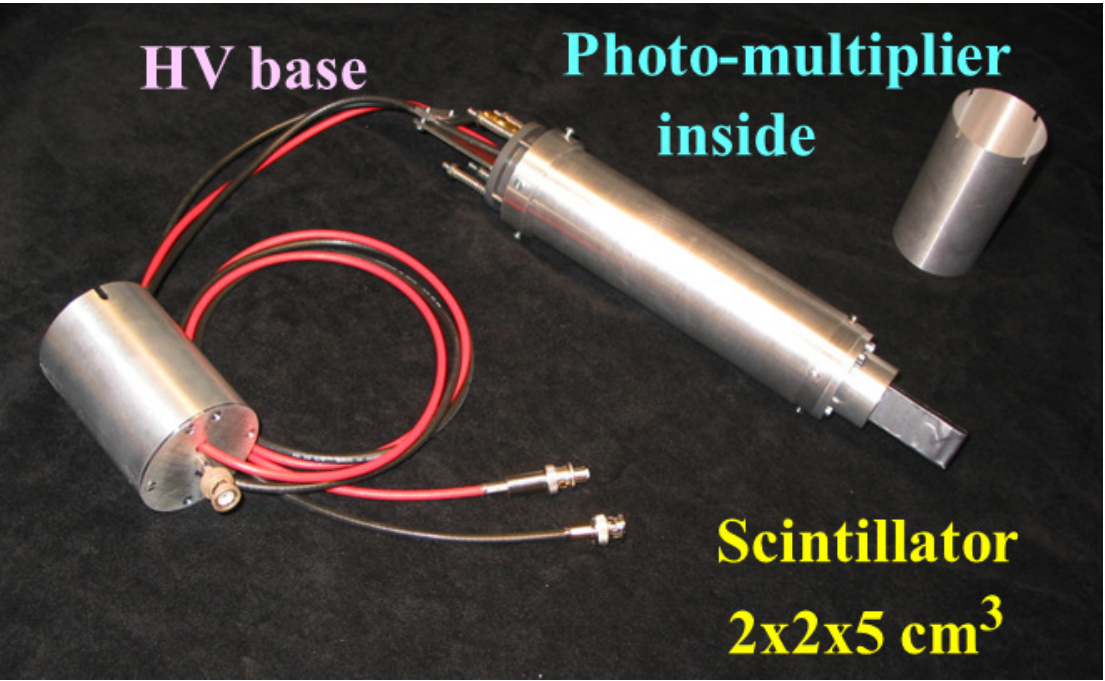}}
\parbox{0.41\linewidth}{\centering
\includegraphics*[width=73mm,angle=0]
{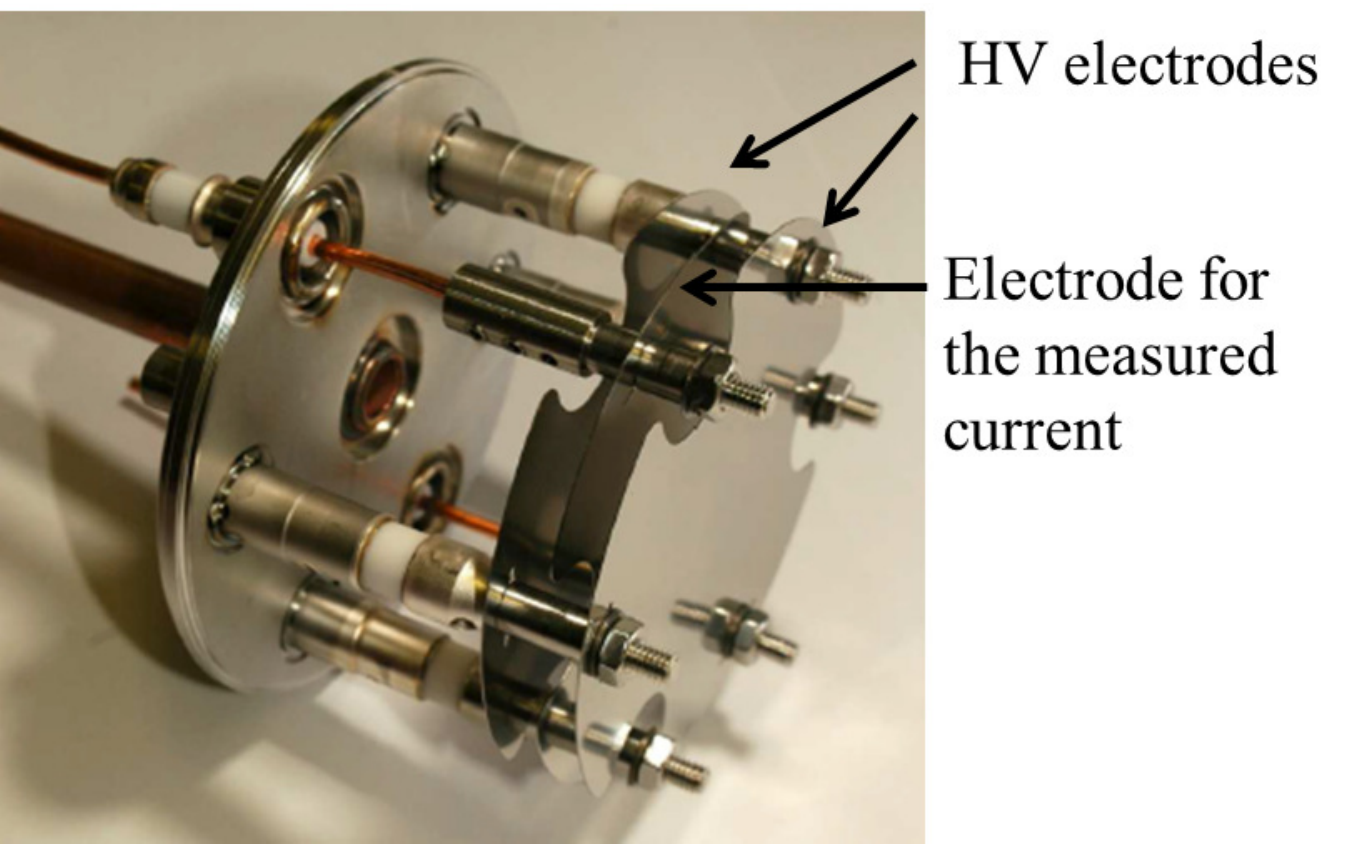} }
\caption{Left: Photo of a typical BLM based on a plastic scintillator connected to a photo-multiplier. Right: Photo of a three plate SEM beam loss monitor \cite{kram-dipac07}.}
\label{fig-sc-loss-phd}
\end{figure}

Plastic scintillators detect charged particles 
owing to the
electronic stopping, as discussed in Section~\ref{kap-current-energyloss}. $\gamma$-rays are detected as they liberate electrons from the molecules via photo effect or Compton scattering; those electrons generate than the optical photons via their electronic stopping. 
Moreover, plastic scintillators are also sensitive to neutrons due to their
elastic scattering on the hydrogen atoms of the~polymers \cite{knoll-buch}.
Due to the elastic scattering process generated by the relativistic neutrons, 
a large momentum transfer to the Hydrogen atoms of the macro-molecules 
is probable, which leads to a fast proton passing 
through the scintillator material and finally generates the 
light emission by its electronic stopping. The~light is guided to a
photomultiplier, converted to photo-electrons and amplified, see Fig.~\ref{fig-sc-loss-phd} (left) 
for a typical realization.
In most cases the scintillation monitors are 
located in the crucial areas, like the~injection or extraction devices, or close to scrapers.
The disadvantage is the
low radiation hardness of the plastic materials due to the complex
chemical composition of the polymers from the plastic matrix.

\subsubsection{Secondary Electron Monitor}
In case the loss rate is very high, the current of secondary electrons
from a surface is high enough to be measured directly. This can be 
realized for a secondary electron monitor by three plates installed in a~small vacuum vessel \cite{kram-dipac07}. 
The outer plates are biased by some +100 V and the
secondary electron current emitted from the central plate is measured
by a trans-impedance amplifier or a current-to-frequency
converter. The inner part of such a monitor is shown in
Fig.~\ref{fig-sc-loss-phd}. Relevant simulations and tests of the~response to different radiation are reported in \cite{kram-dipac07}.

\subsubsection{Ionization chamber}

\begin{figure}
\centering
\includegraphics*[width=90mm,angle=0]
{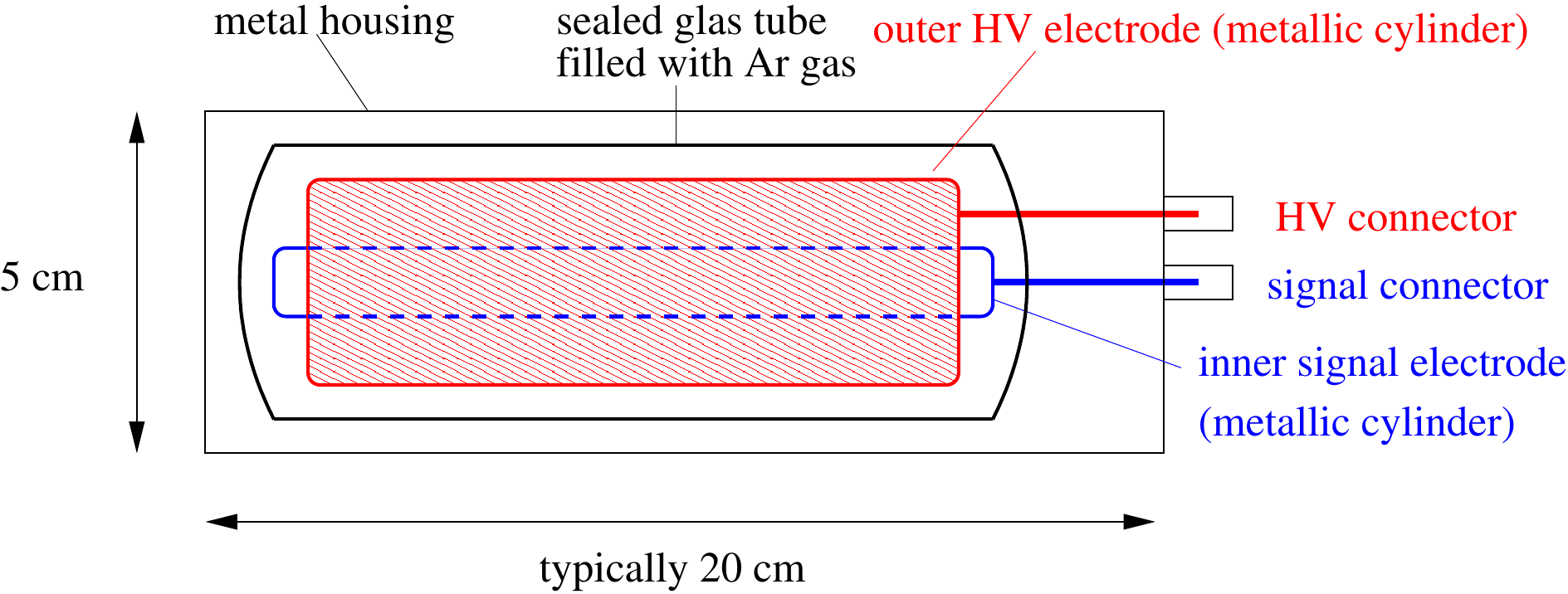} 
\caption{Scheme of a co-axial ionization chamber as used for beam loss detection.}
\label{fig-loss-ic}
\end{figure}

\begin{figure}
\centering
\includegraphics*[width=120mm,angle=0]
{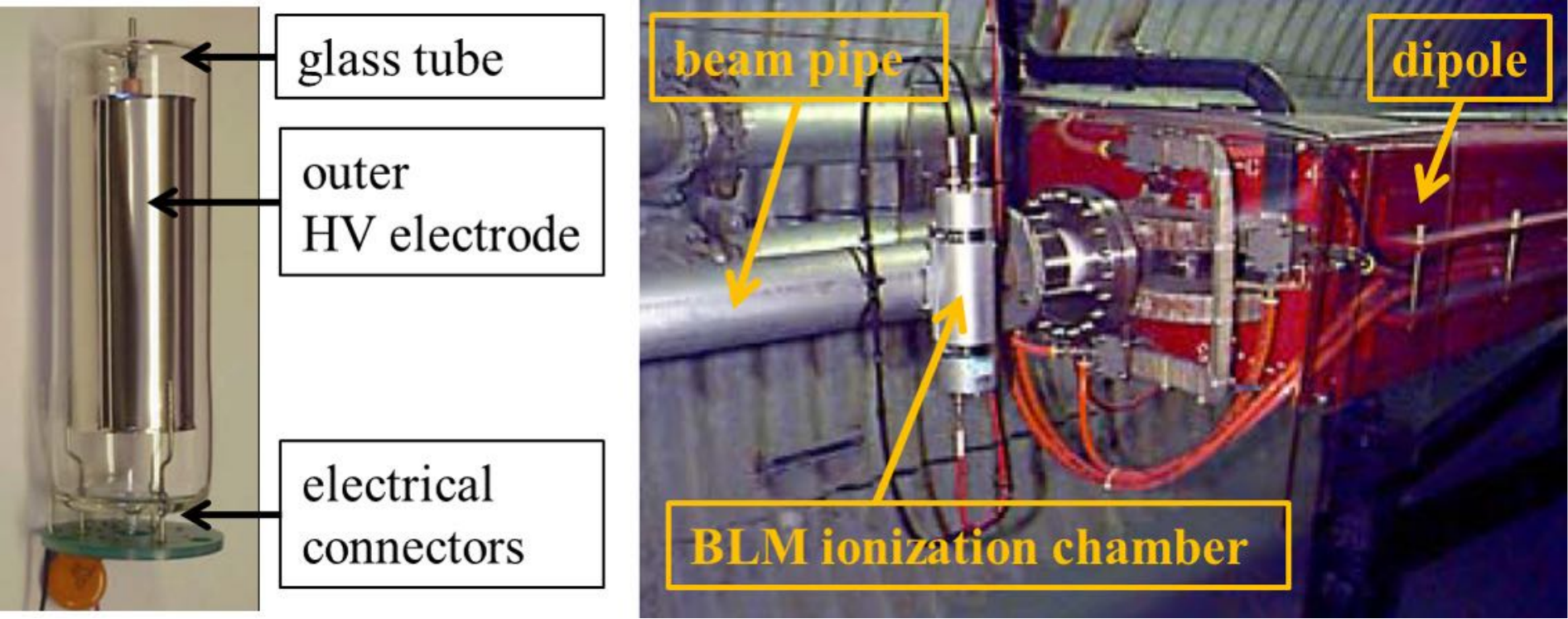} 
\caption{Left: Photo of a 15 cm long co-axial ionization chamber used
as a beam loss monitor. Right: The~installation of such 
a ionization chamber for beam loss detection in a transport line.}
\label{foto-ic-RHICinstallation}
\end{figure}

An Ionization Chamber \textbf{IC} \cite{knoll-buch} measure the number of secondary charges created
in a gas volume. 
Fig.~\ref{fig-loss-ic}
shows such a round ionization chamber filled with Ar or N$_{2}$ gas.
Typically, a sealed glass or metal tube contains $\sim 100 ... 1000$
cm$^3$ of gas between 
an outer high voltage electrode and an inner grounded and readout 
electrode \cite{gassner-biw00}.
The IC is not sensitive to neutrons
and has a low detection efficiency for $\gamma$-rays, mainly charged
hadrons and $e^{\pm}$ are detectable. By definition, the signal strength
gives the absorbed dose directly in Gy. Because inert gases
like Ar or N$_{2}$ are used in the detection volume, the device is very radiation
hard. The signal strength is orders of
magnitude lower than for detectors in a particle counting mode. In
Fig.~\ref{fig-loss-ic}, the scheme of a co-axial IC is depicted. For
this geometry, the IC does not react as fast as the scintillators, because the gas ions, created by the radiation, need $\sim 10$ $\mu$s to reach the~electrode. This time constant leads to a convolution of the primary
signal, which for most applications is acceptable. The readout of the
IC current by the digital electronics is therefore not faster than 1 ms in typical applications.
The essential parameters for such an IC are summarized in
Table~\ref{tab-blm-ic}. 
A typical installation of a co-axial IC at an accelerator beamline is depicted in Fig~\ref{foto-ic-RHICinstallation}.

\begin{figure}
\centering
\includegraphics*[width=75mm,angle=0]
{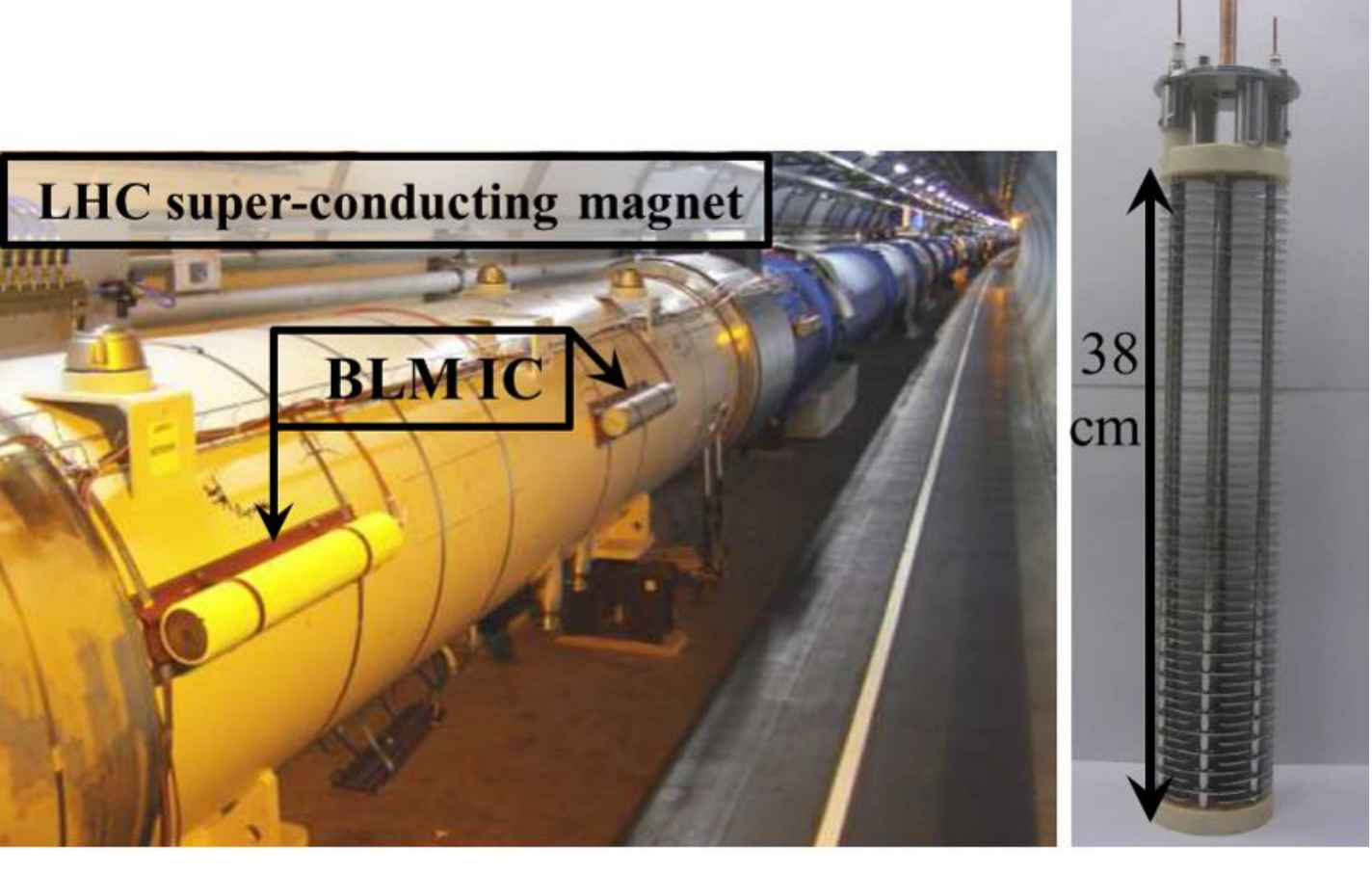} 
\caption{Photo of the installation of a 50 cm long ionization chamber
comprising of 61 parallel plates at CERN LHC \cite{holzer-epac08}.}
\label{foto-ic-LHC}
\end{figure}

The detection threshold of an IC is proportional to the gas volume. To achieve
a fast reaction time even for a large gas volume, the drift time of the
gas ions and electrons should be shortened. This can be realized by a
parallel electrode arrangement with alternating biased and readout
electrodes. Such an~IC is shown in Fig.~\ref{foto-ic-LHC} as installed at CERN
LHC and other facilities. About 4000 BLMs are installed along the 27 km
long LHC corresponding to an average distance of about 6 m. They serve
as the main detectors for beam protection and can trigger a fast beam
abortion. 
The design of such IC for LHC and relevant
simulations concerning the response to various radiation species are
discussed e.g. in \cite{holzer-epac08}.

\begin{table}
\label{tab-blm-ic}
\centering
\caption{Basic parameters for the so called RHIC-type co-axial IC
\cite{gassner-biw00} and
the planar LHC-type IC \cite{holzer-epac08}.}
\begin{tabular}{lll}
\hline \hline
\textbf{Parameter} & \textbf{Co-axial IC} & \textbf{Planar IC}\\ \hline
Outer length \& diameter [cm]& 20 \& 6 & 50 \& 9\\
Active gas volume [l] & 0.11 & 1.5 \\ 
Gas type \& pressure [bar] & Ar at 1.1 & N$_2$ at 1.1 \\ 
Number of electrodes & 2 & 61\\
Distance of electrodes [mm] & \O$_{inner}= 6.3$, \O$_{outer}= 38$ & 5.7 \\
Voltage [kV] & 1 & 1.5 \\
Reaction time [$\mu$s] & $\simeq 3$ & $\simeq 0.3$\\ 
Overall dynamic range & $10^6$ & $10^8$ \\ \hline \hline
\end{tabular}
\end{table}

The application of BLMs for beam operation and machine protection is discussed in the frame of the CAS lecture 'Machine and People Protection'.

\section{Conclusion}

The functionality of various beam instruments is described, and their applicability for different beam parameters are mentioned. For daily accelerator operation mainly current, position, and to some extent transverse profile measurements are executed. It is an advantage if the monitor is non-invasive for the~beam, like current transformers and BPMs, to enable a simultaneous measurement at several locations, archiving of beam parameters, and, if possible for automated beam alignment by a feedback loop. Current transformers of various types serve as non-invasive instruments, but they have a lower detection threshold. For bunched beams, the transverse centre position can be recorded by BPMs. Due to their principle, BPMs are non-invasive but requires a minimum beam current to enable a sufficient precise position determination. The achievable position resolution is influenced significantly on the subsequent electronics: the spatial resolution can be balanced with respect to the observation duration (long time observation results in an improved position resolution). BPMs are the most frequently used monitors for beam alignment and serve in many cases as the sensor for a feedback loop. By exciting the beam to coherent oscillations, important synchrotron parameters, like tune and chromaticity, are determined with BPMs. For profile measurements, various techniques are applied, with scintillation or OTR screens being a favoured technique for regular operation for transfer lines due to their reproduction of the 2-dimensional distribution. SEM-Grids and wire scanners are also often used due to their robustness and large dynamic range. A non-invasive technique for profile measurements is realized by an IPM for medium to high intensity proton beams in transfer lines and synchrotrons. At electron synchrotrons, the~anyhow available synchrotron light is used for transverse and longitudinal diagnostics. At transfer lines behind a LINAC or a cyclotron more complex parameter have to be determined, in particular the transverse and longitudinal emittances. The related measurements are, in many cases performed by experts during dedicated machine development times, which allow for more sophisticated instruments and methods. Bunch length measurement can be performed by recording the temporal structure of synchrotron light. However, at LINAC-based light sources at high time resolution is required, which is realized by electro-optical modulation techniques. A large quantity of beam loss monitors are installed to monitor the correct accelerator functionality permanently and generate an interlock in case of any malfunction to prevent for permanent activation or component destruction.


\end{document}